\documentclass[11pt]{cft}
\usepackage{amssymb,color}

\newcommand{\BC}{\mathbb{C}}

\newcommand{\BN}{\mathbb{N}}

\newcommand{\BQ}{\mathbb{Q}}
\newcommand{\BR}{\mathbb{R}}

\newcommand{\BZ}{\mathbb{Z}}
\newcommand{\bfa}{\raisebox{0pt}{{\boldmath$a$\unboldmath}}}
\newcommand{\bfb}{\raisebox{0pt}{{\boldmath$b$\unboldmath}}}

\newcommand{\bff}{\raisebox{0pt}{{\boldmath$f$\unboldmath}}}
\newcommand{\bfg}{\raisebox{0pt}{{\boldmath$g$\unboldmath}}}

\newcommand{\bfk}{\raisebox{0pt}{{\boldmath$k$\unboldmath}}}
\newcommand{\bfl}{\raisebox{0pt}{{\boldmath$l$\unboldmath}}}
\newcommand{\bfm}{\raisebox{0pt}{{\boldmath$m$\unboldmath}}}
\newcommand{\bfn}{\raisebox{0pt}{{\boldmath$n$\unboldmath}}}

\newcommand{\bfp}{\raisebox{0pt}{{\boldmath$p$\unboldmath}}}

\newcommand{\bfr}{\raisebox{0pt}{{\boldmath$r$\unboldmath}}}

\newcommand{\bfv}{\raisebox{0pt}{{\boldmath$v$\unboldmath}}}
\newcommand{\bfu}{\raisebox{0pt}{{\boldmath$u$\unboldmath}}}
\newcommand{\bfw}{\raisebox{0pt}{{\boldmath$w$\unboldmath}}}
\newcommand{\bfx}{\raisebox{0pt}{{\boldmath$x$\unboldmath}}}
\newcommand{\bfy}{\raisebox{0pt}{{\boldmath$y$\unboldmath}}}

\newcommand{\bfA}{\raisebox{0pt}{{\boldmath$A$\unboldmath}}}
\newcommand{\bfB}{\raisebox{0pt}{{\boldmath$B$\unboldmath}}}

\newcommand{\bfD}{\raisebox{0pt}{{\boldmath$D$\unboldmath}}}
\newcommand{\bfE}{\raisebox{0pt}{{\boldmath$E$\unboldmath}}}

\newcommand{\bfK}{\raisebox{0pt}{{\boldmath$K$\unboldmath}}}

\newcommand{\bfL}{\raisebox{0pt}{{\boldmath$L$\unboldmath}}}

\newcommand{\bfN}{\raisebox{0pt}{{\boldmath$N$\unboldmath}}}

\newcommand{\bfP}{\raisebox{0pt}{{\boldmath$P$\unboldmath}}}

\newcommand{\bfR}{\raisebox{0pt}{{\boldmath$R$\unboldmath}}}
\newcommand{\bfT}{\raisebox{0pt}{{\boldmath$T$\unboldmath}}}
\newcommand{\bfS}{\raisebox{0pt}{{\boldmath$S$\unboldmath}}}

\newcommand{\bfal}{\raisebox{0pt}{{\boldmath$\alpha$\unboldmath}}}

\newcommand{\bfet}{\raisebox{0pt}{{\boldmath$\eta$\unboldmath}}}

\newcommand{\bfxi}{\raisebox{0pt}{{\boldmath$\xi$\unboldmath}}}

\newcommand{\bfrh}{\raisebox{0pt}{{\boldmath$\rho$\unboldmath}}}

\newcommand{\bfta}{\raisebox{0pt}{{\boldmath$\tau$\unboldmath}}}

\newcommand{\bfom}{\raisebox{0pt}{{\boldmath$\omega$\unboldmath}}}

\newcommand{\CA}{\mathcal{A}}
\newcommand{\CB}{\mathcal{B}}
\newcommand{\CC}{\mathcal{C}}
\newcommand{\CD}{\mathcal{D}}

\newcommand{\CF}{\mathcal{F}}

\newcommand{\CH}{\mathcal{H}}
\newcommand{\CI}{\mathcal{I}}

\newcommand{\CL}{\mathcal{L}}
\newcommand{\CM}{\mathcal{M}}
\newcommand{\CN}{\mathcal{N}}
\newcommand{\CO}{\mathcal{O}}
\newcommand{\CP}{\mathcal{P}}

\newcommand{\CR}{\mathcal{R}}

\newcommand{\CT}{\mathcal{T}}

\def\-{~-~}
\def\+{~+~}
\def\={~=~}
\def\eq{~\equiv~}
\def\beq{\begin{equation}}
\def\eeq{\end{equation}}
\def\beqn{\begin{eqnarray}}
\def\eeqn{\end{eqnarray}}
\def\bb{\begin{eqnarray*}}
\def\ee{\end{eqnarray*}}
\newcommand{\calle}[1]{(\ref{#1})}
\newcommand{\ket}[1]{\left| \, {#1} \, \right\rangle }
\newcommand{\bra}[1]{\left\langle \, {#1} \, \right| }
\newcommand{\expectation}[2]{\left\langle\, {#1}\, |\,{#2}\, \right\rangle }
\newcommand{\average}[1]{\left\langle\, {#1}\, \right\rangle }
\newcommand{\residue}[1]{\mathop{\rm Res}\limits_{#1}}
\newcommand\hsp[1] {\mbox{\hspace{#1 em}}}
\def\lefthook      {{\vrule height5pt width0.4pt depth0pt}}
\def\righthook     {{\vrule height5pt width0.4pt depth0pt}}
\def\leftrighthookfill{$\mathsurround=0pt \mathord\lefthook
                   \hrulefill\mathord\righthook$}
\newcommand\lowerpairing[1]{\hsp{.4}\vtop{\ialign{##\crcr$\hfil\displaystyle
                   {\hsp{-.4}#1}\hfil\hsp{-.4}$\crcr
                   \noalign{\kern-1.9pt\nointerlineskip\vskip2pt}
                   \leftrighthookfill\crcr}}\hsp{.4}}
\def\ubrackfill#1{$\mathsurround=0pt
        \kern2.5pt\vrule depth#1\leaders\hrule\hfill\vrule depth#1\kern2.5pt$}
\newcommand{\upperpairing}[1]{\mathop{\vbox{\ialign{##\crcr\noalign{\kern3pt}
        \ubrackfill{3pt}\crcr\noalign{\kern3pt\nointerlineskip}
        $\hfil\displaystyle{#1}\hfil$\crcr}}}\nolimits}
\newcommand{\lp}{\left(}
\newcommand{\lb}{\left\lbrack}
\newcommand{\la}{\left\{ }
\newcommand{\lan}{\left\langle}
\newcommand{\rp}{\right)}
\newcommand{\rb}{\right\rbrack}
\newcommand{\ra}{\right\} }
\newcommand{\ran}{\right\rangle}
\def\boxit#1{\vbox{\hrule\hbox{\vrule\kern3pt
\vbox{\kern3pt#1\kern3pt}\kern3pt\vrule}\hrule}}

\newcommand{\separator}{  \nopagebreak
                          \begin{center}
                          \rule{2.5in}{2mm}
                          \end{center}
                       }

\newcommand{\SQR}{\sqrt{g}}
\newcommand{\half}{ {1\over2} }
\newcommand{\christoffel}[2]{  \Gamma^{#1}_{#2} }
\newcommand{\partition}[2]{  
                            {\scriptstyle #1} \mathop{\square}\limits_{#2}
                           }
\def\iint{\int\kern-8pt\int}
\def\centeronto#1#2{{\setbox0=\hbox{#1}\setbox1=\hbox{#2}\ifdim
\wd1>\wd0\kern.5\wd1\kern-.5\wd0\fi
\copy0\kern-.5\wd0\kern-.5\wd1\copy1\ifdim\wd0>\wd1
\kern.5\wd0\kern-.5\wd1\fi}}
\def\slash#1{\centeronto{$#1$}{$/$}}
\newcommand{\no}{ \nonumber\\ }
\def\counterclockwise{ \rotatebox{90}{\mbox{$\circlearrowleft$}} }

\def\ointleft{\centeronto{$\displaystyle\int$}%
                         {$\counterclockwise$}%
             }
\makeatletter
\input{PART-II.aux}
\makeatother
\begin{document}
\frontmatter

\begin{titlepage}

\begin{center}
       \colorbox{red}{\textcolor{white}{
       \vbox{
       \vspace{0.3cm}
       \centerline{\Large A COLLECTION OF EXERCISES IN}
       \vspace{2mm}
       \centerline{\Large TWO-DIMENSIONAL PHYSICS}
       \vspace{2mm}
       \centerline{\Large PART I}
       \vspace{0.3cm}
            }}}
\end{center}

\vspace{1cm}

\begin{center}
{\bfseries\textcolor{blue}{COSTAS J. EFTHIMIOU}}\\
Newman Laboratory of Nuclear Studies\\
CORNELL UNIVERSITY \& FLORIDA SOUTHERN COLLEGE\\

\vspace{1cm}
{\bfseries\textcolor{blue}{DONALD A. SPECTOR}}\\
Department of Physics\\
HOBART \& WILLIAM SMITH COLLEGES 

\vspace{1cm}
\includegraphics[height=8cm]{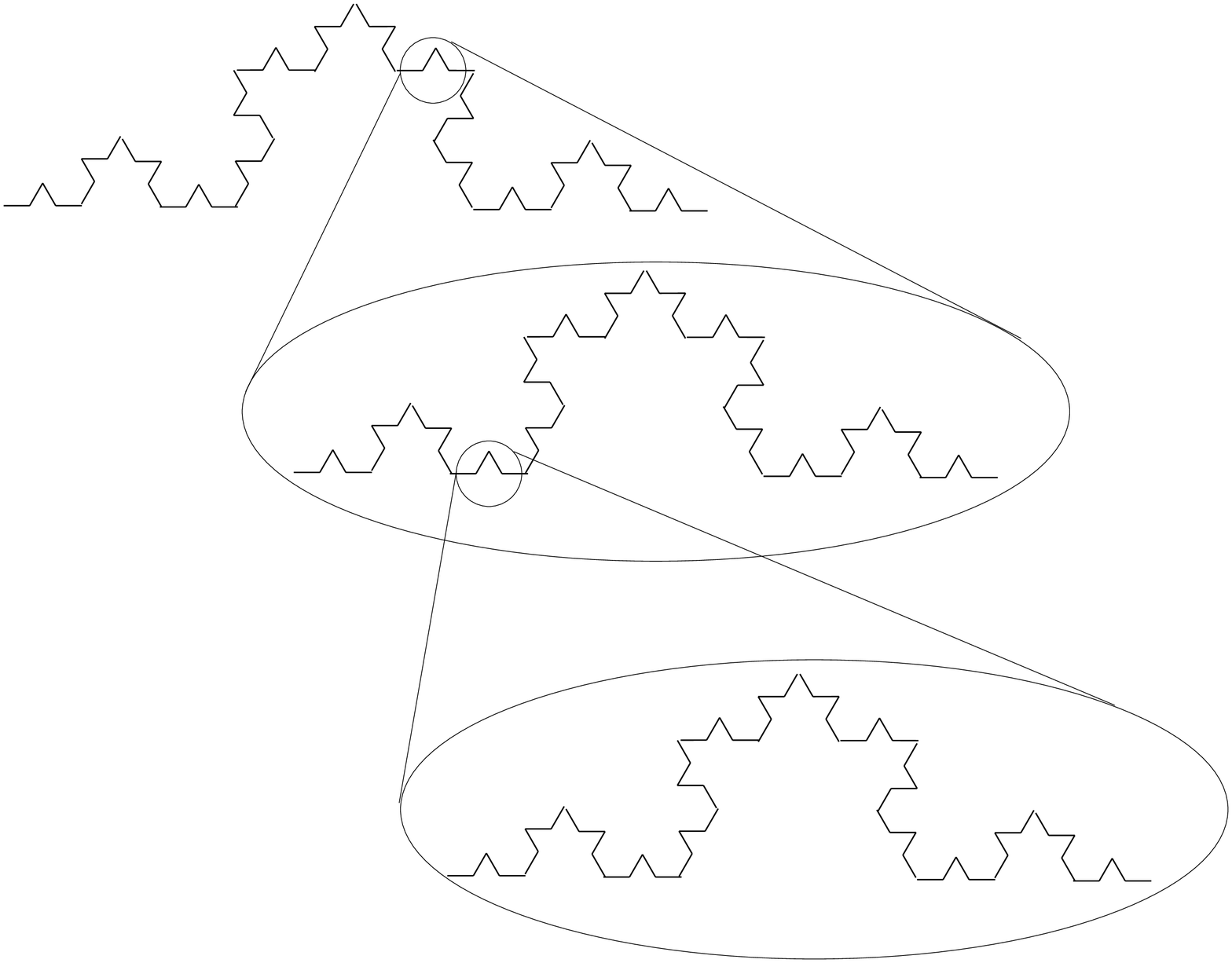}
\end{center}

\vfill
\rightline{\textcolor{red}{RELEASE: 1.0}}
\rightline{Cornell  CLNS 99/1612}

\newpage

\thispagestyle{empty}
\ \vfill
\copyright All rights reserved.

\vfill
This book was typeset in \TeX\
using the \LaTeXe\  Document Preparation System.

\end{titlepage}

\prolegomena
\markboth{\textcolor{blue}{PROLEGOMENA}}{\textcolor{blue}{PROLEGOMENA}}
\begin{center}
  {\bf\Large  A DEFENSE FOR TWO DIMENSIONS}\\
          or \\
  AN EXTRAORDINARY  PREFACE
\end{center}

In his classic book \textsf{Flatland} \cite{Abbott}, E. A. Abbott described
an imaginary two-dimensional world, embedded in three dimensions,
and populated
by two-dimensional figures that think, speak, and have all
the human emotions.
The author, a schoolmaster with a classics background 
who was interested in literature
and theology,
wrote his science fiction fantasy in pure mathematics 
for entertainment.  
He published it under a pseudonym to avoid any negative criticism
resulting from it reflecting poorly on his more formal works.
Apparently, he never
imagined that this work not only would entertain many generations
of physicists and mathematicians, but also 
would contain concepts that future scientists
would work on.

Almost 120 years after \textsf{Flatland} was written\footnote{The first edition
was published in 1880, while a 
second corrected edition was printed in 1884.  Beyond the mathematical
ideas explored, the book reflects
many attitudes that were normative in Western society in the late
nineteenth century, but that today are rejected as offensive and
ill-informed.},
much has happened to our ideas and concepts of spaces with
various dimensions. Relativity placed humans in a 
four-dimensional space-time; more recently, String Theory
claims that the dimensionality of our universe
has to be updated to 10 or 11 dimensions.
We have thus learnt to feel comfortable in a space of many dimensions.

We have also become more comfortable in spaces of
few dimensions. Two-dimensional 
science has grown to  one of the most important, well-controlled, and
well-developed areas of today's physics and mathematics. Abstract
mathematics and theoretical physics have established many breathtaking
results, while
condensed matter physics has devised many systems that behave as
two-dimensional, some of which have great impact on our society.

Of course, despite these two-dimensional applications, many
physicists have argued that `our universe' could not have been
two-dimensional, that life could not have existed in two dimensions,
and that three dimensions is the minimal 
feasible dimensionality for our world. In particular,
S. Hawking in talks and M. Kaku in his best-seller
\textsf{Hyperspace} have argued that the digestive system of living beings
would separate them in two disjoint sets in two dimensions and,
therefore, even this simple argument can rule out a universe in
two dimensions. However, this argument is too quick and facile (although
it does point out that evolution on Earth has selected a digestive system
suitable to three dimensions, and inappropriate for Flatland). 
Indeed, if one puts one's mind to it, a model for life in two spatial
dimensions can be developed, as is done quite thorougly in 
A. K. Dewdney's
book \textsf{Planiverse} \cite{Dewdney}.

So, after all, life in two dimensions might be possible. And there may
even be other universes out there that realize such a possibility. 
We may never discover such places. 
But we can certainly study them. And we will be glad that we did so!

\newpage

\begin{center}
  {\bf\Large CJE'S ORDINARY  PREFACE}
\end{center}

\vspace{5mm}

In contrast to Warren Siegel \cite{Siegel}, when I was a graduate student
I used to love books that contained in the title ``Introduction  to ...".
This is probably because, by the time I was a graduate student,
many good books delivering on this promise\footnote{This has already
become better with the creation of the LANL
archives by P. Ginsparg. Many beautifully written articles
which include ``Introduction  to ..." in their titles
are posted there frequently in a variety of fields. Papers of all
flavors are available there for virtually any taste.}
had been written. 

Even as an undergraduate student,
I recall (with some kind of nostalgia for those years)
that I was lucky enough to read some great introductory
books. Written on the standard  core of physics, those  books usually
would not include in the title the previous phrase.
Among these books, I especially liked 
 \textsf{Problems in Quantum Mechanics} by Constantinescu and Magyari.
The book is nicely written and printed.
It contains a number of problems on Quantum Mechanics (QM), divided thematically
into chapters. Each chapter contains a brief summary of the theory,
a collection of problems, and finally their solutions. I have been influenced
so greatly by this book, that the
architecture of the present
document is modeled after it --- although I 
can only hope that we approach the
quality of that book's presentation.

Although the subject of the present document is distinct
from the subject of its stylistic sibling, the two are
not completely unrelated. What one learns in \cite{CM}, can be used
here. After all,
QM is 1-dimensional Quantum Field Theory (QFT); 
the present document explores
the developments in 2-dimensional QFT.

During the last two decades, we have witnessed an amazing explosion
in the progress of mathematical physics.
String Theory\footnote{According to the standard lore of naming theories
in Field Theory, the name \textit{Quantum Nematodynamics}
(QND) is more appropriate. Perhaps the name \textit{String Theory}
should be used only at the classical level. Another possible
name, given the recent developments 
regarding branes, is \textit{Quantum Branodynamics}
(QBD); however, I personally would vote in favor of QND as I find
\textit{Quantum Nematodynamics} the most euphonic choice.}
has emerged as the
leading candidate for the Theory of Everything (TOE), and has led to
revolutions in our understanding of the principles underlying fundamental
physics. In conjunction with the rise of string theory,
we have witnessed the
discovery of and advancements in many other areas of mathematical
physics:  Conformal 
Field Theory (CFT), Integrable Models (IMs), 2-Dimensional Gravity,
Quantum Groups (QGs),
and Dualities in QFT, to name just a few areas.
The developments in each field have influenced the developments in
the other fields, at times in profound and radical ways, at other times
in subtler and more controlled fashion.

Thus, I embarked on the preparation of this collection of problems,
for those who wish
to study mathematical physics and want to see some solved problems
to whet their appetite.
I hope that people who learn the subjects treated in this
collection of problems will find it
useful. In this sense, it might also
prove useful to people who teach material related to the subjects
explored herein.

In his book, Siegel also writes,
``It is therefore simultaneously the best time 
for someone to read a book and the worst  time  for someone to write one."
This sentence remains true for this collection as well. 
Recent times have proven so
fertile for mathematical physics that many
results are considered `common knowledge' just a few days after they
are posted on the LANL archives. 
As a result, no review or book can cover 
completely such a vast terrain. The present
document is certainly no exception. It covers only a 
fraction of the relevant subjects, and even for those
covered, only a limited number of possible
themes have been touched. 
However, depending on the interest, we hope that in a later edition,
more topics will be added, and more exercises within each area.

We have partitioned the material in \ref{CountChapters}
\space chapters. Although there
are not always sharp boundaries among these chapters, this 
approach was taken to enhance the pedagogical value of this
work.

Finally, we would like to say that for such a subject,
it is impossible to give an exhaustive
list of references. We  primarily cite
review papers and books, as those may be of great help for
the reader who would like to study the material.
However, we have also included many significant
original papers in the bibliography
that might not be cited in the text.
In this spirit, we would like apologize to all people whose works are not
cited; there is no way to be exhaustive, nor have we attempted to be,
and in some cases such omissions are doubtless a result of our ignorance.
If this document serves as an introduction to the field of two-dimensional
physics and its literature, it will have served its purpose.

\newpage

\begin{center}
  {\bf\Large DAS'S ORDINARY  PREFACE}
\end{center}

\vspace{5mm}

The very first paper I published offered new demonstrations of
the integrability
of some non-linear sigma models, with and without supersymmetry, in two
spacetime dimensions.  While that paper predates (by just a little bit)
the revolution in 2-dimensional physics that this manuscript addresses,
I am struck by how much the topics of that paper are echoed here.

If I trace my own personal history of interest in the topics covered
here, there are two pivotal moments, beyond that first paper of mine.
The first moment is the ICGTMP (International Colloquium on
Group Theoretical Methods in Physics) held in Montreal in 1988.  
Those were
heady times.  The conference itself was a cornucopia of string theory, 
conformal field theory, and 
quantum groups, held in a political context that allowed
attendance by an extensive collection of physicists and mathematicians from
around the globe.  How could one not be hooked?

The second moment was my decision to pursue a career in a liberal arts
institution.  These are places not well-understood outside the USA; they
are colleges with no graduate students or postdocs, but that does not mean
that they are void of research.  On the contrary, my colleagues are some
of the most vibrant minds I know.  But it does mean that many of us do
not have the benefit of spending time each day, bumping into colleagues in
the hallway or seminar room, and learning things almost by osmosis.
The value of such a pedagogical document as the present one of course
transcends my own personal context; but this context has made clearer to me
what the value of such a manuscript is.

My co-author Costas Efthimiou has been the driving force behind this
project, and the rationale he presents above is indeed the same rationale
that drew me into this project, and I am glad to have been drawn in.
There is no need for me to repeat the
ideas Costas has expressed above.  But I will express
my hope that this document will find multiple uses, from students
beginning their
explorations of theoretical physics in graduate school, to established
scientists trying to
move into new areas of research, to faculty seeking inspiration for their
courses.

\newpage

\begin{center}
 {\bfseries\Large ACKNOWLEDGEMENTS}
\end{center}

This manuscript and our approach to the topics therein
has benefited from discussions on various occasions
with C. Ahn, M. Ameduri, S. Apikyan, P. Argyres, S. Chaudhuri,
J. Distler, B. Gerganov,
B. Greene, Z. Kakushadze, Y. Kanter, T. Klassen, A. LeClair, G. Shiu, and
H. Tye.  CJE thanks A. LeClair 
in particular, for providing an introduction to
some of the subjects discussed in this document, and
would especially like to thank M. Ameduri who typed some of the problems
from handwritten notes,  providing the momentum needed to
continue on this project. Last, but not least, CJE thanks the Florida
Southern College where some of the final details were written, and in particular
Professor M. Jamshid who gave him the opportunity to visit the college,
while DS acknowledges the support of
NSF Grant PHY-9970771.

\vspace{1cm}
\begin{center}
{\bfseries Comments and criticism are welcomed and greatly encouraged.}
\\  costas@phys.columbia.edu
\\  spector@hws.edu
\end{center}

\newpage

\small
\begin{flushright}
\begin{minipage}{3in}
\textsf{Other mistakes may perchance...await the penetrating glance of
some critical reader, to whom the joy of discovery, and the intellectual
superiority which he will thus discern, in himself, to the author of this
little book, will, I hope, repay to some extent the time and
 trouble its perusal
may have cost him!}
\rightline{\textsc{Lewis Carroll}}
\end{minipage}
\end{flushright}
\normalsize

\vspace{2.5cm}

\centerline{\bf\Large COMMENTS}

\begin{enumerate}
\item
  When you read this document, please keep in mind that the document
  is still in its infancy. Most sections are brief, and the presentation
  at times somewhat abbreviated.
  The reader is also warned that, despite our best efforts,
  no doubt many typos remain. We apologize, too, that different
  conventions may still be used in different parts of the document!
  This is due to the fact that several sections have their origins
  in projects undertaken before conceiving the plan to prepare one comprehensive
  pedagogical collection.  (Some would say this might even be valuable,
  preparing the reader for the array of conventions in the published
  literature!)
  We hope that the reader nonetheless finds this work valuable and
  beneficial.
  If all proceeds according to our
  expectations, when the document reaches its adult stage, it will
  have been cured of all these childhood diseases!

\item
  On the cover page, a release number
  $$
   \mbox{RELEASE $N.n$}
  $$
  is given (see on the cover).
  It should be interpreted as follows. A higher  release number
  of course signals a newer version. A larger $n$ means that 
  \emph{simple} typos
  have been corrected, wording may have been improved,
  conventions and notation may have been uniformized,
  additional references may have been added, but 
  \emph{no essential}
  changes have been made. Reprinting the document is in this
  case \emph{strongly discouraged} --- save the forests!
  A larger $N$
  number means that new material has been added (e.g. new problems
  in previously existing sections, new sections in previously
  existing chapters, or even new chapters) or 
  \emph{conceptual or other important}
  mistakes have been corrected. 
\item
  Of course, there are many topics that could be added
  to the present document. A list of topics that would appear
  as natural extensions to our work would include (see the list
  of abbreviations on page \pageref{table:PRO1}): 
  \begin{itemize}
  \item Background Material in QFT and in Mathematics
  \item Supersymmetric CFT
  \item Higher Genus CFT
  \item CFT in $D>2$ Dimensions
  \item IMs in QM
  \item Classical IMs 
  \item Quantization of Classical IMs 
  \item Bethe Ansatz
  \item Form Factors for IMs
  \item Boundary IMs
  \item Vertex Models
  \item Applications to Condensed Matter
  \item Knot Theory
  \item Matrix Models
  \item Topological Field Theories 
  \item String Theory
  \item Seiberg-Witten (SW) Theory and IMs
  \end{itemize}
  As the release number increases, this list of missing
  items should shorten until it \emph{dissappears} (what
  optimism!!!). However, even if this occurs, it would by no
  means imply that we had achieved a complete coverage of 
  these topics; it should
  be only interpreted as a completion of \textit{our} target, which is to allow
  the reader a foundation for indulging in the
  exploration of mathematical physics and physical mathematics. 
\item
  
Abbreviations are in general defined at the place of 
their first occurrence.
However, especially if you do not read this document sequentially, 
relying on this for definitions may be somewhat
cumbersome.
Certain well-known abbreviations may even not be
expanded in any place
in the document.
Therefore, 
we have included a table of abbreviations (Table \ref{table:PRO1}),
which appears on the following page.
\end{enumerate}

\newpage
\thispagestyle{empty}

\begin{table}[hb!]
\begin{center}
\begin{tabular}{|c|c|}\hline
{\bfseries Abbreviation} & {\bfseries Explanation}\\ \hline\hline
BCFT & Boundary Conformal Field Theory \\ \hline
CFT & Conformal Field Theory \\ \hline
CBC & Conformal Boundary Condition \\ \hline
CGC & Coulomb Gas Construction	 \\ \hline
CGF & Coulomb Gas Formulation \\ \hline
DSZ & Dirac-Schwinger-Zwanziger \\ \hline
GUT & Grand Unified Theory \\ \hline
IM & Integrable Model \\ \hline
LANL & Los Alamos National Laboratory \\ \hline
l.h.s. & left hand side \\ \hline
MM & Minimal Model \\ \hline
OPE & Operator Product Expansion \\ \hline
QCD & Quantum Chromodynamics \\\hline
QED & Quantum Electrodynamics \\\hline
QFT & Quantum Field Theory \\\hline
QG & Quantum Group \\\hline
QM & Quantum Mechanics \\\hline
RCFT & Rational Conformal Field Theory \\\hline
reg & Non-Singular Terms in Operator Product Expansion \\\hline
r.h.s. & right hand side \\ \hline
RSOS & Restricted Solid-on-Solid \\ \hline
SCFT & Supersymmetric Conformal Field Theory \\ \hline
SG  & Sine-Gordon \\ \hline
S-matrix & Scattering Matrix \\ \hline
SOS & Solid-on-Solid \\ \hline
SUSY & Supersymmetry \\ \hline
SUGRA & Supergravity \\ \hline
SW & Seiberg-Witten \\ \hline
TOE & Theory of Everything \\ \hline
UMM & Unitary Minimal Model \\ \hline
W-matrix & Wall (or Reflection) Matrix \\ \hline
w.r.t. & with respect to \\ \hline
WZWN & Wess-Zumino-Witten-Novikov \\ \hline
YBE & Yang-Baxter Equation \\ \hline
\end{tabular}
\end{center}
\caption{Table of abbreviations and acronyms used in this document.}
\label{table:PRO1}
\end{table}

\newpage

\newpage
\ \thispagestyle{empty} 
\newpage
\thispagestyle{empty}
\markboth{\textcolor{blue}{CONTENTS}}{\textcolor{blue}{CONTENTS}}
\noindent{\Huge\bf Contents}

\vspace{2cm}
\noindent{\Large\bf PART I}

\contentsline {chapter}{\numberline {1}CURIOSITIES IN TWO DIMENSIONS}{1}
\contentsline {section}{\numberline {1.1}BRIEF THEORY}{1}
\contentsline {subsection}{\numberline {1.1.1}Statistics in Two Dimensions}{1}
\contentsline {subsection}{\numberline {1.1.2}Bosonization/Fermionization}{3}
\contentsline {subsection}{\numberline {1.1.3}The Conformal Group in Two Dimensions}{5}
\contentsline {subsection}{\numberline {1.1.4}Commonly Used Conformal Transformations}{7}
\contentsline {subsection}{\numberline {1.1.5}Symmetries of the S-Matrix}{10}
\contentsline {section}{\numberline {1.2}EXERCISES}{13}
\contentsline {section}{\numberline {1.3}SOLUTIONS}{15}
\contentsline {chapter}{\numberline {2}GENERAL PRINCIPLES OF CFT}{35}
\contentsline {section}{\numberline {2.1}BRIEF THEORY}{35}
\contentsline {subsection}{\numberline {2.1.1}Basic Notions of CFT}{35}
\contentsline {subsection}{\numberline {2.1.2}Massless Free Boson}{39}
\contentsline {subsection}{\numberline {2.1.3}Massless Free Fermion}{40}
\contentsline {subsection}{\numberline {2.1.4}The $bc$-System}{41}
\contentsline {subsection}{\numberline {2.1.5}Boson with Background Charge}{42}
\contentsline {subsection}{\numberline {2.1.6}Minimal Models of CFT}{43}
\contentsline {subsection}{\numberline {2.1.7}A Prelude to Chapters 10\hbox {} and 15\hbox {}}{44}
\contentsline {section}{\numberline {2.2}EXERCISES}{47}
\contentsline {section}{\numberline {2.3}SOLUTIONS}{50}
\contentsline {chapter}{\numberline {3}CORRELATORS IN CFT}{81}
\contentsline {section}{\numberline {3.1}BRIEF SUMMARY}{81}
\contentsline {subsection}{\numberline {3.1.1}Computation of Correlators}{81}
\contentsline {subsection}{\numberline {3.1.2}The Bootstrap Approach}{82}
\contentsline {subsection}{\numberline {3.1.3}Fusion Rules}{84}
\contentsline {subsection}{\numberline {3.1.4}Local vs. Non-Local Fields}{85}
\contentsline {subsection}{\numberline {3.1.5}Bosonization}{86}
\contentsline {section}{\numberline {3.2}EXERCISES}{88}
\contentsline {section}{\numberline {3.3}SOLUTIONS}{90}
\contentsline {chapter}{\numberline {4}OTHER MODELS IN CFT}{111}
\contentsline {section}{\numberline {4.1}BRIEF SUMMARY}{111}
\contentsline {subsection}{\numberline {4.1.1}Orbifolds}{111}
\contentsline {subsection}{\numberline {4.1.2}WZWN Model}{112}
\contentsline {subsection}{\numberline {4.1.3}Parafermions}{114}
\contentsline {section}{\numberline {4.2}EXERCISES}{116}
\contentsline {section}{\numberline {4.3}SOLUTIONS}{118}
\contentsline {chapter}{\numberline {5}CONSTRUCTING NEW MODELS IN CFT}{137}
\contentsline {section}{\numberline {5.1}BRIEF SUMMARY}{137}
\contentsline {subsection}{\numberline {5.1.1}Coulomb Gas Construction or Formulation}{137}
\contentsline {subsection}{\numberline {5.1.2}Coset Construction}{139}
\contentsline {section}{\numberline {5.2}EXERCISES}{140}
\contentsline {section}{\numberline {5.3}SOLUTIONS}{142}
\contentsline {chapter}{\numberline {6}MODULAR INVARIANCE}{157}
\contentsline {section}{\numberline {6.1}BRIEF SUMMARY}{157}
\contentsline {subsection}{\numberline {6.1.1}Riemann $\vartheta $-function}{157}
\contentsline {subsection}{\numberline {6.1.2}The Modular Group}{159}
\contentsline {subsection}{\numberline {6.1.3}Partition Functions and Modular Invariance}{160}
\contentsline {section}{\numberline {6.2}EXERCISES}{162}
\contentsline {section}{\numberline {6.3}SOLUTIONS}{165}

\vspace{2cm}
\noindent{\Large\bf PART II}\\
\textcolor{red}{\textit{Warning! Part II has not yet
been released.  The contents and page numbers for Part II are therefore
preliminary, and subject to change.}}

\contentsline {chapter}{\numberline {7}FINITE SIZE AND FINITE TEMPERATURE}{193}
\contentsline {section}{\numberline {7.1}BRIEF SUMMARY}{193}
\contentsline {subsection}{\numberline {7.1.1}Main Ideas}{193}
\contentsline {subsection}{\numberline {7.1.2}Casimir Energy}{195}
\contentsline {subsection}{\numberline {7.1.3}Correlation Functions}{195}
\contentsline {section}{\numberline {7.2}EXERCISES}{196}
\contentsline {section}{\numberline {7.3}SOLUTIONS}{197}
\contentsline {chapter}{\numberline {8}BOUNDARY CFT}{211}
\contentsline {section}{\numberline {8.1}BRIEF SUMMARY}{211}
\contentsline {subsection}{\numberline {8.1.1}Boundary States}{211}
\contentsline {subsection}{\numberline {8.1.2}Boundary Partition Function}{213}
\contentsline {subsection}{\numberline {8.1.3}Boundary Green's Functions}{215}
\contentsline {subsection}{\numberline {8.1.4}Method of Images and Boundary Operator}{216}
\contentsline {section}{\numberline {8.2}EXERCISES}{217}
\contentsline {section}{\numberline {8.3}SOLUTIONS}{218}
\contentsline {chapter}{\numberline {9}QUANTUM GROUPS}{227}
\contentsline {section}{\numberline {9.1}BRIEF SUMMARY}{227}
\contentsline {subsection}{\numberline {9.1.1}The Basic Idea of a Quantum Group}{227}
\contentsline {subsection}{\numberline {9.1.2}Hopf Algebras}{230}
\contentsline {subsection}{\numberline {9.1.3}The Quantum Group $U_q(\mathfrak{g})$}{231}
\contentsline {subsection}{\numberline {9.1.4}Deformation of Known Structures}{232}
\contentsline {section}{\numberline {9.2}EXERCISES}{234}
\contentsline {section}{\numberline {9.3}SOLUTIONS}{237}
\contentsline {chapter}{\numberline {10}PERTURBATIONS OF CFT}{255}
\contentsline {section}{\numberline {10.1}BRIEF SUMMARY}{255}
\contentsline {subsection}{\numberline {10.1.1}Deformed CFTs}{255}
\contentsline {subsection}{\numberline {10.1.2}The c-Theorem}{256}
\contentsline {section}{\numberline {10.2}EXERCISES}{257}
\contentsline {section}{\numberline {10.3}SOLUTIONS}{258}
\contentsline {chapter}{\numberline {11}INTEGRABLE MODELS AS PERTURBATIONS OF CFT}{267}
\contentsline {section}{\numberline {11.1}BRIEF SUMMARY}{267}
\contentsline {subsection}{\numberline {11.1.1}Definition of Integrability}{267}
\contentsline {subsection}{\numberline {11.1.2}Conserved Currents}{267}
\contentsline {subsection}{\numberline {11.1.3}Integrable Deformed CFTs}{268}
\contentsline {section}{\numberline {11.2}EXERCISES}{270}
\contentsline {section}{\numberline {11.3}SOLUTIONS}{272}
\contentsline {chapter}{\numberline {12}FACTORIZED SCATTERING}{279}
\contentsline {section}{\numberline {12.1}BRIEF SUMMARY}{279}
\contentsline {subsection}{\numberline {12.1.1}Factorization}{279}
\contentsline {subsection}{\numberline {12.1.2}2-D S-matrix Theory}{281}
\contentsline {subsection}{\numberline {12.1.3}The Method of Non-Local Charges}{283}
\contentsline {subsection}{\numberline {12.1.4}Toda Field Theory}{284}
\contentsline {section}{\numberline {12.2}EXERCISES}{285}
\contentsline {section}{\numberline {12.3}SOLUTIONS}{288}
\contentsline {chapter}{\numberline {13}DIAGONAL SCATTERING}{307}
\contentsline {section}{\numberline {13.1}BRIEF SUMMARY}{307}
\contentsline {subsection}{\numberline {13.1.1}Diagonal S-Matrix}{307}
\contentsline {subsection}{\numberline {13.1.2}Bootstrap and Nuclear Democracy}{309}
\contentsline {subsection}{\numberline {13.1.3}Bootstrap equations}{309}
\contentsline {section}{\numberline {13.2}EXERCISES}{311}
\contentsline {section}{\numberline {13.3}SOLUTIONS}{312}
\contentsline {chapter}{\numberline {14}BOUNDARY IMs}{313}
\contentsline {section}{\numberline {14.1}BRIEF SUMMARY}{313}
\contentsline {subsection}{\numberline {14.1.1}Conserved Currents in the Presence of Boundary}{313}
\contentsline {subsection}{\numberline {14.1.2}Factorization}{314}
\contentsline {subsection}{\numberline {14.1.3}Boundary W-Matrix}{315}
\contentsline {subsection}{\numberline {14.1.4}Hamiltonian Picture}{316}
\contentsline {subsection}{\numberline {14.1.5}Boundary Bootstrap Equations}{318}
\contentsline {section}{\numberline {14.2}EXERCISES}{321}
\contentsline {section}{\numberline {14.3}SOLUTIONS}{322}
\contentsline {chapter}{\numberline {15}STATISTICAL PHYSICS}{329}
\contentsline {section}{\numberline {15.1}BRIEF SUMMARY}{329}
\contentsline {subsection}{\numberline {15.1.1}The spin-$1/2$ Ising Model}{329}
\contentsline {subsection}{\numberline {15.1.2}Spherical Model}{331}
\contentsline {subsection}{\numberline {15.1.3}$\mathbb{Z}_p$ Model}{331}
\contentsline {subsection}{\numberline {15.1.4}Potts Model}{332}
\contentsline {subsection}{\numberline {15.1.5}XYZ Model}{332}
\contentsline {subsection}{\numberline {15.1.6}Ashkin-Teller Model}{332}
\contentsline {subsection}{\numberline {15.1.7}The Transfer Matrix Method}{333}
\contentsline {subsection}{\numberline {15.1.8}Critical Exponents}{333}
\contentsline {subsection}{\numberline {15.1.9}CFT, IMs, and Statistical Physics}{334}
\contentsline {section}{\numberline {15.2}EXERCISES}{335}
\contentsline {section}{\numberline {15.3}SOLUTIONS}{337}

\mainmatter

\newchapter{CURIOSITIES IN TWO DIMENSIONS}
\label{ch:CUR}

\vspace{-2cm}
\small
\begin{flushright}
\begin{minipage}{2.5in}
 \textsf{But to me, proficient though I was in Flatland Mathematics,
 it was by no means a simple matter...although I saw the facts before me, the
 causes were as dark as ever.}
\rightline{\textsc{A. Square}}
\end{minipage}
\end{flushright}
\normalsize

\vspace{2cm}
\footnotesize
\noindent {\bfseries References}:
Good introductions to spin and statistics in two dimensions are found in
\cite{For,MW}. A recent review on generalized statistics in 1+1
dimensions is \cite{Poly}. Bosonization is discussed in \cite{Sto}.
The conformal transformations are discussed in many books on complex analysis,
e.g. \cite{Nehari}.
\normalsize

\section{BRIEF THEORY}

2-dimensional mathematics and physics are
quite special.  The lack of `extra space'
imposes such great constraints that the particles and their
interactions adhere to a highly restrictive set of properties.
This chapter examines some of the basic implications 
of a 2-dimensional world.
This lays the groundwork for what follows in the
subsequent chapters.

\subsection{Statistics in Two Dimensions}

It is well-known that the path integral on a multipy-connected space
$\CC$ decomposes into a sum of path integrals (one for each class of
the first homotopy group), with the contributions from each sector
weighted by a factor dependent 
on the equivalence class only, so that
$$
    Z(x,y)\=\sum_{\alpha\in\pi_1(\CC)} \, \chi(\alpha)\,
    Z_\alpha(x,y)~.
$$
The weights $\chi(\alpha)$ are restricted by two physical requirements:
\\ (a) Physical observables cannot depend on the mesh used to calculate the
       homotopy classes.
\\ (b) The weighted sum must satisfy the standard convolutive property
       $$
           Z(x,y)\=\int dz\,  Z(x,z) Z(z,y)~.
       $$
As a result, the weights satisfy the constraints
\beq
\label{eq:CUR2}
     \left.
     \begin{array}{c}
      |\chi(\alpha)| = 1~,~~~{\rm and}\\
     \chi(\alpha)\chi(\beta) = \chi(\alpha\beta)~.
     \end{array}
    \ra  
\eeq
Therefore the weights $\chi$ provide a 1-dimensional unitary
representation of the fundamental group $\pi_1(\CC)$.

Given $n$ particles in $d$ dimensions, the space of all allowed configurations 
would be $\overline{\CC}_n=\BR^{dn}\smallsetminus D$, where $D$ is the subset
of $\BR^{dn}$ where at least two particles have identical position vectors.
When the particles are of identical kind, configurations that differ by a
permutation are not distinct. To avoid multiple enumeration 
of the configurations,
one must divide by the permutation group for $n$ objects $S_n$. In this
way, the configuration space $\CC_n$ of $n$ identical particles reads
$$
    \CC_n\={\overline{\CC}_n\over S_n}\=
           {\BR^{dn}\smallsetminus D\over S_n}~.
$$

In $d>2$ dimensions, $\CC_n$ is simply-connected and the fundamental group
is $\pi_1(\CC_n)=S_n$. The permutation group has only two unitary abelian
representations; either
$$
    \chi(P)\=1~,~~~~~\forall P\in S_n~,
$$
or
$$
   \chi(P)\=\cases{+1~,& if~$P$~is~even~,\cr
                  -1~,& if~$P$~is~odd~.\cr}
$$
These representations correspond to bosons and fermions, respectively.

In $d=2$ dimensions, $\CC_n$ is multiply-connected and $\pi_1(\CC_n)$ is
more complicated. In fact, $\pi_1(\CC_n)=B_n$, the 
{\bf braid group}\index{group!braid --} of $n$ strings. Combinatorially,
$B_n$ is generated by a set of generators $\sigma_i,~i=1,2,\dots,n$,
obeying the relations
\beqn
\label{eq:CUR3}
    \sigma_i\sigma_{i+1}\sigma_i &=&
    \sigma_{i+1}\sigma_i\sigma_{i+1} ~,\\
\label{eq:CUR4}
    \sigma_i\sigma_j &=& \sigma_j \sigma_i~, ~~~|i-j|>1~.
\eeqn
Intuitevely, we can think of the generator $\sigma_i$
as the operation that twists the $i$-th and $(i+1)$-th strings,
as depicted here:

\begin{center}
\psfrag{1}{$1$}
\psfrag{2}{$2$}
\psfrag{i}{$i$}
\psfrag{i+1}{$i+1$}
\psfrag{n+1}{$n+1$}
\psfrag{n}{$n$}
\includegraphics[height=4cm]{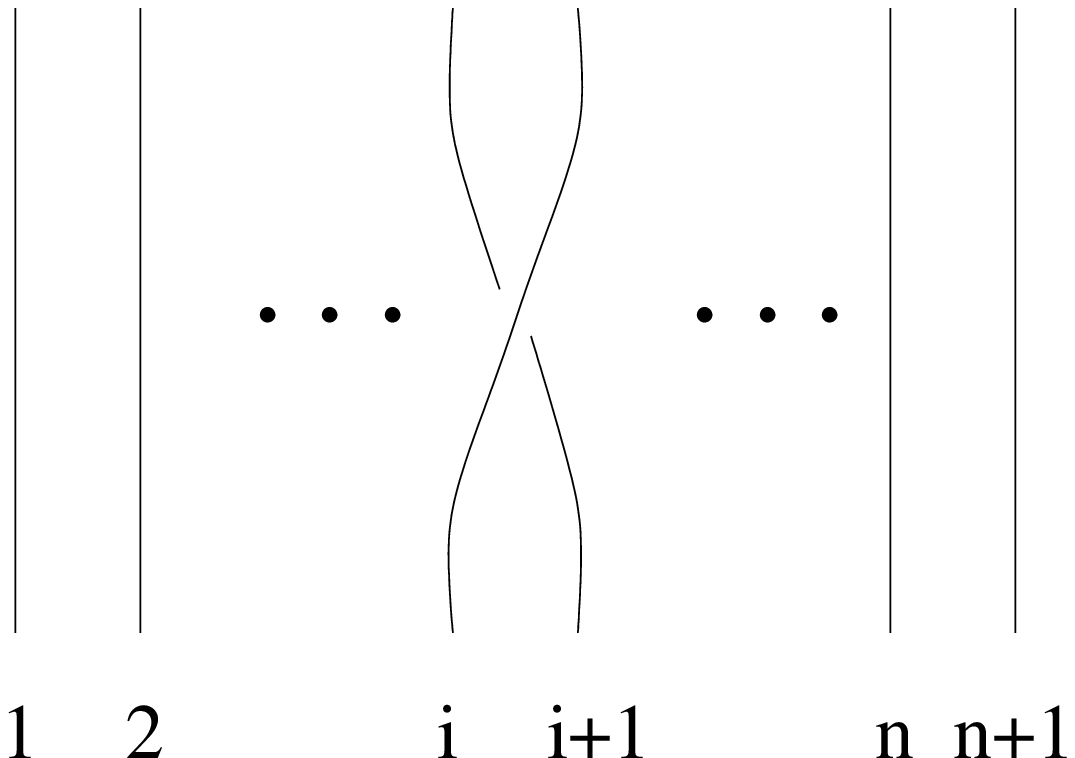}
\end{center}

The inverse  $\sigma_i^{-1}$ of the operator $\sigma_i$ interweaves the
same strings in the opposite way:

\begin{center}
\psfrag{1}{$1$}
\psfrag{2}{$2$}
\psfrag{i}{$i$}
\psfrag{i+1}{$i+1$}
\psfrag{n+1}{$n+1$}
\psfrag{n}{$n$}
\includegraphics[height=4cm]{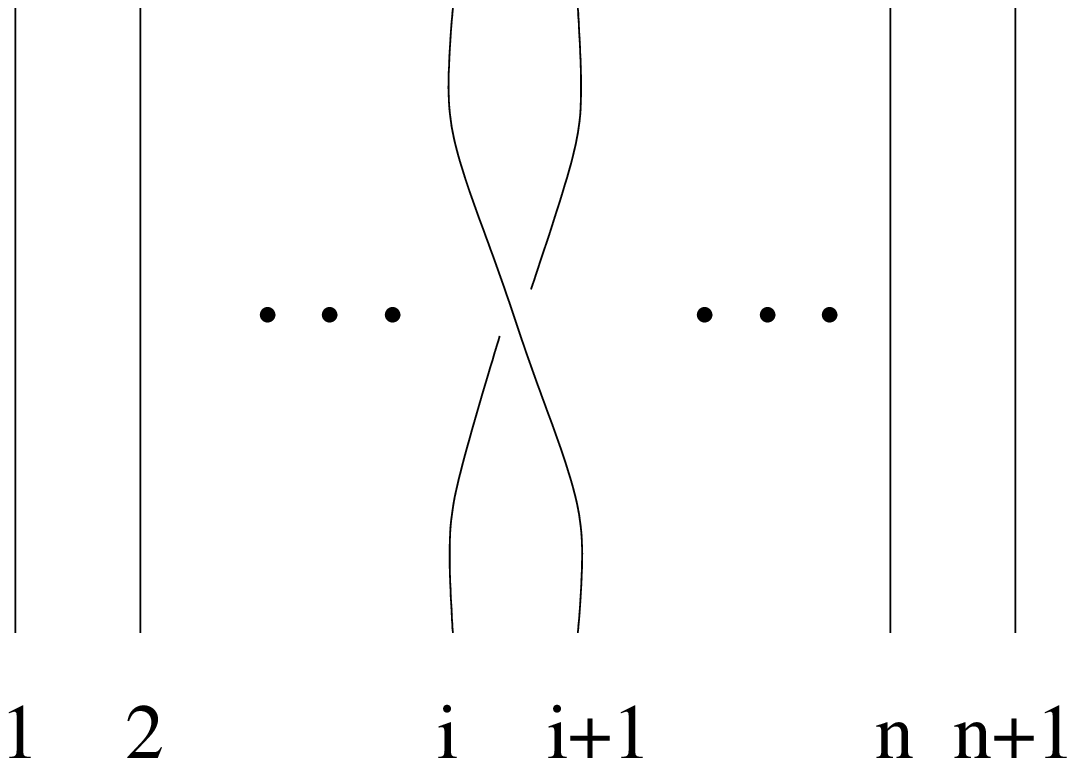}
\end{center}

Any element $\sigma\in B_n$ can be represented as set of $n$ interwoven
strings, that is, as an ordered  product of the generators
$\sigma_i$ and their inverses $\sigma_i^{-1}$. 
Relation \calle{eq:CUR4} says that the twistings of two 
well-separated pairs of strings are independent. Relation \calle{eq:CUR3}
identifies two ways of interweaving three successive strings
that are  equivalent.

Regarding the weights $\chi$, relations \calle{eq:CUR3} and 
\calle{eq:CUR4} require that the phase be universal
$$
   \chi(\sigma_i)\=\chi(\sigma_j)~,~~~~~\forall i,j~.
$$
We  write
$$
    \chi(\sigma_i)\=e^{i\theta}~.
$$
A path  (braid) $\sigma\in\pi_1(\CC_n)$ represented by the product
$$
   \sigma\=\prod_{k=1}^n\, \sigma_{i_k}~,
$$
is thus weighted by the factor
$$
   \chi(\sigma)\=\prod_{k=1}^n\,\chi( \sigma_{i_k})\=
   e^{i\theta\sum_{k=1}^n{\rm sgn}\sigma_{i_k}}~,
$$
where
$$
   {\rm sgn}\sigma_{i_k}
         \=\cases{+1~,& if~$\sigma_{i_k}$~is~a~generator~,\cr
                  -1~,& if~$\sigma_{i_k}$~is~an~inverse~generator~.\cr}
$$

\subsection{Bosonization/Fermionization}

In two dimensions, we 
have seen that particles may obey a variety of statistics,
in contrast with higher dimensions, in which particles must obey
either fermionic or bosonic statistics. However, one might still
expect that each of the choices would be related to a particular
choice of the intrinsic spin of the particles:
$$
  \theta\mbox{-statistics} \leftrightarrow
  \mbox{particles with spin} ~s_\theta~.
$$
Surprisingly, spin is not an intrinsic property in two dimensions,
and statistics are a matter of convention. For example, given any bosonic
operators $b(p)$ and $b^\dagger(p)$ that satisfy the commutation
relations
\bb
   \lb b(p),b^\dagger(p')\rb &=& p^0\delta(p-p')~, \\ 
   \lb b^\dagger(p),b^\dagger(p')\rb\=\lb b(p),b(p')\rb &=& 0~,  
\ee
one can construct operators $f(p)$ and $f^\dagger(p)$, defined by
\beq
   \left.
   \begin{array}{r}
   f^\dagger(p)\=  b^\dagger(p)\,
  e^{-i\pi\int_{p_1}^{+\infty} {dk_1\over k_0} b^\dagger(k)b(k)}~,\\
   f(p)\phantom{^\dagger} \= 
  e^{i\pi\int_{p_1}^{+\infty} {dk_1\over k_0} b^\dagger(k)b(k)}
   \, b(p)~,
  \phantom{{}^{\dagger\dagger-}}
   \end{array} \ra
\label{eq:CUR1}
\eeq
which are fermionic operators that satisfy the anticommutation
relations
\bb
   \la f(p),f^\dagger(p')\ra &=& p^0\delta(p-p')~, \\
   \la f^\dagger(p),f^\dagger(p')\rb\=\la f(p),f(p')\ra &=& 0~.
\ee
Thus there is a 1-to-1 mapping between the states that are symmetric and those
that are antisymmetric under permutations! 

Having no intrinsic spin in two dimensions, we use the term ``spin''
to refer to the Lorentz spin,
i.e., to the eigenvalue of a state under the boost operator. Using
Wigner's classic analysis, the scalar states can be chosen to transform
under boosts as
$$
   U(\alpha)
   \ket{p_0,p_1}\=\ket{p'_0,p_1'}~,
$$
where 
$$
   \lb\matrix{p'_0\cr p'_1\cr}\rb\=
   \lb\matrix{\cosh\alpha & \sinh\alpha\cr \sinh\alpha & \cosh\alpha\cr}\rb
   \lb\matrix{p_0\cr p_1\cr}\rb~.
$$
One can define the states
$$
    \ket{p_0,p_1;s}\eq \lp {p_0-p_1\over m} \rp^s \,
    \ket{p_0,p_1}~,
$$
which then transform by
$$
    U(\alpha)
   \ket{p_0,p_1;s}\= e^{-s\alpha}\,  \ket{p'_0,p'_1;s}~,
$$
i.e., they have Lorentz spin $s$.
Zero mass states have a larger symmetry group --- the conformal group.

The issues we have enountered in this section will arise many times
throughout this document.

\subsection{The Conformal Group in Two Dimensions}

Let $f: M_1\rightarrow M_2$ be a map from the $n$-manifold $M_1$
to the $n$-manifold $M_2$. Then, $f$ is said to be a
{\bf conformal transformation}\index{Transformation!Conformal --} if
$\forall p\in M_1$, given any $\bfv,\bfu\in T_pM_1$,
\bb
   \bfg'_{f(p)}(f_*(\bfv),f_*(\bfu))\= \Omega^2\, 
   \bfg_p(\bfv,\bfu) ~.
\ee
In the above equation,
$\bfg$ and $\bfg'$ are the metrics on $M_1$ and $M_2$ respectively
and $f_*$ is the induced map $f_*:T_pM_1\rightarrow T_{f(p)}M_2$.
When $M_1=M_2$, $f$ is called a 
{\bfseries conformal isometry}\index{Conformal!--Isometry}
of $M_1$.
It is very instructive to write the transformation rule explicitly:
\beq
\label{eq:CUR7}
  \Omega^2(p)\,g_{\mu\nu}(p)\=
  {\partial f^\rho\over\partial x^\mu}\,
  {\partial f^\sigma\over\partial x^\nu}\,
   \,g'_{\rho\sigma}(f(p))~.
\eeq

A conformal transformation leaves invariant
the angle $\theta$ between 
two vectors $\bfv$ and $\bfu$ in the
tangent space, since this angle obeys 
\bb
 \cos\theta\eq 
 {\bfg(\bfv,\bfu)\over\sqrt{\bfg(\bfv,\bfv)\,
   \bfg(\bfu,\bfu)}}~.
\ee
We may  say that a conformal transformation
changes only the size or scale of an object
while keeping its shape fixed; this intuitive visualization
of the transformation is more faithful at small distances.

If for a (pseudo-)Riemannian manifold $M$, there is an atlas  $\{ U_j,\phi_j\}$
such that the metric tensor $\bfg_j$ on each patch is conformally related
to the flat metric, i.e., $\bfg_j=\Omega_j^2\,\bfet$, then the manifold $M$
is said to be {\bf conformally flat}\index{conformally flat}.

Let $f^\mu(\bfx)$ be a conformal isometry of the metric manifold $M$,
i.e., suppose that
\beq
\label{iso1}
   g_{\mu\nu}(\bfx)\= \Omega^2   \,
   g_{\alpha\beta}(\bff(\bfx))\,
   {\partial f^\alpha\over\partial x^\mu}\,
   {\partial f^\beta\over\partial x^\nu}~.
\eeq
Without loss of
generality, we can work with
infinitesimal displacements 
\beq
\label{infinitesimal1}
   f^\mu(\bfx)\=x^\mu+\epsilon\,\xi^\mu(\bfx)~,
   ~~~~~|\epsilon|\ll 1~,
\eeq
since any finite displacement can be made from
an infinite sum of infinitesimal transformations.
For infinitesimal conformal transformations, we must have
\beq
\label{smallOM}
     \Omega^2 \= 1+\epsilon\, \psi +\CO(\epsilon^2) ~.
\eeq
Using \calle{infinitesimal1} and \calle{smallOM},
equation \calle{iso1} becomes
\bb
 (1+\epsilon\,\psi)\,  g_{\mu\nu}\= g_{\mu\nu} + 
 \epsilon\,  g_{\mu\beta}\, \partial_\nu\xi^\beta
   +\epsilon\, g_{\alpha\nu}\, \partial_\mu\xi^\alpha
   +\epsilon\, \partial_\rho g_{\alpha\nu}\, \xi^\rho 
  +\CO (\epsilon^2)~,
\ee
or
\beq
\label{killing1}
  \xi^\rho\,\partial_\rho g_{\mu\nu}
  + \partial_\mu\xi^\rho\,g_{\rho\nu} 
  + \partial_\nu\xi^\rho \,g_{\mu\rho}\=
  \psi\, g_{\mu\nu}~.
\eeq
The scalar $\psi$ is determined by self-consistency.
Multiplying the last equation by $g^{\nu\mu}$ and summing
over the indicated repeated indices, we find
\bb
  \xi^\rho\,g^{\nu\mu}\,\partial_\rho g_{\mu\nu}
  +2\, \partial_\rho\xi^\rho \=
  n\, \psi~.
\ee
Substituting for $\psi$ in equation \calle{killing1}, we finally
obtain
\beq
\label{killing45}
  \xi^\rho\,\partial_\rho g_{\mu\nu}
  + \partial_\mu\xi^\rho\,g_{\rho\nu} 
  + \partial_\nu\xi^\rho \,g_{\mu\rho}\=
{1\over n}\,
  (\xi^\rho\,g^{\kappa\sigma}\,\partial_\rho g_{\sigma\kappa}
  +2\, \partial_\rho\xi^\rho )
  \, g_{\mu\nu}~.
\eeq
This is the
 {\bf conformal  Killing equation}\index{Killing!Conformal -- Equation}.
The
solutions $\bfxi$ of this equation are called the 
{\bfseries conformal Killing vector fields}\index{Killing! Conformal -- Vector 
Fields} associated with the given metric.
The above discussion makes obvious that if $\psi=0$, i.e., if
we are considering isometries\index{Isometry} of $M$,
then the conformal Killing equation 
reduces to the simpler form
\beq
\label{eq:CUR6}
  \xi^\rho\,\partial_\rho g_{\mu\nu}
  + \partial_\mu\xi^\rho\,g_{\rho\nu} 
  + \partial_\nu\xi^\rho \,g_{\mu\rho}\=
   0~.
\eeq
This equation is called simply the 
{\bf Killing equation}\index{Killing!-- Equation};
the corresponding 
solutions $\bfxi$  are called the 
{\bf Killing vector fields}\index{Killing!-- Vector Fields}. 

For a conformally flat
$n$-dimensional manifold, the conformal Killing equation gives
\beq
\label{Killing}
    \partial_\mu \xi_\nu+\partial_\nu \xi_\mu\= {2\over n}\, \eta_{\mu\nu}\,
        \partial_\rho \xi^\rho~.
\eeq
The solutions of this equation generate  the 
{\bfseries conformal group}\index{Conformal! -- Group}.
For $d>2$, the conformal group is finite-dimensional.
This group in
two  dimensions is infinite-dimensional, imposing great restrictions 
on any conformally invariant field theory. In particular, 
the above equation \calle{Killing} reduces to the Cauchy-Riemann
conditions
$$
  {\partial \xi_2\over\partial x^1}\=
  -{\partial \xi_1\over\partial x^2}~,~~~~~
  {\partial \xi_1\over\partial x^1}\=
  {\partial \xi_2\over\partial x^2}~.
$$
Using
complex coordinates
$$
   z\=x^1+ix^2~,~~~~~\overline z\=x^1-ix^2~,
$$
and the function $\xi=\xi_1+i\,\xi_2$, the Cauchy-Riemann conditions
guarantee that $\xi$ is a holomorphic function, i.e., that
$$
  {\partial \xi\over \partial \overline z}\= 0~.
$$
The conformal transformations \calle{eq:CUR7}
are thus realized in two dimensions  by
\beq
   z~\rightarrow~ z'=f(z)~,~~~~~
   \overline z~\rightarrow~ \overline z'=\overline f(\overline z)~.
\eeq
This decoupling of variables allows
one to handle the coordinates $z$ and $\overline z$ as independent;
the reality condition $\overline z=z^*$ can be imposed at the end of the day.

\subsection{Commonly Used Conformal Transformations}
\label{section:conftrans}

In this document, conformal transformations will underlie nearly all that
we do.
Therefore, it is worthwhile to study some of them and become familiar with
their properties.

A fundamental map is
\beq
\label{eq:CUR31}
   w\= e^z
\eeq
from the $z$-plane to the $w$-plane. Decomposing $z$ and $w$ into 
real and imaginary parts via
$z=x+iy$ and  $w=u+iv$,
this map can be expressed in real coordinates as
\bb
   u &=& e^x\, \cos y~, \\
   v &=& e^x\, \sin y~.
\ee
Therefore
\bb
    u^2+v^2 &=& e^{2x}~, \\
    v&=&u\,\tan y~,
\ee
and the line $x=x_0$ is mapped to a circle with radius $\rho=e^{x_0}$.
The mapping \calle{eq:CUR31} transforms lines parallel to the imaginary axis
into circles around the origin.
If $x$ is the temporal coordinate, then time ordering on the $z$-plane
becomes radial ordering on the $w$-plane. A line $y=y_0$ is mapped to the
half-line $v=(\tan y_0) u$. Thus the mapping \calle{eq:CUR31} also  transforms
lines parallel 
to the real axis into rays directed outward from the origin.

\begin{center}
\psfrag{z-plane}{$z$-plane}
\psfrag{w-plane}{$w$-plane}
\psfrag{x}{$x$}
\psfrag{y}{$y$}
\psfrag{v}{$v$}
\psfrag{u}{$u$}
\psfrag{y=y_0}{$y=y_0$}
\psfrag{x=x_0}{$x=x_0$}
\psfrag{r}{$\rho=e^{x_0}$}
\psfrag{f}{$\phi=y_0$}
\psfrag{e}{$w=e^z$}
\includegraphics[width=12cm]{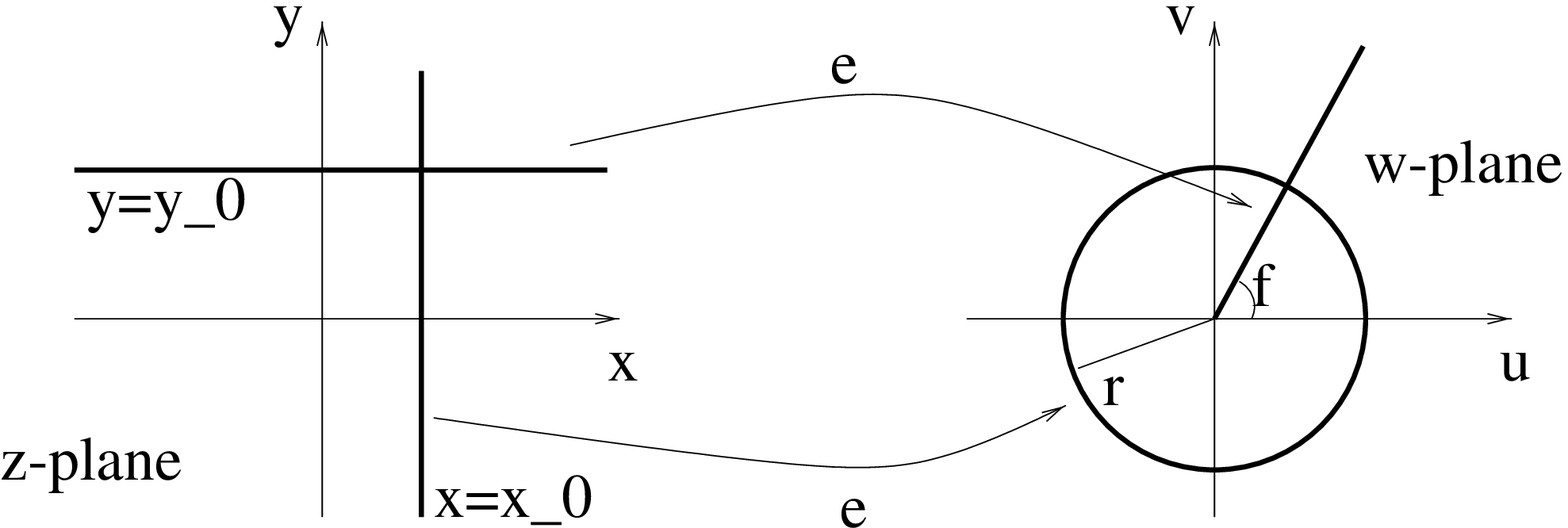}
\end{center}

Additional properties may be derived considering the inverse 
transformation of \calle{eq:CUR31},
\beq
    z\=\ln w~.
\label{eq:CUR32}
\eeq
Using the polar coordinates $\rho$ and $\phi$ on the $w$-plane,
with $w=\rho e^{i\phi}$,
\calle{eq:CUR32} becomes
\beq 
   \left.
   \begin{array}{l} x \= \ln\rho~, \\ y \= \phi~.\end{array}\ra
\label{eq:CUR33}
\eeq

Equation \calle{eq:CUR33} says that an arc $\phi_1\le\phi\le\phi_2$
of the circle $\rho=a$ on the $w$-plane is mapped to a segment
on the line $x=\ln a$. The segment starts at $y=\phi_1$ and ends at
$y=\phi_2$. Also, a segment $a\le\rho\le b$ lying on the ray $\phi=\phi_1$
is mapped to a segment on the line $y=\phi_1$. The segment starts at 
$x=\ln a$ and ends at $x=\ln b$. All this together means that any rectangle
in the $z$-plane becomes an angular sector in the $w$-plane.

\begin{minipage}[c]{6.5cm}

\vspace{4mm}
\psfrag{x}{$x$}
\psfrag{y}{$y$}
\psfrag{u=lna}{$u=\ln a$}
\psfrag{u=lnb}{$u=\ln b$}
\psfrag{f1}{$\phi_1$}
\psfrag{f2}{$-\phi_2$}
\includegraphics[width=7.5cm]{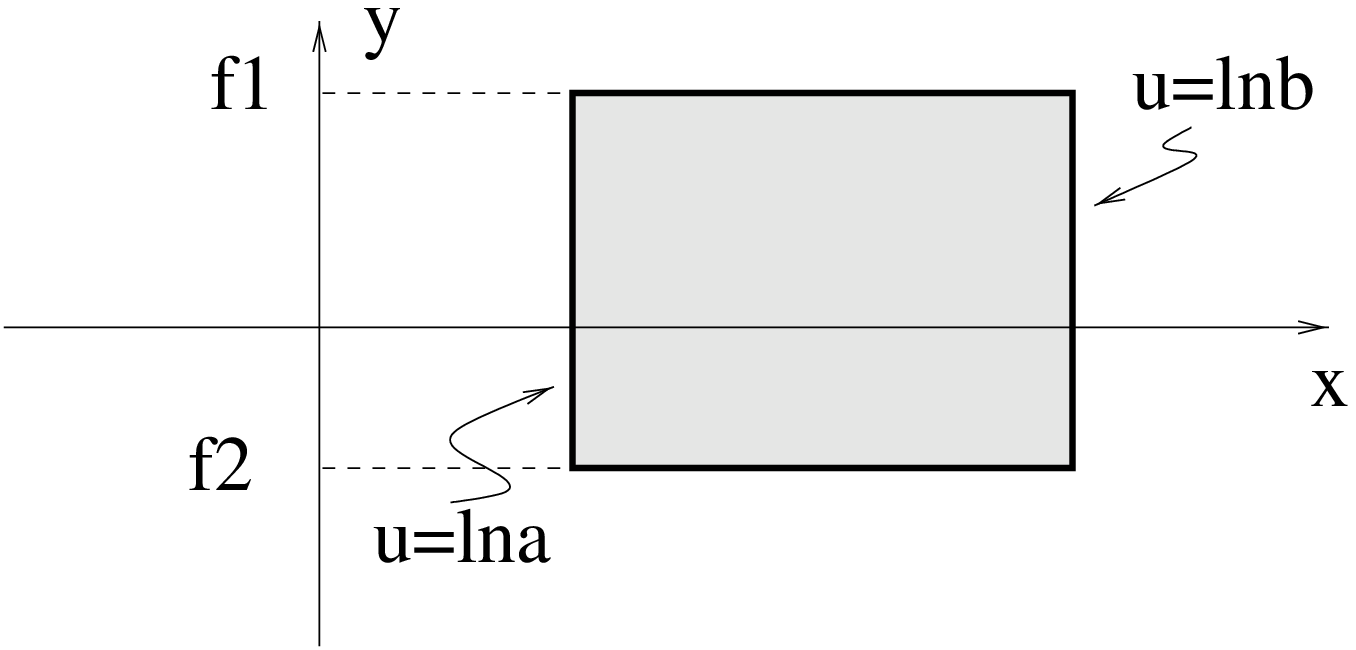}
\end{minipage}
\hspace{1cm}
\begin{minipage}[c]{6.5cm}
\psfrag{x}{$u$}
\psfrag{y}{$v$}
\psfrag{a}{$a$}
\psfrag{b}{$b$}
\psfrag{f1}{$\phi_1$}
\psfrag{f2}{$-\phi_2$}
\includegraphics[width=6.5cm]{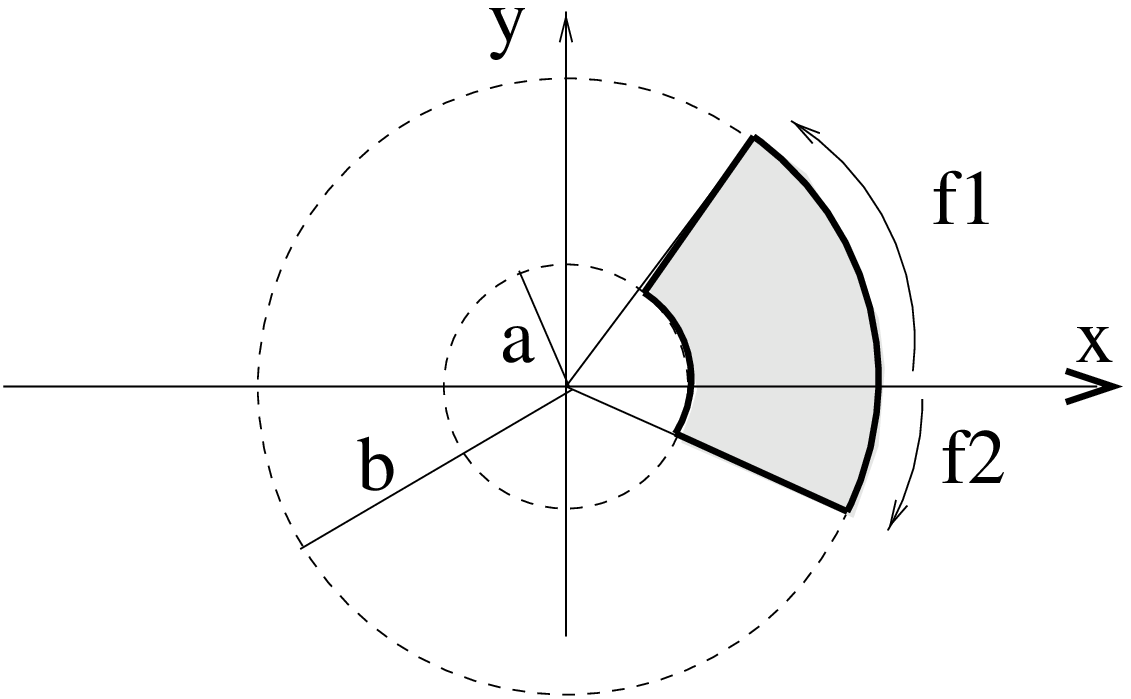} 
\end{minipage}

An important theorem in conformal mapping is the 
{\bfseries Riemann mapping theorem}:

\vspace{3mm}
{\bfseries Theorem} [Riemann] \\
{\small Given two arbitrary simply-connected
domains $D$ and $D'$ whose boundaries contain more than one point,
there always exists an analytic function $w=f(z)$ 
that maps $D$ onto $D'$. The function $f(z)$ depends on three real parameters.
}

\vspace{3mm}
Usually $D'$ is taken to be
the interior of the unit disc. We can restate the
Riemann mapping theorem in this case as:

\vspace{3mm} 
{\bfseries Theorem}\\
{\small Any simply-connected domain $D$ whose boundary 
contains more than one point can be mapped conformally onto the interior
of the unit disc. Moreover, it is possible to choose 
any one point of $D$ and
a direction through this point to 
be mapped to the origin and a direction through
the origin; these choices render the map unique.
}

\vspace{3mm} 
The Riemann mapping theorem, upon careful examination, can be extended
to the case that $D$ is bounded between two closed simple curves, one 
inside the other. In this case, the map is to an annulus.

We can employ similar considerations for various conformal maps.
We tabulate some
results to be used in later sections in table \ref{table:CUR1}.

\begin{table}[p]
\begin{center}
\begin{tabular}{|c|c|}\hline
\multicolumn{2}{|c|}{semi-infinite strip to half-plane~~~ 
$w=\cosh{\pi z\over a}$} \\ \hline
\psfrag{x}{$x$} \psfrag{y}{$y$}\psfrag{a}{$a$}
\psfrag{A}{$A$} \psfrag{B}{$B$} \psfrag{C}{$C$} \psfrag{D}{$D$} \psfrag{E}{$E$}
\includegraphics[width=6.5cm]{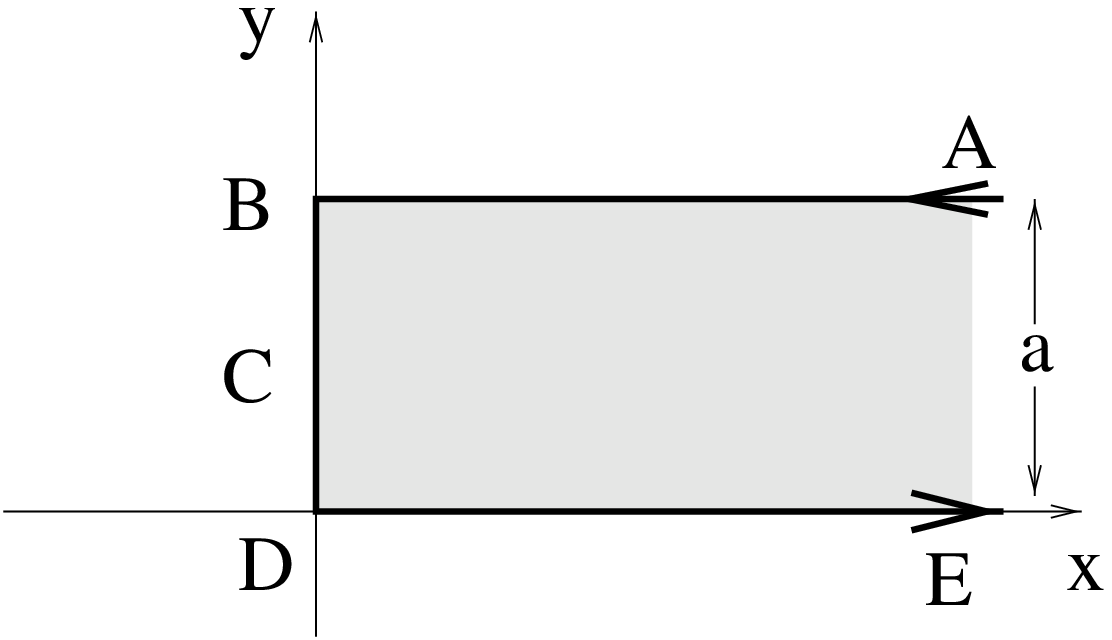}  &
\psfrag{x}{$u$} \psfrag{y}{$v$}
\psfrag{-1}{$-1$} \psfrag{+1}{$+1$}
\psfrag{A}{$A'$}\psfrag{B}{$B'$}\psfrag{C}{$C'$}\psfrag{D}{$D'$}\psfrag{E}{$E'$}
\includegraphics[width=6.5cm]{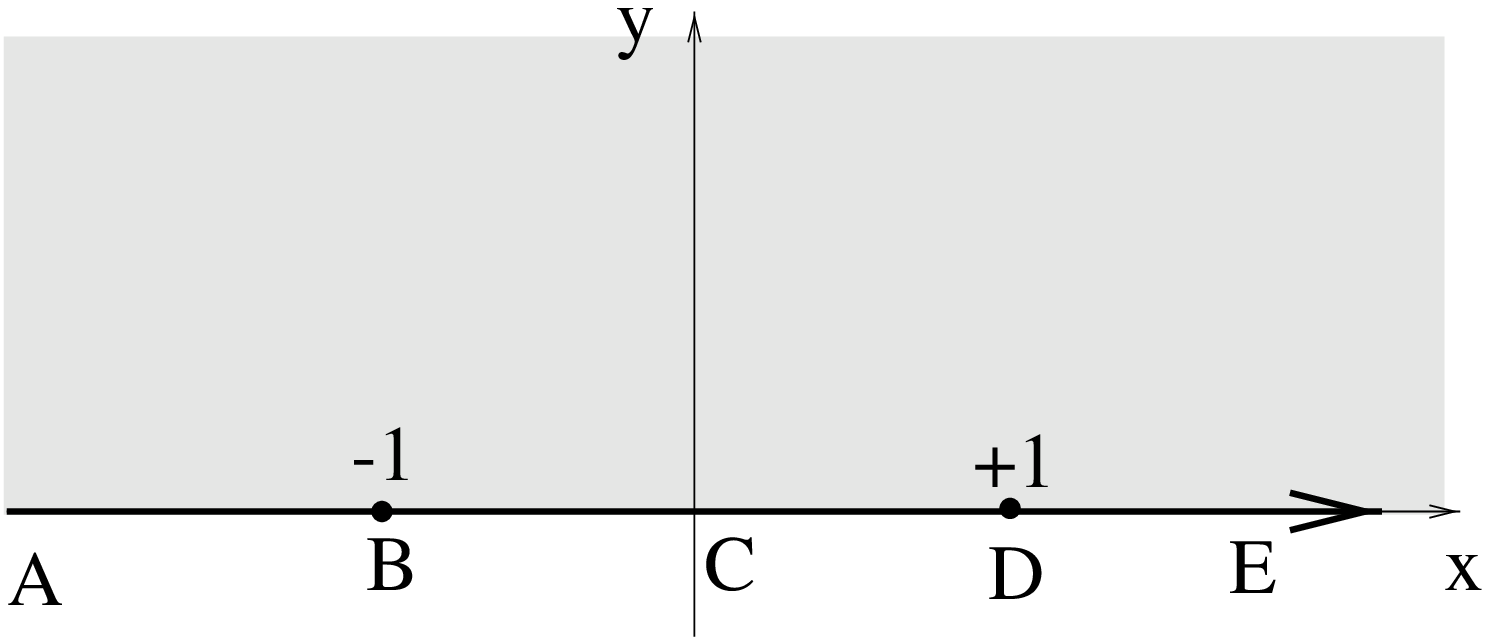}  \\ \hline
\multicolumn{2}{|c|}{infinite strip to half-plane~~~
$w=e^{\pi z\over a}$} \\ \hline
\psfrag{A}{$A$} \psfrag{B}{$B$} \psfrag{C}{$C$} \psfrag{D}{$D$} \psfrag{E}{$E$}
\psfrag{a}{$a$} \psfrag{x}{$x$} \psfrag{y}{$y$}
\includegraphics[width=6.5cm]{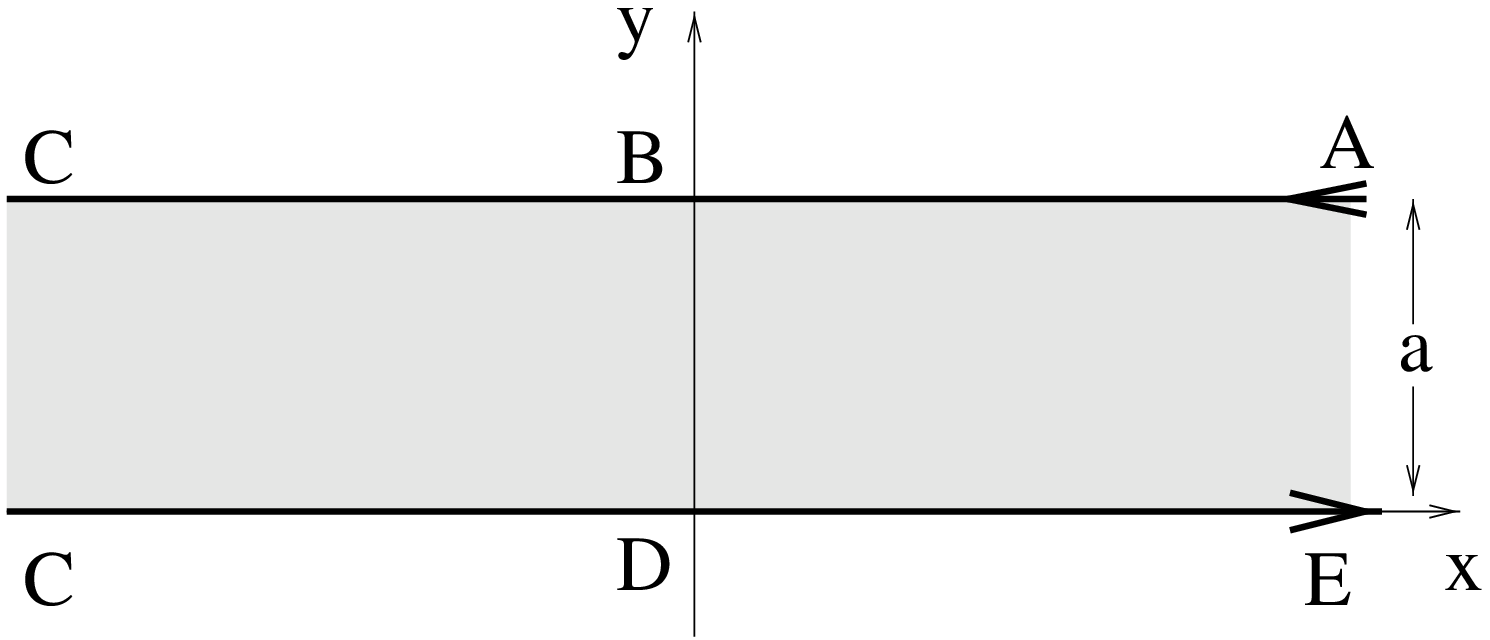}  &
\psfrag{x}{$u$} \psfrag{y}{$v$}
\psfrag{-1}{$-1$} \psfrag{+1}{$+1$}
\psfrag{A}{$A'$}\psfrag{B}{$B'$}\psfrag{C}{$C'$}\psfrag{D}{$D'$}\psfrag{E}{$E'$}
\includegraphics[width=6.5cm]{figures/conf1b.eps}  \\ \hline
\multicolumn{2}{|c|}{infinite wedge to half-plane~~~
$w=z^\alpha~,~~~\alpha\ge1/2$} \\ \hline
\psfrag{x}{$x$} \psfrag{y}{$y$}
\psfrag{f}{$\pi/\alpha$} \psfrag{+1}{$+1$}
\psfrag{A}{$A$}\psfrag{B}{$B$}\psfrag{C}{$C$}\psfrag{D}{$D$}\psfrag{E}{$E$}
\includegraphics[width=6.5cm]{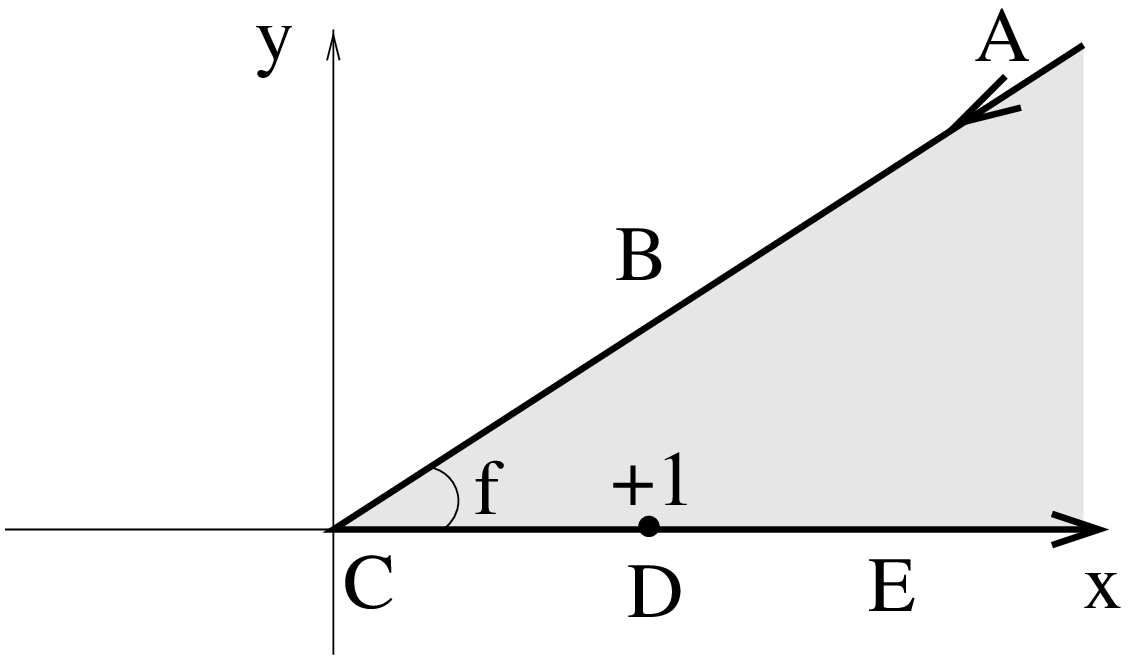} &
\psfrag{x}{$u$} \psfrag{y}{$v$}
\psfrag{+1}{$+1$}
\psfrag{A}{$A'$}\psfrag{B}{$B'$}\psfrag{C}{$C'$}\psfrag{D}{$D'$}\psfrag{E}{$E'$}
\includegraphics[width=6.5cm]{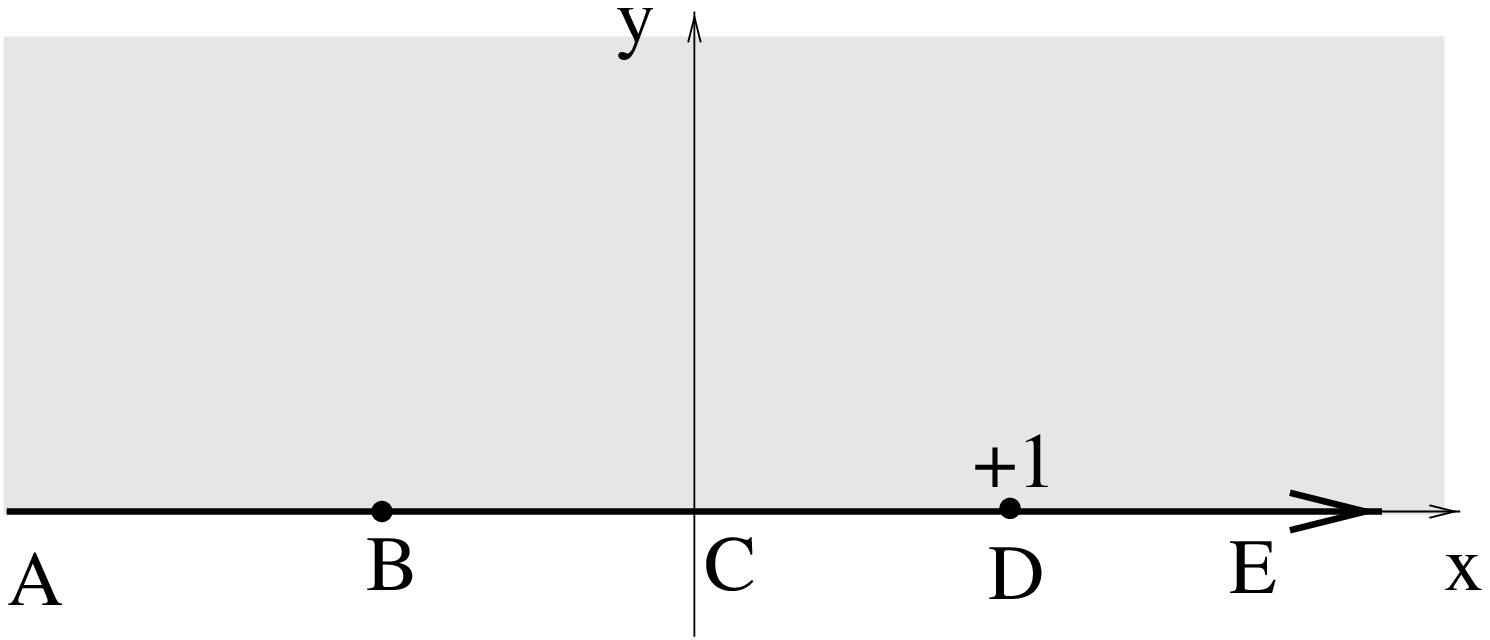}  \\ \hline
\multicolumn{2}{|c|}{angular sector to rectangle~~~
$w=\ln z$}  \\ \hline
\psfrag{A}{$A$}\psfrag{E}{$E$}\psfrag{C}{$C$}
\psfrag{F}{$F$}\psfrag{H}{$H$}\psfrag{G}{$G$}
\psfrag{x}{$x$}
\psfrag{y}{$y$}
\psfrag{a}{$a$}
\psfrag{b}{$b$}
\includegraphics[width=6.5cm]{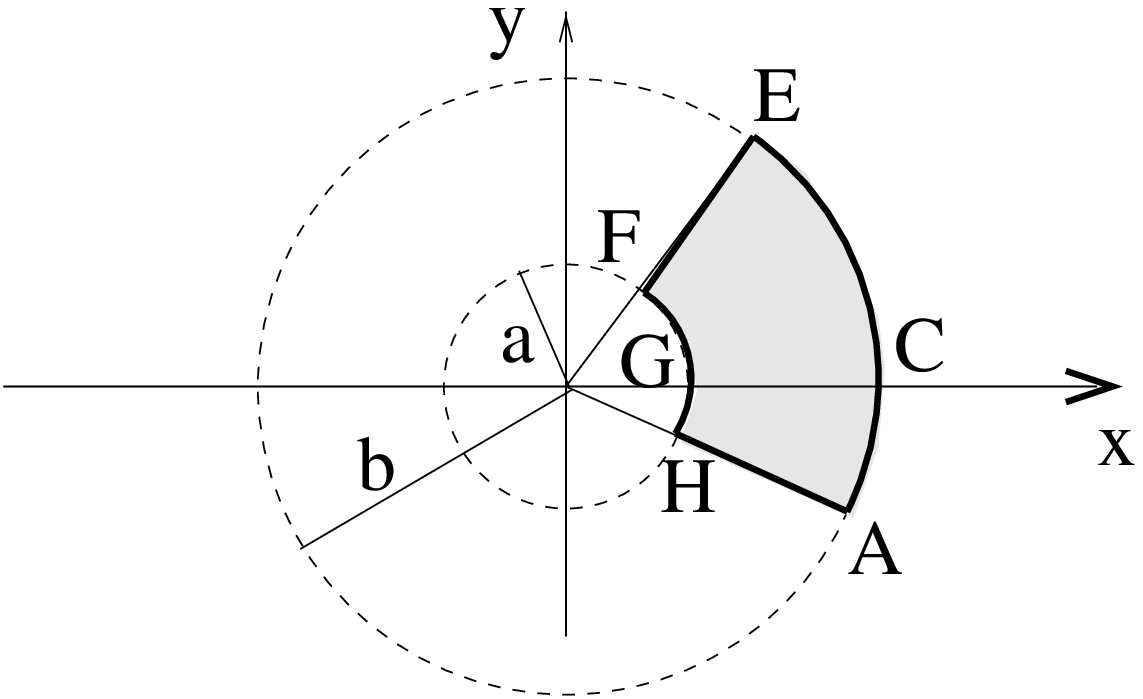} &
\psfrag{A}{$A'$}\psfrag{E}{$E'$}\psfrag{C}{$C'$}
\psfrag{F}{$F'$}\psfrag{H}{$H'$}\psfrag{G}{$G'$}
\psfrag{x}{$u$}
\psfrag{y}{$v$}
\psfrag{u=lna}{$u=\ln a$}
\psfrag{u=lnb}{$u=\ln b$}
\includegraphics[width=6.5cm]{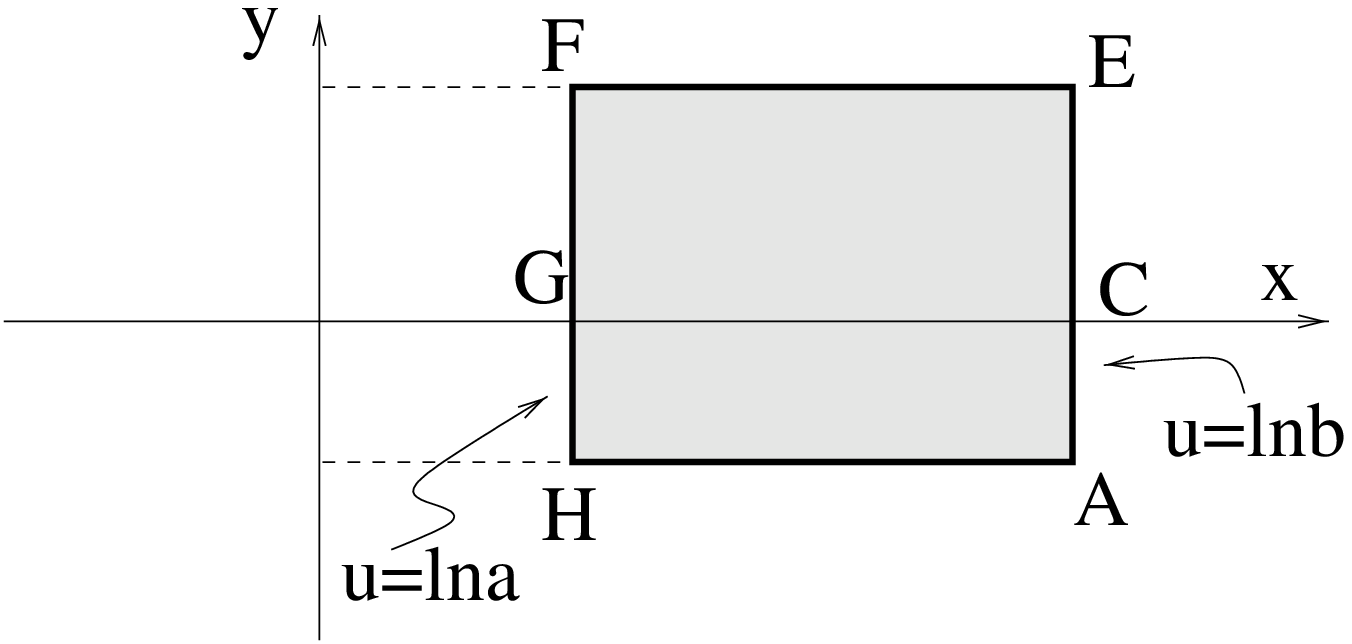}  \\ \hline
\end{tabular}
\end{center}
\caption{Some commonly used conformal transformations.}
\label{table:CUR1}
\end{table}

In the last entry of the table \ref{table:CUR1}, imagine that we take 
$\phi_1=\pi$ and $\phi_2=-\pi$. In this case, the angular sector becomes
a ring, i.e., a domain that is not simply-connected. Its image on the
$w$-plane appears to be a simply-connected rectangle. 

\begin{minipage}[c]{6.5cm}
\psfrag{A}{$A$}\psfrag{E}{$E$}
\psfrag{F}{$F$}\psfrag{H}{$H$}
\psfrag{x}{$x$}
\psfrag{y}{$y$}
\psfrag{a}{$a$}
\psfrag{b}{$b$}
\includegraphics[width=6.5cm]{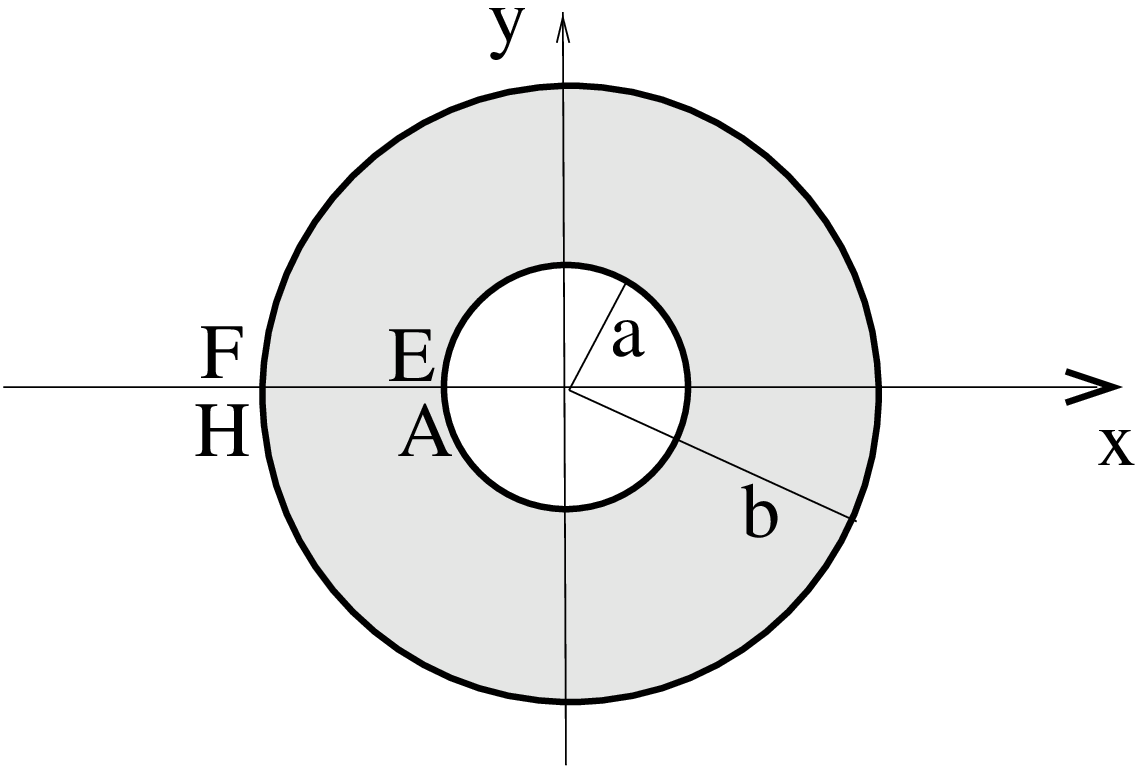}
\end{minipage}
\hspace{2mm}
\begin{minipage}[c]{6.5cm}
\psfrag{p}{$\pi$}
\psfrag{-p}{$-\pi$}
\psfrag{x}{$u$}
\psfrag{y}{$v$}
\psfrag{A}{$A$}\psfrag{E}{$E$}
\psfrag{F}{$F$}\psfrag{H}{$H$}
\includegraphics[width=7.5cm]{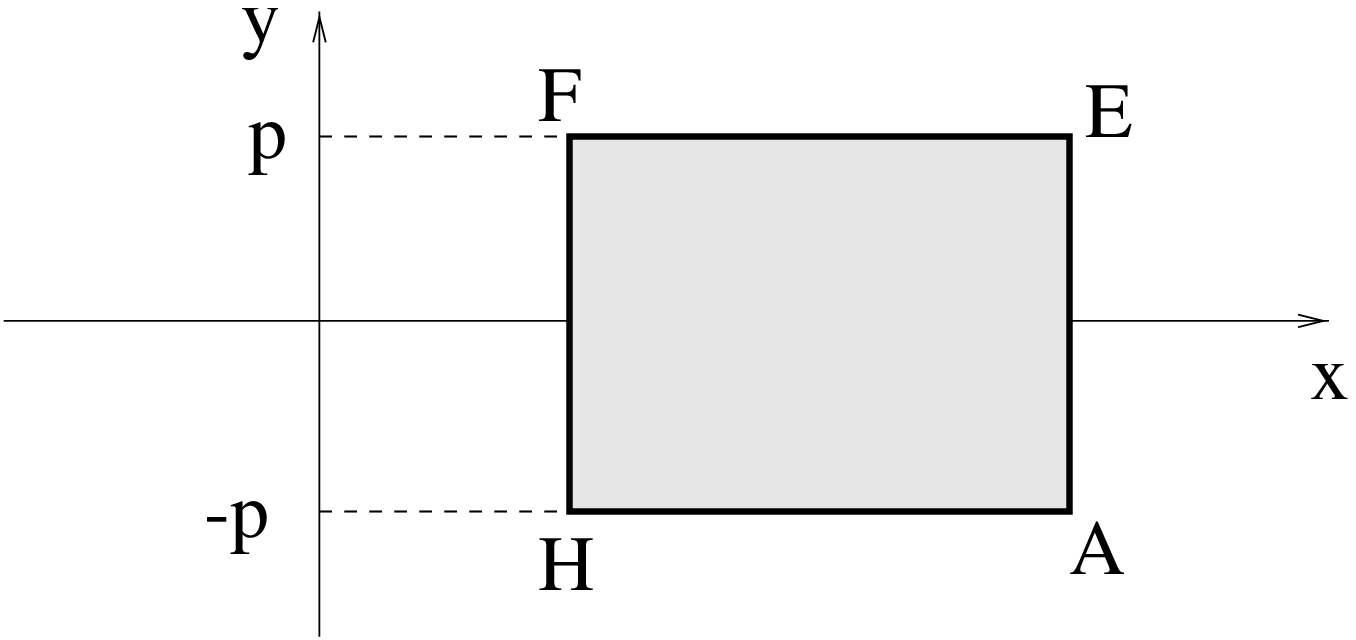}
\end{minipage}

\noindent This seems to contradict
the Riemann mapping theorem. However, one must notice that the
edges $FE$ and $HA$ coincide now on the $z$-plane and therefore,
under the mapping, they should  be identified on the $w$-plane too. This
means that  the rectangle is really a cylinder which is also not a
simply-connected domain.

\begin{center}
\psfrag{A}{$A$}\psfrag{E}{$E$}
\psfrag{F}{$F$}\psfrag{H}{$H$}
\includegraphics[width=11cm]{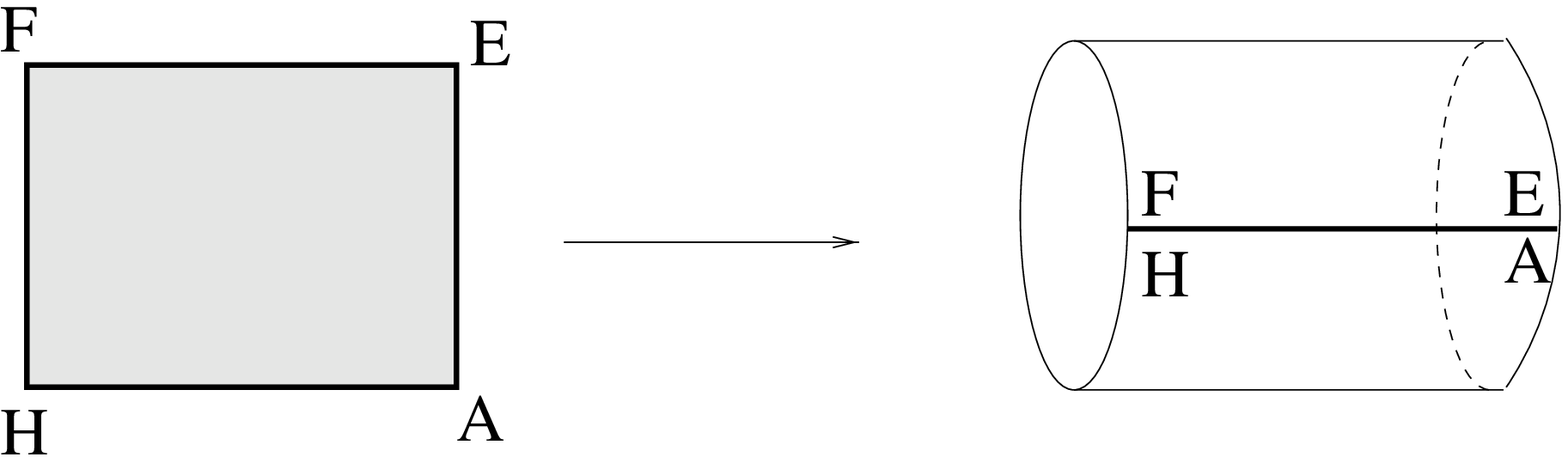}
\end{center}

A final comment is in order here. In applications of the conformal
transformations, special care must be given to all quantities and constants
in the transformation. Sometimes, `innocent' factors are manifestations
of a very subtle situation! For example, in Table \ref{table:CUR2}
we give two transformations that differ only by an
`innocent' factor of 2; nonetheless (see the second and fourth entries in 
Table \ref{table:CUR1}), they are quite different!

\begin{table}
\begin{center}
\begin{tabular}{|c|c|}\hline
plane to cylinder & $w={a\over2\pi}\,\ln z$ \\ \hline
upper half-plane to infinite strip
  & $w={a\over\pi}\,\ln z$ \\ 
{\footnotesize without periodic boundary conditions on the edges} & \\ \hline
\end{tabular}
\caption{Two different conformal transformations.}
\label{table:CUR2}
\end{center}
\end{table}

\subsection{Symmetries of the S-Matrix}

One of the goals  of the machinery of QFT is to devise methods
for determining the S-matrix --- i.e., the array of probabilities for all
possible events --- and to carry out the program and calculate the
S-matrix in particular theories, especially those that
are applicable to the real world.  Clearly, the symmetries of
the S-matrix constrain its possible form.  
(In the Lagrangian framework, all symmetries of the
Lagrangian that are not spontaneously broken appear as symmetries of
the S-matrix.)  Hence, knowing the possible symmetries of the
S-matrix is both useful and informative.

For a QFT theory defined on a flat manifold, one always assumes that
the Poincar\'e group $\CP$ is a symmetry of the S-matrix. 
This is a `must' in the construction of any physical model.
The full symmetry group of the S-matrix must thus include $\CP$ as a
subgroup.  Attempts to find physical symmetry groups that included
$\CP$ in a non-trivial way led ultimately to a no-go theorem known
as the Coleman-Mandula theorem. Under very
mild conditions, Coleman and Mandula proved that that there can be no
exotic space-time symmetries of the S-matrix. 
More exactly, the statement of the theorem is as 
follows \cite{CM}:

\vspace{3mm}
{\bf Theorem} \lbrack Coleman and Mandula\rbrack\\
{\small
Let G be a connected symmetry (Lie) group of the S-matrix, and let
the following five conditions hold:
\begin{enumerate}
\item (Lorentz invariance) G contains a subgroup locally isomorphic to $\CP$.
\item (Particle finiteness) All particle types correspond to positive energy 
      representations of $\CP$. For any finite $M$, there is only a finite number
      of particle types with mass less than $M$.
\item (Weak elastic analyticity) Elastic scattering amplitudes are analytic
      functions of the center-of-mass energy $s$ and invariant momentum
      transfer $t$, in some neighborhood of the physical region, except at
      normal thresholds.
\item (Occurrence of scattering) Let $\ket{p}$ and $\ket{p'}$ be any two
      1-particle momentum eigenstates, and let $\ket{p,p'}$ be the 2-particle 
      state made from these. Then
      $$
          T\ket{p,p'}\ne0~,~~~~~S=1-i(2\pi)^4\delta^4(P-P')\,T~,
      $$
      except perhaps for certain isolated values of $s$. Phrased physically,
      this assumption is
      that at almost all energies, any two plane waves scatter.
\item (An ugly technical  assumption) The kernels of
      the generators of G, considered as
      integral operators in momentum space, are distributions. 
      More precisely: There is a neighborhood of the identity in G such that
      every element in G in this neighborhood lies on some 1-parameter group
      $g(t)$. Further, if $x$ and $y$ are any two states in the the set
      $\CD$ of all 1-particle states whose momentum space wave functions
      are test functions, then
      $$
        -i{d\over dt} \average{x|g(t)|y}\=\average{x|A|y}
      $$
      exists at $t=0$, and  defines a continuous function of $x$ and $y$, linear
      in $y$ and antilinear in $x$.
\end{enumerate} 
Then as long as the S-matrix is not trivial (i.e., as long
as the S-matrix is not that of a set of non-interacting
objects),
G is locally isomorphic to the 
direct product
of an internal symmetry group $\CI$ and the
 the Poincar\'e group:
$$
   \mbox{G}\= \CI\otimes \CP~.
$$ 
}

\vspace{3mm}
A corollary of the Coleman-Mandula theorem is that the generators $I_a$ of $\CI$
must commute with the momentum generators $P_\mu$ and 
the angular momentum generators $J_{\mu\nu}$.
$$
  \lb I_a, J_{\mu\nu} \rb=
  \lb I_a, P_\mu \rb=0~.
$$
Therefore, if we classify the particles in families according to 
representations of
$\CI$, then all particles in the same representation
must have the 
\textit{same mass} and the
\textit{same spin}, since 
$$
  \lb I_a, P^2 \rb=
  \lb I_a, W^2 \rb=0~,
$$
where $W^\mu$ is the Pauli-Lubarski vector
$$
  W_\mu=-{1\over2}
  \varepsilon_{\mu\nu\rho\sigma}J^{\nu\rho}{P^\sigma\over\sqrt{P^2}}~.
$$

There are three important
loopholes in this theorem, two of which were known from
the time the Coleman-Mandula theorem was proven. 

\begin{itemize}
\item
  If assumption 2 is relaxed, the space-time symmetry group
  can be extended to the conformal group, which includes $\CP$ as a subgroup.
  However, if the other assumptions are preserved, the final conclusion
  is almost the same, with the Poincar\'e group replaced by the conformal
  group, as Coleman and Mandula discussed in their original article.
\item
  The theorem makes use of Lie groups, i.e., groups whose algebra is based on
   commutation relations of the form
   $$
     \lb T_i, T_j \rb \= c_{ij}^k\, T_k~.
   $$
  Once  one assumes that generalized commutation relations are possible
  for the generators, the theorem does not apply. In particular, supersymmetry
  (SUSY) includes commutation and anticommutation relations in its algebra,
   \bb
     \lb E_a, E_b \rb &=& f_{ab}^c\, E_c~,\\
     \lb E_a, O_i \rb &=& m_{ai}^j\, O_j~,\\
     \la O_i, O_j \ra &=& c_{ij}^k\, E_k~.
    \ee
  One can re-work the analysis of the Coleman-Mandula theorem in the case
  that  Lie groups and supergroups are allowed \cite{HLS}.
  The results are exactly analogous to those
  of the original theorem, with the conclusion being that the most general
  possible supersymmetry of the S-matrix is as before, except with the
  Poincar\'e (conformal) group replaced by the super-Poincar\'e (superconformal)
  group.
\item
  The theorem assumes that the S-matrix is an analytic function of at least
  two Mandelstam variables $s$ and $t$. However, this can be true only in $d+1$
  space-time dimensions,
  with $d\ge 2$. In $1+1$ dimensions, there is only one Mandelstam
  variable, and therefore this assumption is not valid.  Stated physically,
  in $1+1$ dimensions, there can only be forward or backward scattering,
  and hence there is no prospect of analyticity in the scattering angle.
  Indeed, this is not just a technical observation.  In $1+1$ dimensions,
  theories exist which have a non-trivial S-matrix \textit{and} which also have 
  extended space-time Lie symmetries that contain the Poincar\'e group in
  a non-trivial way.
\end{itemize}

\newpage
\section{EXERCISES}

\begin{enumerate}

\item
Consider Maxwell's equations in $d+1$ dimensions. What is the
electrostatic energy between two charge distributions
$\rho_1(\vec r)$ and $\rho_2(\vec r)$?

\item
Show that the configuration space $\CC_2$ of two
identical particles in $d$
 spatial dimensions is the manifold
$$
    \CC_2 \= \BR^{d+1}\times\BR P^{d-1}~.
$$
Then prove that
$$
   \pi_1(\CC_2)\=\cases{\BZ~, &if $d=2~$\cr
                      \BZ_2~,&if $d\ge3~,$\cr}
$$
and from this deduce that in $d\ge 3$ only fermions and bosons can exist,
while in $d=2$ particles with different statistics are allowed.

\item
(a) Is the posssibility of assigning many spins to the same state in two
    dimensions a new phenomenon compared to four dimensions?

(b) The map \calle{eq:CUR1} can be be defined (in an analogous way) in any
    number of dimensions. Why do we then claim that statistics is a
    matter of convention only in two dimensions?

\item
(a) {\bf Green's theorem}\index{theorem!Green's --}
 on the plane states: If $P$ and $Q$ are continuous 
 functions with
continuous partial derivatives on the domain $D$, then
$$
  \iint_D\lp {\partial Q\over\partial x} - {\partial P\over\partial y}\rp
  \, dxdy\= \ointleft_{\partial D} \lp Pdx+Qdy\rp~.
$$
Rewrite this theorem using complex variables.

(b) Use the result of part (a) to find a representation of the
   $\delta$-function in complex coordinates.

\item
Trying to extend  the Riemann mapping theorem in the case of 
multiply-connected domains, we face the problem that domains of the
same order of connectivity are not necessarily conformally mapped onto
each other \cite{Nehari}. This exercise studies this issue in one simple case.

Consider the circular rings $D=\{ 1<|z|<r\}$ and $D'=\{ 1<|w|<R\}$,
where $r\ne R$. Assume that there is a function $w=f(z)$ which maps
$D$ onto $D'$. Show that this would require $r=R$, which is contrary to
the hypothesis  $r\ne R$.

\item
\label{item:CUR1}
Show that all the 2-dimensional Riemannian manifolds are conformally flat.

\item
Find all Killing vectors of an $n$-dimensional flat manifold.
Compute the algebra they generate (which is
the {\bfseries Poincar\'e algebra}).

\item

Write down the classical Poincar\'e algebra (i.e., using Poisson brackets)
in $1+1$ dimensions, and
find a representation  for a system of $N$
particles interacting via a nearest-neighbor potential \cite{Ruij}.

\item
Equation \calle{eq:CUR6} has been derived for a general metric on
a manifold $M$. Show that if the metric is a Levi-Civita metric,
then the Killing equation  can be brought to the more symmetric
form
\beq
\label{killing3}
    \xi_{\mu,\nu}+\xi_{\nu,\mu}\=2\, \christoffel{\rho}{\mu\nu}\,\xi_\rho~.
\eeq

\item
Using equation \calle{killing3}, find all Killing vectors of the
2-sphere $S^2$ with the standard metric
$$
  \bfg\=\sin^2\theta\,d\phi\otimes d\phi+d\theta\otimes d\theta~.
$$
 Compute the algebra they generate.

\item
Using equation \calle{killing3}, find all Killing vectors of the
Poincar\'e half-plane $H^2$, for which the metric is
$$
  \bfg\= {dx\otimes dx+dy\otimes dy\over y^2}~.
$$
Compute the algebra they generate,
and integrate it to find the corresponding group.

\item
Find all conformal Killing vectors of an $n$-dimensional conformally
 flat manifold.
Compute the algebra they generate.
\item
Recall from Exercise \ref{item:CUR1} that a 2-dimensional manifold is
conformally flat. Thus the result  
of the last problem applies locally in any patch of a
2-dimensional manifold.
From that result, find the global conformal algebra
on the 2-sphere $S^2$.

\item
{\bf Mandelstam variables}: Show that in $D$ space-time dimensions,
for the scattering of $N$ particles (incoming and outgoing all counted),
 we can define
$(D-1)\,N\,-\,{D(D+1) \over 2}$
independent Mandelstam variables. From this result, confirm that in
two space-time dimensions, the 2-particle scattering is determined by one
Mandelstam variable.

\item
{\bfseries A flavor of the Coleman-Mandula theorem}: For a hypothetical model,
assume the existence of a second order tensorial conserved
charge $Q_{\mu\nu}$ different from the angular momentum $J_{\mu\nu}$.
Study the consequences of such
an additional conservation law for
the $n$-to-$n$ elastic scattering process
$$
   a_1(p_1^i) + a_2(p_2^i) + \cdots a_n(p^i_n) \longrightarrow
   a_1(p_1^f) + a_2(p_2^f) + \cdots a_n(p^f_n),
$$
and show that only in 1+1 dimensions is there 
no conflict between this conservation law
and the well-known analyticity
properties of the S-matrix .

\end{enumerate}

\newpage
\section{SOLUTIONS}

\begin{enumerate}

\item
Consier a point charge $q$ in $d$ spatial dimensions.
We can imagine surrounding the charge
by a $(d-1)$-sphere $S^{d-1}$. Then 
Gauss's Law in integral form for electric field
$\bfE$ of the charge $q$ reads
$$
   \oint_{S^{d-1}}  \bfE\cdot d\bfS\= {q\over\varepsilon_0}~. 
$$

The rotational symmetry of the configuration
requires that $\bfE$ be a function of the radial distance
$r$ only, and that it point along the radial direction. Therefore
\bb
    {q\over\varepsilon_0}
           &=&  E(r)\, \oint_{S^{d-1}} dS \\
           &=&  E(r)\, \Omega_{d-1}\, r^{d-1}~,
\ee
where $E$ is the magnitude of the electric field,
and $\Omega_{d-1}$ is the $(d-1)$-dimensional total solid angle,
which is easily computed to be
$$
   \Omega_{d-1}\= {2\pi^{d/2}\over\Gamma(d/2)}~.
$$
Therefore 
$$
    E(r)\= {\Gamma(d/2)\over 2\pi^{d/2-1}\varepsilon_0}\, {q\over r^{d-1}}~.
$$ 
The potential is then given by
$$
    \phi(r) \=
 \cases{-{1\over 2\varepsilon_0}\,q\,r~, &if$~d=1~,$\cr
       -{1\over 2\pi\varepsilon_0}\,q\,\ln r~, &if$~d=2~,$\cr
    -{\Gamma(d/2)\over 4\pi^{d/2}\varepsilon_0}\,{q\over r^{d-2}}~, &if$~d>2~.$\cr}
$$
Consequently, the interaction between the two charge distributions
$\rho_1(\vec r)$ and $\rho_2(\vec r)$ is
\bb
   U\=
 \cases{
 -{1\over 2\varepsilon_0}\,
  \int dr_1dr_2\, \rho_1(r_1)|r_1-r_2|\rho_2(r_2)~, &if$~d=1~,$\cr
 -{1\over 2\pi\varepsilon_0}\,
  \int d^2r_1d^2r_2\, \rho_1(r_1)\ln|r_1-r_2|\rho_2(r_2)~, &if$~d=2~,$\cr
 -{\Gamma(d/2)\over 4\pi^{d/2}\varepsilon_0}\,
  \int d^dr_1d^dr_2\, \rho_1(r_1){1\over r^{d-2}}\rho_2(r_2)~, &if$~d>2~.$\cr}
\ee         

\separator

\item

In $d$ dimensions, the locus of particles is described by their
position vectors $(\bfr_1,\bfr_2)$, with the restriction
$\bfr_1\ne\bfr_2$. As usual, we define the center of mass 
position vector $\bfR$ and the relative position vector $\bfr$
by
$$
  \bfR\={\bfr_1+\bfr_2\over 2}~, ~~~~~
  \bfr\=\bfr_1-\bfr_2~.
$$
In fact, the vector $\bfr$ can furthermore be decomposed into its
magnitude $r$ and a unit vector $\bfu_r$ parallel $\bfr$.
If the particles were distinguishable, the configuration space would
be
\bb
   \overline{\CC}_2 &=&
   \{ (\bfR,\bfr) ~|~ \bfR,\bfr\in\BR^d,~\bfr\ne 0 \} \\ 
   &=& \{(\bfR,\bfu, r) ~|~ \bfR\in\BR^d,~\bfu\in S^{d-1},r\in\BR^*_+\}\\
   &=& \BR^d\times S^{d-1} \times \BR^*_+ \\
   &\sim& \BR^d\times S^{d-1} \times \BR \\
   &=& \BR^{d+1} \times S^{d-1}~.
\ee
For indistinguishable particles,
we must divide by their permutation group
$S_2=\BZ_2$ to avoid double counting.  Thus the
configuration space for two identical particles is
\bb
   \CC_2 &=& { \overline{\CC}_2\over\BZ_2}\=
        \BR^{d+1} \times {S^{d-1}\over\BZ_2} = 
         \BR^{d+1} \times\BR P^{d-1}~,
\ee
since the manifold $S^{d-1}/\BZ_2$ is $\BR P^{d-1}$,
the $(d-1)$-dimensional real projective space. The first two 
such spaces
are $\BR P^1=S^1$ and $\BR P^2\=$SO$(3)$.

Since
$$
   \pi_1(M_1\times M_2)\= \pi_1(M_1)\times \pi_1(M_2)~,
$$
to find the first homotopy group of $\CC_2$, we  notice that
any non-trivial contribution will come from $\BR P^{d-1}$, since 
$\BR^n$ is homotopically trivial. For $d=2$, we thus immediately find that
$\pi_1(\CC_2)=\pi_1(\BZ)=\BZ$. For $d\ge 3$, we must find  
$\pi_1(S^{d-1}/\BZ_2)$. According to a theorem of algebraic topology,
if a manifold $M$ is homotopically trivial, and a discrete group $\Gamma$
acts on $M$ effectively, then $\pi_1(M/\Gamma)=\Gamma$. In our case,
$\pi_1(S^{d-1})=0$ (you cannot lasso the sphere) and $\BZ_2$ acts
effectively on it. Therefore $\pi_1(\BR P^{d-1})=\BZ_2$.

From \calle{eq:CUR2}, we see that the weights $\chi$ are unitary 
1-dimensional representations of $\BZ$ for $d=2$ and of
$\BZ_2$ for $d\ge 3$. Now, $\BZ_2$ has only two such representations,
namely $1$ and $-1$. The first corresponds to bosons and the second to fermions.
On the other hand, $\BZ$ allows for  a continuum of choices parametrized by
a variable $\theta$, with $\chi(n)\mapsto e^{in\theta}$. When $\theta=0$
the statistics are bosonic, and when $\theta=\pi$ the statistics are 
fermionic, while other values of $\theta$ lead to exotic statistics.

\separator

\item
(a) The possibility of assigning many spins to the same state is known in
    four dimensions. A free particle of spin $s$ can be described 
    equivalently by many relativistic wave equations, with fields
    which transform differently
    under the Lorentz group.

(b) In $d$ spatial dimensions, one can certainly define the map
\bb
   f^\dagger(p) &=& b^\dagger(p)\,
  e^{-i\pi\int {d^dk_1\over k_0} b^\dagger(k)b(k)}~,\\
   f(p) &=&
  e^{i\pi\int {d^dk\over k_0} b^\dagger(k)b(k)}
   \, b(p)~.
\ee
However, this map is Lorentz invariant only in two dimensions.   

\separator

\item
Let $A(z,\overline z)$ be a complex function 	
$$
   A(z,\overline z)\= P(x,y)+iQ(x,y)
$$
of $z=x+iy$. Then
\bb
  \ointleft_{\partial D} A(z,\overline z) dz &=&
  \ointleft_{\partial D} (P+iQ)\,(dx+idy) \\
  &=& \ointleft_{\partial D} (Pdx-Qdy)+i\ointleft_{\partial D}(Qdx+Pdy)\\
  &=& -\iint_D\lp {\partial Q\over\partial x}-{\partial P\over\partial y}\rp
  \, dxdy
  +i\iint_D\lp {\partial P\over\partial x}+{\partial Q\over\partial y}\rp
  \, dxdy \\
  &=& -\iint_D\lp {\partial Q\over\partial x}+i{\partial Q\over\partial y}\rp
  \, dxdy
  +i\iint_D\lp {\partial P\over\partial x}+i{\partial P\over\partial y}\rp
  \, dxdy \\
  &=& -\iint_D 2 {\partial Q\over\partial \overline z}\, dxdy
  +i \iint_D 2 {\partial P\over\partial \overline z}\, dxdy \\
   &=& 2i \iint_D {\partial \over\partial \overline z}(P+iQ)\, dxdy\\
   &=& 2i \iint_D {\partial A\over\partial \overline z}\, dxdy~.
\ee
Finally, since
$$
 dxdy\= \left| {\partial(x,y)\over\partial(z,\overline z)}\right|
        \, dz d\overline z
     \= {dz d\overline z\over |-2i|}
     \= {dz d\overline z\over 2}~,
$$
we write
\beq
\label{eq:CUR81}
   \ointleft_{\partial D} A(z,\overline z) {dz\over i}\= 
   \iint_D {\partial A\over\partial \overline z}\, dz d\overline z~.
\eeq

In the same way, we can prove
$$
    \ointleft_{\partial D} B(z,\overline z) {d\overline z\over i}\=
    -\iint_D {\partial B\over\partial  z}\, dz d\overline z~.
$$
Putting the two results together,
$$
  \ointleft_{\partial D} A(z,\overline z) {dz\over i} 
  +B(z,\overline z) {d\overline z\over i}\=
  \iint_D \lp {\partial B\over\partial  z} +
  {\partial A\over\partial\overline z}\rp \,  dz d\overline z~.
$$

(b) Let $A(z)$ be a holomorphic function. Any representation of the
    $\delta$-function must satisfy the equation
$$
    A(x,y)\=\iint_D \delta(x-x')\delta(y-y')\, A(x',y')\, dxdy~,
$$
or
$$
    A(w)\=\iint_D \delta^{(2)}(z-w)\, A(z)\,{dzd\overline z\over 2}~.
$$
We will show that this is the case for the representation
\beq
\label{eq:CUR100}
   \delta^{(2)}(z-w)\={1\over 2\pi}\, \overline\partial{1\over z-w}~.
\eeq
Indeed
\bb
  \iint_D \delta^{(2)}(z-w)\, A(z)\, dz d\overline z &=&
  {1\over 2\pi}\,\iint_D \overline\partial{ A(z)\over z-w}
  \, dz d\overline z \\
  &=& \ointleft_{\partial D} {dz\over 2\pi i}\, { A(z)\over z-w} \\
  &=&  A(w)~.
\ee

In the same way
\bb
  \delta^{(2)}(z-w)\={1\over 2\pi}\,\partial{1\over\overline z-\overline w}~.
\ee

\separator

\item
We consider the analytic function
\bb
  g(z) &=& \ln r \ln f(z) - \ln R \ln z \\
       &=& \ln r \ln\lp |f(z)|\, e^{i\arg f(z)} \rp
          -\ln R \ln\lp |z|\, e^{i\arg z} \rp \\
       &=& \lb \ln r \ln |f(z)| -\ln R \ln|z| \rb
          +i\, \lb \ln r \arg f(z) -\ln R \arg z \rb \\
       &\equiv& h(z) +i\, \alpha(z)~.
\ee
The real part of $g(z)$,
\bb
    h(z) \= \ln r \ln |f(z)| -\ln R \ln|z|~,
\ee
is obviously a harmonic function on $D$. On  the  boundaries
$C_1=\{ |z|=1 \}$ and $C_2=\{ |z|=r \}$, $h(z)$ vanishes:
$$
   h(z)\=0~~~~~\mbox{on}~C_1~{\rm and}~ C_2~.
$$
From Liouville's theorem, then,
$$
   h(z) \= 0~~~~~\mbox{on}~D~.
$$
From this result and the Cauchy-Riemann conditions
$$
  {\partial h\over\partial x}\=
  {\partial \alpha\over\partial y}~~~{\rm and}~~~
  {\partial h\over\partial y}\=-
  {\partial \alpha\over\partial x}~,
$$
we conclude that 
$$
   \alpha(z)\=\mbox{const}\eq a~.
$$
However,
$$
  \alpha(z)\= \ln r \arg f(z) -\ln R \arg z
           \= w\ln r   -\arg z\ln R ~.
$$
When we tranverse the circle $|z|=r$ counterclockwise, we also
tranverse the circle $|w|=R$  counterclockwise. If the map is 1-to-1,
after a full rotation
$$
  z \mapsto z+2\pi i~,~~~~~
  w \mapsto w+2\pi i~,
$$
which implies
$$
 \alpha(z) \mapsto \alpha(z)+ 2\pi \, ( \ln r-\ln R)~.
$$
But if $\alpha(z)$  is going to be a constant, the extra term must
vanish, thus requiring
$$
   \ln r\=\ln R \Rightarrow r\=R~.
$$

This exercise clearly shows that the ``conformal type" of a ring is
described by the ratio $\tau=r_1/r_2$ ($r_2=1$ above). This ratio
is know as the {\bf modulus}\index{modulus!-- of a ring} of the ring.
As long as two rings have different moduli, they are of different
``conformal type" although they have the same connectivity.

\separator

\item
The most general metric in the patch\footnote{All the quantities
in this problem depend on the patch, and so should have a label $j$; for
simplicity, we omit these labels.} $U_j$
of the 2-manifold $M$ has the
form
$$
  \bfg\=g_{11}\,dx\otimes dx+
          g_{12}\,dx\otimes dy+
          g_{21}\,dy\otimes dx+
          g_{22}\,dy\otimes dy~.
$$
We rewrite the above expression in the form
\bb
 \bfg\=\lp \sqrt{g_{11}}\, dx +{g_{12}+i\sqrt{g}\over\sqrt{g_{11}}}\,dy\rp
 \otimes\lp \sqrt{g_{11}}\, dx +{g_{12}-i\sqrt{g}\over\sqrt{g_{11}}}\,dy\rp~.
\ee
Let us concentrate on the 1-form
$$
\sqrt{g_{11}}\, dx +{g_{12}+i\sqrt{g}\over\sqrt{g_{11}}}\,dy~.
$$
Can it be seen as the differential of a function  $z=z(x,y)$?
In other words, is the above expression equal to
\bb
    dz={\partial z\over\partial x}\, dx+
       {\partial z\over\partial y}\, dy
\ee
for some $z(x,y)$?  If it were, this would imply
\bb
       {\partial z\over\partial x}&\=&
\sqrt{g_{11}}~,\\
       {\partial z\over\partial y}&\=&
{g_{12}+i\sqrt{g}\over\sqrt{g_{11}}}\,dy~,
\ee
and since
for a continuous function $z$ with continuous derivatives,
\bb
    {\partial \over\partial y}
    {\partial z\over\partial x}\=
       {\partial \over\partial x}
       {\partial z\over\partial y}~,
\ee
which leads to the constraint
\bb
    {\partial \over\partial y}
    \sqrt{g_{11}}\=
       {\partial \over\partial x}
   {g_{12}+i\sqrt{g}\over\sqrt{g_{11}}}~.
\ee
Obviously this cannot be true in general; the l.h.s. is real,
while the r.h.s. is complex.  This difficulty is overcome
by introducing a complex function $\lambda(x,y)$ such that
\bb
    {\partial \over\partial y}
    (\lambda\,\sqrt{g_{11}}) \=
       {\partial \over\partial x}
   \lp \lambda\, {g_{12}+i\sqrt{g}\over\sqrt{g_{11}}}\rp~.
\ee
This equation in fact determines the function $\lambda$.
Then
\bb
dz\=\lambda\,\lp\sqrt{g_{11}}\, dx +{g_{12}+i\sqrt{g}\over\sqrt{g_{11}}}\,dy
   \rp\eq du+ i  dv~.
\ee
The new coordinates $u=u(x,y)$ and $v=v(x,y)$ introduced above are known as the
{\bf isothermal coordinates}\index{isothermal coordinates}.
In terms of these coordinates
\bb
   \bfg\= \lambda^{-1}\,dz\otimes
   ({\overline\lambda})^{-1}\,
     d\overline z\=
   {1\over |\lambda|^2}\,(du\otimes du+dv\otimes dv)~.
\ee

\separator

\item
In this case $g_{\mu\nu}=\eta_{\mu\nu}$, and so the Killing
equation \calle{killing46} simplifies to
\beq
\label{iso3}
  \partial_\mu\xi_\nu\=-\partial_\nu\xi_\mu~.
\eeq
This equation implies also
\beq
\label{iso4}
  \partial_\rho\partial_\mu\xi_\nu\=-\partial_\rho\partial_\nu\xi_\mu~.
\eeq
In the same way
\beq
\label{iso5}
  \partial_\mu\partial_\rho\xi_\nu\=-\partial_\mu\partial_\nu\xi_\rho~.
\eeq
Notice now that the l.h.s of equations \calle{iso4} and \calle{iso5}
are equal and therefore
$$
  \partial_\rho\partial_\nu\xi_\mu\=\partial_\mu\partial_\nu\xi_\rho~.
$$
Applying equation \calle{iso3} in the r.h.s. of the last equation,
we find
$$
  \partial_\rho\partial_\nu\xi_\mu\=-\partial_\rho\partial_\nu\xi_\mu~,
$$
which necessarily implies that $\bfxi$ is first order in $\bfx$:
$$
\partial_\rho\partial_\nu\xi_\mu\=0~.
$$
Therefore, the allowed Killing vector fields for a flat manifold
are of two kinds:

(i) \underbar{translations}:
$$
    \xi^\mu_{(i)}\=\delta^\mu_i~.
$$
The index $i$ labels the $n$ independent vectors.

(ii) \underbar{rotations}:
$$
    \xi^\mu\= \omega_{\mu\nu}\, x^\nu~.
$$
Notice that equation \calle{iso3} requires 
$$
    \omega_{\mu\nu}\=- \omega_{\nu\mu}~.
$$

Thus, the general Killing vector $\bfxi$ on the Euclidean plane 
has the form
$$
 \bfxi= a_\mu\, \partial^\mu + {1\over 2}\, \omega_{\mu\nu}\,
  (x^\mu\partial^\nu- x^\nu\partial^\mu)~.
$$
We usually define the vector fields
\bb
  \bfP_\mu\eq -i\partial_\mu~,~~~~~
  \bfL_{\mu\nu}\eq i(x_\mu\partial_\nu- x_\nu\partial_\mu)~,
\ee
which are the well-known generators of
translations (linear momentum)  and rotations (angular momentum),
respectively.
The algebra generated by them is denoted $\mathfrak{iso}(n)$,
and 
the corresponding group is denoted
ISO$(n)$. Physicists usually call these the Poincar\'e algebra
and the
{\bf Poincar\'e group}\index{Group!Poincar\'e --}, respectively,
and use the symbol $\CP$ for the group.
One can easily check that the
Poincar\'e algebra has the commutators
\bb
   \lb \bfL_{\mu\nu}, \bfP_\rho \rb &=& i\,
   (\eta_{\nu\rho}\bfP_\mu-\eta_{\mu\rho}\bfP_\nu)~,\\
   \lb \bfL_{\mu\nu}, \bfL_{\rho\tau}\rb &=& i\,
   (\eta_{\mu\tau}\bfL_{\nu\rho}
   +\eta_{\nu\rho}\bfL_{\mu\tau}
   -\eta_{\mu\rho}\bfL_{\nu\tau}
   -\eta_{\nu\tau}\bfL_{\mu\rho})~.
\ee

\separator

\item
In the (1+1)-dimensional case, we have only 2 components of the
linear momentum, 
$P_1$ and $P_2$, and one component of the angular momentum $L\equiv L_{01}$.
From the result of the previous problem (or by a quick computation)
we find that the
classical corresponding Poincar\'e algebra reads:
\beqn
\label{eq:CUR8a}
  \la L, P_0 \ra_{PB} &=& P_1 ~, \\
\label{eq:CUR8b}
  \la L, P_1 \ra_{PB} &=& P_0 ~, \\
\label{eq:CUR8c}
  \la P_0, P_1 \ra_{PB} &=& 0~.
\eeqn
Introducing the operators
$$
   P_\pm \eq P_0\pm P_1~,
$$
we can rewrite \calle{eq:CUR8a}-\calle{eq:CUR8c} in the form
\beqn
\label{eq:CUR9}
   \la L, P_\pm\ra_{PB} &=& \pm\, P_\pm~, \\
\label{eq:CUR10}
   \la P_+, P_-\ra_{PB} &=& 0~.
\eeqn

Now for a system of $N$ particles with coordinates $x_i$ and 
momentum $p_i$, we write
\bb
   \CP_\pm &=& \sum_{i=1}^N e^{\pm p_i}\, V_i(x_1,\dots,x_N)~,\\
   \CL_\pm &=& \sum_{i=1}^N x_i~.
\ee
We will now find the conditions on the functions $V_i$ under which
$\CP_\pm$ and $\CL$ give a representation of the Poincar\'e algebra.
The boost operator $\CL$ has been chosen such that the Poisson 
bracket \calle{eq:CUR9}  is satisfied immediately:
\bb
   \la \CL, \CP_\pm\ra_{PB} &=&
   \sum_{k=1}^N\lp
   {\partial\CL\over\partial x_k}
   {\partial\CP_\pm\over\partial p_k}
   -{\partial\CP_\pm\over\partial x_k}
   {\partial\CL\over\partial p_k}\rp \\
  &=& \sum_{k=1}^N \lp\sum_{i=1}^N\delta_{ik}\rp\,
   \sum_{j=1}^N e^{\pm p_j} (\pm \delta_{jk}) V_j - 0 \\
  &=&  \pm \sum_{i=1}^N  e^{\pm p_i} V_i\\
  &=& \pm \, \CP_\pm~.
\ee
 
We now impose condition \calle{eq:CUR10}:
\bb
   \la\CP_+, \CP_-\ra_{PB} &=&
   \sum_{k=1}^N\lp
   {\partial\CP_+\over\partial x_k}
   {\partial\CP_-\over\partial p_k}
   -{\partial\CP_-\over\partial x_k}
   {\partial\CP_+\over\partial p_k}\rp \\
   &=& \sum_{k=1}^N \lp
       \sum_{i=1}^N e^{p_i} \partial_k V_i 
       \sum_{j=1}^N e^{-p_j}  V_j (-\delta_{jk})
       -\sum_{i=1}^N e^{p_i}  V_i \delta_{ik} 
       \sum_{j=1}^N e^{-p_j}   \partial_k V_j  \rp \\
    &=&  -\sum_{i=1}^N \sum_{j=1}^N
         e^{p_i-p_j} \lp V_i\partial_i V_j +V_j\partial_j V_i\rp \\
    &=&  -\sum_{i\ne j}
         e^{p_i-p_j} \lp V_i\partial_i V_j +V_j\partial_j V_i\rp
         -\sum_{i}
          \lp V_i\partial_i V_i +V_i\partial_i V_i\rp \\
    &=&  -\sum_{i\ne j}
         e^{p_i-p_j} \lp V_i\partial_i V_j +V_j\partial_j V_i\rp
         -\sum_{i} \partial_i V^2_i ~.
\ee
The r.h.s. will vanish for all values of the momenta if and only if
\bb
        V_i\partial_i V_j +V_j\partial_j V_i &=& 0~,~~~~~i\ne j~,\\
        \sum_{i} \partial_i V^2_i &=& 0~.
\ee
The first of these equations is satisfied if the functions $V_i$ have the
functional form:
$$
     V_i \= f(x_{i-1}-x_i)\, f(x_i-x_{i-1})~.
$$
Indeed, if $i$ and $j$ are well-separated, $|i-j|\ge2$, then
none of the points $x_{i-1}, x_i$ and $x_{i+1}$ coincide with
$x_j$, and $\partial_j V_i=0$. If $|i-j|=1$, we may assume without loss of 
generality that $i=j+1$. Then 
\bb
    {\partial_{j+1} V_j\over V_j}+
    {\partial_j V_{j+1}\over V_{j+1}} 
     &=& {\partial_{j+1}f(x_{j-1}-x_{j})\over f(x_{j-1}-x_{j}) } 
       + {\partial_{j+1}f(x_{j}-x_{j+1})\over f(x_{j}-x_{j+1}) } \\ 
       && + {\partial_{j}f(x_{j}-x_{j+1})\over f(x_{j}-x_{j+1}) } 
       + {\partial_{j}f(x_{j+1}-x_{j+2})\over f(x_{j+1}-x_{j+2}) } \\
     &=& 0-  {f'(x_{j}-x_{j+1})\over f(x_{j}-x_{j+1}) } 
          +  {f'(x_{j}-x_{j+1})\over f(x_{j}-x_{j+1}) } 
          +0 \=0~.
\ee

Finally, we must specify $f$ using the last condition
 $S\equiv\sum_i \partial_iV_i^2=0$.
Notice that if we define periodic boundary conditions
$$
    x_0\=x_N~,~~~~~ x_{N+1}\=x_1~,
$$
or if we set
$$
    x_0 \rightarrow -\infty~,~~~~~x_{N+1}\rightarrow+\infty~,
$$
then
\bb
   \sum_{i=1}^N e^{x_i-x_{i+1}} \=  \sum_{i=1}^N e^{x_{i-1}-x_i}~.
\ee
This is a trivial identity that we can rearrange to obtain the function 
$f$! Indeed:
\bb
    0 &=& ab\, \sum_{i=1}^N
    \lp e^{x_i-x_{i+1}}- e^{x_{i-1}-x_i} \rp \\
      &=&  \sum_{i=1}^N
    \partial_i \lp
    a^2+ab\, e^{x_i-x_{i+1}}+ba\, 
    +ab\, e^{x_{i-1}-x_{i}}
    +b^2\, e^{x_{i-1}-x_{i+1}}\rp\\
 &=&  \sum_{i=1}^N
    \partial_i \lb\lp a+be^{x_{i-1}-x_{i}}\rp
               \lp a+be^{x_{i}-x_{i+1}}\rp\rb~,
\ee
which shows that the function
\bb
   f^2(y)\=  a+b\, e^y
\ee
is a solution to our problem. 

The system thus obtained has the Hamiltonian
\bb
   \CP_0\=
   \sum_{i=1}^N \cosh p_i\, 
    \sqrt{a+be^{x_{i-1}-x_i}}\,
    \sqrt{a+be^{x_i-x_{i+1}}}~
\ee
and is known as the 
{\bfseries relativistic Toda system}\index{Toda!relativistic -- system}
\cite{Ruij}.

\separator

\item
For a Levi-Civita metric
$$
   \christoffel{\kappa}{\rho\nu} \= {1\over2}\, g^{\kappa\lambda}\,
   \lp \partial_\rho g_{\nu\lambda}
      +\partial_\nu g_{\lambda\rho}
      -\partial_\lambda g_{\rho\nu} \rp~.
$$
Thus
\bb
   \christoffel{\kappa}{\rho\nu}
    g_{\kappa\mu} &=& 
   {1\over 2}\, g^{\kappa\lambda}\, g_{\kappa\mu}\,
   \lp \partial_\rho g_{\nu\lambda}
      +\partial_\nu g_{\lambda\rho}
      -\partial_\lambda g_{\rho\nu} \rp\\
   &=& {1\over2}\,\delta^\lambda_\mu\,
   \lp \partial_\rho g_{\nu\lambda}
      +\partial_\nu g_{\lambda\rho}
      -\partial_\lambda g_{\rho\nu} \rp\\
     &=& {1\over2}\, \lp \partial_\rho g_{\nu\mu}
      +\partial_\nu g_{\mu\rho}
      -\partial_\mu g_{\rho\nu} \rp~.
\ee
Interchanging $\mu$ and $\nu$, we obtain
\bb
   \christoffel{\kappa}{\rho\mu} g_{\kappa\nu}&=& 
   {1\over2}\, \lp \partial_\rho g_{\mu\nu}+\partial_\nu g_{\nu\rho}
      -\partial_\nu g_{\rho\mu} \rp~.
\ee
Summing the two preceding results yields
$$
   \christoffel{\kappa}{\rho\nu}  g_{\kappa\mu} 
   +\christoffel{\kappa}{\rho\mu} g_{\kappa\nu} 
   \= \partial_\rho g_{\mu\nu}~,
$$
where the symmetry of the metric tensor was used. We finally substitute
this result in \calle{killing46}, and find
\bb
   0 &=& \xi^\rho\, \lp \christoffel{\kappa}{\rho\nu}g_{\kappa\mu} 
   +\christoffel{\kappa}{\rho\mu}g_{\kappa\nu}\rp
  + \partial_\mu \xi^\kappa\, g_{\kappa\nu}
  + \partial_\nu \xi^\kappa\, g_{\kappa\mu} \\
  &=& 
   g_{\kappa\mu}\,\lp \christoffel{\kappa}{\rho\nu}\xi^\rho
                      + \partial_\nu \xi^\kappa \rp
   +g_{\kappa\nu}\,\lp \christoffel{\kappa}{\rho\mu}\xi^\rho
                      + \partial_\mu \xi^\kappa \rp \\
   &=&  g_{\kappa\mu}\, \xi^\kappa_{~;\nu} 
       + g_{\kappa\nu}\, \xi^\kappa_{~;\mu} \\ 
   &=&  \lp g_{\kappa\mu}\, \xi^\kappa\rp_{;\nu} 
       + \lp g_{\kappa\nu}\, \xi^\kappa\rp_{;\mu} \\ 
   &=&  \xi_{\mu;\nu} 
       + \xi_{\nu;\mu} \\ 
   &=&  
   (\partial_\nu\xi_\mu-
   \christoffel{\rho}{\mu\nu}\,\xi_\rho)+
   (\partial_\mu\xi_\nu-
   \christoffel{\rho}{\mu\nu}\,\xi_\rho) \\
   &=&
    \xi_{\mu,\nu}+\xi_{\nu,\mu}-2\, \christoffel{\rho}{\mu\nu}\,\xi_\rho~,
\ee   
where we have used the properties
\bb
   g_{\kappa\mu;\rho}&=&0~,\\
   \xi_{\mu;\nu} &=&
   \partial_\nu\xi_\mu-
   \christoffel{\rho}{\mu\nu}\,\xi_\rho~.
\ee

\separator

\item
For the standard metric of the sphere, 
the Christoffel symbols are
\bb
     \phantom{\christoffel{\phi}{\phi\,\phi}=}
     \christoffel{\phi}{\phi\,\phi}=
     \christoffel{\phi}{\theta\,\theta} = 0~,&&~~~
     \christoffel{\phi}{\theta\,\phi}=
     \christoffel{\phi}{\phi\,\theta}= \cot\theta~,\\
     \christoffel{\theta}{\theta\,\theta}=
     \christoffel{\theta}{\theta\,\phi}=
     \christoffel{\theta}{\phi\,\theta}=0~,&&~~~
     \christoffel{\theta}{\phi\,\phi}=-\sin\theta\cos\theta~.
\ee
Using these expressions in   
\calle{killing3}, we find 
\beqn
\label{killing4}
  \partial_\theta\xi_\theta &=& 0~,\\
\label{killing5}
  \partial_\phi\xi_\phi +\sin\theta\,\cos\theta\,\xi_\theta &=& 0~,\\
\label{killing6}
  \partial_\phi\xi_\theta +
  \partial_\theta\xi_\phi -
  \sin\theta\,\cos\theta\,\xi_\theta &=& 0~.
\eeqn
This system can be solved in straightforward fashion.
Equation \calle{killing4} requires that $\xi_\theta$ be a function of
$\phi$ only, and so
\beq
\label{killing7}
    \xi_\theta\= \Phi(\phi)~.
\eeq
Substituting this result in equation \calle{killing5}, we can also
solve for $\xi_\phi$, and find
\beq
\label{killing8}
    \xi_\phi\= -\sin\theta\,\cos\theta\,\tilde\Phi(\phi)+\Theta(\theta)~,
\eeq
where $\Theta(\theta)$ is an arbitrary function of $\theta$ and
\bb
  \tilde \Phi(\phi) \eq \int^\phi\, d\phi'\, \Phi(\phi')~.
\ee
Finally, equation \calle{killing6} gives
\bb
  \partial_\phi\Phi +\tilde \Phi \=
  -\partial_\theta\Theta +2\, \cot\theta\, \Theta~.
\ee
The $\phi$-dependent terms are now separated from the 
$\theta$-dependent terms and therefore each term must be
constant:
\beqn
\label{killing11}
  {d\Phi\over d\phi} +\tilde \Phi &=& c_1~,\\
  -{d\Theta \over d\theta} +2\, \cot\theta\, \Theta &=& c_1~.
   \nonumber
\eeqn
The second equation is a first-order differential equation and can be
solved with well-known techniques. In particular, multiplying
it by $1/\sin^2\theta$, we find
\beq
\label{killing13}
  {d\over d\theta}\lp {\Theta \over \sin^2\theta}\rp \= -
   {c_1\over\sin^2\theta} ~\Rightarrow~
   \Theta(\theta)\= (c_1\, \cot\theta + c_2)\, \sin^2\theta~.
\eeq
Now, differentiating equation \calle{killing11}, we arrive at a
simple equation,
\bb
   {d^2\Phi\over d\phi^2}+\Phi\=0 ~\Rightarrow~
   \Phi(\phi)\= a_1\, \sin\phi - a_2\, \cos\phi ~.
\ee
From this 
\bb
   \tilde\Phi(\phi) \= - {d\Phi\over d\phi} +c_1 \=
   a_1\, \cos\phi + a_2\, \sin\phi + c_1 ~.
\ee
The components of the Killing vector are thus
\bb
  \xi^\theta &=& g^{\theta\theta}\,\xi_\theta\= 
  a_1\, \cos\phi + a_2\, \sin\phi~, \\
  \xi^\phi &=& g^{\phi\phi}\, \xi_\phi \=
  -a_1\, \cot\phi\, \sin\phi + a_2\, \cot\theta\, \cos\phi +c_2~.
\ee
The Killing vector $\bfxi=\xi^\theta\,\partial_\theta+\xi^\phi\,
\partial_\phi$ is usually written in the form
\bb
  \bfxi \= -a_1\, \bfL_1 + a_2\, \bfL_2 + c_2\, \bfL_3~,
\ee
where
\bb
   \bfL_1 &=& -\cos\phi\,\partial_\theta +\cot\theta\,\sin\phi\,
   \partial_\phi~,\\
   \bfL_2 &=& \sin\phi\,\partial_\theta +\cot\theta\,\cos\phi\,
   \partial_\phi~,\\
   \bfL_3 &=& \partial_\phi~.
\ee
These vectors generate the ${\frak su}(2)$ algebra:
\bb
   \lb \bfL_1, \bfL_2\rb &=& \bfL_3~, \\
   \lb \bfL_2, \bfL_3\rb &=& \bfL_1~, \\
   \lb \bfL_3, \bfL_1\rb &=& \bfL_2~.
\ee

\separator

\item
From the given metric, one finds that the Christoffel symbols are
\bb
     \Gamma^x_{xx}=
     \Gamma^x_{yy}=0~,~~~&&
     \Gamma^x_{xy}=
     \Gamma^x_{yx}=-{1\over y}~,\\
     \Gamma^y_{xy}=
     \Gamma^y_{yx}=0~,~~~&&
     \Gamma^y_{xx}=
     -\Gamma^y_{yy}={1\over y}~.
\ee
Then, the Killing equations \calle{killing3} for the Poincar\'e half-plane
are
\beqn
\label{killing25}
 \partial_x\xi_x -{1\over y}\, \xi_y &=& 0~, \\
\label{killing26}
 \partial_y\xi_y +{1\over y}\, \xi_y &=& 0~, \\
\label{killing27}
 \partial_x\xi_y + \partial_y \xi_x + {2\over y}\, \xi_x &=& 0~.
\eeqn
Equation \calle{killing26} is solved easily, to wit,
\bb
 \partial_y(y\,\xi_y)\=0 ~\Rightarrow~ 
 \xi_y \= {X(x)\over y}~,
\ee
where $X(x)$ is a function of $x$ only.
Substituting this in equation \calle{killing25}, we find
\bb
  \xi_x \= {\tilde X(x)\over y^2} + Y(y)~,
\ee
where $Y(y)$ is a function of $y$ and
\bb
    \tilde X(x) \eq \int^x \, dx'\, X(x')~.
\ee 
Equation \calle{killing27} also gives
\bb
  {dX(x)\over dx} \= - y\, {dY(y)\over dy} -2\, Y(y)~.
\ee
In this formula, the  variables $x$ and $y$ are separated, and
therefore each term has to be a constant:
\bb
  {dX(x)\over dx} &=& 2a ~, \\
  - y\, {dY(y)\over dy} -2\, Y(y) &=& 2a~.
\ee
The solutions are
\bb
   X(x) &=& 2a\, x + b~, \\
   Y(y) &=& -a + {c\over y^2}~.
\ee
The components of the Killing vector are
\bb
  \xi^x &=& g^{xx}\, \xi_x \= a\, (x^2-y^2) +b\, x +c~, \\
  \xi^y &=& g^{yy}\, \xi_y \= 2a\, xy + b\, y~.
\ee
Thus, in general here we have
\bb
   \bfxi \= c\, \bfL_1 + b\, \bfL_2 + a\, \bfL_3~,
\ee
where
\bb
   \bfL_1 &=& \partial_x~, \\
   \bfL_2 &=& x\,\partial_x + y\,\partial_y~, \\
   \bfL_3 &=& (x^2-y^2)\,\partial_x +2xy\, \partial_y~.
\ee
The above vector fields generate the algebra 
\bb
   \lb \bfL_1, \bfL_2\rb &=& -\bfL_1~, \\
   \lb \bfL_2, \bfL_3\rb &=& - \bfL_3~, \\
   \lb \bfL_3, \bfL_1\rb &=& 2\,\bfL_2~.
\ee
 
Now we would like to integrate the algebra to the corresponding
group. To make contact with customary notation, let us use complex variables.
In these variables, the vector vields take a nice symmetric form, namely
\bb
   \bfL_1 &=& {\partial\over\partial z}+
           {\partial\over\partial \overline z}\eq 
          \bfl_1+\overline{\bfl}_1~,\\
   \bfL_2 &=& z\,{\partial\over\partial z}+
           \overline z\,{\partial\over\partial \overline z}\eq 
          \bfl_2+\overline{\bfl}_2~,\\
   \bfL_3 &=& z^2\,{\partial\over\partial z}+
           \overline z^2\,{\partial\over\partial \overline z}\eq 
          \bfl_3+\overline{\bfl}_3~.
\ee   
To integrate the algebra, let us work with the holomorphic part
$\bfl_1, \bfl_2, \bfl_3$ only.
The flow generated by $\bfl_1$ is 
$$
    {dz\over dt} \= a\, \bfl_1(z) \= a ~\Rightarrow~ z\= z_0 +a~,
$$
where $z_0$ and $a$ are two complex constants.
In the same way, the flow generated by $\bfl_2$ is 
$$
    {dz\over dt} \= b\, \bfl_2(z) \= b\,z ~\Rightarrow~ z\= z_0\, e^{bz} ~.
$$
Again, $z_0$ and $b$ are two complex constants. 
Finally, the flow generated by $\bfl_3$ is 
$$
    {dz\over dt} \= c\, \bfl_3(z) \= c\, z^2 ~\Rightarrow~ z\= 
   {z_0\over z_0 t-1}  ~,
$$
where $z_0$ and $c$ are two complex constants. The generic combination
of the above transformations
is represented by the overall transformation
\beq
    z ~\rightarrow~ w \= {\alpha z +\beta\over \gamma z +\delta}~,
  ~~~~~\alpha\delta-\beta\gamma\not=0~.
\label{eq:CUR5}
\eeq
The set of tranformations \calle{eq:CUR5} constitute the 
{\bf M\"obius group}\index{group!M\"obius}.

\separator

\item
We are interested in 
determining all possible conformal Killing vector fields 
of an 
$n$-manifold that it is conformally flat:
$$
   g_{\mu\nu}=\eta_{\mu\nu}~.
$$ 
In this case, equation \calle{killing45} simplifies to
\beq
\label{eq:CUR18}
  \partial_\mu\xi_\nu+\partial_\nu\xi_\mu\=
   {2\over n}\, \partial^\rho\xi_\rho\, \eta_{\mu\nu}~.
\eeq
Acting with $\partial^\nu$, we obtain
\beq
\label{eq:CUR19}
  \left( 1 -{2\over n}\right)\, \partial_\mu(\partial\cdot\xi)
  + \square \xi_\mu \= 0~.
\eeq
We notice that the behavior of this equation is different for
$n\ge 3$ and $n=2$, since in the latter case the first term
drops out of the equation. Therefore, we 
will examine the two cases 
separately.

\vspace{2mm} $\bullet$ Case I:  $n>2$

Acting with $\partial_\nu$ in \calle{eq:CUR19}, one obtains
\beq
\label{eq:CUR20}
  \left( 1 -{2\over n}\right)\, 
  \partial_\nu\partial_\mu(\partial\cdot\xi)
  + \square \partial_\nu\xi_\mu \= 0~.
\eeq
Contracting the indices $\nu$ and $\mu$ in the last equation,
we arrive at the result
\beq
\label{killing46}
   \square \partial\cdot\xi \= 0~.
\eeq
Now, we rewrite equation \calle{eq:CUR20} with the indices $\nu$ 
and $\mu$ interchanged, and we add the resulting equation to the
original one, yielding
$$
  2\,\left( 1 -{2\over n}\right)\, 
  \partial_\nu\partial_\mu(\partial\cdot\xi)
  + \square (\partial_\nu\xi_\mu 
  +  \partial_\mu\xi_\nu) \= 0~.
$$
However, using equations \calle{eq:CUR18} and \calle{killing46}
successively, this is transformed to
\beq
\label{killing48}
 \partial_\nu\partial_\mu\partial\cdot\xi \= 0~.
\eeq
Using this in conjunction with \calle{eq:CUR18} yields
$$
  \partial_\rho\partial_\sigma\partial_\mu\xi_\nu
  +\partial_\rho\partial_\sigma\partial_\nu\xi_\mu\=0~.
$$
Playing with the indices, we find that
$$
  \partial_\rho\partial_\sigma\partial_\mu\xi_\nu
  \=0~.
$$
This is a very strong result. For $n\ge 3$, the
allowed conformal Killing vector
fields are at most quadratic functions of the coordinates.
The four possibilities are:

(i) \underbar{translations}:
$$
    \xi^\mu_{(i)}\=\delta^\mu_i~.
$$
The index $i$ labels the $n$ independent vectors.

(ii) \underbar{rotations}:
$$
    \xi^\mu\= \omega^\mu_{~\nu}\, x^\nu~.
$$
Notice that equation \calle{eq:CUR18} requires 
$$
    \omega_{\mu\nu}\=- \omega_{\nu\mu}~.
$$

(iii) \underbar{dilatations}:
$$
    \xi^\mu\= \lambda\, x^\nu~.
$$

(iv) \underbar{special conformal transformations}:
$$
    \xi^\mu\= b^\mu\, x^2 - 2\, b^\nu\,x_\nu x^\mu~.
$$

For the above transformations, we define the respective generators as follows:
\bb
  \bfP_\mu &\equiv& -i\partial_\mu~,\\
  \bfL_{\mu\nu} &\equiv& i(x_\mu\partial_\nu- x_\nu\partial_\mu)~,\\
  \bfD &\equiv& -ix^\mu\partial_\mu~,\\
  \bfK_\mu &\equiv& i(2x_\mu x^\rho\partial_\rho-x^\rho x_\rho\partial_\mu)\=
     -2x_\mu\bfD+x^\rho x_\rho\bfP_\mu~.
\ee
Then the algebra they generate, the 
{\bfseries conformal Poincar\'e algebra}\index{Algebra!Conformal Poincar\'e --}
in $d>2$ dimensions, is given by:
\bb
   \lb \bfL_{\mu\nu}, \bfP_\rho \rb &=& i\,
   (\eta_{\nu\rho}\bfP_\mu-\eta_{\mu\rho}\bfP_\nu)~,\\
   \lb \bfL_{\mu\nu}, \bfL_{\rho\tau}\rb &=& i\,
   (\eta_{\mu\tau}\bfL_{\nu\rho}
   +\eta_{\nu\rho}\bfL_{\mu\tau}
   -\eta_{\mu\rho}\bfL_{\nu\tau}
   -\eta_{\nu\tau}\bfL_{\mu\rho})~, \\
  \lb \bfL_{\mu\nu}, \bfK_\rho \rb &=& i\,
   (\eta_{\nu\rho}\bfK_\mu-\eta_{\mu\rho}\bfK_\nu)~,\\
   \lb \bfD, \bfP_\rho \rb &=& +i\, \bfP_\mu~,\\
   \lb \bfD, \bfK_\rho \rb &=& +i\, \bfK_\mu~,\\
   \lb \bfP_\mu, \bfK_\nu \rb &=& 2i\, (\eta_{\mu\nu}\bfD+\bfL_{\mu\nu})~.
\ee

\vspace{2mm} $\bullet$ Case II:  $n=2$

In two dimensions, equation \calle{eq:CUR18} becomes:
\bb
     \partial_1\xi_1 &=& \partial_2\xi_2 ~,\\
     \partial_1\xi_2 &=& \partial_2\xi_1 ~.
\ee
But these are exactly the Cauchy-Riemann conditions! In other
words, in 2-dimensions, any holomorphic
 function
$\xi=\xi_1+i\xi_2$ of $z=x^1+ix^2$ is a conformal
transformation.  Recalling that any holomorphic function 
can be expanded in a Laurent series
$$
   \xi(z) \= - \sum_n\, \xi_{(n+1)}\, z^{n+1}~,
$$
we recognize that any transformation 
$$
 z\rightarrow z'
   \= z - \epsilon\, \xi_{(n+1)}\, z^{n+1}
$$
is a conformal transformation, and thus we have an infinite dimensional
algebra $\CL$.
In particular, the gererators $\bfl_n$,
with
\bb
    \bfl_n = - z^{n+1}\, \partial_z~,
\ee 
satisfy the algebra
\beq
\label{classicalL}
   \lb  l_n, l_m \rb \= (n-m)\, l_{n+m}~.
\eeq
The algebra $\CL$ given by equation \calle{classicalL}
is called the 
{\bfseries classical Virasoro algebra}\index{Algebra!Classical Virasoro --}.

Of course, similar considerations are valid for
$\overline\xi(\overline z)=\xi_1-i\xi_2$ and the corresponding
gererators $\overline{\bfl}_n$. It is customary to study only
half of the symmetry, keeping in the back of one's head that
the full symmetry is $\CL\times\overline\CL$. Moreover, although 
the anti-holomorphic part of the theory is the complex conjugate of 
the holomorphic part, it is usually treated as
an  independent piece of the theory. This piece is related
to the holomorphic piece only if the transformation
$(z,\overline z)\mapsto(\overline z,z)$ is a symmetry of the theory.

\separator

\item

The 2-sphere $S^2$ is equivalent to the compactified plane
\bb
    S^2 \= \BC \cup \{ \infty \},
\ee
One can put coordinates on this manifold using
two patches, $U_1$ which excludes the
point at infinity, and $U_2$ which excludes the origin. 
If $z$ and $w$ are the (complex) coordinates on these
patches, then the equation
$$
    w\={1\over z}
$$
is the transition function from one patch to the other.

The maximal subalgebra that is globally defined on $S^2$
must be spanned by vector fields that are smooth
on the whole of $S^2$. 
To find this maximal subalgebra, 
we consider the local algebra $\CL$ on one of the patches, and 
then attempt to extend this to the entire other patch.
Obviously, the only points at which
subtleties may arise are at 0 and $\infty$.

From their very definition,
$$
   \bfl_n = - z^{n+1}\, \partial_z~,
$$
it is clear that not all vector fields 
are well defined at the point $z=0$. In particular, when $n< -1$, the
corresponding vector fields $\bfl_n$ are singular at $z=0$.
Thus, only the
vector fields $\bfl_n,~n\ge -1$ have a chance of being globally defined.

To examine now what happens at the point $z=\infty$, we  use the substitution
$w=1/z$, and examine what happens at $w=0$.  Since under this
transformation
\bb
   \bfl_n \mapsto \bfl_n \= w^{-n+1}\, \partial_w~,
\ee
we see that the vector fields $\bfl_n$ for $n>1$ have a singularity at
$w=0$. 
    
Therefore, the only the vector fields $\bfl_n$ that are globally
defined are those with $-1\le n \le 1$, i.e.,
\bb
   \bfl_{-1} &=&  \partial_z~, \\
   \bfl_0 &=& z\, \partial_z~, \\
   \bfl_1 &=& z^2\, \partial_z~.
\ee
These fields generate an ${\frak sl}\,(2)$ algebra, with
commutation relations
\bb
   \lb \bfl_{-1} , \bfl_0  \rb  &=& -\bfl_{-1}~,\\
   \lb \bfl_0 , \bfl_1  \rb  &=& -\bfl_1~,\\
   \lb \bfl_1 , \bfl_{-1}  \rb &=& 2\bfl_0~,
\ee
which can be integrated
to the  M\"obius group\index{Group!M\"obius --}.

\separator

\item
For a collision in which $N$ particles (the total number of incoming
particles plus the total number of outgoing particles) are involved,
we have $ND$ momentum components,
but not all of them are independent.
Using
Lorentz invariance we can fix $D(D-1)/2$
of them, and using translational invariance we can eliminate another
set of $D$ of them.
In addition, the on--shell conditions $p^2_i\=m^2_i$
reduce the independent components further by an amount $N$. 
So, finally, we have
$$
    ND\, -\,{D(D-1) \over 2}\,D\,-N\= (D-1)\,N\,-\,{D(D+1) \over 2}~
$$
independent momentum components in this collision.
For the particular case of two particle scattering in two
dimensions, that is for $D=2$ and $N=4$ (two particles in plus two particles
out), we thus have
only one independent Mandelstam variable, which is given by
$$
  s~\equiv ~ (p_1+p_2)^2\=m_1^2+m_2^2+2(\epsilon_1\epsilon_2-k_1 k_2)~.
$$

\separator

\item
Without loss of generality, we will assume that $Q_{\mu\nu}$ is traceless,
since if it is not, we can define
$$
   Q'_{\mu\nu} \eq Q_{\mu\nu} - \delta_{\mu\nu}\, Q^\rho_\rho~
$$
and work with this new tensor.

For a 1-particle state $\ket{p}$, the expectation value of
the charge $Q_{\mu\nu}$ must be of the form
$$
 \langle p | Q_{\mu\nu} | p \rangle =
 A(p^2)\, p_\mu \, p_\nu +
 B(p^2) \, \eta_{\mu\nu}~.
$$
Taking into account the vanishing trace of $Q_{\mu\nu}$, we find
$$
 \langle p | Q_{\mu\nu} | p \rangle =
 A(p^2)\, \lp  p_\mu \, p_\nu - {1\over D}\, p^2 \,
               \eta_{\mu\nu} \rp ~.
$$
For an asymptotic state of many particles,
$\ket{p_1, p_2, \dots, p_n; {\rm asym}}=\ket{p_1}\, \ket{p_2} \dots 
\ket{p_n}$, factorization means that $Q_{\mu\nu}$ acts additively,
which means 
$$
   \bra{p_1, p_2, \dots, p_n; {\rm asym}}
   Q_{\mu\nu}
   \ket{p_1, p_2, \dots, p_n; {\rm asym}}
   ~=~
   \sum_{k=1}^n \, A(p^2_k)\, \lp  p_{k,\mu} \, p_{k,\nu} 
                   - {1\over D}\, p^2_k \, \eta_{\mu\nu} \rp ~.
$$

The conservation of $Q_{\mu\nu}$ ensures that its values before and after
the collision will be identical, given by the condition
$$
  \sum_{k=1}^n \, A(m_k)\, \lp  p^i_{k,\mu} \, p^i_{k,\nu}
                   - {1\over D}\, m^2_k \, \eta_{\mu\nu} \rp
  ~=~
  \sum_{k=1}^n \, A(m_k)\, \lp  p^f_{k,\mu} \, p^f_{k,\nu}
                   - {1\over D}\, m^2_k \, \eta_{\mu\nu} \rp~,
$$
or, equivalently,
$$
    \sum_{k=1}^n \, A(m_k)\,  p^i_{k,\mu} \, p^i_{k,\nu} ~=~
    \sum_{k=1}^n \, A(m_k)\,   p^f_{k,\mu} \, p^f_{k,\nu}~.
$$
The above equation is true for all $n$-to-$n$ elastic reactions.
In particular, it is true for the reaction in which all particles
are identical (and therefore have the same rest mass). Consequently,
either $A=0$ (an utterly trivial case), or
$$
 \{   p^i_1, p^i_2, \dots, p^i_n \} ~=~
 \{   p^f_1, p^f_2, \dots, p^f_n \} ~.
$$
This says that for an elastic reaction, if there is this
additional conserved quantity $Q_{\mu\nu}$, there can be only
a reshuffling of the momenta of the particles; momenta
cannot change values in any other way. For the 2-to-2 scattering
this means that there can be only forward or backward scattering
as seen in figure \ref{fig:FB}.

\begin{figure}[htb]
\begin{center}
\includegraphics[height=3cm]{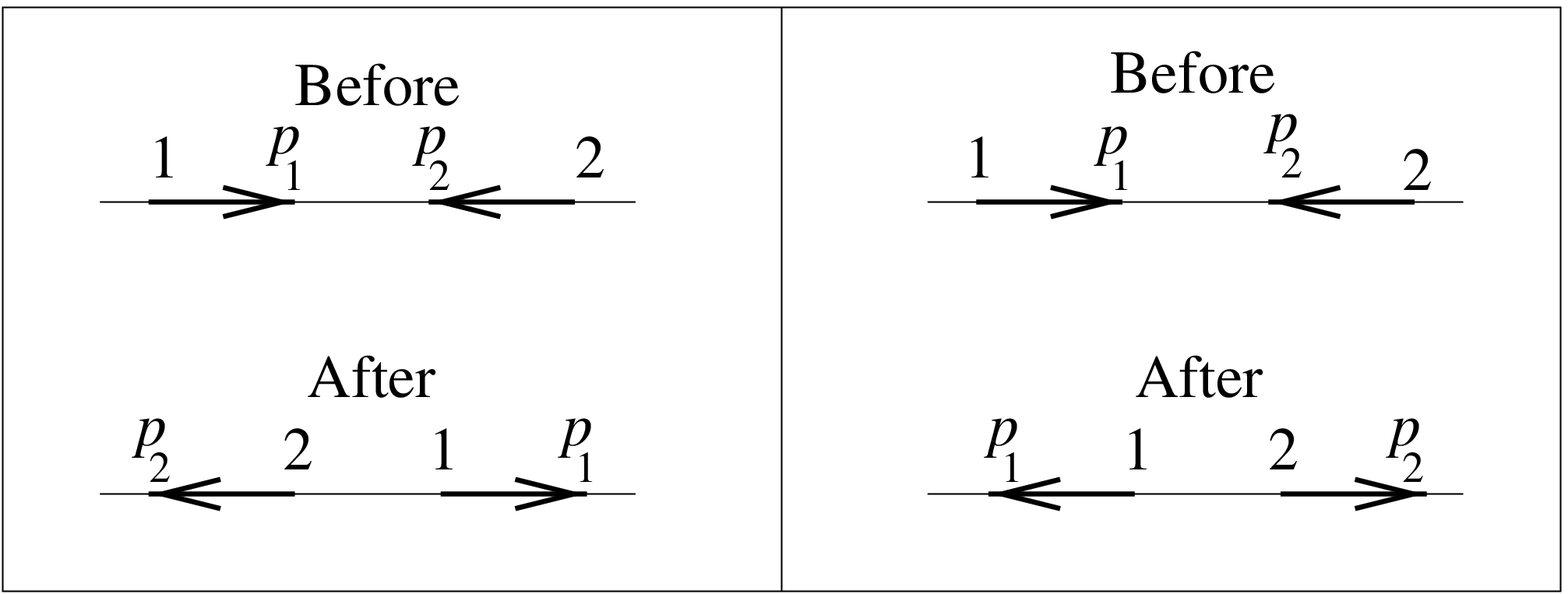}
\end{center}
\caption{Forward  and backward  scattering. These are the
analogues of transmission and reflection of a wavepacket
in Quantum Mechanics.}
\label{fig:FB}
\end{figure}
 
Any S-matrix has to be analytic in the Mandelstam variables
$s$ and $t$ (i.e. analytic on the complex $s$-$t$ plane).
However, this is impossible if only backward and forward scattering
are allowed. This rules out the presence of the non-trivial conserved
charge $Q_{\mu\nu}$ described here when there are at least two
Mandelstam variables (except in the case of a non-interacting
theory).  However, as we saw in an earlier problem,
in 1+1 dimensions, there is only one Mandelstam variable, and consequently
no analyticity constraint.  (Indeed, in 1+1-dimensions, even without
an exotic conserved charge, only backward and forward scattering are
possible.)  Therefore, in 1+1 dimensions it is feasible to have
non-trivial models in which there is an extra spacetime symmetry with
an associated conserved tensorial charge.

Incidentally, let us point out the this analysis
can be used to show another loophole in the Coleman-Mandula
Theorem. Our discussion made use of the fact that the conserved
charge $Q_{\mu\nu}$ is additive when acting on  asymptotic
multi-particle states. To formulate this a little more mathematically,
let
$$
  \ket{a(p_a)} \otimes \ket{b(p_b)}
$$
be a 2-particle state in the Hilbert space $\CH\otimes\CH$. Additivity
of the charge means that the action of $Q_{total}^{\mu\nu}$ on
$\CH\otimes\CH$
is given by
$$
   Q_{total}^{\mu\nu} \=
   Q_{\mu\nu}\otimes I + I \otimes Q_{\mu\nu}~.
$$
There are symmetries (but not Lie group symmetries) that do not
satisfy this condition, and hence are possible in non-trivial
physical theories.  Supersymmetry and quantum groups
(see chapter \ref{ch:qgroups}) are two
such possibilities.  The analogous analysis determines what
range of possible symmetries can arise in these generalized
cases.

\end{enumerate}

\newchapter{GENERAL PRINCIPLES OF CFT}
\label{ch:GP}

\vspace{-2cm}
\small
\begin{flushright}
\begin{minipage}{2.5in}
 \textsf{Big fleas have little fleas upon their backs to bite them,
 and little fleas have lesser fleas, and so ad infinitum.}
\rightline{\textsc{Swift}}
\end{minipage}
\end{flushright}
\normalsize

\vspace{2cm}

\vspace{2cm}

\footnotesize
\noindent {\bfseries References}:
It all started with \cite{BPZ}.
A standard reference for CFT is \cite{Ginz}. However, 
now there are a few books on the subject \cite{DiFMS,Ketov},
as well as a few shorter reviews for the impatient reader
\cite{Banks,BN,Dixon}. A recent review on CFT from an algebraic perspective
is \cite{Gaberdiel}.
One can also find a review of irrational CFTs,
which are not examined in the
present document, in \cite{HKOC}.
\normalsize

\section{BRIEF THEORY}

\subsection{Basic Notions of CFT}

The basic object of study of QFT
is the set of Green's functions of the local fields
of the theory.  These Green's functions or correlation
functions are denoted
$$
   \langle \CT( A_1(x_1)A_2(x_2)...A_n(x_n) ) \rangle ~. 
$$
Usually,
the time ordering operator $T$ is not written explicitly.
In the Langrangian 
approach, the above Green's functions can be expressed as functional
integrals
\bb
   \average{A_1(x_1)A_2(x_2)...A_n(x_n)} \=
   {1\over Z}\, \Big\lmoustache\, d\mu\lb\phi\rb \, 
    {A_1(x_1)A_2(x_2)...A_n(x_n)} \, e^{-{i\over\hbar}S\lbrack\phi\rbrack}~,
\ee
where 
$S\lbrack\phi\rbrack$ is the action for the theory, expressed in
terms of a set of fundamental fields
$\phi$. 

Any physical QFT must be translationally and rotationally invariant. 
Translational invariance results, via Noether's Theorem, in the 
energy-momentum conservation law
\beq
\label{emcons1}
   \partial^\mu T_{\mu\nu}=0~,
\eeq
where $ T_{\mu\nu}$ is the energy-momentum tensor defined by
\beq
    T_{\mu\nu} \equiv - {4\pi\over\sqrt{g}}
     \frac{\delta S}{\delta g^{\mu\nu}}~.
\label{eq:GP55}
\eeq
It can always be defined 
such that it is symmetric, with $ T_{\mu\nu}= T_{\nu\mu}$.
In general, QFTs are not scale invariant. Scale invariance
is restored at the points where the beta function $\beta(\lambda)$
of the coupling constant $\lambda$ vanishes.
At these points, the dilatation current $D_{\mu}=T_{\mu\nu}\,x^\nu$
is conseved due to scale invariance, i.e.,
$$
   \partial^\mu D_{\mu}=\partial^\mu\bigl(T_{\mu\nu}\,x^\nu\bigr)=0~.
$$
Since the energy-momentum tensor is itself conserved, 
this conservation law for $D_\mu$ in turn 
implies that the energy momentum tensor of the theory is traceless
\beq
\label{trace}
       \Theta\eq T^\mu_\mu\=0~.
\eeq
Applying condition \calle{trace} in conjunction with
the locality of the energy-momentum tensor
implies invariance under conformal transformations, too, that is,
invariance under coordinate
transformations $y^\mu=y^\mu(x)$ for which the metric is invariant up
to an overall factor $\Omega$:
$$
   ds^2\=g_{\mu\nu}(x)\,dx^\mu dx^\nu
   \=\Omega(y)\,g_{\mu\nu}(y)\,dy^\mu dy^\nu~.
$$
To see this, we define the conformal currents
$$
   J_\mu\eq T_{\mu\nu}\,f^\nu~,
$$
which obviously generalize the translation current $T_{\mu\nu}a^\nu$
and the dilatation current $T_{\mu\nu}x^\nu$. Clearly
$$
 \partial_\mu J^\mu\={1\over2}\, T^{\mu\nu}\, (\partial_\mu f_\nu+
            \partial_\nu f_\mu )~.
$$
In $d$ space-time dimensions, these currents will be conserved if
\bb
    \partial_\mu f_\nu+\partial_\nu f_\mu\= {2\over d}\, \eta_{\mu\nu}\, 
        \partial_\rho f^\rho~.
\ee
This is simply the conformal Killing equation,
which in two dimensions is solved by
\bb
   z~\rightarrow~ z'=f(z)~,~~~~~
   \overline z~\rightarrow~ \overline z'=\overline f(\overline z)~.
\ee
This decoupling of variables allows
one to handle the coordinates $z$ and $\overline z$ as independent; 
the reality condition $\overline z=z^*$ can be imposed at the end of the day. 

In  complex coordinates, the energy-momentum conservation law \calle{emcons1}
becomes
\beq
\label{emcons2}
   \overline\partial T_{zz}=0~,~~~~~
    \partial T_{\overline z\overline z}=0~,
\eeq
where $\partial=\partial_z,~
\overline\partial=\partial_{\overline z}$. 
Equation \calle{emcons2} implies that
\beq
\label{emcons3}
    T_{zz}=T(z)~,~~~~~
    T_{\overline z\overline z}=\overline T(\overline z)~,
\eeq
and thus one can expand the components of the energy-momentum tensor
in Laurent series
\bb
 T(z)&\=&\sum_{n=-\infty}^{+\infty}\,{L_n\over z^{n+2}}~,\\ 
 \overline T(\overline z)&\=&\sum_{n=-\infty}^{+\infty}\,
 {\overline L_n\over \overline z^{n+2}}~.
\ee
The coefficients $L_n$ and $\overline L_n$ satisfy two independent 
{\bfseries Virasoro algebras}:
\beqn
\label{Vira1}
 \lb L_n, L_m \rb &\=& (n-m)\,L_{n+m}\+{c\over 12}\,n(n^2-1)\,
 \delta_{n+m,0}~,\\ 
\label{Vira2}
 \lb \overline L_n, \overline L_m \rb &\=& 
   (n-m)\,\overline L_{n+m}\+{\overline c\over 12}\,n(n^2-1)\,
  \delta_{n+m,0}~,\\ 
\label{Vira3}
 \lb  L_n, \overline L_m \rb &\=& 0~. 
\eeqn
The numbers $c$ and $\overline c$ are called {\bfseries central charges}. 

The Operator Product Expansion (OPE) plays a central role in the study
of CFT. According to
the OPE assumption, in any QFT, the product of local operators
acting at points that are sufficiently close to each other may
be expanded in terms of the local fields of theory, as in
$$
    A_i(x)A_j(y)~\sim~\sum_k\, C_{ij}^k(x-y)\, A_k(y)~.
$$
This statement
is usually an asymptotic statement in QFT.
In  CFT though, it becomes an exact statement,
\beq
\label{eq:GP7}
  A_i(x)A_j(y)\=\sum_k\, C_{ij}^k(x-y)\, A_k(y)~.
\eeq
In qualitative terms, the reason this statement becomes exact 
is that 
scale invariance prevents the appearance of any length 
parameter $l$ in the theory.  Consequently, there is no
parameter to control the expansion, and thus terms like
$e^{-l/|x-y|}$ which would signal
the breakdown of the exactness of the expansion \calle{eq:GP7}
cannot arise. Note that in two dimensions
the l.h.s. of equation \calle{eq:GP7} is time-ordered:
\bb
  \CT(A_i(x)A_j(y))\=\sum_k\, C_{ij}^k(x-y)\, A_k(y)~.
\ee         
When using complex variables, the time-ordering becomes radial-ordering
as has been pointed out in section \ref{section:conftrans}:
\bb
  \CR(A_i(z,\overline z)A_j(w, \overline w))\=\sum_k\, 
  C_{ij}^k(z-w,\overline z-\overline w)\, A_k(w, \overline w)~.
\ee    
By abuse of the notation, it is customary to omit\footnote{We shall often use
this convention, although not always.} the symbol $\CR$.

One can translate the Virasoro algebra \calle{Vira1}-\calle{Vira3}
into OPEs:
\bb
 T(z)T(w)&=&{c/2\over (z-w)^4}+ {2T(w)\over (z-w)^2}
    +{\partial T(z_2)\over z-w}+{\rm reg}~,\\ 
 \overline T(\overline z)\overline T(\overline w)&=&
{\overline c/2\over (\overline z-\overline w)^4}
 + {2\overline T(\overline w)
 \over (\overline z-\overline w)^2}
    +{\overline \partial \overline T(\overline w)\over 
  \overline z-\overline w}+{\rm reg}
  ~,\\ 
  T(z)\overline T(\overline w)&=&
  {\rm reg}
  ~, 
\ee
where ``reg'' refers to non-singular terms that do not blow up as
$z\rightarrow w$.

One of the most exciting results of two dimensional
CFT is that one can classify all possible theories. 
To this end, we will introduce the notion of a primary field.
First, recall that a classical tensor $T_{z...z\overline z...\overline z}$ 
in two dimensions
would transform under a  change of variables as
$$
T_{\underbrace{z...z}_n\underbrace{\overline z...\overline z}_{\overline n}}
(z,\overline z)~\mapsto~
T'_{z...z\overline z...\overline z}(z',\overline z')\=
\lp \partial z\over\partial z'\rp^n\,
\lp \partial \overline z\over\partial \overline z'\rp^{\overline n}\,
T_{z'...z'\overline z'...\overline z'}(z,\overline z)~.
$$
Taking into account that in quantum field theory, fields may acquire
anomalous dimensions, we define a primary field to be a field
that under a change of variables obeys the transformation law
\beq
\label{primary}
\phi(z,\overline z)~\mapsto~
\phi'(z',\overline z')\=
\lp \partial z\over\partial z'\rp^\Delta\,
\lp \partial \overline z\over\partial \overline z'\rp^{\overline \Delta}\,
\phi(z,\overline z)~.
\eeq
The real numbers $\Delta$ and $\overline \Delta$
are known, respectively, as the {\bfseries left} and {\bfseries right conformal
weights} of the field $\phi$. In particular, $D=\Delta+\overline \Delta$ is the usual
anomalous scale dimension of $\phi$, and $s=\Delta-\overline\Delta$ is the spin
of $\phi$.

The decoupling of the  holomorphic and anti-holomorphic degrees of
freedom will allow us frequently to concentrate on the holomorphic 
part of the theory;
however, we must keep in mind that the full theory also has an
anti-holomorphic part, with analogous properties.

The OPE of a primary field \calle{primary} with the energy-momentum tensor 
is
\bb
 T(z)\phi(w,\overline w)= {\Delta\over (z-w)^2}\,\phi(w,\overline w)
    +{\partial_w \phi(w,\overline w)\over z-w}+{\rm reg}~.
\ee

By a simple inspection of the $T(z)T(w)$ OPE, we can immediately conclude that
the energy-momentum tensor is not a primary field, due to the extra term with
the central charge. Only when $c=0$ 
is $T(z)$ is a primary field,  with weight 2.
The extra term in the OPE of the energy-momentum tensor with itself induces
a corresponding term in the `tensorial' transformation law of $T(z)$:
\bb
 T(z)~\mapsto~ T'(w)\=
 \lp \partial z\over\partial w\rp^2\, T(z) +{c\over12}\, S(w,z)~.
\ee
This equation may be considered as the definition of the anomalous term
$S(w,z)$ which can be found to be
\bb
   S(w,z)\= {z'''\over z'}-{3\over2}\,\lp {z''\over z'}\rp^2~,
\ee
where $z'=dz/dw$. This expression is known in the mathematics 
literature as the {\bfseries Schwarzian derivative}.

The vacuum $\ket{\emptyset}$ of any CFT satisfies
\bb
   L_n\ket{\emptyset}\=0~,~~~n\ge -1~.
\ee
For a field $\phi$ with conformal weight $\Delta$, we also define the state 
\bb
    \ket{\Delta}\eq\phi(0)\,\ket{\emptyset}~,
\ee
which has the properties
\bb
   L_0\ket{\Delta}\=\Delta\, \ket{\Delta }~,~~~~~L_n\ket{\Delta}\=0~,~~~n>0~.
\ee
The space of states
\bb
 \lb\phi\rb\eq\{ ~L_{-n_1}L_{-n_2}...L_{-n_l}\ket{\Delta}~,~
   n_1\ge n_2\ge ...\ge  n_l>0~ \}~
\ee
is known as the {\bfseries Verma module} built over the field $\phi$.
The state $L_{-n_1}L_{-n_2}...L_{-n_l}\ket{\Delta}$ is called 
a {\bf descendant}\index{field!descendant --} of $\ket{\Delta}$, 
and the quantity $\sum_{j=1}^l n_j$ 
is the {\bf level}\index{level} of such a descendant.
Obviously, the Virasoro algebra maps the  Verma module $\lb\phi\rb$ into
itself; this implies that  this space is
a representation space of the Virasoro algebra.

\subsection{Massless Free Boson}

The simplest and one of the most
important examples of CFT is the massless free boson, for
which the action is
\beq
\label{boson}
   S={g\over 4\pi}\,\iint\,d^2z\, \partial\Phi\overline\partial\Phi~,
\eeq
where $g$ is a normalization parameter. (Often we will take $g=1/2$.)
The propagator is the solution of the equation
\bb
  \overline\partial\partial\average{\Phi(z,\overline z)
  \Phi(w,\overline w)}\= -{2\pi\over g}\,\delta^{(2)}(z-w)~.
\ee
Therefore,
\beq
\label{propa2}
  \average{\Phi(z,\overline z)
  \Phi(w,\overline w)}\= -{1\over 2g}\,\ln|z-w|^2~.
\eeq
The free boson can be decomposed into a holomorphic part and
an anti-holomorphic part, producing the decomposition
$$
      \Phi(z,\overline z)\= \phi(z)\+\overline \phi(\overline z)~.
$$
The result \calle{propa2} for the propagator immediately implies
\beqn
\label{Prop1}
  \langle \phi(z)
  \phi(w) \rangle &=& -{1\over 2g}\,\ln(z-w)~,\\
\label{Prop2}
  \average{\overline\phi(\overline z)
  \overline\phi(\overline w)}&=&-{1\over 2g}\,\ln(\overline z-\overline w)~,\\
  \average{\partial_z\phi(z)
  \partial_w\phi(w)}&=& -{1/2g\over (z-w)^2}~,\nonumber \\
  \average{\partial_{\overline z}\overline \phi(\overline z)
  \partial_{\overline w}\overline \phi(\overline w)}&=& -{1/2g\over 
   (\overline z-\overline w)^2}~,\nonumber\\
  \average{\partial_z\phi(z)
  \partial_{\overline w}\overline\phi(\overline w)}&=& 
  {\pi\over 2g}\,\delta^{(2)}(z-w)~. \nonumber
\eeqn

The energy-momentum tensor for the massless free boson is 
\beq
\label{bT}
  T(z)\=-g\,:\partial\phi
  \partial\phi:~.
\eeq
From this, it follows that
\beq
\label{bTT}
  T(z)T(w)\={{1/ 2}\over (z-w)^4}+
  {2T(w)\over (z-w)^2}+
  {\partial T(w)\over z-w}+{\rm reg}~.
\eeq
The central charge for the massless free boson is $c=1$.
One can also easily calculate that
\bb
  T(z)\partial_w\phi(w)\={1\over (z-w)^2}\,\partial_w\phi+
  {1\over z-w}\, \partial_w (\partial_w\phi)+{\rm reg}~,
\ee
i.e., the free boson field acquires no anomalous scale dimension.

Other interesting  primary operators are the normal ordered
exponentials 
\beqn
\label{vert1}
      V_\alpha(z)&\=&:e^{i\alpha\phi(z)}:~,
\eeqn
usually referred to as {\bfseries vertex operators}.
One can easily derive the OPE
\beq
\label{TV}
  T(z)V_\alpha(w)\={{\alpha^2}\over (z-w)^2}\,V_\alpha(w)+
  {1\over z-w}\, \partial_w V(w)+{\rm reg}~.
\eeq
In other words, the conformal weights of the vertex operator 
\calle{vert1}
are $(\alpha^2/4g,0)$.

\subsection{Massless Free Fermion}

Another important example of CFT is the  massless free 
Majorana-Weyl fermion, for which the action is
\beq
\label{fermion}
   S={\lambda\over 2\pi}\,\iint\,d^2z\, \Psi\slash\partial\Psi~,
\eeq 
where $\lambda$ is a normalization parameter (which we often we will set
equal to 1).
The 2-dimensional  spinor $\Psi$ has two components:
$$
   \Psi\=\lb\matrix{\psi\cr\overline\psi}\rb~.
$$
The action can then be written in the equivalent form
\beq
\label{fermion2}
   S={\lambda\over 2\pi}\,\iint\,d^2z\,
  \lp \psi\overline\partial\psi
      + \overline\psi\partial\overline\psi\rp~.
\eeq
The propagator is the solution of the equation 
\bb
  \overline\partial\average{\psi(z)\psi(w)}&=&{2\pi\over\lambda}\,\delta(z-w)~, \\
  \partial\average{\overline\psi(\overline z)\psi(\overline w)}&=&
   {2\pi\over\lambda}\,\delta(\overline z- \overline w)~.
\ee
The result for the propagator is:
\bb
  \langle \psi(z)
  \psi(w) \rangle &=& {1/\lambda\over z-w}~,\\
  \average{\overline\psi(\overline z)
  \overline\psi(\overline w)}&=& {1/\lambda\over \overline z-\overline w}~.
\ee

The energy-momentum tensor for the massless free fermion is 
\bb
  T(z)\={\lambda\over 2}\,:\psi
  \partial\psi:~.
\ee
From this it follows that
\bb
  T(z)T(w)\={{1/ 4}\over (z-w)^4}+
  {2T(w)\over (z-w)^2}+
  {\partial T(w)\over z-w}+{\rm reg}~.
\ee
The central charge for the massless free fermion is $c=1/2$.
One can also easily verify the OPE result
\bb
  T(z)\psi(w)\={1/2\over (z-w)^2}\,\psi(w)+
  {1\over z-w}\, \partial_w\psi(w)+{\rm reg}~.
\ee
Like the free boson, the free fermion field acquires no anomalous
 scale dimension.

\subsection{The $bc$-System}

A generalization of the free fermion system is the so-called
{\bfseries $bc$-system}\index{$bc$-System}
which includes two anticommuting fields $b(z,\overline z)$ and 
$c(z,\overline z)$
of weights $j$ and $1-j$, respectively, so that
\bb
   T(z)b(w)\= {j\, b(w)\over (z-w)^2}+{\partial_w b(w)\over z-w}+\mbox{reg}~,\\
   T(z)c(w)\= {(1-j)\, c(w)\over (z-w)^2}+{\partial_w c(w)\over z-w}+\mbox{reg}~.
\ee
The action for the $bc$-system is
\beq
\label{eq:GP6}
   S={1\over 2\pi}\,\iint\,d^2z\,
  \lp b\overline\partial c
      + \overline b\partial\overline c\rp~.
\eeq
The holomorphic propagators of the theory are then
\bb
  \average{b(z) c(w)}  &=& {1\over z-w}~,\\
  \average{c(z) b(w)}  &=& {1\over z-w}~.
\ee
The energy-momentum tensor is given by 
\bb
  T(z)\= -j\, :b(z)\, \partial c(w): + (1-j)\,:\partial b(z) \, c(z):~,
\ee
from which one finds
\bb
  T(z)T(w)\={{-2(6j^2-6j+1)/2}\over (z-w)^4}+
  {2T(w)\over (z-w)^2}+
  {\partial T(w)\over z-w}+{\rm reg}~.
\ee
The central charge for the $bc$-system  is
$$
      c\= -2\, (6j^2-6j+1)~.
$$

A variation of the  $bc$-system is the 
{\bfseries $\beta\gamma$-system}\index{$\beta\gamma$-System},
which includes two commuting fields $\beta(z,\overline z)$ and 
$\gamma(z,\overline z)$
of weights $j$ and $1-j$, respectively, that obey 
the action:
\beq
\label{eq:GP12}
   S={1\over 2\pi}\,\iint\,d^2z\,
  \lp \beta\overline\partial \gamma
      + \overline \beta\partial\overline \gamma\rp~.
\eeq
Since the `statistics' of the theory has changed, some of the signs
in various formul\ae\  will be reversed.
In particular, the holomorphic propagators of the theory are
\bb
  \average{\beta(z) \gamma(w)}  &=& {1\over z-w}~,\\
  \average{\gamma(z) \beta(w)}  &=& {-1\over z-w}~.
\ee
As a result, the central charge for the $\beta\gamma$-system is
$$
      c\= 2\, (6j^2-6j+1)~.
$$

\subsection{Boson with Background Charge}

An interesting modification of the free boson theory is a model
that includes a coupling between the boson and the scalar 
curvature\footnote{Notice
that the normalization factor
has nothing to do with the determinant of the metric tensor which is
also typically denoted by $g$.  Here, we are momentarily denoting
this determinant as $\tilde g$ to avoid confuction.}:
$$
   S\= {g\over4\pi}\,\iint\, d^2z\, \partial\Phi\overline\partial\Phi
     +{ie_0\over 4\pi}\, \iint\, d^2z\,\sqrt{\tilde g} R\, \Phi~.
$$
Incidentally, notice that this coupling is non-minimal.
The constant $e_0$ is known as the
{\bfseries background charge}\index{Charge!Background --}.

The (holomorphic) energy-momentum tensor for this theory reads:
\beq
\label{aT}
  T(z)\=-g\,:\partial\phi
  \partial\phi:\+ ie_0\partial^2\phi~,
\eeq
Such a total derivative  will not affect the status of $T(z)$ as a 
generator of
conformal transformations, but it does modify the value of the central
charge, which is now
\beq
\label{ceff}
   c\=1-{6\over g}\, e_0^2~.
\eeq

It also modifies the conformal weights of the vertex operators
 \calle{vert1}; the new 
weights are
\beq
\label{eq:GP31}
    \Delta_\alpha\= {\alpha(\alpha-2e_0)\over 4g}~.
\eeq

 Consistency requires
\bb 
     V_\alpha^\dagger\= V_{2e_0-\alpha}~,
\ee
such that
$$
  \average{ V_\alpha^\dagger(z) V_\alpha(w)}\={1\over (z-w)^{2\Delta_\alpha}}
  ~.
$$

The introduction of the background charge modifies the transformation
properties of the boson such that it is consistent with \calle{eq:GP31}.
In particular,
$$
  \phi(z) ~{\buildrel z\mapsto w \over \mapsto} ~ \phi(w)+{ie_0\over 2g}\,
  \ln{dw\over dz}~,
$$
and so
\beq
\label{eq:GP32}
  e^{i\alpha\phi(z)}  ~{\buildrel {z\mapsto w} \over \mapsto} ~
  \lp{dw\over dz}\rp^{\alpha^2/4g}\,  e^{i\alpha\phi(w)}\,
  e^{-{\alpha e_0\over 2g}\ln{dw\over dz}}\=
  \lp{dw\over dz}\rp^{\alpha(\alpha-2e_0)\over 4g}\, e^{i\alpha\phi(w)}~.
\eeq

\subsection{Minimal Models of CFT}

An important subclass of CFTs is that of the so-called 
{\bfseries Minimal Models} (MMs).
These models are characterized
by two relatively prime positive integers $r$ and $s$,
and the model corresponding to
a particular pair of such integers is denoted
$M_{rs}$. Each such model has only a finite number of primary fields $\phi_{mn}~,
~1\le m \le r-1,~1\le n \le s-1.$ The corresponding 
conformal weights are given by
\beq
\label{weights}
     \Delta_{mn}\={(mr-ns)^2-(r-s)^2\over 4rs}~.
\eeq
The central charge of such a model is
\beq
\label{cMM}
     c\=1-{6(r-s)^2\over rs}~.
\eeq
Not all the MMs are unitary; some contain negative norm states. 
Calculating the norm of the the state $L_{-1}\ket{\phi}$,
we see that
\beqn
\label{norm}
  ||L_{-1}\ket{\phi}||^2&\=&\bra{\phi}L_{-1}^\dagger\,L_{-1}\ket{\phi}\=
  \bra{\phi}L_1\,L_{-1}\ket{\phi} \nonumber\\
  &\=&\bra{\phi}\lb L_1\, ,\,L_{-1}\rb\ket{\phi}\= 
  \bra{\phi}2L_0\ket{\phi}\= 2\Delta ||\ket{\phi}||^2~,
 \nonumber
\eeqn
where we used the fact that $L^\dagger_{-n}=L_n$.
Therefore, for any CFT (not just a minimal model) to be
unitary, it must only have primary fields with positive conformal weights.
Among the MMs, then, the only unitary models are those with
$$
          s=r+1~.
$$
The central charge of such models is thus
$$
     c\=1-{6\over r(r+1)}~,~~~~~r=3,4,5,\dots~.
$$
We call these models the Unitary Minimal Models (UMMs).

\subsection{A Prelude to Chapters \ref{ch:PT} and \ref{ch:STAT}}

We have pointed out that the left and right sectors of a CFT are studied
independently. There are several reasons\footnote{For example,
requiring cancellation of local gravitational anomalies in order to
allow a system to consistently couple to 2-dimensional gravity
translates to the condition $c=\overline c$ \cite{AlvWit}.} to impose
the condition $c=\overline c$; here, though, we will simply use
the condition, rather than justifying it.

The first model of the UMMs has $(r,\overline r)=(3,3)$, or
$(c, \overline c)=(1/2,1/2)$. This model has three primary operators
$\Phi_{11}$, $\Phi_{21}$, and $\Phi_{12}$ with conformal weights (0, 0),
(1/2, 1/2), and (1/16, 1/16), respectively.

CFT provides a classification of models according to
critical behavior in 2-dimensions, and
so one can identify the particular CFT model that matches
the critical behavior of a known statistical mechanical system.

Consider a mechanical system with an order parameter $\sigma$. For high
temperature, the 2-point correlation function of this field falls off 
exponentially
\beq
  \average{\sigma(z,\overline z)\sigma(0,0)}\propto r^{-r/\xi}~,
\label{eq:GP91}
\eeq
where $z=re^{i\phi}$, and $\xi$ is called the correlation length, 
which is a function of the temperature $T$. 
The quantity $m=\xi^{-1}$ has the
dimensions of mass, and it is a measure of how far the system is
from criticality (where $m=0$). When $T$ equals a critical temperature
$T_c$, the correlation function \calle{eq:GP91} takes the form
$$
 \average{\sigma(z,\overline z)\sigma(0,0)}\propto r^{-\eta}~,
$$
where $\eta$ is called the
{\bf critical exponent\index{Exponent!Critical --} of the order parameter}. 
Similarly, there are critical exponents for other quantities of the
system.

The Ising system is defined on a lattice such that at each site $i$ of the
lattice there is a spin variable $\sigma_i$ taking values in $\{1,-1\}$.
The Hamiltonian of the system is
$$
  H\=h\, \sum_i \sigma_i -J\, \sum_i \sigma_i \sigma_{i+1}~.
$$
In the continuum limit, it takes the form of a free Dirac fermion
$$
  H\=\int dx\lb\Psi i\gamma_5{d\Psi\over dx} + m\,\overline\Psi\Psi
     +\mbox{h.c.}\rb~.
$$
The critical exponent $\eta$ for the 2-dimensional Ising model is
$\eta=1/4$, i.e.
$$
 \average{\sigma_i\sigma_j}\propto |i-j|^{-1/4}~.
$$
We notice the similarity with the 2-point correlation function 
$\average{\Phi_{12}\Phi_{12}}$ of the UMM(3). This is an indication
that the UMM(3) might describe the critical behavior of the Ising model.
To draw this conclusion, one must find a complete map for a rest
quantities
of the two models. This can, in fact, be done, establishing the
desired identification. In
particular, the energy operator of the Ising model $\varepsilon_i$
has a 2-point correlation function
$$
 \average{\varepsilon_i\varepsilon_j}\propto |i-j|^{-2}~,
$$
and therefore it can be identified with the primary operator $\Phi_{21}$
in
the UMM(3).

\begin{table}[h!]
\begin{center}
\begin{tabular}{|c|c||c|c|}\hline\hline
\multicolumn{2}{|c||}{UMM(3)}&\multicolumn{2}{|c|}{Ising Model}\\ \hline\hline 
$\Phi_{11}$ & (11)-primary operator & $1$ & identity \\
\hline
$\Phi_{12}$ & (12)-primary operator & $\sigma$ & spin variable \\
\hline
$\Phi_{21}$ & (21)-primary operator & $\varepsilon$ & energy variable \\
\hline
\end{tabular}
\end{center}
\caption{The identification of the critical behavior of the Ising model
with the UMM(3).}
\end{table}

A final comment is in order here. At the critical point the free fermion
becomes massless with energy operator
$\varepsilon\sim\psi(z)\overline\psi(\overline z)$. Moving away from
criticality is equivalent to adding a perturbation $\Phi_{pert}=\delta
m\, \psi(z)\overline\psi(\overline z)=\delta m\, \Phi_{21}$ to the
kinetic term (conformal action).

The Ising model is, of course, only a single case among many models for which
the critical behavior has been succesfully identified with a particular
CFT; some examples are listed in the following table.

\begin{table}[h!]
\begin{center}
\begin{tabular}{|c|c|}\hline\hline
 {\bf model} & $c$ \\ \hline\hline
 Ising & 1/2 \\ \hline
 tricritical Ising & 7/10 \\ \hline
 3-state Potts & 4/5 \\ \hline
  tricritical 3-state Potts & 6/7 \\ \hline
\end{tabular}
\end{center}
\caption{Some statistical models  for which the critical behavior has been
identified with particular CFTs given by the value of the central charge.}
\end{table}

\newpage \

\newpage
\section{EXERCISES}

\begin{enumerate}

\item
Consider a classical CFT on a Riemann surface $M$ that has genus $h$
and no boundary. Usually,
we study tensorial objects
transforming as
$$
   T(z) \mapsto 
     \lp{dz\over dw}\rp^m\,
 T(z)~, ~~~~~m\in\BZ~.
$$
Is it possible to define objects that transform in a similar way, but
 with fractional values of $m$,
i.e. $m\in\BQ$?

\item
Show that if $\phi(z)$ is a primary field, then in general
$\partial_{z}\phi(z)$ is not primary.  Is there any exception?

\item
Consider a possible central extension of the classical Virasoro algebra
\calle{classicalL}
\beq
\label{eq:GP8}
 \lb L_n,L_m\rb\=(n-m) \, L_{n+m}+c_{nm}~~,
\eeq
where
$$
 \lb c_{nm}\, , \, L_{r}\rb \=0 ~, ~~~\forall n,m,r~.
$$

(a) From the Jacobi relations deduce the constraints on $c_{nm}$.

(b) Use the freedom to shift the $L_n$'s by a constant (i.e., 
to make the change
$ L_n\rightarrow L_n +a_n $)
to set $c_{n,0}=c_{0,n}=c_{1,-1}=0$.

(c) Find the most general solution for $c_{nm}$ satisfying the conditions
in part (b).

\item
\label{item:GP2}
The Kac formula predicts a null state of dimension $\Delta_{1,3}+3$ in the
module of highest weight $\Delta_{1,3}$. Find the explicit form of this
null state as a descendant, i.e., in the form
$$
 \ket{\chi}\=(L_{-3}\,+\cdots)\,\ket{\Delta_{1,3}}~.
$$

\item
For the free boson theory described by the action \calle{boson},
show that the Euclidean propagator is
$$
 \langle \,\Phi(z,\overline{z})\,\Phi(0)\,\rangle\=-{1\over 2g}\,\ln|z|^{2}~.
$$

\item
\label{item:GP1}
For the free boson theory, show that: 
\bb
 \CR \lp \partial_{z}\phi(z)\,:e^{i\alpha\phi(w)}:\rp
  \= : \partial_{z}\phi(z)e^{i\alpha\phi(w)}:
      \- \frac{i\alpha}{2g}\,\frac{1}{z-w}\,
     :e^{i\alpha\phi(w)}: ~.
\ee
Generalize the above identity to the case
$$
  \CR \lp (\partial_{z}\phi)^k \,:e^{i\alpha\phi(w)}: \rp~.
$$

\item
Prove the following identity for a free boson:
\beq
 \CR \lp :e^{i\alpha\phi(z)}: ~:e^{i\beta\phi(w)}:\rp \=
   (z-w)^{\alpha\beta/2g}
    ~  :e^{i\alpha\phi(z)}
    e^{i\beta\phi(w)}: ~.
\eeq
\item
For the free boson theory, the scalar field
can be expanded as
$$
  i\partial_{z}\phi\,=\,\sum_{n}\alpha_{n} \, z^{-n-1}~~,~~~
 n \, \in \, \BZ ~.
$$
Compute the commutation relations of
the modes $\alpha_{n}$.

\item
\label{item:GP4}
Construct the Virasoro generators $L_n$'s in terms of the modes
$\phi_n$ for the bosonic action \calle{aT}.

\item
\label{pr:compactboson}
It is often relevant to consider the CFT of a free boson
{\bf compactified on a circle} of radius $r$, so that
$$
   \phi(\sigma+\beta, t)
  \sim \phi(x, t) + 2\pi r\, w~, ~~~w\in\BZ~.
$$
This relation is to be interpreted as follows. If the spatial
coordinate is compact (i.e. a circle of length $\beta$),
then the boson field is a multi-valued function with its
values at any one point differing by an amount $2\pi r$.

Discuss the CFT of the compactified boson.

\item
For the free fermion theory
$$
  S\= {\lambda\over 2\pi}\, \iint\, d^2z \, (\psi\overline\partial\psi
  +\overline\psi\partial\overline\psi)~,
$$
show that the Euclidean propagator is
$$
   \average{\psi(z)\psi(w)}\={1\over\lambda}\, {1\over z-w}~.
$$

\item
In the free fermion theory, consider expanding 
the fermion field as
$$
 \psi(z)\,=\,\sum_{r}\psi_{r} \, 
 z^{-r-1/2}~~,~~~r\,\in \, \BZ + 1/2~.
$$

Compute the anticommutation relations of
the modes $\psi_{r}.$

\item
Let $\psi(z)$ be a dimension 1/2 free fermion field. Compute the first 
regular term $L_{-2}(\psi(z))$ (i.e., express it in terms of the
field $\psi$) in the OPE
$$
 {\cal R}(T(w)\psi(z))\=\frac{1}{2}\frac{\psi(z)}{(w-z)^2}+
 \frac{\partial_{z}\psi(z)}{w-z}+L_{-2}(\psi(z)) +\dots
$$
two different ways:

(a) using 
$T(w) = \frac{1}{2} :\psi(w) \partial_{w} \psi(w):$ and calculating
the OPE explicitly; and

(b) using the null state predicted by the Kac formula.

\item
We can treat the $bc$-system and the $\beta\gamma$-system in a unified
fashion by defining a quantity  $\varepsilon$ (which we call the
{\bfseries signature}\index{Signature})
that takes the value
$+1$ for a  $bc$-system  and the value $-1$ for a $\beta\gamma$-system.
Then the generic $BC$-system
$$
    S\={1\over 2\pi}\, \iint d^2z\,
    (B\overline\partial C 
    +\overline B\partial\overline C)~,
$$
can give us either a $bc$-system or a $\beta\gamma$-system, depending
on whether we have commutating or anticommuting fields (with $\epsilon$
identifying which situation we have).

Using this unified notation, compute, for both the $bc$-system
and the $\beta\gamma$-system,

(a) the energy-momentum tensor $T(z)$; and 

(b) the central charge $c$.

\item
{\bf A puzzle}: From the OPE $\CR(T(z)T(w))$ show that
$$
 \average{\partial_z\Theta(z)T(w)}\={2c\pi\over3}\,
 \partial_z^2\partial_w\delta(z-w)~.
$$
This implies that $\Theta(z)$ cannot vanish identically for a CFT.
However, recall that if  $\Theta(z)$ is not identically zero the theory
cannot be conformally invariant. How can these two statements both be
true?

\item
{\bf The Liouville action}: In two dimensions, dimensional arguments 
require that the trace of the energy-momentum tensor of a theory $S$ on a
Riemann surface with scalar curvature $R$ be given by
\beq
\label{eq:traceanomaly}
  \Theta \= a\, R +b~,
\eeq
where $a$ and $b$ are two constants (which can be computed exactly using well-known
methods of QFT).
The non-vanishing of this trace is known as the 
{\bf conformal anomaly}\index{anomaly!conformal --}
(or {\bf trace anomaly}\index{anomaly!trace --}).

Using equation \calle{eq:traceanomaly}, find an action 
$S_L=S\lb\sigma\rb$, such that the energy-momentum tensor of the
action 
$$
   S'\= S+S_L
$$
has vanishing trace. The field $\sigma$ is called the
{\bf Liouville field}\index{field!Liouville --} and $S_L$ is called
the {\bf Liouville action}\index{action!Liouville --}.

\item
\label{item:CNM1}
For a boson on a curved two-dimensional manifold $M$ with action
$$
   S\= \alpha\, \iint_M\, d^2\xi\, \sqrt{g}g^{ab}\partial_a\Phi
       \partial_b\Phi
       +\beta\,\iint_M\, d^2\xi\, \sqrt{g} R\, \Phi~,
$$
compute the energy-momentum tensor as  defined by the equation
\calle{eq:GP55}.

\end{enumerate}

\newpage
\section{SOLUTIONS}

\begin{enumerate}

\item
The solution of this problem requires the use of some topology --- in 
particular, it makes use of the Chern (characteristic) classes for
the Riemann surface $M$.

The integral of the tangent bundle over $M$ gives
the Euler characteristic 
$$
     \chi(M) \= 2\,(1-h)\=\int_M \, c_1(TM)~.
$$
The analogous integral of any other line bundle $E$ over $M$
must be an integer, as well,
$$
     \int_M \, c_1(E) \= n~.
$$

Recall that a tensor on a manifold $M$ is a section of
the tangent bundle $TM$ (a holomorphic tangent  bundle in
our case). Higher order tensors are sections of other
bundles. If an object  $T(z)$ belongs to a line
bundle $E$  of $M$ formed as the $m$-th power of
the holomorphic tangent bundle,
\bb
  T(w) \=
     \lp{dz\over dw}\rp^m\,  T(z)~, ~~~m\in\BQ~, 
\ee
then
$$
   c_1(E) \= m \, c_1(TM)~.
$$
Integrating this relation over $M$, we find a relation that
restricts the allowed values of $n$:
$$
   n \= 2\,m\,(1-h)~,
$$

Notice that, consequently,
the line bundle with $m=1/2$ is always acceptable.
This line bundle (which is, loosely speaking, the square root
of $TM$) is called the 
{\bf canonical bundle}\index{Bundle!Canonical --}  of $M$,
and it gives rise to fermions on the surface $M$.

For other values of $m$, the corresponding bundle  can be constructed only
in special cases, namely if $h-1$ is a multiple of $m$. 
For example, for $m=1/(2k)$~, $k\in\BZ$, the bundle $E$ can be constructed
only on Riemann surfaces for which $h-1$ is a multiple of $k$.

\separator

\item
Using the relation
$$
 T(z)\phi(w)\=\frac{\Delta}{(z-w)^{2}}
 \, \phi(w)+ \frac{1}{z-w}\,\partial_{w}
 \phi(w) + \mbox{reg}  
$$
for the OPE of a primary field, we find
$$
 T(z)\partial_{w}\phi(w)\= \frac{2\Delta}{(z-w)^{3}} \,\phi(w)\,+\,
 \frac{\Delta+1}{(z-w)^{2}}\,\partial_{w}\phi(w) \,+\, \frac{1}{z-w}
 \,\partial_{w} (\partial_{w}\phi(w))  + \mbox{reg} ~.
$$
A piece proportional to $\phi/(z-w)^{3}$ has appeared in the OPE, and
therefore $\partial_{w} \phi$ is not primary. However, notice that if $\phi$
is a primary field of conformal weight zero, that is with $\Delta=0$, 
then this problematic term drops out, and
then in this case $\partial_{w}
\phi$ {\it is} a primary field of conformal weight 1. (As an added
exercise, compare 
this with what you know about the free massless boson.)

\separator

\item
(a) We start
by imposing the Jacobi identity for $L_l$, $L_m$, and $L_n$:
$$
 \lb L_l,\lb L_m, L_n\rb\rb 
 +\lb L_n,\lb L_l,L_m\rb\rb+\lb L_{m},\lb L_{n},L_{m}\rb\rb
 \=0~~.
$$
Substituting 
the proposed algebra \calle{eq:GP8}, we obtain the equation
\beqn
 (m-n)(l-m-n)L_{l+m+n}\,+\,(m-n)c_{l,m+n} & & \nonumber \\
 +\,(l-m)(n-l-m)L_{l+m+n}\,+\,(l-m)c_{n,n+l} & & \nonumber \\
 + \, (n-l)(m-n-l)L_{l+m+n}\,+\,(n-l)c_{m,n+l}& = & 0~.
\label{eq:GP9}
\eeqn
This is the main restriction on the form of $c_{nm}$. However,
 since
$$
 \lb L_{n},L_{m}\rb \=-\lb L_{m},L_{n}\rb~,
$$
there is an additional  constraint on the $c$'s, which is
$$
 c_{mn}\=-c_{nm}~.
$$

(b) Our algebra is isomorphic to another one, which we find
by making the 
transformation
$$
 L'_{n}\= L_{n}\,+\,\alpha_{n}~,
$$
for any $n$.  We will choose specific values for the constants
$\alpha_{n}$ as we proceed. Explicitly,
$$
 \lb L'_{n},L'_{m} \rb \=(n-m)L'_{n+m}\,+\,c'_{n+m} ~,
$$
where
\beq
\label{eq:GP10}
 c'_{nm}\=c_{nm}-(n-m)\alpha_{n+m}~.
\eeq
We wish to set $c'_{-1,1}=0$ and $c'_{n0}=c'_{0n}=0$.
To set $c'_{n0}=0$, we must simply choose
\bb
\alpha_n = \frac{1}{n}c_{n0}~,~~~~~n\ne 0~.
\ee
Of course, $c'_{00}=0$ automatically, and once $c'_{n0}=0$,
then $c'_{0n}=0$.

We still have $\alpha_0$ to specify, and we can use this
freedom to set $c'_{-1,1}=0$; this is achieved by choosing
\bb
\alpha_0 = \frac{1}{2}c_{-1,1}~.
\ee

Thus, by setting
\bb
 \alpha_n &=& \frac{1}{n}\,c_{n0}~,~~~~~n\ne 0~,\\
 \alpha_0 &=&  \frac{1}{2}\,c_{1,-1}~,
\ee
the algebra of the $L'_{n}$ has $c'_{n0}=c'_{0n}=0~~\forall n$
and $c'_{-1,1}=0$.

(c) For $l=0$, \calle{eq:GP9} applied to the $L'_{n}$ gives
$$
 (m-n)\,c'_{0,m+n}\,-\,(n+m)\,c'_{nm}\=0~.
$$
This is true if and only if
$$
 (n+m)\,c'_{nm}\=0~.
$$
Therefore
$$
 c'_{nm}\=0~, ~~~n+m\ne 0~,
$$
while $c'_{nm}$ is left undetermined if $n+m=0$. We can thus write
$$
 c'_{nm} \eq c_{n}\,\delta_{n,-m} ~.
$$
Now if we consider the case $l=-(m+1)$ and $n=1$, \calle{eq:GP9}
yields
$$
 (m-1)\,c'_{-(m+1),(m+1)}\,+\,(-2m-1)\,c'_{1,-1}\,+\,(m+2)\,c'_{m,-m}\=0~.
$$
This in turn yields the recursion relation
$$
 c_{m+1}\= \frac{m+2}{m-1}\,c_{m}~,
$$
which implies that
$$
 c_{m+1}\=c_2\, \prod_{k=2}^m\,\frac{k+2}{k-1}\=
 \frac{m(m+1)(m+2)}{6}\,c_2~.
$$
Defining $c \equiv 2c_2$, we can write
$$
 c_{m} \,=\, \frac{m(m^{2}-1)}{12} \,c~.
$$
Our final result is thus
$$
 c'_{nm}\,=\,\frac{m(m^{2}-1)}{12}\,c\,\delta_{n,-m}~.
$$

\separator
\item
At level 3, the descendant states are:
\begin{eqnarray*}
 \ket{i} & \, \equiv \, & L_{-3} \, \ket{ \Delta }\\
 \ket{j} & \, \equiv \, & L_{-1} \, L_{-2} \, \ket{ \Delta }\\
 \ket{k} & \, \equiv \, & L_{-1}^{3} \, \ket{ \Delta }\\
\end{eqnarray*}
Let $\ket{\chi}$ be a null state constructed from them:
\beq
\label{eq:GP11}
 0 \,=\, \ket{\chi }\,\equiv \,\alpha\, \ket{i}\,+\,\beta \,\ket{j} \,+\,
 \gamma \, \ket{k}~.
\eeq
One of the coefficients $\alpha$, $\beta$, 
and $\gamma$ is irrelevant, but we
are going to keep all three of them just for {\ae}sthetic reasons! 
By scalar multiplication
of \calle{eq:GP11} with $\ket{i}$, $\ket{j}$, and $\ket{k}$,
respectively, we get the following
system of equations:
\beq
\label{eq:dasGP11a}
 \left.
 \begin{array}{r}
 \alpha \, \expectation{i}{i} +
 \beta \, \expectation{i}{j} +
 \gamma \, \expectation{i}{k}
   \=0\\
 \alpha \, \expectation{j}{i} +
 \beta \, \expectation{j}{j} +
 \gamma \, \expectation{j}{k}
   \=0\\
 \alpha \, \expectation{k}{i} +
 \beta \, \expectation{k}{j} +
 \gamma \, \expectation{k}{k}
   \=0
 \end{array} \ra
\eeq
Notice that we have a homogeneous linear system to be solved for
$\alpha$, $\beta$, and $\gamma$. Non-trivial solutions exist if the
determinant of the coefficients is zero:
$$
 D\eq \left| \begin{array}{ccc}
 \expectation{i}{i} & \expectation{i}{j} & \expectation{i}{k}\\
 \expectation{j}{i} & \expectation{j}{j} & \expectation{j}{k}\\
 \expectation{k}{i} & \expectation{k}{j} & \expectation{k}{k}\\
 \end{array} \right| \= 0~.
$$
It is easy to calculate the elements of the above matrix using the
Virasoro algebra commutators,
$$
 \lb L_{n}, \, L_{m} \rb \,=\, (n \,-\, m) \, L_{n+m} \,+\,
 \frac{c}{12} \, n \, (n^{2} \,-\,1) \, \delta_{n,-m} ~.
$$
One finds the inner products take the following forms:
\bb
 \expectation{i}{i}
 & = & \bra{\Delta} \, L_{3} \, L_{-3} \, \ket{\Delta} \,=\,
       \bra{\Delta} \, \lb L_{3}, \, L_{-3} \rb \,
                 \ket{\Delta} \,=\, 6\Delta+2c ~, \\
 \expectation{j}{j}
 & = & \bra{\Delta} \, L_{2} \, L_{1} \, L_{-1} \, L_{-2} \,
       \ket{\Delta} \,=\, \bra{\Delta} \, \lb L_{2} \, L_{1} , \,
                 L_{-1} \, L_{-2} \rb \, \ket{\Delta}\\
       & \,=\, & \ldots \,=\, 34 \, \Delta \,+\, \Delta \, c \,+\, 8 \,
                 \Delta^{2} \,+\, 2 \, c~, \\
 \expectation{k}{k}
 & = & \ldots \,=\, 48 \, \Delta^{3} \,+\, 64 \, \Delta^{2}
                 \,+\, 20 \, \Delta ~, \\
 \expectation{i}{j} & = & \, \expectation{j}{i}\,=\,
                          16 \Delta \,+\,2 \, c~, \\
 \expectation{i}{k} & \,=\, & \expectation{k}{i}
               \,=\, 24 \, \Delta~, \\
 \expectation{j}{k} & \,=\, & \expectation{k}{j}
               \,=\, 6 \, \Delta \, (6 \, \Delta \,+\, 3)~.
\ee
Now it is straightforward, though a little messy, to solve 
the system of equations \calle{eq:dasGP11a}. After some calculations,
one finds
$$
 D\=
 \Delta^{2} \, ( \Delta \,-\, \Delta_{1}^{+} ) \, ( \Delta \,-\,
 \Delta_{2}^{+} ) \, ( \Delta \,-\, \Delta_{1}^{-} ) \, ( \Delta \,-\,
 \Delta_{2}^{-} ) ~,
$$
where
\bb
 \Delta_{1}^{\pm} & \,=\, & \frac{1 \,-\, c}{96} \, \left[ \left(
 3 \, \pm \, \sqrt{ \frac{25 \,-\, c}{1 \,-\, c}} \right)^{2} \,-\, 4
 \right] \,=\, \frac{1}{16} \, \left[ (5 \,-\, c) \, \pm \, \sqrt{
 (1 \,-\, c) \, (25 \,-\, c)} \, \rb ~, \\
 \Delta_{2}^{\pm} & \,=\, & \frac{1 \,-\, c}{96} \, \left[ \left(
 4 \, \pm \, 2 \, \sqrt{ \frac{25 \,-\, c}{1 \,-\, c}} \right)^{2} \,-
 \, 4\right] \,=\, \frac{1}{6} \, \left[ (7 \,-\, c) \, \pm \, \sqrt{
 (1 \,-\, c) \, (25 \,-\, c)} \, \rb~.
\ee
Setting $D=0$, we find that $\Delta = 0$, $\Delta_1^{\pm}$, or $\Delta_2^{\pm}$.
The solution $\Delta=0$ gives a null state at level 1, while $\Delta=\Delta_1^{\pm}$
are exactly the values that give a null state at level 2.
Thus the solution $\Delta=\Delta_{2}^{\pm}$ is the one corresponding
to the null state at level 3. Using these values
and two of the three equations of the system \calle{eq:dasGP11a}, 
we can solve for
$\beta / \alpha$ and $\gamma / \alpha$ (or for $\beta$ and $\gamma$, if we
accept $\alpha \,=\, 1$). We  find:
$$
 \frac{\beta}{\alpha} \,=\, - \, \frac{2}{\Delta \,+\, 2} ~~,~~~
 \frac{\gamma}{\alpha} \,=\, \frac{1}{(\Delta \,+\, 1)\, (\Delta
 \,+\, 2)}~,
$$
so that the level 3 null state is
$$
 \ket{\chi} \,=\, \left[ L_{-3} \,-\, \frac{2}{\Delta \,+\, 2} \,
 L_{-1} \, L_{-2} \,+\, \frac{1}{(\Delta \,+\, 1)\, (\Delta 
 \,+\, 2)} \, L_{-1}^{3} \, \right] \, \ket{\Delta}
 ~.
$$

\separator

\item
The boson propgator $G(x,x')$ is the solution of the differential equation
\beq
\label {greenf}
 g \partial^{\mu} \partial_{\mu} G(x, \, x') \,=\,
 - 2 \pi \delta^{(2)}(x - x') ~~.
\eeq
One can proceed in several ways to solve this equation.

\underbar{First Solution}: The first way of calculating the boson propagator
is by far the simplest one, once one realizes the analogies with Poisson's
equation in electromagnetism. In particular, recall that Poisson's equation
for the potential $\phi$ is
\beq
\label{Poisson}
   \nabla^2\phi \= - 4\pi\, \rho~,
\eeq
where $\rho$ is the electric charge density.
Equation \calle{greenf} becomes identical to \calle{Poisson}
for a {\em linear} charge distribution of infinite length
and density 
$$
   \rho = -{1\over 2g}\, \delta^{(2)}(x-x')~.
$$
However, in this geometry, the symmetry of problem allows us to
perform the calculation in a very simple way. Using Gauss's 
law,
one can calculate the electric field at distance 
$r$ from the infinite line, obtaining
$$
    E_r \= {1\over g}\, {1\over r}~.
$$
From this, the calculation of the potential $\phi$ is straightforward,
since $E_r = -{\partial \phi\over \partial r}$, and thus
one determines that
$$
   \phi(r) \=-{1\over g}\, \ln r + {\rm constant}~.
$$
Using the convention that $\phi(r_0)=0$, this fixes
$$
    \phi(r)\=-{1\over g}\, \ln {r\over r_0}~.
$$
This is of course the result for the boson propagator, too.
Noting that $r = |z|$, we finally get the desired result 
expressed in complex coordinates,
$$
  G(z,\overline{z}) \=
  -\frac{1}{2g}\,\ln|z|^{2} ~.
$$

\underbar{Second Solution}:
Even without the analogy to the electric field, one can
still calculate the boson propagator quite easily by
making use of the symmetries  in Green's equation \calle{greenf}.
First, observe that  equation \calle{greenf}
is translationally and Lorentz invariant. This implies
that the Green's function is a function of the radius $r$
only,
$$
   G(x,x') \= G(r)~.
$$
Writing equation \calle{greenf} in polar coordinates,
\beq
\label{greenf2}
  {1\over r}\, {d\over dr} \lp r {dG\over dr} \rp \= 0~,
\eeq
one immediately finds
$$
   G(r)\= a\, \ln r + b~,
$$
where $a$ and $b$ are two constants of integration. To calculate these constants,
we need two boundary conditions.

Notice that equation \calle{greenf2} is singular for $r=0$; therefore,
we will derive a boundary condition to handle the singularity
at this point.
To this end, recall Green's Theorem on the plane,
$$
   \ointleft_{\partial D} P dx + Q dy \=
   \iint_D  \lp {\partial Q\over \partial x}-
   {\partial P\over\partial y} \rp dx dy ~,
$$
where $P(x,y)$ and $Q(x,y)$ are two functions with continuous derivatives
and $D$ is the region of the plane interior to the curve $\partial D$.
For the case
$$
   Q\={\partial G\over \partial x}~,~~~~~
   P\=-{\partial G\over \partial y}~,
$$
Green's Theorem tell us that
$$
   \ointleft_{\partial D} \nabla G \cdot \vec n \, dl \=
   \iint_D  \nabla^2 G \, dS~,
$$
where $\vec n$ is the outward-pointing normal vector on $\partial D$.
Taking the domain $D$ to be a circular region of radius 
$\varepsilon\rightarrow 0$, we see that
$$
   {dG \over dr}\Bigg|_{r=\varepsilon} \,\cdot\, 2\pi\varepsilon 
   \longrightarrow -{2\pi\over g}~.
$$
Comparing this with
$$
  {dG \over dr} \= a\, {1\over r}~,
$$
we conclude that $a=-1/g$.

The constant $b$ is calculated by introducing a normalization 
condition, which we choose arbitrarily to be $G(r_0)=0$ for
some distance $r_0$.

Thus, finally, we obtain
$$
   G(r) \= -{1\over g}\, \ln{r\over r_0}~,
$$
which, as we saw in the first solution to this problem,
is exactly the desired propagator when put in complex
form.

\underbar{Third Solution}:
Our last way of calculating the boson propagator is simply to grind
through the calculation.  It is straightforward, in the sense that
no analogy or trick is used; however, it does require
a relatively higher level of comfort with the calculation of integrals.

The Fourier transform of $G(x-x')$ is given by
\bb
 \tilde{G}(k)\=\iint d^2x\,e^{-ik \cdot (x-x')}\, G(x-x')~.
\ee
The inverse transformation is
\begin{equation}
\label{fou2}
 G(x-x')\=\iint \frac{d^2k}{(2\pi)^2} \, e^{+ik \cdot (x-x')}
 \, \tilde{G}(x-x')~.
\end {equation}
We also know that
\begin {equation}
\label{delta}
 \delta (x-x')\= \iint \frac{d^2k}{(2\pi)^2} e^{+ik \cdot (x-x')}~.
\end {equation}
Substituting equations \calle{fou2} and \calle{delta} into the definitions of
the propagator, we find
\bb
 \tilde{G}(k)\=\frac{2\pi}{g}\,\frac{1}{k^{2}}~.
\ee
Notice that this propagator exhibits the well-known $1/k^{2}$ behaviour 
for a massless boson in two
dimensions. Now we can do the integrations in (\ref{fou2}) using polar
coordinates:
\beqn
\label{calc1}
 \tilde{G}(x-x') & = & \frac{2\pi}{g}\, \iint \frac{d^{2}k}{(2\pi)^{2}} \, 
                       \frac{e^{ik \cdot (x-x')}}{k^{2}} \nonumber \\
     & = &  \frac{1}{2\pi g}\,\int_0^{+\infty} dk \int_0^{2\pi} d\theta
            \, \frac{e^{ik r \cos \theta }}{k} \nonumber \\
     & = &  \frac{1}{g}\, \int_{0}^{+ \infty} dk \, \frac{J_{0}(kr)}{k} ~~,
\eeqn
where we have set set $r \equiv x-x'$,
and where $J_{0}$ is a Bessel function of the
first kind.
Since the small $x$ behavior of $J_{0}(x)$ is
$$
 J_{0}(x) \= 1 - \frac{x^{2}}{4} + \ldots ~~,
$$
the integral in (\ref{calc1}) diverges; therefore we regulate its
infrared behavior introducing a reference point $r_{0}$:
$$
 G_{{\rm reg}}(r) \= G(r) \,-\, G(r_{0}) ~~.
$$
The regulated propagator is then
\beq
\label{Greg}
 G_{{\rm reg}}(r) \= \frac{1}{g} \, \int_{0}^{+ \infty} dk \,
    \frac{J_{0}(kr) - J_{0}(kr_{0})}{k} ~.
\eeq
The integral in this equation can be evaluated in many ways.
At the end of this problem, we present one method, which
applied here shows that
$$
    \int_{0}^{+ \infty} dk \,
    \frac{J_{0}(kr) \,-\, J_{0}(kr_{0})}{k}\= -\ln{r\over r_0}~,
$$ 
thus leading once more to the result
$$
   G(z,\overline{z})\=
     -\frac{1}{2g}\,\ln|z|^{2} ~.
$$

\footnotesize
{\bf APPENDIX}

The integral in (\ref{Greg}) is a case of a broad category of
integrals known as {\bf Frullani integrals}, which are given by
$$
 I \lb f(x) \rb \, \equiv \, \int_{0}^{+\infty} dx \,
       \frac{f(bx) \,-\, f(ax)}{x} ~~.
$$
We  now evaluate  the Frullani integral for functions $f(x)$ with properties
similar to $J_0(x)$, i.e.
   
(i) $\lim\limits_{x\rightarrow +\infty}f(x)={\rm finite}$,

(ii) $f'(x)$ is continuous and integrable on $\lb 0,+\infty\rp$.

The above two conditions imply
\bb
    \int_{0}^{+\infty}\,dx\, f'(x)\=f(+\infty)-f(0)~.
\ee

We now define the integral
\beq
\label{Ia1}
    I(\alpha)\=\int_{0}^{+\infty}\,dx\, f'(ax)~.
\eeq
The integration can be trivially performed to give
\beq
\label{Ia2}
    I(\alpha)\={f(+\infty)-f(0)\over\alpha}~.
\eeq
Now let us calculate the integral
$$
     \int_{a}^{b}\,d\alpha\, I(\alpha)
$$
in two ways: first using the definition \calle{Ia1},
and second based on the result \calle{Ia2}.

The definition \calle{Ia1} gives
\beqn
     \int_a^b\,d\alpha\, I(\alpha)&\=&
    \int_a^b\,d\alpha\,\int_0^{+\infty}\,dx\, f'(ax)\nonumber\\
    &\=&\int_0^{+\infty}\,dx\,\int_a^b\,d\alpha\, f'(ax)\nonumber\\
    &\=&\int_0^{+\infty}\,dx\,{f(bx)-f(ax)\over x}~.
\label{Ia3}
\eeqn
Using equation \calle{Ia2}, on the other hand, we find
\beqn
     \int_a^b\,d\alpha\, I(\alpha)\=
    \lb f(+\infty)-f(0)\rb\,\int_a^b\,{d\alpha\over\alpha}
    \= \lb f(+\infty)-f(0)\rb\,\ln{b\over a}~.
\label{Ia4}
\eeqn

Comparing the two results, \calle{Ia3} and \calle{Ia4}, we conclude
that
\bb
    \int_0^{+\infty}\,dx\,{f(bx)-f(ax)\over x}
    \=\lb f(+\infty)-f(0)\rb\,\ln{b\over a}~.
\ee

\normalsize

\separator

\item
Expanding the exponential in a power series, we have
$$
 \partial_{z}\phi(z)\, :e^{i\alpha\phi(w)}:\,=\, \partial_{z} \phi(z)
 \, \sum_{n=0}^{+\infty}\, :\frac{\lb i\alpha\phi(w)\rb^{n}}{n!}:\, 
  ~~.
$$
Using Wick's theorem \cite{itzy}, this becomes
$$
 \partial_{z}\phi(z)\, :e^{i\alpha\phi(w)}:\= 
 :\partial_{z}\phi(z)e^{i\alpha\phi(w)}: \+ 
 \partial_{z}\lowerpairing{\phi(z)\phi}(w)
 \, \sum_{n=1}^{+\infty}\, \frac{(i\alpha)^{n}:\lb\phi(w)\rb^{n-1}:}{n!}\,
 n ~.
$$
Notice the factor $n$ which appears as a result of all the possible
contractions of $\partial_{z}\phi(z)$ with $\phi^n(w)$.
 Using \calle{Prop1}, we can now write
\bb
 \partial_{z}\phi(z)\, :e^{i\alpha\phi(w)}: \,
 &=& :\partial_{z}\phi(z)e^{i\alpha\phi(w)}: 
       \-\frac{i \alpha}{2 g}\, \partial_{z}\ln(z-w)\,
        \sum_{n=1}^{+\infty}
       :\frac{\lb i\alpha\phi(w)\rb^{n-1}}{(n-1)!}:
        \\
 &=& :\partial_{z}\phi(z)e^{i\alpha\phi(w)}: 
       \-\frac{i\alpha}{2g} \,\frac{1}{z-w}\, 
       \sum_{m=0}^{+\infty}
       \, :\frac{\lb i\alpha\phi(w)\rb^{m}}{m!}:\\
 &=& :\partial_{z}\phi(z)e^{i\alpha\phi(w)}:  
       \-\frac{i\alpha}{2g}\,\frac{1}{z-w}\,:e^{i\alpha\phi(w)}:
        ~,
\ee
which is the desired result.

Having worked the simple case and mastered the ideas, we can now proceed to the
general case. For simplicity, let us set
$$
  A\eq\partial_z\phi(z)~~~{\rm and}~~~
  B\eq i\alpha\phi(w)~.
$$
According to Wick's theorem,  the OPE
$$
 \CR\lp :A^k: \, :e^B:\rp
 \= :A^k: 
 \, \sum_{n=0}^{+\infty}\, {:B^n:\over n!}
$$
is calculated by adding all possible contributions for one contraction,
two contractions, three contractions, etc., up to $k$ contractions.

We notice that  
the term $B^n$ can contribute $l$ contractions if $n\ge l$. In particular,
when the inequality $n\ge l$ is satisfied $:A^k:\, :B^n:$ contributes
$$
 (\lowerpairing{AB})^l\,{k(k-1)\cdots(k-l+1) \, n(n-1)\cdots(n-l+1)\over l!}
 \= (\lowerpairing{AB})^l\,{k!n!\over (k-l)!l!(n-l)!}
$$ 
since there are $k(k-1)\cdots(k-l+1)$ ways to chose the $l$ $A$'s and
$ n(n-1)\cdots(n-l+1)$ ways to choose the $l$ B's. 
The number is divided by
$l!$ since the pairs $\lowerpairing{AB}$ are indistinguishable.

In this way, we find
\bb
 \CR\lp :A^k: \, :e^B:\rp \=  :A^k ~ e^B:
      &+& (\lowerpairing{AB}) :\,A^{k-1}
      \sum_{n=1}^{+\infty}\, {B^{n-1}:\over n!}\,
       {n! \, k!\over (k-1)!\, 1! (n-1)!} \\
      &+& (\lowerpairing{AB})^2 \,: A^{k-2}
      \sum_{n=2}^{+\infty}\, {B^{n-2}:\over n!}\,
       {n! \, k!\over (k-2)!\, 2! (n-2)!} \\
  &+&  \cdots \\
  &+&  (\lowerpairing{AB})^k \,: A^{k-k}
      \sum_{n=k}^{+\infty}\, {B^{n-k}:\over n!}\,
       {n! \, k!\over (k-k)!\, k! (n-k)!} \\
  \= :A^k ~ e^B: 
  &+& \sum_{l=1}^k
      {(\lowerpairing{AB})^l\over l!} 
      ~ :A^{k-l}
      \sum_{n=l}^{+\infty}\, {B^{n-l}:\over (n-l)!}\,
        {k!\over (k-l)!}\\
  \= :A^k ~ e^B: 
  &+& \sum_{l=1}^k \pmatrix{k\cr l\cr}\,
      (\lowerpairing{AB})^l
      ~ :A^{k-l}
      \sum_{m=0}^{+\infty}\, {B^m:\over m!}\\
  \= :A^k ~ e^B: 
  &+& \sum_{l=1}^k \pmatrix{k\cr l\cr}\,
      (\lowerpairing{AB})^l
      ~ :A^{k-l} e^B: ~,
\ee
or
\beq
\label{eq:GP87}
   \CR\lp :A^k: \, :e^B:\rp \= 
   \sum_{l=0}^k \pmatrix{k\cr l\cr}\,
      (\lowerpairing{AB})^l
      ~ :A^{k-l} e^B:~.
\eeq
Substituting the values of $A$ and $B$ in terms of the scalar fields,
we arrive at our final result: 
\bb
  \CR\lp:(\partial_z\phi(z))^k:\,: e^{i\alpha\phi(w)}:\rp
  \= \sum_{l=0}^k \pmatrix{k\cr l\cr} \, 
      {(-i\alpha/2g)^l\over (z-w)^l}
       ~ :(\partial_z\phi(z))^{k-l} \, e^{i\alpha\phi(w)}:  ~.
\ee

\separator

\item
We prove this identity employing the same method we used in the 
solution to the preceding problem.  
Using
Wick's theorem, the initial product of the two exponentials
\bb
 :e^{i\alpha\phi(z)}:\,:e^{i\beta\phi(w)}: & = &
 \sum_{n=0}^{+\infty}\,:\frac{\lb i\alpha\phi(z)\rb^{n}}{n!}:\,
 \sum_{n=0}^{+\infty}\,:\frac{\lb i\beta\phi(w)\rb^{m}}{m!}:
\ee
equals a sum of terms, each of which takes into
account all possible zero, single, double, triple, etc. contractions.
In particular, the term with zero contractions is
\bb
   \sum_{n=0}^{+\infty} \sum_{m=0}^{+\infty} \,
 :\frac{\lb i\alpha\phi(z)\rb^{n} \, \lb i\beta\phi(w)\rb^{m}}{n!m!}: \qquad,
\ee
the term with one contraction is
\bb
 \sum_{n=1}^{+\infty} \sum_{m=1}^{+\infty}\, -\alpha\beta
 \lowerpairing{\phi(z)\phi}(w) \, :\frac{\lb i\alpha\phi(z)\rb^{n-1}
 \lb i\beta\phi(w)\rb^{m-1}}{
 n! \, m!}:  \, {n\, m\over 1!} \qquad,
\ee
the term with two contractions is
\bb
 \sum_{n=2}^{+\infty}\sum_{m=2}^{+\infty}\, \lb -\alpha\beta
 \lowerpairing{\phi(z)\phi}(w)\rb^2 \, :
  \frac{\lb i\alpha\phi(z)\rb^{n-2}\, \lb i\beta\phi(w)\rb^
 {m-2}}{n!\,m!}: \, {n (n-1)\, m (m-1)\over 2!} \qquad,
\ee
and so on. The factors in the r.h.s. of the above expressions
count all possible ways that the corresponding contractions can 
be performed.
Therefore 
\bb
 :e^{i\alpha\phi(z)}:\,:e^{i\beta\phi(w)}: & = &
 \sum_{n=0}^{+\infty}\,:\frac{\lb i\alpha\phi(z)\rb^{n}}{n!}:\,
 \sum_{n=0}^{+\infty}\,:\frac{\lb i\beta\phi(w)\rb^{m}}{m!}:\\
 &=& \phantom{+}\sum_{n=0}^{+\infty} \sum_{m=0}^{+\infty} \,
 :\frac{\lb i\alpha\phi(z)\rb^{n} \, \lb i\beta\phi(w)\rb^{m}}{n!m!}:\\
 & & +\sum_{n=1}^{+\infty} \sum_{m=1}^{+\infty}\, \frac{-\alpha\beta
 \lowerpairing{\phi(z)\phi}(w)}{1!} \, :\frac{\lb i\alpha\phi(z)\rb^{n-1}
 \lb i\beta\phi(w)\rb^{m-1}}{
 (n-1)! \, (m-1)!}\\ 
 & & +\sum_{n=2}^{+\infty}\sum_{m=2}^{+\infty}\, \frac{\lb -\alpha\beta
 \lowerpairing{\phi(z)\phi}(w)\rb^{2}}{2!}:\,
  \frac{\lb i\alpha\phi(z)\rb^{n-2}\, \lb i\beta\phi(w)\rb^
 {m-2}}{(n-2)!\,(m-2)!}:
 +~ \cdots \\
 &=& \sum_{n=0}^{+\infty} \sum_{m=0}^{+\infty} \,
 :\frac{\lb i\alpha\phi(z)\rb^{n} \, \lb i\beta\phi(w)\rb^{m}}{n!m!}:\\
 & & +\frac{-\alpha\beta\lowerpairing{\phi(z)\phi}(w)}{1!} \,
       \sum_{n=0}^{+\infty} \sum_{m=0}^{+\infty} \,
 :\frac{\lb i\alpha\phi(z)\rb^{n} \, \lb i\beta\phi(w)\rb^{m}}{n!m!}:\\
 & & +\frac{\lb-\alpha\beta \lowerpairing{\phi(z)\phi}(w)\rb^{2}}{2!}:\,
      \sum_{n=0}^{+\infty} \sum_{m=0}^{+\infty} \,
 :\frac{\lb i\alpha\phi(z)\rb^{n} \, \lb i\beta\phi(w)\rb^{m}}{n!m!}:
  +~ \cdots \\
 &=&  \sum_{k=0}^{+\infty} \,
 \frac{\lb-\alpha\beta\lowerpairing{\phi(z)\phi}(w)\rb^k}{k!}
 ~ \lb \sum_{n=0}^{+\infty} \sum_{m=0}^{+\infty} \, 
 :\frac{\lb i\alpha\phi(z)\rb^{n}\lb i\beta\phi(w)\rb^{m}}{n!m!}: \rb~.
\ee
The last expression is just a product of exponentials, and thus this
equation becomes
\bb
 :e^{i\alpha\phi(z)}: ~ :e^{i\beta\phi(w)}: 
 &=& \exp\lp-\alpha\beta\,\lowerpairing{\phi(z)\phi}(w)\rp\, 
   :e^{i\alpha\phi(z)} \,e^{i\beta\phi(w)}: \\
 &=& (z-w)^{\alpha\beta/2g} ~
   :e^{i\alpha\phi(z)} \,e^{i\beta\phi(w)}: ~.
\ee

Another way to prove this result is by direct use of formula
\calle{eq:GP87} with
$$
   A\=i\alpha\phi(z)~,~~~~~
   B\=i\beta\phi(w)~.
$$
In this way, we find
\bb
   \CR(e^A e^B) &=& \sum_{k=0}^{+\infty}{:A^ke^B:\over k!}\\
   &{\calle{eq:GP87}\atop=}& \sum_{k=0}^{+\infty}{1\over k!}
   \sum_{l=0}^k \pmatrix{k\cr l\cr}\,
      (\lowerpairing{AB})^l  ~ :A^{k-l} e^B: \\
   &=& \sum_{k=0}^{+\infty}
   \sum_{l=0}^k 
      {(\lowerpairing{AB})^l\over l!}  ~ {:A^{k-l} e^B:\over (k-l)!} ~.
\ee
Changing the summation indices from $k, l$ to
$l, N$, where $N=k-l$ (as is represented pictorially
in the figure that follows), we can decouple the
two sums in the series.
\begin{center}
\psfrag{k=l}{$k\!=\!l$}
\psfrag{k-l=0}{$k\!-\!l\!=\!0$}
\psfrag{k-l=1}{$k\!-\!l\!=\!1$}
\psfrag{k-l=2}{$k\!-\!l\!=\!2$}
\psfrag{k-l=3}{$k\!-\!l\!=\!3$}
\psfrag{k-l=4}{$k\!-\!l\!=\!4$}
\psfrag{k=1}{$k\!=\!1$}
\psfrag{k=2}{$k\!=\!2$}
\psfrag{k=3}{$k\!=\!3$}
\psfrag{k=4}{$k\!=\!4$}
\psfrag{k}{$k$}
\psfrag{l}{$l$}
\psfrag{0}{$0$}
\psfrag{1}{$1$}
\psfrag{2}{$2$}
\psfrag{3}{$3$}
\psfrag{4}{$4$}
\includegraphics[width=10cm]{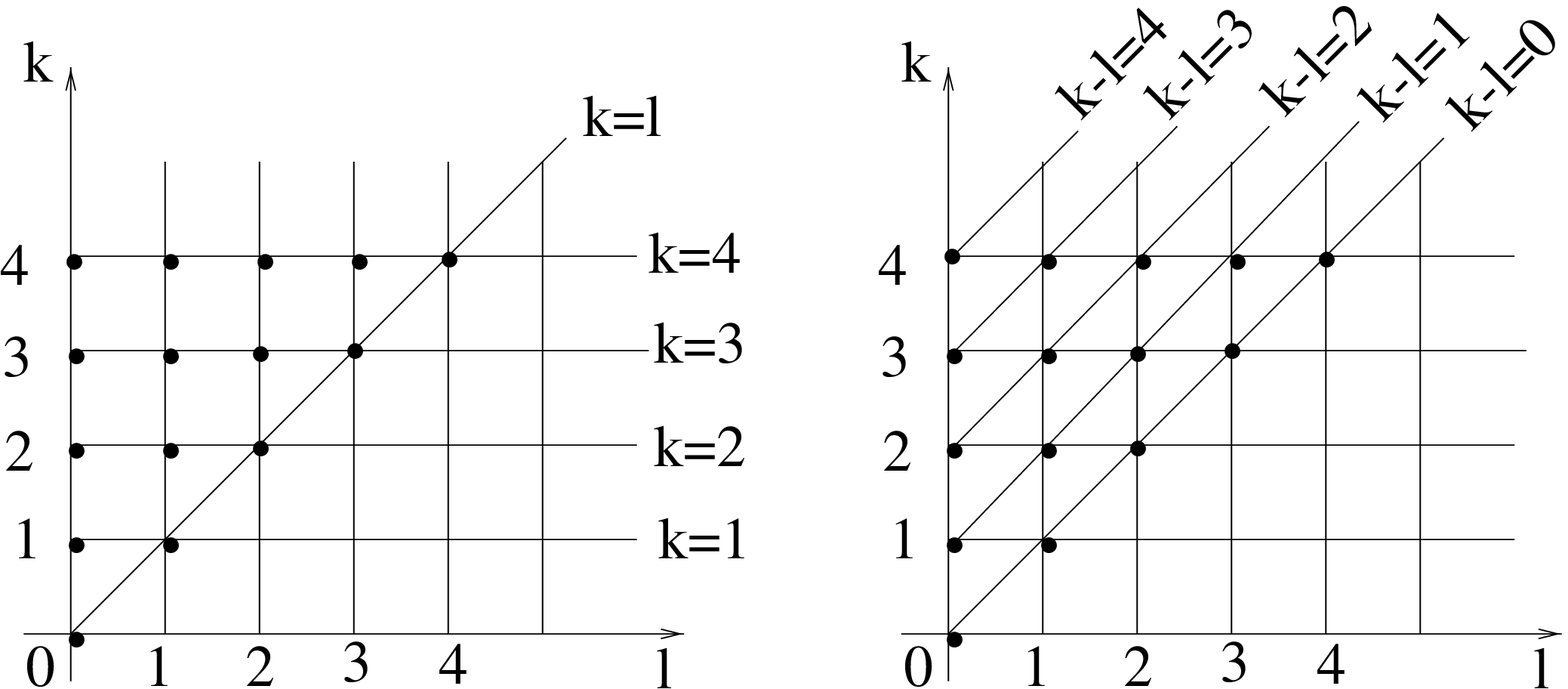}
\end{center}
This re-writing yields
\bb
   \CR(e^A e^B) &=& 
    \sum_{l=0}^{+\infty} {(\lowerpairing{AB})^l\over l!}
   \sum_{N=0}^{+\infty}{:A^N e^B:\over N!}
    \= \exp\lp\lowerpairing{AB}\rp\,:e^Ae^B:~.
\ee

\separator

\item
We begin by making a Laurent expansion for $i \partial_z \phi (z)$,
$$
 i \partial_z \phi (z) \eq \sum_n\, \alpha_nz^{n-1}~.
$$
From the theory of complex variables, we have the relation
$$
 \alpha_n\= \ointleft_C \frac{dz}{2 \pi i}\, z^{n}\,i\,\partial \phi(z)~,
$$
and thus
$$
 \lb \alpha_{n},\alpha_{m}\rb\,=\,i^{2}\,\lb \ointleft_{C(z)}\,\frac{dz}{
 2\pi i}\,,\, \ointleft_{C(w)}\,\frac{dw}{2\pi i} \rb \, z^{n}\,
 \partial_{z} \phi(z) \, w^{m} \, \partial_{w} \phi(w)~.
$$

\begin{figure}[h!]
\begin{center}
\psfrag{C1}{$C_1$}
\psfrag{C2}{$C_2$}
\includegraphics[height=4.5cm]{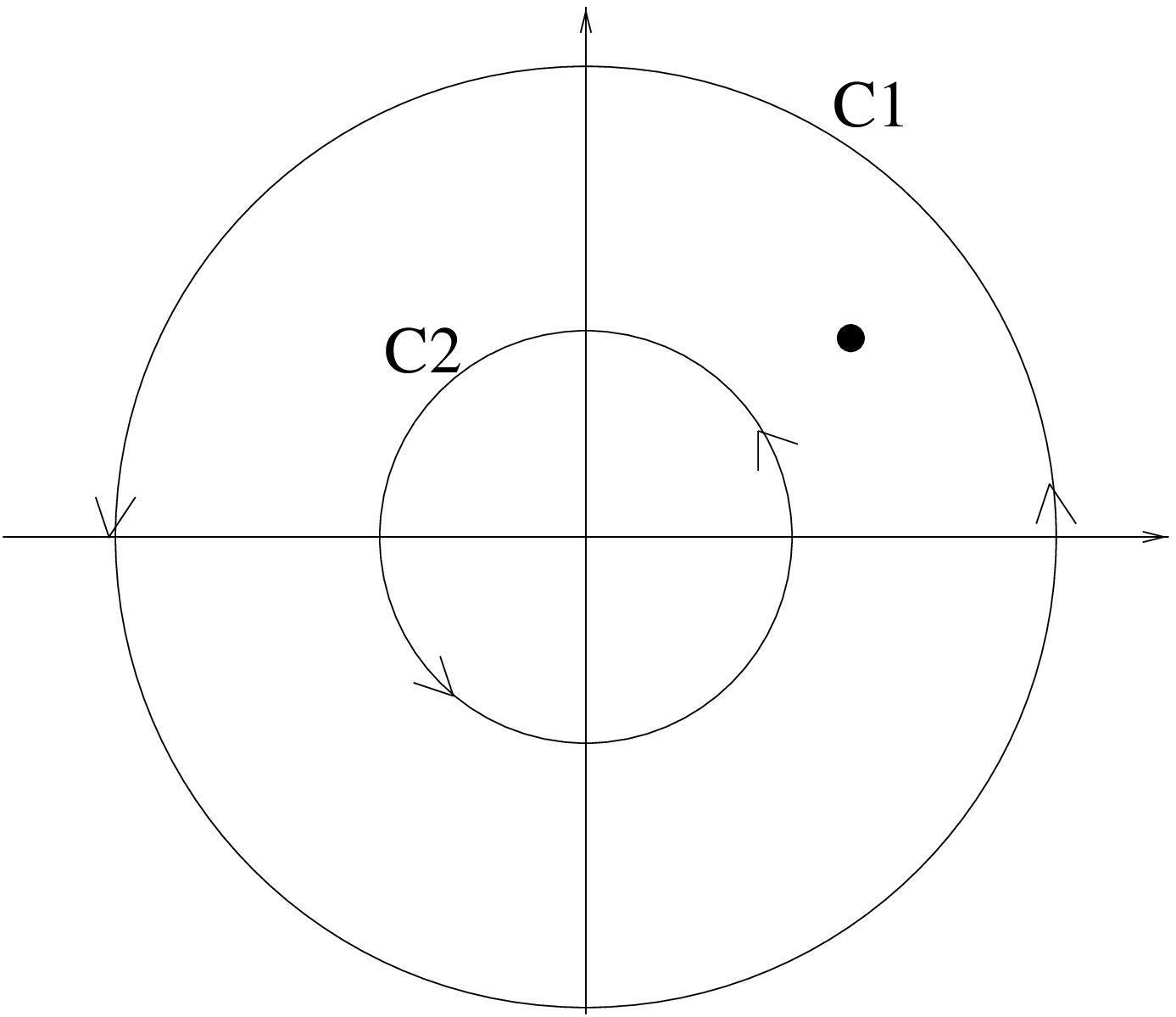}
\end{center}
\end{figure}

Notice that $C(z)$ and $C(w)$ must be closed curves around the origin of the
complex plane; other than that, there is no other restriction on these curves.
Therefore we can keep 
$C(z)$ fixed and take $C(w)=C_{2}$ for the first term of the commutator
and $C(w)=C_{1}$ for the second term (see the diagram).
Doing so, the above equation becomes
\bb
 \lb \alpha_n, \alpha_m\rb &=& - \,\ointleft_{C(z)}\, \frac{dz}{2 \pi i}
 \,\lp \, \ointleft_{C_{2}} - \ointleft_{C_{1}}\rp \, \frac{dw}{2 \pi i} \,
 z^{n}w^{m} \, \partial_z \phi(z) \, \partial_{w} \phi(w)\\
 &=& - \, \ointleft_{C(z)} \, \frac{dz}{2 \pi i} \, \ointleft_{\gamma} \, 
 \frac{dw}{2 \pi i} \, z^n w^m \, 
 \CR \lp \partial_z \phi(z) \partial_w
 \phi(w)\rp ~,
\ee
where $\CR$ indicates radial ordering of the fields (analogous
to temporal ordering) and the curve $\gamma$ encircles the point
$z$.

\begin{figure}[h!]
\begin{center}
\psfrag{-C1}{$-C_1$}
\psfrag{C2}{$C_2$}
\psfrag{z}{$z$}
\psfrag{-c}{$-\gamma$}
\includegraphics[height=5cm]{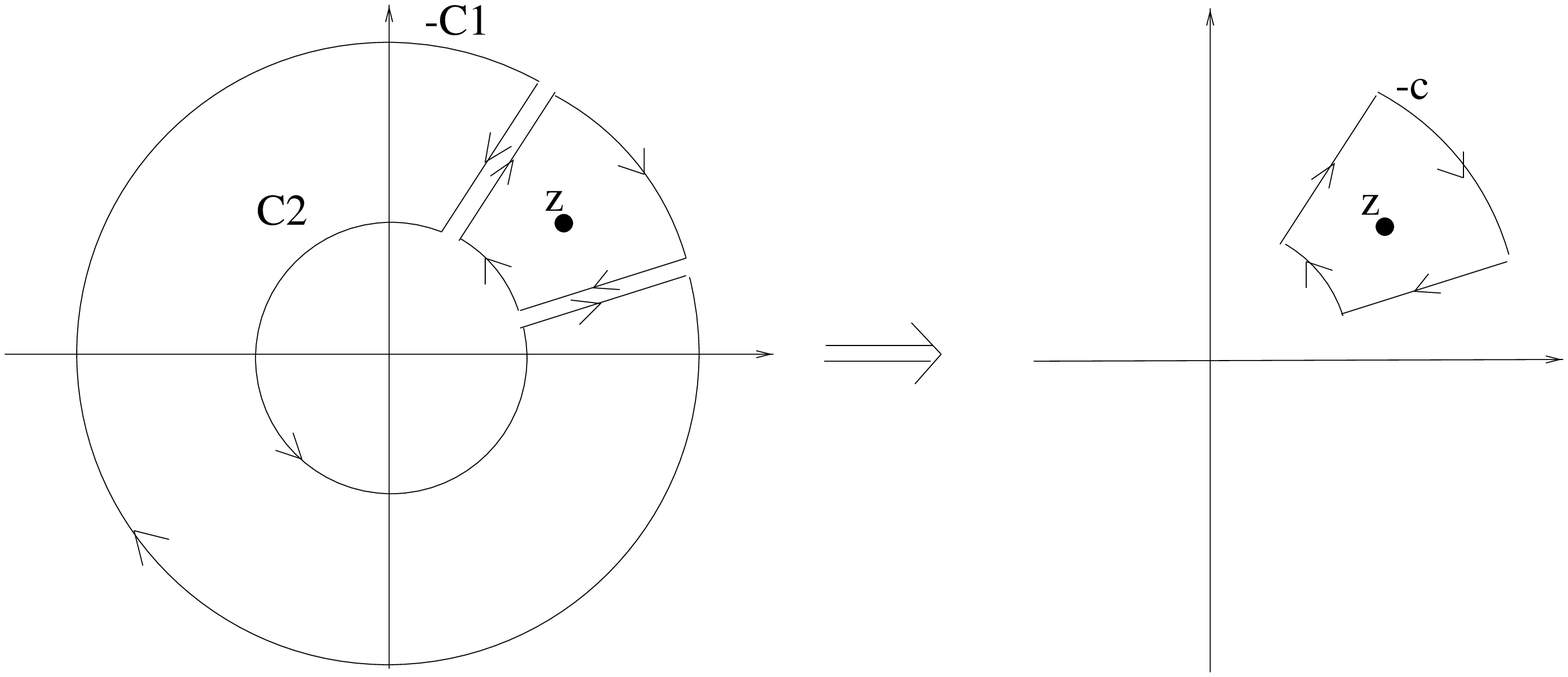}
\end{center}
\end{figure}
 
Since
$$
 \CR\lp\phi(z) \phi(w)\rp \= - \frac{1}{2g} \, \ln(z-w)
 +  \mbox{reg}~,
$$
we see that
$$
 \CR\lp \partial_{z} \phi(z) \partial_{w} \phi(w)\rp \=
 -\, \frac{1}{2g} \,
 \frac{1}{(z-w)^{2}}  + \mbox{reg}~ .
$$
and the commutation relation between two $\alpha$'s becomes:
\bb
 \lb\alpha_{n}, \alpha_{m}\rb\, & = &  -\, \frac{1}{2g} \, \ointleft_{C(z)} \,
 \frac{dz}{2 \pi i} \, \ointleft_{\gamma} \, \frac{dw}{2 \pi i} \,
 \frac{z^{n} w^{m}}{(z-w)^{2}}\\
 & = & \, - \,\frac{1}{2g} \ointleft_{C(z)} \, \frac{dz}{2 \pi i} \, z^n\,
 mz^{m-1}\\
 & = &  - \, \frac{m}{2g} \, \delta_{n,-m}~,
\ee
or
$$
 \lb\alpha_{n}, \alpha_m\rb \= \frac{n}{2g} \, \delta_{n,-m}  ~.
$$

\separator

\item
Substituting the expression
$$
   i\partial\phi(z)\eq \sum_n {\phi_n\over z^{n+1}}~~,
$$
which defines the modes of the boson in the presence of the
$e_0$ term in \calle{aT},  we find:
\bb
  T &=& g\, \lp\sum_n{\phi_n\over z^{n+1}}\rp^2
        +e_0\, \partial\sum_n {\phi_n\over z^{n+1}} \\
    &=& g\, \sum_{n,m} {\phi_n\phi_m\over z^{n+m+2}}
        -e_0\,\sum_n { (n+1)\,\phi_n\over z^{n+2} } \\
    &=& g\, \sum_{k,m} {\phi_{k-m}\phi_m\over z^{k+2}}
        -e_0\,\sum_k { (k+1)\,\phi_k\over z^{k+2}} \\
    &=& \sum_k z^{-(k+2)} \,
        \lb  g\,\sum_m \phi_{k-m}\phi_m -e_0\, (k+1)\,\phi_k\rb~.
\ee
From the last expression, it is obvious that
$$
  L_k\= g\,\sum_m \phi_{k-m}\phi_m -e_0\, (k+1)\,\phi_k~.
$$
The modes $\phi_n$ obey the commutation relations
$$
   \lb \phi_n,\phi_m\rb \= {n\over 2g}\,\delta_{n,-m}~.
$$
Using these results,
\bb
    L_0 &=& g\, \sum_{m=-\infty}^{+\infty}\phi_{-m}\phi_m -e_0\, \,\phi_0\\
        &=& g\, \sum_{m=1}^{+\infty}\phi_{-m}\phi_m 
            +g\, \sum_{m=-1}^{-\infty}\phi_{-m}\phi_m 
            +g\, \phi_0^2 
            -e_0\, \phi_0\\
        &=& g\, \sum_{m=1}^{+\infty}\phi_{-m}\phi_m 
            +g\, \sum_{m=1}^{+\infty}\phi_{m}\phi_{-m} 
            +g\, \phi_0^2 
            -e_0\, \phi_0\\
        &=& g\, \sum_{m=1}^{+\infty}\phi_{-m}\phi_m 
            +g\, \sum_{m=1}^{+\infty} \lp\phi_{-m}\phi_{m}+{m\over2g}\rp 
            +g\, \phi_0^2 
            -e_0\, \phi_0\\
        &=& 2g\, \sum_{m=1}^{+\infty}\phi_{-m}\phi_m 
            +{1\over2}\, \sum_{m=1}^{+\infty} m
            +g\, \phi_0^2 
            -e_0\, \phi_0~,
\ee
or, in other words,
$$
  L_0\= 2g\, \sum_{m=1}^{+\infty}\phi_{-m}\phi_m+g\, \phi_0^2 -e_0\, \phi_0
        + c_0~,
$$
where $c_0=\sum_m m/2$.
The constant $c_0$ can be set to zero by normal ordering.

\underline{Comment}: In some instances, we will need to introduce 
the $\zeta$-function renormalization.  Then
$$
  c_0\={1\over2}\, \sum_{m=1}^{+\infty} m\={1\over2}\,\zeta(-1)
     \=-{1\over24}~.
$$
This last equation is explained in the appendix that follows.

\footnotesize
{\bf APPENDIX}
  
The sum $\sum_m m/2$ often is denoted by $\zeta(-1)/2$ in analogy with
the $\zeta$-function\index{function!$\zeta$-function} of number
theory,
$$
   \zeta(s)\= \sum_{m=1}^{+\infty} {1\over m^s}~.
$$
As defined, this series converges for $Re(s)>1$. The series
does not converge for other values of $s$, but for these values,
the function $\zeta(s)$ is defined uniquely
by analytic continuation.

In the rest of the document, we will meet the  $\zeta$-function
on several occasions.
Here are two of the properties of the $\zeta$-function:
\bb
\Gamma(s)\zeta(s) &=& \int_0^{+\infty} {x^{s-1}e^{-x}\over 1-e^{-x}}\, dx
 ~,~~~{\rm and} \\
 \zeta(s) &=& 2(2\pi)^{s-1}\,\Gamma(1-s)\, \sin{\pi s\over2}\, \zeta(1-s)~.
\ee

Setting $s=-1$ in the second identity, and noticing that
$$
  \zeta(2)\=\sum_{n=1}^{+\infty} {1\over n^2}\={\pi^2\over6}~,
$$
we find
\bb
 \zeta(-1)&=& 2(2\pi)^{-2}\,\Gamma(2)\, \sin{-\pi s\over2}\, \zeta(2)\\
          &=& {2\over 4\pi^2}\, 1! \, (-1)\, {\pi^2\over6}\= -{1\over12}~.
\ee
         
\normalsize

\separator

\item
We consider the free boson defined
by the Lagrange density
$$
  \CL = {1\over8\pi}\,  (\partial_{\mu}\Phi)^2~,
$$
where 
$\Phi$ is compactified on a circle of radius $R$.
From the Lagrangian, we derive the equation of motion
\beq
\label{eq:bosoneqm}
  \partial_t^2\Phi -\partial^2_\sigma\Phi \= 0~.
\eeq
The canonical momentum is given by
$$
  \Pi \= {\partial\CL\over\partial\dot\Phi}\=
  {1\over 4\pi} \, \dot{\Phi}~.
$$
$\Phi$ and $\Pi$ satisfy the canonical commutation relations
\bb
 \lb\Phi(\sigma, t),\Phi(\sigma',t)\rb\ &=& 
 \lb\Pi(\sigma, t),\Pi(\sigma',t)\rb\ \= 0~, \\
 \lb\Phi(\sigma, t),\Pi(\sigma',t)\rb\ &=& i\delta(\sigma-\sigma')~.
\ee
The Hamiltonian density for this model is
\beq
\label{eq:GP53}
  \CH \eq \Pi\dot\Phi-\CL 
      \= {1\over8\pi}\, \lb (\partial_t\Phi)^2+(\partial_\sigma\Phi)^2\rb~.
\eeq

As is conventional, to solve equation \calle{eq:bosoneqm}, 
we use the method of separation of variables, setting
$$
  \Phi(\sigma,t)\= \Sigma(\sigma)\, T(t)~.
$$
Substituting this in \calle{eq:bosoneqm}, we find
$$
 {\ddot T\over T} \=
 {\Sigma'' \over \Sigma} \= -k^2~,
$$
where $k^2$ is a constant. When $k\ne 0$,
these last equations give
\bb
   T(t) &=& a\, e^{ikt}+b\, e^{-ikt}~,\\
   \Sigma(\sigma) &=& c\, e^{ik\sigma}+d\, e^{-ik\sigma}~,
\ee
while for $k=0$, these equations yield
\bb
   T(t) &=& a + b\, t~,\\
   \Sigma(\sigma) &=& c + d\, \sigma~.
\ee
Therefore, a free boson can consist of
\bb
   \Phi_0(\sigma,t) &=&
   \phi_0 +  b_0\, \sigma + c_0\, t + d_0\,\sigma t~,
\ee
and
\bb
   \Phi_k(\sigma,t) &=& \tilde a_{-k}\, e^{ik(\sigma+t)}
                       +\tilde a_k\, e^{-ik(\sigma+t)}
       +  a_k\, e^{ik(\sigma-t)}+ a_{-k}\, e^{-ik(\sigma-t)}~,~~~k\ne0~.
\ee
The Hermiticity condition $\Phi=\Phi^\dagger$ 
requires
$$
     a_{-k}\=a_k^\dagger~,~~~~~
     \tilde a_{-k}\=\tilde a_k^\dagger~,
$$
and that $\phi_0$, $b_0$, $c_0$, $d_0$, and $k$ be real.
The periodicity condition 
$$
   \Phi(\sigma+\beta, t)
  \= \Phi(\sigma, t) + 2\pi R\, w~, 
$$
can be rewritten as
\bb
  \Phi_0(\sigma+\beta,t)&=& \Phi_0(\sigma, t) + 2\pi R\, w~,\\
  \Phi_k(\sigma+\beta, t)&=&\Phi_k(\sigma, t) ~.
\ee
The first equation implies that 
\bb
  d_0 \= 0~,~~~~~ b_0\={2\pi R\over\beta}\, w~,
\ee
while the second imposes integer momentum modes:
\bb
  k \= {2\pi n\over\beta}~,~~~~~n\in\BZ^*~.
\ee
Therefore
\bb
  \Phi(\sigma,t)&=&\Phi_0(\sigma,t)+\sum_{n\in\BZ^*} \Phi_n(\sigma,t)\\
  &=& \phi_0+{2\pi R\over\beta}w\sigma+{4\pi\over\beta}pt
  +i\,\sum_{n\ne0}\lb {\alpha_n\over n}\, e^{i{2\pi\over\beta}n(\sigma-t)}
    -{\overline\alpha_{-n}\over n}\,  e^{i{2\pi\over\beta}n(\sigma+t)}\rb\\
  &=& \phi_0+{2\pi R\over\beta}w\sigma+{4\pi\over\beta}pt
  +i\,\sum_{n\ne0}\lb {\alpha_n\over n}\, e^{i{2\pi\over\beta}n(\sigma-t)}
    +{\overline\alpha_n\over n}\,  e^{-i{2\pi\over\beta}n(\sigma+t)}\rb~,
\ee
where we redefined
$$
  \alpha_n\eq i{a_n\over n}~,~~~~~
  \overline \alpha_n\eq i{\tilde a_{-n}\over -n}~,~~~~~
  c_0\eq {4\pi\over\beta}p~.
$$
Differentiating w.r.t. time, we get
\bb
  \Pi\={1\over 4\pi}\, \dot\Phi(\sigma,t)&=&
   {1\over\beta}p+{1\over2\beta}\,
  \sum_{n\ne0}\lb \alpha_n\, e^{i{2\pi\over\beta}n(\sigma-t)}
    -\overline\alpha_n\,  e^{i{2\pi\over\beta}n(\sigma+t)}\rb~.
\ee
From the canonical commutation rules, we see that
\bb
    \lb \phi_0, p \rb &=& i~, \\
    \lb \alpha_n,\alpha_m\rb &=&\lb\overline \alpha_n,\overline\alpha_m\rb
    \=n\, \delta_{n+m,0}~.
\ee
The rest of the commutators vanish. Notice that $p$ has the interpretation
of momentum, and therefore has eigenvalues $p=n/R$.

After the analytic continuation $t\mapsto -i\tau$, we define
$$
  z\eq e^{-i{2\pi\over\beta}(\sigma+i\tau)}\=
       e^{{2\pi\over\beta}(-i\sigma+\tau)}~, ~~~~~
  \overline z\= e^{{2\pi\over\beta}(i\sigma+\tau)}~.
$$
The series describing the boson thus gives
\bb 
  \Phi(z)&=&
  \phi_0+{wR\over-2i}(\ln z-\ln \overline z)-ip(\ln z+\ln \overline z)
  +i\,\sum_{n\ne0}\lb {a_n\over n}\, z^{-n}
    +{\overline a_n\over n}\,  \overline z^{-n}\rb \\
   &=& \phi_0+i\lp {wR\over2}-p\rp\, \ln z
         +i\lp -{wR\over2}-p\rp\, \ln \overline z
  +i\,\sum_{n\ne0}\lb {a_n\over n}\, z^{-n}
    -{\overline a_{-n}\over n}\,  \overline z^{-n}\rb~.
\ee
The boson field may be decomposed into two chiral components:
\bb
  \phi (z) &=&
  \frac{\phi_0}{2} -i\alpha_0\, \ln  z
 +  i\sum_{n\ne0} {\alpha_n\over n z^n}~,  
\\
  \overline\phi ( \overline z) &=&
  \frac{\phi_0}{2}  -i\overline\alpha_0\, \ln \overline z
  +  i\sum_{n\ne0} {\overline\alpha_n\over n\overline z^n}    ~,
\ee
where we have defined
\bb
  \alpha_0 \eq p-{wR\over2}~~~{\rm and}~~~
  \overline\alpha_0 \eq p+{wR\over2}~.
\ee
Evaluating the expression \calle{eq:GP53}, one 
determines that the Hamiltonian for the
free compactified boson is
\bb
 H \= \frac{2 \pi}{\beta} \lp L_0+\overline L_0\rp~,
\ee
where
\bb
    L_0 &=&     \lp p- {Rw\over2} \rp^2 
              + \sum_{n=1}^{+\infty} \alpha_{-n} \alpha_n
              - \frac{1}{24}~, \\
    L_0 &=&     \lp p+{Rw\over2} \rp^2  
              + \sum_{n=1}^{+\infty} \overline\alpha_{-n} \overline\alpha_n
              - \frac{1}{24}~.
\ee
The vacuum state $\ket{\emptyset; w,k}$ is labeled by the winding number
$w$ and the momentum integer $k$. These are also heighest weight states 
with conformal weights
\bb
 \Delta_{kw}&=&{1\over2}\,\lp {k\over R}-{Rw\over2}\rp^2-\frac{1}{24}~,\\  
 \overline\Delta_{kw}&=&{1\over2}\,\lp {k\over R}+{Rw\over2}\rp^2-\frac{1}{24}~.
\ee

\separator

\item
The fermion propagator $G$ is the solution of the equation
$$
   \overline\partial G \= {2\pi\over\lambda}\, \delta^{(2)}(z-w)~.
$$
Using the representation \calle{eq:CUR100} of the $\delta$-function
$$
 \delta^{(2)}(z-w)\={1\over 2\pi }\, \overline\partial{1\over z-w}~,
$$
the above equation reads
$$
 \overline\partial G \= {1\over\lambda}\,  \overline\partial{1\over z-w}~,
$$
from which we can immediately infer that
$$
    G \= {1\over\lambda}\,{1\over z-w}~.
$$

\separator

\item

Working in the same way as in the case
of the free boson, we define
$$
 i\psi(z)  \eq \sum_n \, \psi_n \,z^{-n-1/2}~,
$$
for $n\in\BZ+1/2$. Then
$$
 \psi_{n} \= \ointleft \, \frac{dz}{2 \pi i} \, z^{n-1/2} \, i\psi(z)~.
$$
We now recall that, due to the fermionic nature of the field $\psi$,
changing the order of two $\psi$'s gives an extra negative sign.
Thus we can write the anticommutator in terms of contour
integrals as follows:
\bb
 \{ \psi_n, \psi_m \} \,
 &=& \, i^{2} \,\lb\ointleft \, \frac{dz}{2 \pi i},
       \, \ointleft \, \frac{dw}{2\pi i}
       \rb \, z^{n-1/2} \, \psi(z) \, w^{m-1/2} \, \psi(w)\\
 &=& \, \ointleft_{C(z)} \, \frac{dz}{2 \pi i} \, \ointleft_{\gamma} \,
          \frac{dw}{2 \pi i} \, z^{n-1/2} \, w^{m-1/2} \,
          \CR\lp\psi(z) \psi(w)\rp\\
 &=& \, \ointleft_{C(z)} \, \frac{dz}{2 \pi i} \, \ointleft_{\gamma} \,
         \frac{dw}{2 \pi i} \, z^{n-1/2} \, w^{m-1/2} \, \frac{-1}{z-w}~,
\ee
which gives us
$$
 \{ \psi_{n}, \psi_{m} \} \= \delta_{n,-m}~.
$$

\separator

\item
(a) Using Wick's theorem, we obtain
\bb
 \CR \lp T(w) \psi(z)\rp 
 & = & \, \CR \lp \frac{1}{2} \, :\psi(w) \partial_{w} \psi(w):
          \psi(z)\rp\\
 & = & \, \frac{1}{2} \, :\psi(w) \partial_{w} \psi(w) \psi(z): -
          \frac{1}{2} \, \lowerpairing{\psi(z)\psi}(w) \, \partial_{w}\psi(w)\\
 &   &  + \frac{1}{2} \, \partial_w\lowerpairing{\psi(w)\psi}(z) \, \psi(w)\\
 & = & - \frac{1}{2} \, : \psi(w) \partial_w \psi(w) \, \psi(z):
         + \frac{1/2}{w-z} \, \partial_w \psi(w)  +
          \frac{1/2}{(w-z)^2} \, \psi(w)~.
\ee
Then Taylor expanding the field $\psi(w)$ around $z$ gives
\bb
  \psi(w)\= \psi(z)+\partial_z \psi(z) (w-z)
   +\frac{1}{2} \partial_z^2 \psi(z) (w-z)^2 + \cdots~,~~~{\rm and}\\
  \partial_w\psi(w)\= \partial_z \psi(z) 
   + \partial_z^2 \psi(z) (w-z) + \cdots~,
\ee
and so we find
\bb
 \CR \lp T(w) \psi(z)\rp &=&
 - \frac{1}{2} \, :\lb \psi(z) + \partial_{z}\psi(z)(w-z)
        + \ldots\rb
  \lb\partial_z \psi(z)
          + \partial_z^2 \psi(z) (w-z)+ \ldots\rb
          \psi(z):\\
 && + {1/2\over w-z} \,  \lb \partial_z
          \psi(z)+ \partial_z^2 \psi(z)
          (w-z) + \ldots\rb \\
 && + {1/2\over (w-z)^2} \, \lb\psi(z) \, + \, \partial_{z} \psi(z)(w-z)
          \, + \, \frac{1}{2!} \partial_{z}^{2} \psi(z) (w-z)^{2} \, +
          \, \ldots \rb \\
 &=& \frac{1/2}{(w-z)^{2}} \, \psi(z) \,+ \, \frac{1}{w-z} \,
          \partial_{z} \psi(z) \, + \, \frac{3}{4} \, \partial_{z}^{2}
          \psi(z) \, + \, \ldots~,
\ee
where we have used the fact that
$$
 : \psi(z) \, \Bigl(\partial_{z} \psi(z) \Bigr) \, \psi(z): \= 0~.
$$
From the result of the  calculation above and the definition
\beq
\label{eq:GP52}
    \CR \lp T(w) \psi(z)\rp \= \sum_n \, {L_n\psi(z)\over(w-z)^{n+2}}~,
\eeq
it is thus obvious that
$$
 L_{-2} \psi(z) \= \frac{3}{4} \, \partial_{z}^{2} \psi(z)~.
$$

(b) Kac's formula on states
$$
 (L_{-2} \, - \, \frac{3}{4} \, L_{-1}) \, \ket{1/2} \= 0
$$
can be written as a differential equation on fields:
\beq
\label{eq:GP21}
 L_{-2}(\psi(z)) \, -\, \frac{3}{4} \, L_{-1}^{2} (\psi(z)) \= 0~.
\eeq
From the definition \calle{eq:GP52}, we find that
$$
 L_{-n}(\psi(w)) \=\ointleft  {dz\over 2 \pi i} \, (z-w)^{-n+1} \,
 \CR\lp T(z) \psi(w\rp)~.
$$
In particular
\bb
 L_{-1} (\psi(w)) \,
 & = & \ointleft  \frac{dz}{2 \pi i} \, \CR\lp T(z) \psi(w)\rp \\
 & = & \ointleft  \frac{dz}{2 \pi i} \, \lb\frac{1/2}{(z-w)^{2}} \,
          + \, \frac{1}{z-w} \partial_{z}\psi(z) \,+\, \ldots\rb~.\\
\ee
Therefore
$$
 L_{-1}(\psi(w)) \= \partial_{w} \psi(w)~.
$$
Then \calle{eq:GP21} gives
$$
 L_{-2} (\psi(z)) \= \frac{3}{4} \, \partial_{z}^{2} \psi(z)~.
$$

\separator

\item

(a) For a generic $BC$-system with signature $\varepsilon$,
the OPEs between
the two fields can be written as
\bb
    B(z)C(w) &=& {1\over z-w}~,\\
    C(z)B(w) &=& {\varepsilon\over z-w}~.
\ee

The energy-momentum tensor may now be derived easily  using the following
argument. Since it is a field of weight 2, it should be a linear combination
of all weight 2 products made out of $B$ and $C$, and so it must be of the
form
$$
   T(z)\= \alpha_1 \, :B(z)\, \partial_z C(z):
          +\alpha_2 \, :\partial_zB(z) \, C(z):
   ~.
$$
The coefficients $\alpha_1$ and $\alpha_2$ can be 
determined by demanding that
$T(z)$ leads to the correct conformal weights for $B$ and $C$.
In particular, we must have
\beq
\label{eq:GP5}
   T(z)B(w)\= {j\, B(w)\over (z-w)^2}+{\partial_w B(w)\over z-w}+\mbox{reg}~.
\eeq
On the other hand, using the expression for $T(z)$ in terms of the
fields $B$ and $C$, along with the OPEs for these fields,
we see that
\bb
    T(z)B(w) &=& \lb \alpha_1 \, :B(z)\, \partial_z C(z):
                +\alpha_2 \, :\partial_zB(z) \, C(z):\rb B(w) \\
    &=& 
       \alpha_1 \,B(z)\,\partial_z
        \lowerpairing{C(z)B}(w) +
       +\alpha_2 \,\partial_zB(z) \, \lowerpairing{C(z)B}(w) +\mbox{reg}\\
    &=&
       \alpha_1 \,B(z)\,\partial_z{1\over z-w}
     +\alpha_2 \,{\partial_zB(z)\over z-w} +\mbox{reg}\\
    &=&
       -\alpha_1 \, {B(z)\over (z-w)^2}  +\alpha_2 \,{\partial_zB(z)\over z-w}
    +\mbox{reg} ~.
\ee
We can now expand $B(z)$ and $\partial_zB(z)$ in  Taylor series around $w$:
\bb
    T(z)B(w)
     &=&
   -\alpha_1 \, {B(w)+(z-w)\,\partial_w B(w)+\CO((z-w)^2)\over (z-w)^2}  
   +\alpha_2 \,{\partial_wB(w)+\CO(z-w)\over z-w}
    +\mbox{reg} \\
    &=&
   -\alpha_1 \, {B(w)\over (z-w)^2} + 
   {(-\alpha_1+\alpha_2)\partial_w B(w)\over z-w} + \mbox{reg}~. 
\ee
Comparing the last expression with \calle{eq:GP5}, we see that
$\alpha_1=-j$ and $\alpha_2=1-j$. Therefore, the energy-momentum
tensor is given by
$$
   T(z)\= -j \, :B(z)\, \partial_z C(z):
          +(1-j) \, :\partial_zB(z) \, C(z):
   ~.
$$

(b) To find the central charge $c$, we need to compute the
    OPE for $T(z)T(w)$, using the energy-momentum tensor we 
    obtained in the previous part. In particular, it is enough to
    focus only on the most singular term of the OPE,
    the one that is obtained upon two contractions among the
    fields.  The calculation is straightforward, and we present
    it here:
\bb
   T(z)T(w) &=&
   \lb-j \, :B(z)\, \partial_z C(z):+(1-j) \, :\partial_zB(z) \, C(z):\rb\\
   &&\lb-j \, :B(w)\, \partial_w C(w):+(1-j) \, :\partial_wB(w) \, C(w):\rb\\
   &=&
    +j^2 \, :B(z)\, \partial_z C(z):~:B(w)\, \partial_w C(w): \\
   &&-j(1-j) \, :B(z)\, \partial_z C(z):~:\partial_wB(w) \, C(w):\\
   &&
   -j(1-j) \,  :\partial_zB(z) \, C(z):~:B(w)\, \partial_w C(w): \\
  &&+(1-j)^2 \,  :\partial_zB(z) \, C(z):~ :\partial_wB(w) \, C(w):\\
   &=&
   j^2 \,
 :\lowerpairing{B(z)\,\partial_z \lowerpairing{C(z):~:B}(w)\,\partial_w C}(w):\\
   &&-j(1-j) \, 
 :\lowerpairing{B(z)\, \partial_z \lowerpairing{C(z):~:\partial_wB}(w) \, C}(w):
   \\
  &&  -j(1-j) \,
  :\partial_z\lowerpairing{B(z)\,\lowerpairing{C(z):~:B}(w)\,\partial_w C}(w):\\
  &&+(1-j)^2 \,  
  :\partial_z\lowerpairing{B(z) \, \lowerpairing{C(z):~ :\partial_wB}(w)\,C}(w):
  +\cdots   \\
  &=&
  j^2 \,\partial_w{1\over z-w}\, \partial_z{\varepsilon\over z-w}
  -j(j-1)\,{1\over z-w}\,\partial_w\partial_z{\varepsilon\over z-w} \\
  &&
   -j(j-1)\,\partial_w\partial_z{1\over z-w}\,{\varepsilon\over z-w}
   +(1-j)^2 \,\partial_z{1\over z-w}\,\partial_w{\varepsilon\over z-w}
   +\cdots \\
  &=&
  j^2\, {1\over (z-w)^2}\,{-\varepsilon\over (z-w)^2}
  -j(j-1)\, \, {1\over z-w}\,{-2\varepsilon\over (z-w)^3} \\
  &&
  -j(j-1)\, ,{-2\over (z-w)^3}\,{\varepsilon\over z-w}
  +(j-1)^2\,  {1\over (z-w)^2}\,{\varepsilon\over (z-w)^2}
   +\cdots \\
  &=& \varepsilon\,  {-j^2+2j(j-1)-(j-1)^2\over (z-w)^4} +\cdots~.
\ee  
Thus we conclude that the central charge of the generic $BC$-system is
$$
  c\= -\varepsilon\, (6j^2-6j+1)~.
$$

\separator

\item
Starting  from the OPE 
\bb
 T(z_1)T(z_2)&=&{c/2\over (z_1-z_2)^4}+ {2T(z_2)\over (z_1-z_2)^2}
    +{\partial T(z_2)\over z_1-z_2}+{\rm reg}~,
\ee
we see that
\bb
    \average{T(z,\overline z) T(w,\overline w)}\=
    {c/2\over (z-w)^4} ~.
\ee
We rewrite this result in the form
\bb
    \langle T(z) T(w) \rangle  \= -{c\over 12}\, \partial_z^2\partial_w
   {1\over z-w}~.
\ee
Differentiating with respect to $\overline z$, we get
\bb
    \langle \partial_{\overline z} T(z) T(w) \rangle &=&
   -{c\over 12}\, \partial_z^2\partial_w\partial_{\overline z}{1\over z-w}
   \delta(z-w) \\
   &=& -{c\pi\over 6}\, \partial_z^2\partial_w
   \delta(z-w)~,
\ee
where we have made use of equation \calle{eq:CUR100}. 
Using the  conservation of the energy momentum tensor 
$$
  \overline\partial T_{zz}+\partial T_{\overline zz}\=
  \overline\partial T+{1\over4}\,\partial\Theta\=0
  ~\Rightarrow~ \overline\partial T\=- {1\over4}\,\partial\Theta~,
$$
we finally find the relation sought:
\beq
\label{eq:notraceless}
    \langle \partial_z \Theta(z) T(w) \rangle  \=
   {2c\pi\over 3}\, \partial_z^2\partial_w
   \delta(z-w)~.
\eeq

The confusion that creates the contradictory natury of equations 
\calle{trace} and \calle{eq:notraceless} is eliminated if 
we say that the energy-momentum tensors that appear in the two equations
are different! Therefore, the  puzzle examined in the problem is
only superficial, and arises by (bad?) notation.

Let us try to refine our notation. We call the energy-momentum
tensor defined by \calle{trace} the {\bf physical energy-momentum
tensor} $T^{(phys)}_{\mu\nu}$.
On the other hand, we call the tensor of equation \calle{eq:notraceless}
the {\bf conformal energy-momentum tensor}
$T^{(conf)}_{\mu\nu}$.
How are the two tensors related?
Simply, any acceptable conformally invariant theory is  a sum
of CFTs, each contributing its own \textit{conformal} piece to
the \textit{physical} energy-momentum tensor:
\bb
   T^{(phys)}_{\mu\nu} = \sum_{CFTs}
   T^{(conf)}_{\mu\nu} ~.
\ee
The total theory has $c=0$, and therefore 
$$
   \langle \partial_z \Theta(z) T(w) \rangle  \= 0~
$$
Thus no paradox is present.

\separator

\item
The energy-momentum tensor  \calle{eq:GP55},
can be defined through
\beq
\label{eq:GP56}
   \delta S \= -{1\over 4\pi} \, \iint\, d^2 x\, 
    \sqrt{g}\, T_{\mu\nu}\, \delta g^{\mu\nu}~.
\eeq
We can write $g_{\mu\nu}$
in terms of some reference metric $\tilde g_{\mu\nu}$
by defining
$$
  g_{\mu\nu} \= e^\sigma \, \tilde g_{\mu\nu} ~\Rightarrow~
  g^{\mu\nu} \= e^{-\sigma} \, \tilde g^{\mu\nu}~.
$$
Then
$$
    \delta g^{\mu\nu}= - \delta\sigma \,  g^{\mu\nu}~,
$$
and equation \calle{eq:GP56} gives 
\bb
   \delta S &=& {1\over 4\pi} \, \iint\, d^2 x\,
    \sqrt{g}\, T_{\mu\nu}\, g^{\mu\nu}\, \delta \sigma \\
            &=& {1\over 4\pi} \, \iint\, d^2 x\,
    \sqrt{g}\, \Theta\, \delta \sigma ~.
\ee
Since the total theory $S'$ has vanishing trace ($\Theta'=0$),
the equation just found, applied to the total action $S'$,
implies that $\delta S'=0$. On the other hand, $\delta S'=\delta S+\delta S_L$.
We thus conclude that
\bb
   \delta S_L &=& -\delta S\\
    &=&  -{1\over 4\pi} \, \iint\, d^2 x\, 
                \sqrt{g}\, \Theta\, \delta \sigma \\
    &=&  -{1\over 4\pi} \, \iint\, d^2 x\, 
                \sqrt{g}\,  (a R +b)
  \, \delta \sigma ~.
\ee
In terms of the reference metric $\tilde g_{\mu\nu}$
(see appendix):
$$
 \sqrt{g}\=e^\sigma\, \sqrt{\tilde g}~,~~~~~
 R\= e^{-\sigma}\, \lp\tilde R+\tilde\triangle\sigma\rp~.
$$
Then
\bb
   \delta S_L   &=&  -{1\over 4\pi} \, \iint\, d^2 x\, 
               \sqrt{\tilde{g}}\, 
 \lb a\lp \tilde R+\tilde\Delta\sigma\rp
   +b\, e^\sigma \rb \, \delta \sigma~. 
\ee
This result can be integrated easily:
\bb 
   S_L =  -{1\over 4\pi} \, \iint\, d^2 x\, 
               \sqrt{\tilde g}\, 
 \lb a \lp \tilde R\, \sigma
  +{1\over2}\sigma\tilde\Delta\sigma\rp
   +b\, e^\sigma \rb ~. 
\ee
This action is  the Liouville action sought.
Usually, the parametrization $\tilde g_{\mu\nu}=e^\phi\,\delta_{\mu\nu}$
is used. Then  
$$
   \sqrt{\tilde g}\= e^\phi~,~~~
   \tilde R\=-e^{-\phi}\,\partial_z\partial_{\overline z}\phi~,~~~
   \tilde \triangle\sigma\=-e^{-\phi}\,\partial_z\partial_{\overline z}\sigma~,
$$
and
\bb
   S_L =  {1\over 4\pi} \, \iint\, d^2 x\, 
 \lb a \lp \sigma\partial_z\partial_{\overline z}\phi
          +\sigma\partial_z\partial_{\overline z}\sigma\rp
    -b\, e^{\phi+\sigma} \rb ~. 
\ee

\footnotesize
{\bf APPENDIX}

As we have proved in Exercise \ref{item:CUR1} of Chapter \ref{ch:CUR},
the Riemann surfaces are
conformally flat. A metric can then always be chosen such that
\beq
\label{eq:GP60}
   g_{\mu\nu} \= \rho\,\delta_{\mu\nu}~.
\eeq
Using the equation
$$
  R\= -{2\over\sqrt{g_{11}g_{22}}}\lb
  {\partial\over\partial x^1}\lp{1\over\sqrt{g_{11}}}
  {\partial\over\partial x^1}\sqrt{g_{22}}\rp+
  {\partial\over\partial x^2}\lp{1\over\sqrt{g_{22}}}
  {\partial\over\partial x^2}\sqrt{g_{11}}\rp\rb~,
$$
which gives the scalar curvature of a 2-dimensional manifold,
we can easily find that for the metric \calle{eq:GP60},
$$
   R\= -{1\over\rho}\, (\partial_1^2+\partial_2^2)\ln\rho\=
     -{1\over\rho}\, \partial_z\partial_{\overline z}\ln\rho~.
$$

Given a second metric 
$$
  \tilde g_{\mu\nu} \= \tilde\rho\,\delta_{\mu\nu}~,
$$
the corresponding scalar curvature would be
$$
 \tilde R\=-{1\over\tilde\rho}\,\partial_z\partial_{\overline z}\ln\tilde\rho~.
$$
  
We can easily find the relation between the
two curvatures. If we set $\rho=e^\sigma\,\tilde\rho$, then
$$
    R\= e^{-\sigma}\, \lp \tilde R -{1\over\tilde\rho}\,
        \partial_z\partial_{\overline z}\sigma \rp~.
$$
Notice that the Laplacian of the metric $\tilde g_{\mu\nu}$ would be:
$$
 \tilde\triangle\sigma\=-{1\over\sqrt{\tilde g}}\,\partial_\mu\lp\sqrt{\tilde g}
 \tilde g^{\mu\nu}\partial_\nu\sigma\rp\=
 -{1\over\tilde\rho}\,\partial_\mu\lp\tilde\rho\,
   {1\over\tilde\rho}\delta^{\mu\nu}\,\partial_\nu\sigma\rp\=
   -{1\over\tilde\rho}\,(\partial_1^2+\partial_2^2)\sigma~.
$$
Therefore, we may write
$$
  R\= e^{-\sigma}\, \lp \tilde R +\tilde\triangle\sigma\rp~.
$$

\normalsize

\separator

\item
We need to calculate the variation of $S$ with respect the
variation of $g_{ab}$. Through the calculations, we shall make use of the
well-known identities (for a reference on identities for
metrics, Christoffel symbols, and related topics, see any standard 
reference on general relativity, such as \cite{LL,MTW,Weinberg}):
\bb
   \delta g_{mn} &=& -g_{ma} g_{nb} \delta g^{ab}~, \\
   \delta g  &=&  g\,  g^{mn}\, \delta  g_{mn}~,  \\
   \delta \SQR  &=& \half\, \SQR\,  g^{mn}\, \delta  g_{mn}~,  \\
    \delta( \sqrt{g} g^{ab} ) &=& \sqrt{g}\,
    (\delta^a_m \delta^b_n - {1\over 2}\, g_{mn} g^{ab} )\,
    \delta g^{mn}~,~~~{\rm and}\\
     g^{mn} \delta g_{ms} &=& - g_{ms} \delta g^{mn}~.
\ee

Now we calculate $\delta S$, and get
\bb
  \delta S &=& 
  \alpha\iint_M\,d^2\xi\,  \delta(\sqrt{g}\, g^{ab})\,
          \partial_a\Phi\partial_b\Phi
  + \beta\iint_M\,d^2\xi\,  \delta(\sqrt{g}\,g^{ab}\, R_{ab} )\, \Phi
   \\
           &=&
  \alpha\iint_M\,d^2\xi\,  \delta(\sqrt{g}\, g^{ab})\,
          \partial_a\Phi\partial_b\Phi
  + \beta\iint_M\,d^2\xi\,  \delta(\sqrt{g}\,g^{ab})\, R_{ab} \, \Phi
  + \beta\iint_M\,d^2\xi\,  \sqrt{g}\,g^{ab}\, \delta R_{ab} \, \Phi
   \\
           &=&
  \alpha\iint_M\,d^2\xi\, \SQR  \,(\delta^a_m\delta^b_n-\half g_{mn}g^{ab})\,
          \partial_a\Phi\partial_b\Phi
  + \beta\iint_M\,d^2\xi\,  (\delta^a_m\delta^b_n-\half g_{mn}g^{ab})\,
   R_{ab} \, \Phi
   \\
           && 
  \phantom{\iint_M\,d^2\xi\, \SQR  \,(\delta^a_m\delta^b_n-\half 
  g_{mn}g^{ab})\, \partial_a\Phi\partial_b\Phi}
  + \beta\iint_M\,d^2\xi\,  \sqrt{g}\,g^{ab}\, \delta R_{ab} \, \Phi
   \\
           &=&
 \alpha\iint_M\,d^2\xi\, \SQR  \,(\partial_m\Phi\partial_n\Phi
 -\half g_{mn}g^{ab}\,
 \partial_a\Phi\partial_b\Phi)\, \delta g^{mn}
  + \beta \iint_M\,d^2\xi\,  (R_{mn}-\half g_{mn}R)
   \, \Phi \, \delta g^{mn}
   \\
           &&
  \phantom{\iint_M\,d^2\xi\, \SQR  \,
   (\partial_m\Phi\partial_n\Phi
 -\half g_{mn}g^{ab}\,
 \partial_a\Phi\partial_b\Phi)\, \delta g^{mn}}
  + \beta\iint_M\,d^2\xi\,  \sqrt{g}\,g^{ab}\, \delta R_{ab} \, \Phi
  ~.
\ee 
Therefore
$$
  T_{mn}\=-4\pi\alpha(\partial_m\Phi\partial_n\Phi-\half g_{mn}\,
            \partial_a\Phi\partial^a\Phi)
          -4\pi\beta (R_{mn}-\half g_{mn}R)
          -4\pi\beta\,{1\over\sqrt{g}} {\delta A\over\delta  g^{mn}}~,
$$
where
$$
  A\eq  \iint_M\,d^2\xi\,  
  \sqrt{g}\,g^{ab}\, \delta R_{ab} \, \Phi~.
$$

One must still rearrange the last term in $T_{mn}$ in a convenient form. This 
requires some work. Before we undertake this task, 
the reader should keep in mind the following remarks:
\begin{itemize}
\item As well-known, the metric is covariantly constant, i.e.,
      $g_{mn;r}=0$. We can thus freely move it in and out of a
      covariant derivative.
\item If $A_a$ is a vector, then its divergence is given by
     \beq
     \label{eq:CNM5}
     A^a_{~;a} = {1\over \sqrt{g}}\, \partial_a(\sqrt{g}A^a)~.
     \eeq
\item We will drop all terms of the form
      $$
        \iint d^2\xi (\cdots)_{,n}
      $$
      since, by Gauss' theorem, they become boundary terms and thus vanish.
\end{itemize}
We now continue with the calculations.

Using the Palatini identity \calle{variation:Rij}, we  write
\bb
   A &=&  \iint_M\,d^2\xi\,
  \SQR\,g^{mn}\, (\delta\Gamma_{ma;n}^a-\delta\Gamma^a_{mn;a}) \, \Phi
   \\
   &=&
   \underbrace{\iint_M\,d^2\xi\,
  \SQR\,g^{mn}\, \delta\Gamma_{ma;n}^a \, \Phi}_{B}
   - \underbrace{\iint_M\,d^2\xi\,
  \SQR\,g^{mn}\, \delta\Gamma^a_{mn;a} \, \Phi}_{C}~.
\ee
Note that here and in what follows,
the variation symbol ``$\delta$'' acts only on the object 
immediately following it.

We begin re-arranging the integrand in the first term $B$:
\bb
  B &=& \iint_M\,d^2\xi\,
  \SQR\,g^{mn}\, \delta\Gamma_{ma;n}^a \, \Phi
  \nonumber \\
  &=& \iint_M\,d^2\xi\,
  \SQR\,(g^{mn}\, \delta\Gamma_{ma}^a)_{;n} \, \Phi
  \\
  &{\calle{eq:CNM5}\atop =}& \iint_M\,d^2\xi\,
  (\SQR\, g^{mn}\, \delta\Gamma_{ma}^a)_{,n} \, \Phi
  \\
 &=& \iint_M\,d^2\xi\,
  (\SQR\, g^{mn}\, \delta\Gamma_{ma}^a\, \Phi)_{,n}
 - \iint_M\,d^2\xi\,
  \SQR\, g^{mn}\, \delta\Gamma_{ma}^a \, \Phi_{,n}
  \\
 &{\calle{eq:CNM6}\atop =}& 
 - \iint_M\,d^2\xi\,
  \SQR\, g^{mn}\, \partial_m\delta(\ln\SQR) \, \Phi_{,n}
  \\
 &=& 
 - \iint_M\,d^2\xi\,
  \partial_m(\SQR\, g^{mn}\, \delta(\ln\SQR) \, \Phi_{,n})
 +\iint_M\,d\xi^1 d\xi^2\,
  \partial_m(\SQR\, g^{mn}\,  \Phi_{,n})\, \delta(\ln\SQR)
   \\
 &=&
  \iint_M\,d^2\xi\,
  \partial_m(\SQR\, g^{mn}\,  \Phi_{,n})\,{\delta\SQR\over\SQR}
 \\
 &{\calle{eq:CNM5}\atop =}&
  \iint_M\,d^2\xi\, \delta\SQR
   \, \Phi^{,m}_{~;m}
 \\
 &=&
  -\half\, \iint_M\,d^2\xi\,
   \SQR\, \delta g^{mn} g_{mn} \, \Phi^{,r}_{~;r}
  ~.
\ee
We now continue with the $C$ term: 
\footnotesize
\bb
  C &=&\iint_M\,d^2\xi\,
  \SQR\,g^{mn}\, \delta\Gamma^a_{mn;a} \, \Phi
   \\
 &=& \iint_M\,d^2\xi\,
  \SQR\, (g^{mn}\, \delta\Gamma^a_{mn})_{;a} \, \Phi
   \\ 
 &{\calle{eq:CNM5}\atop =}& \iint_M\,d^2\xi\,
  (\SQR\, g^{mn}\, \delta\Gamma^a_{mn})_{,a} \, \Phi
   \\ 
  &=& \iint_M\,d^2\xi\,
  \partial_a(\SQR\, g^{mn}\, \delta\Gamma^a_{mn} \, \Phi)
  -\iint_M\,d^2\xi\,
  \SQR\, g^{mn}\, \Phi_{,a}\,
  \delta\Gamma^a_{mn}
   \\ 
 &=&  -\iint_M\,d^2\xi\,
  \SQR\,  \Phi_{,a}\,
  g^{mn}\, \delta\Gamma^a_{mn} \\
  &=&  \iint_M\,d^2\xi\,
  \SQR\,  \Phi_{,a}\,
   \lb - \delta \lp g^{mn}\Gamma^a_{mn}\rp +\delta g^{mn} \Gamma^a_{mn}\rb\\
  &{\calle{eq:CNM7}\atop=}&
   \iint_M\,d^2\xi\, \SQR\,  \Phi_{,a}\,
   \delta\lp {1\over\SQR}\partial_s(\SQR g^{as})\rp
  +\iint_M\,d^2\xi\, \SQR\,  \Phi_{,a}\,\delta g^{mn} \Gamma^a_{mn}\\
  &=& \iint_M\,d^2\xi\, \SQR\,  \Phi_{,a}\,
    {\delta\SQR\over -g}\partial_s(\SQR g^{as})
    +\iint_M\,d^2\xi\, \SQR\,  \Phi_{,a}\,
   {1\over\SQR}\partial_s\delta(\SQR g^{as})
   +\iint_M\,d^2\xi\, \SQR\,  \Phi_{,a}\,\delta g^{mn} \Gamma^a_{mn}\\
  &=& -\iint_M\,d^2\xi\,  \Phi_{,a}\,
    {\delta\SQR\over \SQR}\partial_s(\SQR g^{as})
    +\iint_M\,d^2\xi\,   \Phi_{,a}\,
   \partial_s\delta(\SQR g^{as})
   +\iint_M\,d^2\xi\, \SQR\,  \Phi_{,a}\,\delta g^{mn} \Gamma^a_{mn}\\
  &=& -\iint_M\,d^2\xi\,  \Phi_{,a}\,
    {\delta\SQR\over \SQR}\partial_s(\SQR g^{as})
    +\iint_M\,d^2\xi\,   
    \lb\partial_s\lp\Phi_{,a}\delta(\SQR g^{as})\rp
     -\partial_s\Phi_{,a}\delta(\SQR g^{as})\rb
   +\iint_M\,d^2\xi\, \SQR\,  \Phi_{,a}\,\delta g^{mn} \Gamma^a_{mn}\\
  &=& -\iint_M\,d^2\xi\,  \Phi_{,a}\,
    {\delta\SQR\over \SQR}\partial_s(\SQR g^{as})
    -\iint_M\,d^2\xi\,   
     \partial_s\Phi_{,a}\delta(\SQR g^{as})
   +\iint_M\,d^2\xi\, \SQR\,  \Phi_{,a}\,\delta g^{mn} \Gamma^a_{mn}\\
    &{\calle{eq:CNM7}\atop=}&
   \iint_M\,d^2\xi\,  \Phi_{,a}\,
    \delta\SQR\, g^{mn} \Gamma^a_{mn}
    -\iint_M\,d^2\xi\,
     \partial_s\Phi_{,a}\delta(\SQR g^{as})
   +\iint_M\,d^2\xi\, \SQR\,  \Phi_{,a}\,\delta g^{mn} \Gamma^a_{mn}\\
   &=& \iint_M\,d^2\xi\,  \Phi_{,a}\,
    \delta\SQR\, g^{mn} \Gamma^a_{mn}
    -\iint_M\,d^2\xi\,
     \partial_s\Phi_{,a}\lp\delta\SQR\, g^{as}+\SQR\,\delta g^{as}\rp
   +\iint_M\,d^2\xi\, \SQR\,  \Phi_{,a}\,\delta g^{mn} \Gamma^a_{mn}\\
   &=& -\iint_M\,d^2\xi\,  
    \delta\SQR\, g^{mn} \,\lp\partial_m\Phi_{,n} -\Gamma^a_{mn} \Phi_{,a}\rp
   -\iint_M\,d^2\xi\, \SQR\,  \delta g^{mn}\lp\partial_m\Phi_{,n}
    - \Gamma^a_{mn}\Phi_{,a}\rp\\
   &=& -\iint_M\,d^2\xi\,  
    \delta\SQR\, g^{mn} \,\Phi_{;nm} 
   -\iint_M\,d^2\xi\, \SQR\,  \delta g^{mn} \, \Phi_{;nm}\\
   &=& -\iint_M\,d^2\xi\,
    \delta\SQR\,\Phi^{,a}_{~;a}
   -\iint_M\,d^2\xi\, \SQR\,  \delta g^{mn} \, \Phi_{;nm}\\
    &=& \iint_M\,d^2\xi\,
    {1\over2}\SQR\, g_{mn}\, \delta g^{mn}\,\Phi^{,a}_{~;a}
   -\iint_M\,d^2\xi\, \SQR\,  \delta g^{mn} \, \Phi_{;nm}~.
\ee\normalsize
Combining the results for $B$ and $C$ we arrive at
\bb
   A\= 
   \iint_M\,d^2\xi\, \SQR\,  \delta g^{mn} \, 
   \lp  \Phi_{;nm}
        - g_{mn}\, \Phi^{,a}_{~;a} \rp ~.
\ee

Therefore
$$
   {1\over\sqrt{g}} {\delta A\over\delta  g^{ij}}\=
    \Phi_{;ik}   -\Phi^{,a}_{~;a}\,
    g_{ik}~,
$$
and the energy-momentum tensor is 
\bb
  T_{ij} &=& -4\pi\alpha\,
  \lp\partial_i\Phi\partial_j\Phi
  - \half g_{ij}  \partial_a\Phi\partial^b\Phi\rp  
  - 4\pi\beta\,\lp R_{ij}
  -\half g_{ij} R\rp  
  -4\pi\beta\, \lp  \Phi_{;ij} - g_{ij}\, \Phi_{~;a}^{,a}\rp
   ~.
\ee

\footnotesize
{\bfseries APPENDIX}

For the Christoffel symbols, it is known that
\beqn
\label{eq:CNM6}
    \Gamma^a_{ma} &=& \partial_m\ln\sqrt{g} \=
   {\partial_m\sqrt{g}\over\sqrt{g}}
\eeqn
and 
\beqn
\label{eq:CNM7}
    g^{jk}\, \Gamma^a_{jk} &=& - {1\over\SQR}\,
    \partial_s (\sqrt{g}\, g^{as} ) 
      ~.
\eeqn

Since the latter formula is not as well-known, we derive it here.
Let $A^{ij}$ be the cofactors of $g_{ij}$, i.e.,
$$
   g = g_{ij} A^{ij}
$$
From this, we see that 
\bb
    {\partial g\over\partial g_{ij}} = A^{ij}
   \Rightarrow 
   g_{ij} {\partial g\over\partial g_{ij}} = g_{ij}  A^{ij} = g
   \Rightarrow 
   {\partial g\over\partial g_{ij}}  = g g^{ij}~.
\ee
On the other hand, from the chain rule,
\bb
  {\partial g\over \partial x^l}=
  {\partial g\over \partial g_{ij}}
  \,{\partial g_{ij}\over \partial x^l} 
  = g g^{ij} \,{\partial g_{ij}\over \partial x^l}
  ~.
\ee
Consequently,
\bb
   g^{jk} \Gamma^i_{jk} &=& \half \, g^{jk} g^{il}\,
   (\partial_j g_{lk} +\partial_k g_{lk} -\partial_l g_{jk} )
   \\
  &=&
   \half \, g^{jk} g^{il}\,
   (2\partial_j g_{lk}  -\partial_l g_{jk} )
   \\
  &=&
    g^{jk} g^{il}\,
   \partial_j g_{lk}  -
   \half \, g^{jk} g^{il}\,
   \partial_l g_{jk} 
   \\
  &=&
    -\partial_j g^{jk} g^{il}\,
    g_{lk}  -
   \half \,  g^{il}\,
    ( g^{jk} \partial_l g_{jk}) 
   \\
  &=&
   - \partial_j g^{ji} -{1\over 2g}\, g^{il} \, \partial_l g
   \\
  &=&  -{1\over\SQR}\, (\SQR\, \partial_j g^{ji}
   +\partial_j \SQR\, g^{ij} )
   \\
  &=&
   - {1\over\SQR}\, \partial_j (\SQR\, g^{ij})~.
\ee

For the Christoffel symbols
\bb
    \Gamma^m_{ik} = \half\, g^{mj}\, 
    (  \partial_k g_{ij}
    +  \partial_i g_{jk}
    -  \partial_j g_{ik} )~,
\ee
the variation is
\beqn
     \delta\Gamma^m_{ik} &=& \half\, \delta g^{mj}\,
     (  \partial_k g_{ij}
    +  \partial_i g_{jk}
    -  \partial_j g_{ik} ) + \half g^{mj}\,
    (  \partial_k \delta g_{ij}
    +  \partial_i \delta g_{jk}
    -  \partial_j \delta g_{ik} )
\nonumber \\
&=& -\half\, \delta g_{ln}\, g^{ml}\, g^{nj}\,
     (  \partial_k g_{ij}
    +  \partial_i g_{jk}
    -  \partial_j g_{ik} ) + \half g^{mj}\,
    (  \partial_k \delta g_{ij}
    +  \partial_i \delta g_{jk}
    -  \partial_j \delta g_{ik} )
\nonumber \\
 &=& - \delta g_{ln}\, g^{ml} \, \Gamma^n_{ik}
    + \half g^{mj}\,
    (  \partial_k \delta g_{ij}
    +  \partial_i \delta g_{jk}
    -  \partial_j \delta g_{ik} )
\label{variation:Gamma}
    \\
&=& \half\, \delta g^{mj}\,
    \Big\lbrack
       \partial_k \delta g_{ij} - \Gamma^n_{ki}\, \delta g_{nj}
                                - \Gamma^n_{kj}\, \delta g_{in}
    +  \partial_i \delta g_{jk} - \Gamma^n_{ij}\, \delta g_{nk}
                                - \Gamma^n_{ik}\, \delta g_{jn}
\nonumber \\
&&\phantom{ -\half\, \delta g^{mj}\,\Bigg\lbrack}
    -  \partial_j \delta g_{ik}- \Gamma^n_{ji}\, \delta g_{nk}
                                - \Gamma^n_{jk}\, \delta g_{in}
     \Big\rbrack
\nonumber \\
    &=&{1\over 2}\, g^{mj}\,
    ( \delta g_{ij;k}
    + \delta g_{jk;i}
    - \delta g_{ik;j} )~.
\label{variation:Gamma2}
\eeqn

Finally, we study the variation of the Ricci tensor
\bb
    -R_{ij} = \partial_a\Gamma^a_{ij}
            - \partial_j\Gamma^a_{ia}
            + \Gamma^b_{ij} \, \Gamma^a_{ba} 
            - \Gamma^b_{ia} \, \Gamma^a_{bj} ~.
\ee
We have
\beqn
   -\delta R_{ij} &=&
              \delta\partial_a\Gamma^a_{ij}
            - \delta\partial_j\Gamma^a_{ia}
            + \delta\Gamma^b_{ij} \, \Gamma^a_{ba} 
            + \Gamma^b_{ij} \, \delta\Gamma^a_{ba} 
            - \delta\Gamma^b_{ia} \, \Gamma^a_{bj} 
            - \Gamma^b_{ia} \, \delta\Gamma^a_{bj} 
\nonumber \\ 
                 &=&
              \partial_a\delta\Gamma^a_{ij}
            - \partial_j\delta\Gamma^a_{ia}
            + \delta\Gamma^b_{ij} \, \Gamma^a_{ba} 
            + \Gamma^b_{ij} \, \delta\Gamma^a_{ba} 
            - \delta\Gamma^b_{ia} \, \Gamma^a_{bj} 
            - \Gamma^b_{ia} \, \delta\Gamma^a_{bj} 
\nonumber \\ 
                 &=&
            \phantom{-}  \partial_a\delta\Gamma^a_{ij}
            - \Gamma^b_{ia}\, \delta\Gamma^a_{bj}  
            - \Gamma^b_{ja} \, \delta\Gamma^a_{ib} 
            + \Gamma^a_{ba} \, \delta\Gamma^b_{ij} 
\nonumber \\ 
                 &&
            - \partial_j\delta\Gamma^a_{ia}
            + \Gamma^b_{ij} \, \delta\Gamma^a_{ba} 
            + \Gamma^b_{ja} \, \delta\Gamma^a_{ib}
            - \Gamma^a_{bj} \, \delta\Gamma^b_{ia} 
\nonumber \\ 
                 &=&
           \delta\Gamma^a_{ij;a}
           -\delta\Gamma^a_{ia;j}~.
\label{variation:Rij}
\eeqn
This last formula is known as the {\bfseries Palatini identity}.

\normalsize

\end{enumerate}

\newchapter{CORRELATORS IN CFT}
\label{ch:CMCC}

\footnotesize
\noindent {\bfseries References}:
The standard
references for the material in this chapter are 
the same as in the previous chapter, namely
\cite{Ginz,DiFMS,Ketov}.
Bosonization is summarized in \cite{Sto}, which also reprints
many original papers.
The application of bosonization to 2-dimensional QCD is reviewed in
\cite{FriSon}.
\normalsize

\section{BRIEF SUMMARY}

\subsection{Computation of Correlators}

In any QFT, the {\bf Ward identities} can be used in the
computation of the correlation functions
$$
  G= \langle\emptyset| \Phi(x_1)\cdots\Phi(x_N) |\emptyset\rangle~.
$$
In CFT, there are three holomorphic 
{\bf conformal Ward identities}, given by
\beq
\label{eq:cWI}
  \langle\emptyset|\lb L_n, \Phi(x_1)\cdots\Phi(x_N)\rb |\emptyset\rangle
  \= 0~,~~~~~n=-1,0,1~;
\eeq
likewise, there are three antiholomorphic conformal
Ward identities associated with the generators
$\overline L_n,~n=-1,0,1$.

Although the identities \calle{eq:cWI} are true for correlation
functions of any fields, they are
most useful for correlation functions
of primary fields. In this case, the conformal Ward identities
read
\bb
  \sum_{i=1}^N \, \partial_{z_i}
  \langle\emptyset| \Phi(x_1)\cdots\Phi(x_N) |\emptyset\rangle
  &=& 0~, \\
  \sum_{i=1}^N \, (z_i\partial_{z_i}+\Delta_i)
  \langle\emptyset| \Phi(x_1)\cdots\Phi(x_N) |\emptyset\rangle
  &=& 0~, \\
  \sum_{i=1}^N \, (z_i^2\partial_{z_i}+2z_i\Delta_i)
  \langle\emptyset| \Phi(x_1)\cdots\Phi(x_N) |\emptyset\rangle
  &=& 0~.
\ee

All correlation functions containing secondary fields can be
obtained by
the action of differential operators on correlation functions
containing primary fields only.
For example, suppose that in the correlation function 
$$
  G= \langle\emptyset| \Phi_1(x_1)\cdots\Phi_{N-1}(x_{N-1})\Phi_N(x_N)
      |\emptyset\rangle~,
$$
all fields are primary except
$$
  \Phi_N(x_N)=L_{-k_1}\dots L_{-k_l}\Phi(z_N,\overline z_N)~.
$$
Then
$$
  G = \CL_{-k_1}\dots \CL_{-k_l}
  \langle\emptyset|\Phi_1(x_1)\cdots\Phi_{N-1}(x_{N-1})
  \Phi(x_N)|\emptyset\rangle~,
$$
where
$$
  \CL_k \= -\sum_{j=1}^{N-1}\lb
  {(1-k)\Delta_j\over (z_j-z_N)^k}
  +{1\over (z_j-z_N)^{k-1}} \, {\partial\over\partial z_j} 
  \rb~.
$$
One can perform similar calculations on 
correlation functions that contain more
descendant fields.

\subsection{The Bootstrap Approach}

We can treat the set of OPEs of a QFT
as the {\em fundamental} information of the theory. Then, knowing the OPEs,
one can reconstruct the
whole theory. This approach, which we discuss below, is called the
{\bfseries bootstrap approach}.

Let $\CA$ be the space of all fields $\{\Phi_i\}$ of the
QFT. This space can be decomposed, as usual, into a direct
sum of two subspaces, a fermionic subspace $\CF$ and a bosonic
subspace $\CB$:
\beq
\label{decompose}
   \CA \= \CF \oplus \CB~.
\eeq
The fields that belong in these respective subspaces satisfy
the following commutation and anti-commutation relations:
\beqn
\label{com1}
  \Phi^{(B)}(x)
  \Phi^{(B)}(y)
  &=& \phantom{-}\Phi^{(B)}(y)
  \Phi^{(B)}(x)~, \\
\label{com2}
  \Phi^{(B)}(x)
  \Phi^{(F)}(y)
  &=& \phantom{-}\Phi^{(F)}(y)
  \Phi^{(B)}(x)~, \\
\label{com3}
  \Phi^{(F)}(x)
  \Phi^{(F)}(y)
  &=& -\Phi^{(F)}(y)
  \Phi^{(F)}(x)~,
\eeqn
where $\Phi^{(F)}(x)\in\CF$ and $\Phi^{(B)}(x)\in\CB$.
As a result, the selection rules in $\CA$ are
\bb
  \Phi^{(B)}
  \Phi^{(B)} \in \CB~,\\
  \Phi^{(F)}
  \Phi^{(F)} \in \CB ~, \\
  \Phi^{(B)}
  \Phi^{(F)} \in \CF~.
\ee

The basic assumptions  in this (non-Lagrangian) formulation 
are the following:
\begin{enumerate}
\item
     $\CA$ is {\em complete}, i.e., it contains all fundamental fields, as well
     as all composite fields.
\item 
    $\CA$ is an {\em infinite dimensional} space that admits a 
    {\em countable basis}.
\item
    $\CA$ is an {\em associative algebra}.
\end{enumerate}    
 
We can now spell out explicitly the equations emerging from the previous
assumptions. Let
$ \{ A_a(x),~a=1,2,\dots\} $ be a basis in $\CA$; notice that
this is in concordance with the requirement that we have a
countably infinite basis. Then any product
$A_a(x)A_b(y)$
 in the algebra can be decomposed in terms of the basis
fields.  It is therefore enough to consider the 
products\footnote{Notice that, for convenience,
we take the OPEs to be time-ordered.}
of the basis fields among themselves:
\beq
\label{ope1}
    \CT\Big(A_a(x)A_b(y)\Big)~=~\sum_k\, C_{ab}^c(x-y)\, A_c(y)~,
\eeq
where $C_{ab}^c(x-y)$ are c-numbers which are called 
{\bf structure coefficients}\index{Structure!-- Coefficients}.

In this approach to QFT, equation \calle{ope1}
and the commutation rules \calle{com1}-\calle{com3} are 
understood as
constraints on the correlation functions.
In particular, if $\Phi(x)$ is some field in the QFT, then
\bb
    \langle \CT\lp\Phi(x) A_a(x_1)A_b(x_2)\rp \rangle
   \=\sum_k\, C_{ab}^c(x_1-x_2)\, 
      \lan \CT\lp\Phi(x) A_c(x_2)\rp\ran ~.
\ee
Hence, equation \calle{ope1} allows us to calculate all
correlation functions of the theory recursively, reducing them
ultimately to the 2-point correlation fuctions
\bb
    \lan  \CT\lp A_a(x_1)A_b(x_2)\rp \ran
   \eq \Delta_{ab}(x_1, x_2)~.
\ee

Obviously, then, if we know the 
structure coefficients, we know the QFT.
In this way, {\bf the problem of the classification of QFTs is reduced
to the problem of classifying all possible structure coefficients}.
One can use the associativity of the algebra $\CA$ to aid in the calculation
of the structure coefficients. The correlation function
$$
    \langle \CT\lp A_a(x_1)A_b(x_2)
             A_c(x_3)A_d(x_4)\rp\rangle
$$
can be calculated in two different ways by applying the expansion
\calle{ope1}.  One can take the OPEs of
    $A_a(x_1)A_b(x_2)$ and
             $A_c(x_3)A_d(x_4)$, 
or one can instead take the OPEs of
             $A_a(x_1)A_c(x_3)$ and
             $A_b(x_2)A_d(x_4)$. Consistency requires that these two 
computations lead to identical results.  This is referred to as
{\bfseries crossing symmetry}\index{Symmetry!-- Crossing}.
Mathematically, this takes the form
$$
   \sum_{nm}\, C_{ab}^n(x_1-x_2) \Delta_{nm}(x_2-x_4) C_{cd}^m(x_3-x_4)
   \= 
   \sum_{nm}\, C_{ac}^n(x_1-x_3) \Delta_{nm}(x_3-x_4) C_{bd}^m(x_2-x_4)~.
$$
This equation is known as the 
{\bf bootstrap equation}\index{Equation!Bootstrap --}, and finding
the sets of structure coefficients that satisfy this equation would amount
to a classification of QFTs.  As mentioned earlier, this whole program
is known as the bootstrap appraoch. 
Making progress using the bootstrap approach has proven to be
very hard in more than two dimensions. However, in two dimensions it becomes
easier to implement. For example,
for 2-dimensional CFTs, conformal invariance
severely restricts the 
2-point and 3-point correlation functions; they 
must be of the form
\bb
    \average{\Phi_i(z_i,\overline z_i) \Phi_j((z_j,\overline z_j)} &=&
    {\delta_{ij} \over z_{ij}^{2\Delta_i} \,
    \overline z_{ij}^{2\overline\Delta_i}} ~, ~~~{\rm and} \\
    \average{\Phi_i(z_i,\overline z_i) \Phi_j((z_j,\overline z_j)
             \Phi_k((z_k,\overline z_k)} &=&
    {C_{ijk} \over z_{ij}^{\gamma_{ij}} \,
    \overline z_{ij}^{\overline\gamma_{ij}} 
     z_{jk}^{\gamma_{jk}}
    \overline z_{jk}^{\overline\gamma_{jk}}
     z_{ki}^{\gamma_{ki}} 
    \overline z_{ki}^{\overline\gamma_{ki}} } ~,
\ee 
where $z_{ij}=z_i-z_j$ and $\gamma_{ij}=\Delta_i+\Delta_j-\Delta_k$ (and 
similarly for $\overline z_{ij}$ and $\overline\gamma_{ij}$). 
Only a set of
  {\bfseries structure constants}\index{Structure!-- Constants}
 in the OPEs remains undetermined:
$$
  \Phi_i(z_i,\overline z_i) 
  \Phi_j(0,0) \= \sum_k \,
  C_{ijk}\, z^{\Delta_k-\Delta_i-\Delta_j} \,
   \overline z^{\overline\Delta_k-\overline\Delta_i-\overline\Delta_j} \,
 \lb \Phi_k (0,0)\rb~.
$$
where the sum in r.h.s. is over the whole conformal tower.
Correlation functions that contain descendants may be
computed using the OPEs of $T(z)\Phi_i(w,\overline w)$.
Finally, the structure constants are the solutions of the bootstrap
equation
\bb
  \sum_p C_{ijp}\, C_{lmp}\, \CF^{lm}_{ij}(p|x)
  \overline{\CF}{}^{lm}_{ij}(p|\overline x)\=
  \sum_q C_{ilq}\, C_{jmq}\, \CF^{jm}_{il}(q|1-x)
  \overline{\CF}{}^{jm}_{il}(q|1-\overline x)~,
\ee
where $\CF^{lm}_{ij}(p|x)\overline{\CF}{}^{lm}_{ij}(p|\overline x)$ is
the contribution of the conformal family $\lb\Phi_p\rb$ to the 4-point
correlation function.

\subsection{Fusion Rules}

The  OPEs for any  two fields of any conformal families are determined by the OPEs
of the corresponding  primaries.  This fact allows one to write down the so-called
{\bf fusion rules}
$$
    \lb \Phi_i\rb \times \lb\Phi_j\rb \= N_{ij}^k \, \lb \Phi_k\rb
$$
as a shorthand for the OPEs of the conformal families. More concretely, the
l.h.s. of the above equation represents the OPE between a field of the conformal
family $\lb\Phi_i\rb$ and a field of the conformal family $\lb\Phi_j\rb$.
Then the r.h.s.  indicates which conformal families  $\lb\Phi_k\rb$ may have
their members appearing in the OPE.

The numbers $N_{ij}^k$  in the fusion rules are integers that count the
number of occurrences of the family $\lb\Phi_k\rb$ in the OPE.  Said in another way,
they count the number of independent ways to create the field  $\Phi_k$
from the original fields $\Phi_i$ and $\Phi_j$.  Due to this interpretation,
the numbers $N_{ij}^k$ immediately satisfy the symmetry condition
$$
       N_{ij}^k\= N_{ji}^k~.
$$
Also,  due to the associativity of the algebra of the fields, they satisfy a
quadratic constraint. Using the  matrix notation $\bfN_i=\lb N_{ij}^k\rb$,
this quadratic constraint takes the form
$$
   \bfN_i \, \bfN_l \= \bfN_l \, \bfN_i ~.
$$
Therefore, the matrices $\bfN_i$ form a commutative associative representation
of the fusion rules.  If the theory contains an infinite number of fields, then 
these matrices are infinite dimensional. However, there is a special set of
CFTs, known as {\bf rational CFTs}\index{CFT!rational --} (RCFTs), that contain
only a finite number of primary fields. The MMs discussed in Chapter \ref{ch:GP}
are special cases of RCFTs. The  RCFTs can be analyzed completely.
In  these theories, the matrices $\bfN_i$ can be diagonalized simultaneously,
and their
eigenvalues form 1-dimensional representations of the fusion rules.

\subsection{Local vs. Non-Local Fields}

 The selections rules \calle{com1}-\calle{com3} are 
 not the most general possible
 in two dimensions, as we saw in Chapter \ref{ch:CUR}.
 Thus in two dimensions, it is
 appropriate to expand the space of fields $\CA$ to
 include generalized selection rules.

 In particular,
 let us denote by $\pi_\CC(x)$ the operation of analytic
 continuation of the point $x$ around the point $y$ on the
 contour $\CC$. In general,\\
 \begin{minipage}{4in}
 $$
     A_a(\pi_\CC (x) ) A_b(y) =
     R_{ab}^{\tilde a\tilde b} \, A_{\tilde a}(x) A_{\tilde b}(y)~.
 $$ 
 \end{minipage}
 \hspace{2cm}
 \begin{minipage}{2in}
 \psfrag{x}{$x$}
 \psfrag{y}{$y$}
 \psfrag{C}{$C$}
 \includegraphics[height=1.5cm]{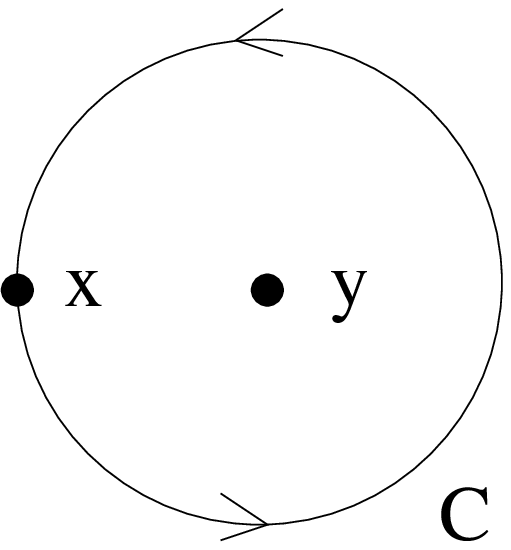}
 \end{minipage}
 In the course of this collection of problems, we will find many
 realizations of this equation.

 An interesting case that we shall study in a later chapter
 is the case of {\bf para-fermions}:
 \beq
 \label{pigamma2}
     A_a(\pi_C (x) ) A_b(y) \=
    e^{2\pi i \varphi_{ab}}\, A_a(x) A_b(y)~.
 \eeq
 The phase $\varphi_{ab}$ is a real number in the interval
 $[0,1)$; it is called the {\bf exponent of relative
 locality} of the fields $A_a(x)$ and $A_b(x)$.
 The two fields $A_a(x)$ and $A_b(y)$ are called {\bf mutually local}
 if $\varphi_{ab}=0$; a field $A_a(x)$ is called {\bf local} if
 $\varphi_{aa}=0$. According to this definition, bosons are
 local fields, while a boson and a fermion are mutually local.

 Let $A_d$ be a field occurring in the OPE of $A_a$ and $A_b$:
 $$
   A_a(x) A_b(y) \= \dots + C_{ab}^d\, A_d + \dots~.
 $$
 Then the exponent of relative locality of $A_d$ with respect
 a field $A_c$ is
 $$
   \varphi_{dc} \= \la \varphi_{ac} + \varphi_{bc} \ra~,
 $$
 where $\{ r \}$ denotes the fractional part of $r$. 
 From this result, we conclude that the correlation function
 $$
    \langle A_d(x) A_c(y) \rangle
 $$
 can be non-vanishing only if $\varphi_{dc}=0$, that is, if
 \beq
 \label{eq:CMCC31}
    \varphi_{ac} \= 1- \varphi_{bc}~.
 \eeq
   
 Consider the special case that
 the algebra can be generated by the two fields
 $\psi_1$ and $\psi_1^\dagger\equiv\psi_{-1}$, with the
 additional condition, motivated by \calle{eq:CMCC31}, that
 $$
    \varphi_{1,1}
    \=1-\varphi_{1,-1}\eq \varphi~.
 $$
 In this model, we see that we can define a discrete
 charge $Q$ such that the space $\CA$ is decomposed into
 sectors of definite charge:
 \beq
 \label{decompose2}
  \CA \= \bigoplus_{n=-\infty}^{+\infty}\, \CA_n~,
  ~~~Q\CA_n\=n\, \CA_n~.
 \eeq
 In particular,
 $$
  \psi_1\in \CA_1~,~~~
  I \in \CA_0~,~~~{\rm and}~~~
  \psi_1^\dagger\in \CA_{-1}~.
 $$
 For two fields $\Phi_n\in\CA_n$ and $\Phi_m\in\CA_m$,
 the selection rule implies that
 $$
   \Phi_n \, \Phi_m \in \CA_{m+n}~.
 $$
 Consequently,
 the exponent of mutual locality of $\Phi_n$ and $\Phi_m$ is
 \bb
   \varphi_{nm} \= \{ nm\varphi\}~,
 \ee
since $\Phi_n$ is in the same charge sector as a product 
of $n$ $\psi_1$ fields.
 Finally, the spins of the fields $\Phi_n\in\CA_n$ are
 calculated trivially from the exponent $\gamma_n$.
 Noticing that
  $$\Phi_n(\pi_\CC(x))\Phi_n(y)\=
   e^{2\pi in^2\varphi}\,
  \Phi_n(x)\Phi_n(y)~,
  $$
 and comparing this result 
 with the definition of Lorentz spin $s$, namely that 
 under a rotation of $\theta=2\pi$,
  $$
    \Phi(x) \rightarrow e^{is\theta} \Phi(x)~,
  $$
 we conclude that
  \bb
     s_n \= {1\over 2}\, \{ n^2\varphi \}
    +{1\over 2}\, n_k \= {1\over 2}\, \varphi_{nn}
    +{1\over 2}\, n_k~,
  \ee
 where the $n_k$ are integers not determined by the information
 we have specified.
  
 If $\varphi$ is rational, we can make a little more progress.
 Let us find $k,N\in  \BN$
 such that $\varphi=k/N$. Then
 the decomposition \calle{decompose2} is finite since the
 exponents of mutual  locality are periodic with period $N$:
 $$
 \varphi_{n+N,m} \=
 \{ (n+N)m\varphi \} \=
 \{ nm\varphi+mk \} \=
 \{ nm\varphi \} \= \varphi_{nm}~.
 $$
 In other words, if $\varphi$ is rational, then
 the charge $Q$ is a $\BZ_N$ charge.

\subsection{Bosonization}

As has been explained in Chapter \ref{ch:CUR}, in two dimensions, statistics
is a matter of convention, and a map can be established between fields
of different
statistics. 
One often uses this, changing the initial representation
to a bosonic representation, in order to calculate the
correlation functions.

In the language of CFT,  the free boson with $c=1$ is equivalent to
two spinors of $c=1/2$. The bosonization map is established by
$$
   \psi_1+i\psi_2 \= \sqrt{2}\, e^{i\phi}~.
$$
This is only a special case of more general constructions. 

The $bc$-system  can be bosonized using a boson 
with background charge:
$$
   T(z)\=-{1\over2}\, (\partial\phi)^2+ie_0\partial^2\phi~.
$$
The central charge of this bosonic system is
$$
    c\=1-12e_0^2~.
$$
By adjusting the coefficient $e_0$, one can match the central charge
of the $bc$-system. 

For example, the case of $j=2$ of the $bc$-system, which is of importance to
string theory, is bosonized by a theory with $e_0=3/2$ boson.
In this case
\bb
     b(z) &=& e^{-i\phi(z)}~,\\
     c(z) &=& e^{i\phi(z)}~.
\ee
A  $\beta\gamma$-system with $c=11$ also appears in string theory.
Its bosonization proceeds via a $j=0$ $bc$-system of $c=-2$ and a
boson with background charge of  $c=13$. The latter has a background
charge of $e_0=i$. The map is given by
\bb
    \beta &=& \partial b\, e^{-\phi}~,\\
    \gamma &=& c\, e^{\phi}~.
\ee

\newpage
\section{EXERCISES}

\begin{enumerate}

\item 
Given any four points $x_i, x_j, x_k, x_l$ we can define
their {\bfseries cross-ratio}\index{cross-ratio} by
$$
  z_{ijkl}\= { |x_i-x_j| \, |x_k-x_l| \over
               |x_i-x_l| \, |x_k-x_j|}~.
$$
Clearly, the cross ratios are invariant under conformal
transformations.  Find the number
of independent variables of
the $n$-point correlation function
$\average{\phi_1(x_1)\phi_2(x_2)\dots\phi_n(x_n)}$
for a conformally invariant theory in $D=2$ dimensions
by explicitly demonstrating how such a correlation
fucntion
can be expressed in terms of independent cross ratios.

\item
(a) Explain as precisely as possible why, given the Virasoro algebra,
   we can write Ward identities only for the three generators
   $L_{-1}$, $L_0$, and $L_1$.

(b) Derive the Ward identities of the operators $L_{-1}$, $L_0$,
and $L_1$.

\item
(a) A function $f(x_1,\dots,x_n)$ is called 
homogeneous\index{function!homogeneous --} of degree $r$ if
\beq
\label{def-homo}
   f(\lambda x_1,\dots,\lambda x_n) \= \lambda^r\,
   f(x_1,\dots,x_n) ~.
\eeq
For a homogeneous function, prove {\bf Euler's identity}:
$$
   \sum_{i=1}^n {\partial f\over \partial x_i}\, x_i
    \= r \, f~.
$$

(b) Show that the Ward identity of $L_0$ for a correlation function $G^{(n)}$
can be translated to the following statement: $G^{(n)}$ is a homogeneous
function of degree $\sum_{i=1}^n \Delta_i$.

\item
\label{item:CMCC4}

Show that any correlation function in CFT has the form
$$
  G \= \prod_{i=1}^N \prod_{j=i+1}^N z_{ij}^{-\gamma_{ij}}\,
        Y~,
$$
where $z_{ij}=z_i-z_j$ and $Y$ is a function of the cross-ratios
(see the solution to Exercise 1).
Find the equations satisfied by the $\gamma_{ij}$.

\item
Show that it is always possible to choose the 2-point functions such that
\beq
\label{eq:CMCC1}
  \average{\phi_i(z_1)\phi_j(z_2)} \=
  {\delta_{\Delta_i\Delta_j}\over (z_1-z_2)^{2\Delta_i} }~.
\eeq

\item
\label{item:CMCC3}
Let $\phi(z)$ be a field of conformal weight $\Delta$ with
Fourier modes $\phi_n$.
Show that
$$
   \phi_n^\dagger \= \phi_{-n}~.
$$

\item
Show that a 3-point correlation function 
$$
  \average{\phi_i(z_1)\phi_j(z_2)\phi_k(z_3)} 
$$
is fully determined up to a constant. Then relate this constant to
the OPE of the given fields.

\item
\label{item:CMCC5}
Determine the functional form of an arbitrary 4-point correlation function, 
and show that it depends on
an arbitrary function of a single variable.

\item
For a free boson $\phi(z)$, show that the correlation function
$$
 \average{ \prod_{j=1}^n \, e^{ia_j\phi(z_j)} }
$$
must vanish unless $\sum_{j=1}^n a_j = 0$.

\item
\label{item:CMCC1}
Calculate the correlation function of the previous excercise.

\item
\label{ising1}

(a) Derive the differential equation satisfied by the correlation function
$$
 \average{\sigma(z_1,\overline z_1) \sigma(z_2,\overline z_2)
 \dots \sigma(z_{2M-1},\overline z_{2M-1})
 \sigma(z_{2M},\overline z_{2M}) }
$$
of $2M$ spin fields in the Ising model.

(b) Solve the differential equation obtained in part (a) for the case
   of four spins and show that the solution can be expressed in terms
    of the hypergeometric
     function.

\item
\label{item:CMCC2}
Consider the generic OPE
$$
 \CR\lp \phi_{n}(z, \overline{z}) \, \phi_{m}(0,0)\rp \=
   \sum_{p} \, \sum_{ \{ k \} } \, \sum_{ \{ \overline{k} \} } \,
   c_{nm}^{p; \{ k \} \{ \overline{k} \} } \, \frac{\phi_{p}^{
   \{ k \} \{ \overline{k} \} }}{z^{\Delta_{n} + \Delta_{m} - \Delta_{p}
   \sum k_{i}} \, \overline{z}^{ \overline{\Delta}_{n} +
   \overline{\Delta}_{m} - \overline{\Delta}_{p} - \sum \overline{k}_{i}}}
   ~,
$$
with
\beq
\label{eq:CMCC16}
 c_{nm}^{p; \{ k \} \{ \overline{k} \} } \= c_{nm}^{p} \,
  \beta_{nm}^{p; \{ k \} } \, \overline{\beta}_{nm}^{p; \{ \overline{k} \} }
 ~.
\eeq
Describe a method that can be used to calculate the coefficients
$c_{nm}^{p; \{ k \} \{ \overline{k} \} }$.

\item
In the theory of open strings, the lowest order contribution
to the
tachyon-tachyon scattering amplitude
is given by the CFT correlation function
$$
   A \= \int\, dz_3 \,
        \average{c(z_1)c(z_2)c(z_4)V_1V_2V_3V_4}~,
$$
where $c(z)$ is a weight $j=2$ $c$-field of a $bc$-system (which in
this context are ghosts), and
$$
     V_i \= e^{i k_i^\mu X_\mu(z_i)}~
$$
is the vertex operator 
for a tachyon, which has $k^2=2$. 

Compute the amplitude $A$ in terms of standard functions of mathematical
physics.

\end{enumerate}
\newpage
\section{SOLUTIONS}

\begin{enumerate}

\item
 Translation invariance requires that $G^{(n)}$ depend only
 on the differences $x_i-x_j,~i<j$. Since
 $$
  x_i-x_j\=(x_i-x_1)-(x_j-x_1)~,
 $$
 $G^{(n)}$ depends only on the $n-1$ vectors $x_i-x_1$, $i=2,3,\dots,n$,
 i.e. it depends on $2(n-1)$ variables.

  Furthermore, rotational invariance  implies that $G^{(n)}$
  depends only on the magnitudes $r_{ij}$ of $x_i-x_j, ~\forall i,j$
  and the relative orientations of these vectors, and not on
  their absoluate orientations.  Since
  $$
    r_{ij}^2\= r^2_{i1}+r^2_{j1}-2r_{i1} r_{j1} \cos\theta_{ij}~,
  $$
  where $\theta_{ij}$ is the angle between $x_i-x_1$ and $x_j-x_1$,
  and since
  in two dimensions, all these angles are determined once one
  knows $\theta_{i1}$, the Green function can be written in terms 
  of the $n-1$ magnitudes $r_{i1}$ and the $n-2$ angles $\theta_{i2}$.
  (One has $i>1$ for the magnitudes, and $i>2$ for the angles.)
  One can see this represented in figure \ref{fig:CMCC1}.

\begin{figure}[htb!]
\begin{center}
\psfrag{a}{$x_1-x_2$}
\psfrag{b}{$x_1-x_3$}
\psfrag{c}{$x_1-x_4$}
\psfrag{n}{$x_1-x_n$}
\psfrag{q1}{$\theta_{23}$}
\psfrag{q2}{$\theta_{24}$}
\psfrag{qn}{$\theta_{2n}$}
\includegraphics[height=6cm]{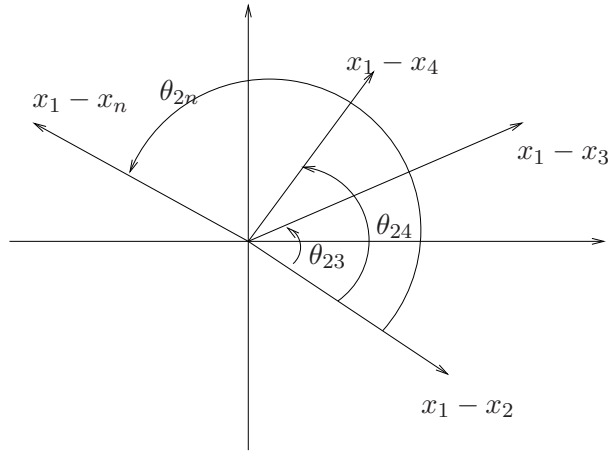}
\end{center}
\caption{In $D=2$ dimensions, knowlege of the angles $\theta_{2i}$,
  $i=3,4,\dots,n$ is enough to determine the relative orientation
of the vectors $x_1-x_i$.}
\label{fig:CMCC1}
\end{figure}

Continuing, scale invariance requires that $G^{(n)}$ depends only
on ratios $r_{ij}/r_{kl}$ and the angles (and in our case, we can
represent the angles using magnitudes of vectors).
Since any arbitrary ratio
$r_{ij}/r_{kl}$ can be written in the form
$$
    {r_{ij}\over r_{kl}}\=
    {r_{ij}\over r_{12}}\,
    {r_{12}\over r_{kl}}~,
$$
we can consider $G^{(n)}$ as a function of
$$
     u_{ij}\=r_{ij}/r_{12}~.
$$
Obviously, we have now one variable less
since $u_{12}=1$. Therefore, so far,  $G^{(n)}$ depends on $2(n-2)$
variables
$u_{1i}, u_{2i}$ $i=3,4,\dots,n$.

Finally, special conformal invariance requires that
$G^{(n)}$  depends only on the cross-ratios
$$
 {r_{ij} r_{kl}\over r_{il} r_{kj} }\=
 {u_{ij}\over  u_{il}}\,\lp {u_{kj}\over  u_{kl}}\rp^{-1}~.
$$
We can thus consider $G^{(n)}$ as a function of
$$
   {u_{ij}\over  u_{il}}~.
$$
Aparently we have $2(n-3)$ such quantities:
\bb
   {u_{1i}\over  u_{13}}~,~~~
   {u_{2i}\over  u_{23}}~,~~~ i=4,\dots,n~.
\ee

Therefore,  the number of
independent variables the correlation function can depend upon
is $2(n-3)$.

\separator

\item
(a) Let $G$  be a group of transformations with generators $T_a$, i.e.,
   if $U\in G$ then $U=e^{i\alpha^aT_a}$ for some $\alpha^a$.

  The fields of the theory  transform as
$$
  \phi \rightarrow \phi'=U\phi U^{-1} \= 
  e^{i\alpha T} \phi e^{-i\alpha T}~.
$$
 For infinitesimal transformations, i.e., transformations
 with infinitesimal $\alpha$, the fields transform as
$$
  \phi \rightarrow \phi'\=\phi+i\alpha\,\lb T,\phi\rb~.
$$

The Ward identities are obtained from the group elements that
leave the vacuum invariant:
\bb
     U \ket{\emptyset} &=& \ket{\emptyset} ~, \\
     \bra{\emptyset}U^{-1} &=& \bra{\emptyset} ~,
\ee
which is equivalent to saying that
the Ward identities are obtained using those generators
of $G$ that annihilate the vacuum state, i.e., those $T_a$ such that
$$
   \bra{\emptyset} T_a \= T_a \ket{\emptyset} \= 0~.
$$
For such generators,  
the correlation  function
$$
     G^{(n)}\=\bra{\emptyset}\phi_1(x_1)\phi_2(x_2)\dots
              \phi_n(x_n)\ket{\emptyset}~,
$$
can be re-written as follows:
\bb
     G^{(n)} &=& 
     \bra{\emptyset}
     U^{-1}U
     \phi_1(x_1)U^{-1} U\phi_2(x_2)U^{-1}U \dots
     UU^{-1}\phi_n(x_n)U^{-1}U\ket{\emptyset}  \\
     &=& \bra{\emptyset}
     (U\phi_1(x_1)U^{-1}) (U\phi_2(x_2)U^{-1}) \dots
     (U\phi_n(x_n)U^{-1})\ket{\emptyset}  \\
     &=& \bra{\emptyset}
     \phi'_1(x_1)\phi'_2(x_2) \dots
     \phi'_n(x_n)\ket{\emptyset}  \\
    &\simeq&
     \bra{\emptyset}
     \phi_1(x_1)\phi_2(x_2) \dots
     \phi_n(x_n)\ket{\emptyset}  +  i\alpha^a
     \bra{\emptyset}
     \lb T_a, \phi_1(x_1)\phi_2(x_2) \dots
     \phi_n(x_n)\rb \ket{\emptyset}  \\
    &=&
     G^{(n)}  +  i\alpha^a
     \bra{\emptyset}
     \lb T_a, \phi_1(x_1)\phi_2(x_2) \dots
     \phi_n(x_n)\rb \ket{\emptyset}  ~.
\ee
From the above result we infer that
\bb
     \bra{\emptyset}
     \lb T_a, \phi_1(x_1)\phi_2(x_2) \dots
     \phi_n(x_n)\rb \ket{\emptyset}\=0  ~,
\ee
for all generators $T_a$ that annihilate the vacuum. 

Thus, given the Virasoro algebra, we will have a Ward identity
for each generator that annihilates the vacuum. Therefore,
we must first establish which generators do so. Not all
Virasoro operators annihilate the vacuum, since if this were true,
that is, if
$$
    L_n\ket{\emptyset}\=0~, ~~~\forall n~,
$$
then we would have
$$
    \lb L_n, L_m\rb \ket{\emptyset}\=0~,
    ~~~\forall n,m~,
$$
which is not true, because of the central charge in the algebra.

However, we 
do notice now that the central term cancels if $n,m\ge 0$. Thus we can
consistently set
$$
    L_n\ket{\emptyset}\=0~, ~~~\forall n\ge 0~.
$$
Moreover, the central term still vanishes when $L_{-1}$ is added to the
list of operators that annihilate the vacuum, and so it is consistent
to have
$$
    L_{-1}\ket{\emptyset}\=0~.
$$
Using the Hermiticity condition $L_n^\dagger=L_{-n}$, we see that
$$
   \bra{\emptyset} L_{-n} \=0~, ~~~\forall n\ge -1~.
$$
Therefore, we can see that the three operators
$$
     L_{-1}~,L_0~,L_1
$$
constitute the maximal set of Virasoro generators
that can annihilate the vacuum from both the left and right.

(b) The Ward identities for a generic conformal field theory can 
now be derived in straightforward fashion. We have, for 
$j=-1$, $0$, and $1$,
\bb
    0 &=&  \bra{\emptyset}
     \lb  L_j, \phi_1(x_1)\phi_2(x_2) \dots
     \phi_n(x_n)\rb \ket{\emptyset}   \\
     &=&  
     \sum_{i=1}^n
     \bra{\emptyset}
     \phi_1(x_1)\phi_2(x_2) \dots
     \lb L_j, \phi_i \rb  \dots
     \phi_n(x_n) \ket{\emptyset}   \\
     &=& 
     \sum_{i=1}^n
     \bra{\emptyset}
     \phi_1(x_1)\phi_2(x_2) \dots
       \CL_j\phi_i  \dots
     \phi_n(x_n) \ket{\emptyset}   ~,
\ee
where
\bb
     \CL_{-1}\phi_i(x_i) &=& {\partial\over\partial z_i} ~, \\
     \CL_{0}\phi_i(x_i) &=& z_i{\partial\over\partial z_i}+\Delta_i ~, \\
     \CL_{1}\phi_i(x_i) &=& z_i^2{\partial\over\partial z_i}+
     2x_i\Delta_i ~.
\ee

\separator

\item
(a) Differentiating the defining relation \calle{def-homo}
of a homogeneous function with respect to $\lambda$, we find
$$
  {d\over d\lambda}f(u_1, \dots, u_n)
  \=
  r \, \lambda^{r-1}\, f(x_1,\dots,x_n)~,
$$
where we have defined $u_i=\lambda x_i$. Setting $\lambda=1$ gives
$$
  {d\over d\lambda}f(u_1, \dots, u_n)\Big|_{\lambda=1}
  \=
  r \,  f(x_1,\dots,x_n)~.
$$
The derivative appearring in
the l.h.s. is calculated using the chain rule:
\bb
  {d\over d\lambda}f(u_1, \dots, u_n) &=& 
  \sum_{i=1}^n \, {\partial f(u_1, \dots, u_n)\over\partial u_i}  
  \, {du_i\over d\lambda} \\
  &=& \sum_{i=1}^n \, {\partial f(u_1, \dots, u_n)\over\partial u_i}  
  \, x_i ~.
\ee
When $\lambda=1$, we have $u_i=x_i$, and therefore 
$$
  {d\over d\lambda}f(u_1, \dots, u_n)\Big|_{\lambda=1} \= 
  \sum_{i=1}^n \, {\partial f(x_1, \dots, x_n)\over\partial x_i}  
  \, x_i ~,
$$
thus completing the proof of Euler's identity.

(b) Recall now that any correlation function $G^{(n)}$ is a function of
   the $n(n-1)/2$ differences $z_{ij}=z_i-z_j,~i<j$:
$$
     G=G(z_{ij})~.
$$
It is now easy to see that the $L_0$ Ward identity takes the form
$$
    \sum_{i=1}^n \sum_{j>i}^n \,
    z_{ij}\, {\partial G\over \partial z_{ij} }
    \= -\sum_{i=1}^n \, \Delta_i \, G~.
$$
The desired result now follows by a direct comparison with
Euler's equation.

\separator

\item
The Ward identity of $L_{-1}$ for the correlation function $G$ is
$$
  \lp {\partial\over\partial z_1}
     + \dots
     + {\partial\over\partial z_N}
      \rp G \=0~,
$$
which implies that $G$ is a function of the differences
$z_{ij}$ only, i.e.,
$$
  G\=G(z_{ij})~,~~~~~i<j~.
$$

The $L_0$ Ward identity requires that $G$ be a homogeneous function
of degree
$(-\sum_{i=1}^N \, \Delta_i)$. We thus write
\bb
   G \= \prod_{i=1}^N \prod_{j=i+1}^N z_{ij}^{-\gamma_{ij}}\,
        Y~,
\ee
with
$$
    \sum_{i=1}^N\sum_{j=i+1}^N\gamma_{ij}\=
    \sum_{i=1}^N \, \Delta_i~.
$$
Therefore, $Y$ is a homogeneous function of degree zero;
it will be a function of the independent cross ratios.

Now, under the transformations
$$
   z \mapsto f(z)={az+b\over cz+d}
   \Rightarrow
   z_{ij}\mapsto {z_{ij}\over(cz_i+d)(cz_j+d)}~,
$$
the correlation function, as written above, transforms as
\bb
    G \mapsto  
     \prod_{i=1}^N \prod_{j=i+1}^N (cz_i+d)^{\gamma_{ij}}(cz_j+d)^{\gamma_{ij}}
     G
     ~.
\ee
This should be compared with the way the correlation should 
transform under these transformations:
$$
   G \mapsto \prod_{i=1}^N (cz_i+d)^{2\Delta_i}\, G~.
$$
Therefore, straighforwardly, we find that the numbers
$\gamma_{ij}$ must satisfy the equations:
$$
   \sum_j{}'\gamma_{ij}\=2\Delta_i~,
$$
where we have defined $\gamma_{ij}=\gamma_{ji}$.

\separator

\item
Let
$$
    G(z_1,z_2)\=\bra{\emptyset}\phi_1(z_1)\phi_2(z_2)\ket{\emptyset}~.
$$

Substituting in the Ward identity for $L_{-1}$, we find
$$
   {\partial G\over\partial z_1}
   + {\partial G\over\partial z_2} \= 0~,
$$
which implies that $G$ is a funtion only of the difference $z_1-z_2$.

The $L_0$ Ward identity for $G$ gives
$$
    z_1{\partial G\over\partial z_1}
     + z_2{\partial G\over\partial z_2}
     + (\Delta_1 +\Delta_2) G \= 0 ~.
$$
Setting $z=z_1-z_2$, the above equation takes the simple form
$$
    z{dG\over dz}
     + (\Delta_1 +\Delta_2) G \= 0 ~,
$$
which gives
$$
      G(z)={ A\over  z^{\Delta_1+\Delta_2} }~,
$$
where $A$ is a constant.

Finally, we make use of the $L_1$ Ward identity for $G$,
which states that
$$
    z_1^2{\partial G\over\partial z_1}
     + z_2^2{\partial G\over\partial z_2}
     + (2z_1\Delta_1 +2z_2\Delta_2) G \= 0 ~.
$$
Inserting the expression we had just found for $G$,
we arrive at the algebraic equation
$$
    z\, 
      (\Delta_1 -\Delta_2)\, A  \= 0 ~,
$$
from which we conclude that
$A=0$ if $\Delta_1\ne\Delta_2$. When $\Delta_1=\Delta_2$,
we can rescale the fields in the theory such that
$A=1$. So, finally,
$$
      G(z)={ 1\over  z^{2\Delta_1} }~,
$$
which is the desired result.

Incidentally, let us make the following observation. For $z=re^{i\theta}$,
the full 2-point  correlation function takes the form
(including the holomorphic and the antiholomorphic dependence)
$$
     G={ e^{-2i\theta(\Delta-\overline\Delta)}\over
         r^{2(\Delta_1+\overline\Delta_2)}  } ~.
$$
Under rescaling $r\rightarrow\lambda r$, the correlation function
scales as
$$
    G\rightarrow\lambda^{-2(\Delta+\overline\Delta)} \, G~.
$$
{}From this, we read the scaling dimension of the fields to be 
$h=\Delta+\overline\Delta$.
Under rotations, $\theta\rightarrow\theta+\alpha$, the correlation
function changes as
$$
    G\rightarrow e^{-2i\alpha(\Delta-\overline\Delta)} \, G~.
$$
From this, the spin of the fields is seen to be 
$s=\Delta-\overline\Delta$.

\separator

\item
The expansion of the field $\phi(z)$ in terms of its modes is
$$
  \phi(z)\=\sum_n \, \phi_n \, z^{-n-\Delta}~.
$$
Then 
\beq
\label{eq:CMCC4}
   \phi^\dagger(z)\=\sum_n \, \phi^\dagger_n \, z^{-n-\Delta}~.
\eeq
The Hermitian conjugate $\phi^\dagger(z)$
of the field is constrained by the condition 
$$
  \average{\phi|\phi}\=1~.
$$
This implies that
\beq
\label{eq:CMCC3}
   \bra{\phi}\eq \lim_{z\rightarrow0}\bra{\emptyset}\phi^\dagger(z)
    \= \lim_{w\rightarrow\infty}
   \bra{\emptyset}\phi(w)w^{2\Delta}~.
\eeq
To check this, notice that under the transformation  $w=1/z$,
$$
  \phi(w)\= \lp{dw\over dz}\rp^{-\Delta}\, \phi(1/z)\=
             z^{-2\Delta}\, \phi(1/z)~.
$$
Then
$$
   \average{\phi|\phi}\=\lim_{z\rightarrow0}
                        \lim_{w\rightarrow\infty}
                    \bra{\emptyset}\phi(w)w^{2\Delta}\phi(z)\ket{\emptyset}~,
$$
or using \calle{eq:CMCC1}
$$
    \average{\phi|\phi}\=
    \lim_{z\rightarrow0}\lim_{w\rightarrow\infty}
    \bra{\emptyset} {w^{2\Delta}\over(w-z)^{2\Delta}} \ket{\emptyset}
    \=1~.
$$

Now, equation \calle{eq:CMCC3} implies
$$
  \lim_{z\rightarrow0}\bra{\emptyset}\phi^\dagger(z)
  \= \lim_{z\rightarrow0}\bra{\emptyset}\phi(1/z)z^{-2\Delta}~,
$$
which in turn means that
\beqn
   \phi^\dagger(z)&=& 
             z^{-2\Delta}\, \phi(1/z) \nonumber\\
         &=& z^{-2\Delta}\,\sum_n \, \phi_n \, z^{n+\Delta} \nonumber\\
         &=& \sum_n \, \phi_n \, z^{n-\Delta} \nonumber\\
         &{(n=-m)\atop =}& \sum_m \, \phi_{-m} \, z^{-m-\Delta}~.
\label{eq:CMCC2}
\eeqn
Note that in the last line, we have relabeled using $m=-n$.
  Then, comparing equations \calle{eq:CMCC4} and \calle{eq:CMCC2},
we arrive immediately at the result  
$$
   \phi_n^\dagger\=\phi_{-n}~.
$$

\separator

\item
The Ward identity from $L_{-1}$ for the correlation function $G$ is
$$
  \lp {\partial\over\partial z_1}
     + {\partial\over\partial z_2}
     + {\partial\over\partial z_3} \rp G \=0~,
$$
which implies that $G$ is a function of the differences
$z_1-z_2$, $z_2-z_3$, and $z_3-z_1$ only, i.e.,
$$
  G\=G(z_1-z_2,~z_2-z_3,~z_3-z_1)~.
$$
For simplicity, we will set
$$
   \alpha\=z_1-z_2~,~~~
   \beta\=z_2-z_3~,~~~{\rm and}~~~
    \gamma\=z_3-z_1~,
$$
so we can write $G$ as a function $G(\alpha,\beta,\gamma)$.
Subsituting this expression in the Ward identity for $L_0$,
we find
$$
  \alpha {\partial G\over\partial\alpha}+
  \beta {\partial G\over\partial\beta}+
  \gamma {\partial G\over\partial\gamma}\= 
  -(\Delta_1+\Delta_2+\Delta_3)\, G~.
$$
We can solve the last equation using the method of separation
of variables:
$$
     G\= A(\alpha) B(\beta) \Gamma(\gamma)~.
$$
Then
$$
  {\alpha\over A}\, {\partial A\over\partial\alpha}+
  {\beta\over B}\, {\partial B\over\partial\beta}+
  {\gamma\over\Gamma}\, {\partial \Gamma\over\partial\gamma}\= 
  -(\Delta_1+\Delta_2+\Delta_3)~,
$$
from which we conclude that
\bb
  {\alpha\over A}\, {\partial A\over\partial\alpha} &=& a ~,\\
  {\beta\over B}\, {\partial B\over\partial\beta} &=& b ~, \\
  {\gamma\over\Gamma}\, {\partial \Gamma\over\partial\gamma} &=& c~,
\ee
where $a$, $b$, and $c$ are three constants such that
$$
  a+b+c\=
  -(\Delta_1+\Delta_2+\Delta_3)~.
$$
Therefore
\bb
     A \= ({\rm constant})\,\alpha^a~, ~~~
     B \= ({\rm constant})\,\beta^b~, ~~~
     \Gamma \= ({\rm constant})\, \gamma^c~,
\ee
and so
$$
    G \= ({\rm constant})\, \alpha^a\, \beta^b\, \gamma^c~.
$$
Finally, to calculate the constants $a$, $b$, and $c$, we substitute the
last expression in the Ward identity of $L_1$, which gives:
$$
    (a+c+2\Delta_1) z_1 +
    (a+b+2\Delta_2) z_2 +
    (b+c+2\Delta_3) z_3 \=0~.
$$
Since $z_1$, $z_2$, and $z_3$ are independent, we conclude that
$$
    a+c+2\Delta_1 \=
    a+b+2\Delta_2 \=
    b+c+2\Delta_3 \=0~.
$$
These algebraic equations are solved easily, yielding
$$
   a=\Delta_3-\Delta_1-\Delta_2~,~~~
   b=\Delta_1-\Delta_2-\Delta_3~,~~~{\rm and}~~~
   c=\Delta_3-\Delta_1-\Delta_2~.
$$
So our final result reads
$$
   G \= {C_{ijk}\over
   (z_1-z_2)^{\Delta_1+\Delta_2-\Delta_3}\,
   (z_2-z_3)^{\Delta_2+\Delta_3-\Delta_1}\,
   (z_1-z_2)^{\Delta_1+\Delta_2-\Delta_3}} ~,
$$
where $C_{ijk}$ is a constant that depends only on the fields that
enter in the correlation function. In fact, it follows that
$C_{ijk}$ is related to the OPE of the given fields by
$$
  \phi_i(z)\phi_j(w) \=
  C_{ijk} \phi_k +\dots
$$

\separator

\item
From the result of Exercise \ref{item:CMCC4}
\bb
    G\= {1\over z_{12}^{\gamma_{12}}
                z_{13}^{\gamma_{13}}
                z_{14}^{\gamma_{14}}
                z_{23}^{\gamma_{23}}
                z_{24}^{\gamma_{24}}
                z_{34}^{\gamma_{34}} } \, Y(z_{ij})~,
\ee
where 
\bb
   \gamma_{12}+\gamma_{13}+\gamma_{14}&=&2\Delta_1~,\\
   \gamma_{21}+\gamma_{23}+\gamma_{24}&=&2\Delta_2~,\\
   \gamma_{31}+\gamma_{32}+\gamma_{34}&=&2\Delta_3~,\\
   \gamma_{41}+\gamma_{42}+\gamma_{43}&=&2\Delta_3~,
\ee
and $\gamma_{ij}=\gamma_{ji}$.

The function $Y$ must be homogeneous of degree $0$; therefore,
it can be written as a function of the cross ratios.
There are only two independent
cross ratios; without loss of generality, we choose
$$
 \frac{z_{12} \, z_{34}}{z_{13} \, z_{24}} ~~~{\rm and}~~~
 \frac{z_{14} \, z_{23}}{z_{13} \, z_{24}}~.
$$
It is a quick exercise to see that these two cross ratios sum to unity,
that is, if we define
$$
 \frac{z_{12} \, z_{34}}{z_{13} \, z_{24}} \eq x
$$
 then
$$
 \frac{z_{14} \, z_{23}}{z_{13} \, z_{24}} \= 1- x ~.
$$
Thus, $Y=Y(x)$. The correlation function depends on the function $Y(x)$.

\separator

\item
We define the chiral current
$$
   J(z) \eq i\,\partial\phi(z)~,
$$
which has conformal dimensions (1,0).
Notice that
$$
   \langle J(z)J(0) \rangle \= {1\over z^2}~,
$$
and therefore for large $z$, 
$$
   J(z) ~\sim~ {1\over z^2}~.
$$
We will use this information shortly.

Now observe that
\bb
 \average{ J(z) \prod_{j=1}^n e^{ia_j\phi(z_j)} } &=&
 \sum_k \average{ \lowerpairing{J(z)e}^{ia_k\phi(z_k)}  
           \prod_{j\ne k} e^{ia_j\phi(z_j)} } \\  
 &=& \sum_k \, {\alpha_k\over z-w_k} \,
 \average{ \prod_{j=1}^n e^{ia_j\phi(z_j)} } ~,
\ee
where we made use of Exercise \ref{item:GP1} in Chapter \ref{ch:GP}.

For large $z$, the preceding result reads
\bb
 0\, {1\over z} +{1\over z^2}\, G 
 +\CO\lp{1\over z^3}\rp
 &=& \sum_k \, {\alpha_k\over z} 
  \lp 1-{w_k\over z} \rp^{-1} \,
  G \\
  &=& \sum_k \, {\alpha_k\over z}
  \lp 1-{w_k\over z} \rp^{-1} \,
  G  \\
  &=& \sum_k \, {\alpha_k\over z}
  \lb 1+{w_k\over z}+\lp{w_k\over z}\rp^2+\cdots\rb \,
  G \\
  &=& {1\over z}\, \lp\sum_k \,\alpha_k\rp \, G +
 \CO\lp{1\over z^2}\rp~.
\ee
Comparing the two sides, we conclude that
$$
   \lp\sum_k \,\alpha_k\rp \, G \= 0~.
$$
Therefore $G$ can only be non-vanishing when
$$
   \sum_k \,\alpha_k  \= 0~.
$$

\separator

\item
We first notice that
\bb
  \partial_{z_i}G &=& \average{ e^{ia_1\phi(z_1)} 
                       \cdots ia_1:\partial_{z_i}\phi(z_i)e^{ia_i\phi(z_i)}:
                       \cdots e^{ia_n\phi(z_n)} } \\
                  &=& a_i\, \average{ e^{ia_1\phi(z_1)} 
                          \cdots :J(z_i)e^{ia_i\phi(z_i)}:
                          \cdots e^{ia_n\phi(z_n)} } \\
                  &=& \sum_{j\ne i} {a_ia_j\over z_i-z_j} \, G~.
\ee
This first order differential equation is solved easily by multiplying
both sides by
$$
  p \= \exp\lp -\sum_{j\ne i}\int{a_ia_j\over z_i-z_j}dz_i\rp
    \= \prod_{j\ne i} (z_i-z_j)^{-a_ia_j}~.
$$
Then the differential equation reads
$$
   \partial_{z_i} (pG)\=0~,~~~~~\forall i~,
$$
which gives
$$
    G \= A\, \prod_{i<j} (z_i-z_j)^{a_ia_j}~,
$$
where $A$ is a constant.
To calculate the constant, we consider the 2-point correlation function
$\average{e^{ia\phi(z)}e^{ib\phi(w)}}$, which can be calculated 
easily from standard identities in QFT:
$$
 \average{e^{ia\phi(z)}e^{ib\phi(w)}}
 \= e^{-ab\average{\phi(z)\phi(w)}}
 \= e^{ab\ln(z-w)}
 \= (z-w)^{ab}~.
$$
Therefore, $A=1$, and thus
\beq
\label{eq:manye}
    G \=  \prod_{i<j} (z_i-z_j)^{a_ia_j}~.
\eeq

\separator

\item

(a) Let $g^{(4)}$ be the holomorphic part of the correlation function.
 Given the state $\ket{\sigma}=\sigma(0)\ket{\emptyset}$,
one can construct a null state
$$
   \ket{\chi}\=\lp L_{-2}-{4\over3} L_{-1}^2 \rp \, \ket{\sigma}~.
$$
Then
\bb
   0 &=& \average{\chi(z_1)\sigma(z_2)\dots\sigma(z_{2M})} \\
     &=& \average{\lp L_{-2}-{4\over3} L_{-1}^2 \rp\sigma(z_1)
          \sigma(z_2)\dots\sigma(z_{2M})} \\
     &=& \lp \CL_{-2}-{4\over3} \CL_{-1}^2 \rp\average{\sigma(z_1)
          \sigma(z_2)\dots\sigma(z_{2M})} \\
     &\equiv&   \lp \CL_{-2}-{4\over3} \CL_{-1}^2 \rp  g^{(4)} ~, 
\ee
with
\bb
   \CL_{-2} &=& \sum_{j=2}^{2M}\lp
   {1/16\over z^2_{1j}}+{1\over z_{1j}}\,{\partial\over\partial z_j}\rp~,{\rm and}\\
   \CL_{-1} &=& -\sum_{j=2}^{2M} {\partial\over\partial z_j}~.
\ee
From the $L_{-1}$ Ward identity, we know that
\bb
  \sum_{j=1}^{2M} {\partial g^{(4)}\over\partial z_j}\=0
  \Rightarrow
  {\partial g^{(4)}\over\partial z_1}\=-\sum_{j=2}^{2M} 
  {\partial g^{(4)}\over\partial z_j}
  \= \CL_{-1}g^{(4)} \Rightarrow  {\partial \over\partial z_1}\=\CL_{-1}~.
\ee
Inserting this last result in the previous expression, we find that
$$
  {4\over3}  {\partial^2 g^{(4)}\over\partial z_1^2}
  - \sum_{j=2}^{2M}\lp
  {1/16\over z^2_{1j}}+{1\over z_{1j}}\,{\partial\over\partial z_j}\rp\,g^{(4)} 
   \= 0~.
$$

(b)  Let
$$
 G^{(4)}(z_{i}, \, {\overline z}_{i}) \, \equiv \, \average{
 \sigma(z_{1},\,{\overline{z}}_{1}) \, \sigma(z_{2},\,{\overline{z}}_{2})
 \, \sigma(z_{3},\,{\overline{z}}_{3}) \, \sigma(z_{4},\,{\overline{z}}_{4})}
 ~.
$$
As we have discussed in Exercise \ref{item:CMCC5},
the conformal Ward identities determine the functional form of $G^{(4)}$
to be
$$
 G^{(4)}(z_{i}, \, {\overline z}_{i}) \= 
 \frac{1}{z_{12}^{\gamma_{12}} \, z_{13}^{\gamma_{13}} \,
          z_{14}^{\gamma_{14}} \, z_{23}^{\gamma_{23}} \,
          z_{24}^{\gamma_{24}} \, z_{34}^{\gamma_{34}}} \,
 \frac{1}{ {\overline z}_{12}^{{\overline \gamma}_{12}} \,
           {\overline z}_{13}^{{\overline \gamma}_{13}} \,
           {\overline z}_{14}^{{\overline \gamma}_{14}} \,
           {\overline z}_{23}^{{\overline \gamma}_{23}} \,
           {\overline z}_{24}^{{\overline \gamma}_{24}} \,
           {\overline z}_{34}^{{\overline \gamma}_{34}}} \,
  Y(x,\overline x)~,
$$
where $z_{ij}\equiv z_i-z_j$, 
$$
  x\eq  \frac{z_{12} \, z_{34}}{z_{13} \, z_{24}}~,
$$
and similar definitions for the conjugate 
(i.e., overbarred) quantities. 
In addition, we have
\bb
 \sum_j{}' \, \gamma_{ij} &=& 2 \, \Delta_{i} \= 1/8~, {\rm and} \\
 \sum_j{}' \, {\overline \gamma}_{ij}
 &=& 2 \, {\overline \Delta}_{i} \= 1/8 ~.
\ee
We can rewrite $G^{(4)}$ as
\beqn
 G^{(4)}(z_{i}, \, {\overline z}_{i}) &=&
 \lp \frac{z_{13} \, z_{24}}{z_{12} \, z_{34}} \right)^{1/8} \,
 \frac{1}{z_{23}^{1/8} \, z_{41}^{1/8}} \,
 \lp \frac{{\overline z}_{13} \, {\overline z}_{24}}{
              {\overline z}_{12} \, {\overline z}_{34}} \right)^{1/8} \,
 \frac{1}{{\overline z}_{23}^{1/8} \, {\overline z}_{41}^{1/8}} \,
 F(x, \, {\overline x}) \no\no 
 &=& \left( \frac{1}{x \, z_{23} \, z_{41}} \right)^{1/8} \,
 \lp \frac{1}{{\overline x} \, {\overline z}_{23} \, {\overline z}_{41}}
 \rp^{1/8} \, F(x, \, {\overline x})
 ~.
\label{eq:G4b}
\eeqn
The holomorphic part $g^{(4)}$
of $G^{(4)}$ satisfies the differential equation
\beq
 -{4\over3} \, \frac{\partial^2g^{(4)}}{\partial z_1^2} + 
 \sum_{j=2}^3 \, \lp {1/16\over z_{1j}^2} +
 \frac{1}{z_{1j}} \, \frac{\partial}{\partial z_{j}} \rp \,
 g^{(4)}
 \= 0~.
\label{eq:G4c}
\eeq
There is a similar equation for the antiholomorphic part
$\overline g^{(4)}(\overline x)$ of $G^{(4)}$.

If we use \calle{eq:G4b} to substitute for 
$$
   g^{(4)} \= (x \, z_{23} \, z_{41})^{-1/8}\, f(x)
$$
in \calle{eq:G4c}, we find the second
order differential equation
\beq
 x \, (1 \,-\, x) \, \frac{d^{2} f}{dx^{2}} \,+\, \left(\frac{1}{2} \,-\,
 x \right) \, \frac{df}{dx} \,+\, \frac{1}{16} \, f \,=\, 0
 ~,
\label{eq:G4d}
\eeq
which is the differential equation sought.

(b) Compare\footnote{One can solve \calle{eq:G4d} by making 
            the substitution $x=\sin^2\theta$,
            in which case one finds
            $$ {d^2f\over d^2\theta}+{1\over4}f~=~0~.$$
            This equation has the obvious solution
            $$f~=~A\cos{\theta\over2}+B\sin{\theta\over2}~. $$ 
             However, we present yet another, longer but completely
             direct approach to solve \calle{eq:G4d} since, if one does not
             know the final answer, it would require a terrible amount
             of imagination and experience
             to write down this transformation.}
the last equation \calle{eq:G4d}
 with the well-known hypergeometric differential equation
$$
 x \, (1 \,-\, x) \, \frac{d^{2}}{dx^{2}} \,+\, \lb c \,-\,
 (1 \,+\, a \,+\, b) \, x \rb \, \frac{dF}{dx} \,-\, a \, b \, F
 \,=\, 0 ~.
$$
We immediately see that \calle{eq:G4d} is a special case of 
the hypergeometric equation for which
$$
 c = 1/2 ~~~{\rm and}~~~ a=-b = 1/4~.
$$
 Therefore, the solution to our differential equation
 can be expressed in terms of the hypergeometric
function by
\bb
 f(x) &=& A \, F(a, \, b; \, c; \, x) \,+\, B\, x^{1-c} \, 
 F(a+1-c, \, b+1-c; \, 2-c; \, x) \\
 &=& A \, F \lp \frac{1}{4}, \, - \frac{1}{4}; \, \frac{1}{2}; \,
            x \right) \,+\, B \, x^{1/2}\,
            F \left( \frac{3}{4}, \, \frac{1}{4}; \, \frac{3}{2}; \, x
            \rp~,
\ee
where $A$ and $B$ are constants. Recall that the hypergeometric function 
can be expanded as
$$
 F(a, \, b; \, c; \, x) \, \equiv \, \sum_{n=0}^{+ \infty} \,
 \frac{1}{n!} \, \frac{(a)_{n} \, (b)_{n}}{(c)_{n}} \, z^{n}
 ~,
$$
where $(a)_{n}$ is the {\bf Barnes symbol}\index{symbol!Barnes --}:
$$
 (a)_{n} \, \equiv \, a \, (a\,+\, 1) \, \ldots \, (a \,+\, n \,-\, 1)
 \,=\, \frac{\Gamma(a+n)}{\Gamma(a)}~.
$$
Two standard properties of the hypergeometric function \cite{AbrSte} are
\bb
 F \left( \frac{n}{2}, \, - \frac{n}{2}; \, \frac{1}{2}; \, \sin^{2}
      \theta \right) &=& \cos \, n \theta ~,\\
 F \left( \frac{n+1}{2}, \, - \frac{n-1}{2}; \, \frac{3}{2}; \,
     \sin^{2} \theta \right) &=& \frac{\sin \, n \theta}{n \,
     \sin \theta} ~.
\ee
In particular, setting $n=1/2$,
\bb
  F \left( \frac{1}{4}, \, - \frac{1}{4}; \, \frac{1}{2}; \,
            \sin^{2} \theta \right) &=& \cos \, \frac{\theta}{2}~,~~{\rm and}\\
  F \left( \frac{3}{4}, \, \frac{1}{4}; \, \frac{3}{2}; \, \sin^{2}
            \theta \right) &=& \frac{1}{\cos \, \frac{\theta}{2}} ~.
\ee
If we write
$$
   \sin^{2} \theta \, \equiv \, x~,
$$
 then
$$
  f(x) \,=\, A\, \cos \frac{\theta}{2} \,+\, B' \, \sin \frac{\theta}{2}~.
$$
Notice that
\bb
 \cos\,\frac{\theta}{2} &=& \sqrt{1 \,-\, \sin^{2} \frac{\theta}{2}}
 \,=\, \sqrt{1 \,-\, \frac{1 - \cos \theta}{2}} \,=\, \frac{1}{\sqrt{2}}
 \, \sqrt{1 \,+\, \cos \theta}\\
 &=& \frac{1}{\sqrt{2}} \, \sqrt{1 \, + \, \sqrt{1 \,-\, \sin^{2}
       \theta}} \,=\, \frac{1}{\sqrt{2}} \, \sqrt{1 \,+\, \sqrt{1 \,-\, x}}
\ee
and
$$
 \sin \frac{\theta}{2} \,=\, \frac{1}{\sqrt{2}} \, \sqrt{1 \,-\,
 \sqrt{1 \,-\, x}} ~.
$$
Therefore
$$
   f(x) \= \alpha \, f_{1}(x) \,+\, \beta \, f_{2}(x)~,
$$
where
$$
 f_{1}(x) \, \equiv \, \sqrt{1 \,+\, \sqrt{1 \,-\, x}} ~,~~~{\rm and}~~~
 f_{2}(x) \, \equiv \, \sqrt{1 \,-\, \sqrt{1 \,-\, x}} ~.
$$
If we restore  the antiholomorphic part, $G^{(4)}$ can be written as
$$
 G^{(4)}(z_{i}, \, {\overline z}_{i}) \,=\, \left| \frac{1}{
  x \, z_{23} \, z_{41}} \right|^{1/4} \, \sum_{i,j=1}^{2} \,
  \alpha_{ij} \, f_{i}(x) \, f_{j}( {\overline x})
 ~.
$$
Since the field $\sigma (z, {\overline z})$ is local,
$ G^{(4)}(z_{i}, \, {\overline z}_{i}) $ must be single-valued.
This condition is better seen using the $\theta$, ${\overline
\theta}$ variables. Under the substitution
\beq
\label{eq:thetamonodromy}
 \theta \rightarrow - \theta ~~,~~~ 
 {\overline \theta} \rightarrow - {\overline \theta}
  ~,
\eeq
$ G^{(4)}(z_{i}, {\overline z}_{i}) $ is single-valued if there
are not cross terms, {\it i.e.}
$$
 G^{(4)}(z_{i}, \, {\overline z}_{i}) \,=\, \left| \frac{1}{x \,
 z_{23} \, z_{41}} \right|^{1/4} \, \left( a \, |f_{1}(x)|^{2}
 \,+\, b \, |f_{2}(x)|^{2} \right)~~.
$$
Actually, the substitution \calle{eq:thetamonodromy}
 takes into account the analytical
continuation only around the point $(x=0,{\overline x}=0)$.
However, the differential equation
\calle{eq:G4d} has another singular point, namely $(x=1,{\overline x}=1)$,
which corresponds to $(\theta=\pi/2,{\overline \theta}=\pi/2)$.
Single-valuedness here implied that $a=b$, i.e. that
\beq
 G^{(4)}(z_{i}, \, {\overline z}_{i}) \,=\, \left| \frac{1}{x \,
 z_{23} \, z_{41}} \right|^{1/4} \, a \, \lp |f_{1}(x)|^{2}
 \,+\, |f_{2}(x)|^{2} \rp~.
\label{eq:G4f}
\eeq

Finally, to calculate the coefficient $a$ we proceed as follows:
For $z_{1} \, \simeq \, z_{2}$ and $z_{3} \, \simeq \, z_{4}$, 
equation \calle{eq:G4f} gives
\begin{equation}
\label{eq:G4g}
 G^{(4)}(z_{i}, \, {\overline z}_{i}) \, \sim \, \frac{2 a}{
 \left| z_{12} \right|^{1/4} \, \left| z_{34} \right|^{1/4}}
  ~.
\end{equation}
On the other hand, using the OPE
$$
 \sigma (z_{1}, \, {\overline z}_{1}) \, 
 \sigma (z_{2}, \, {\overline z}_{2}) \, \sim \, \frac{1}{
 \left| z_{12} \right|^{2}}~,
$$
we can find the singular part of $G^{(4)} = \average{\sigma \sigma
 \sigma \sigma}$ independently, obtaining
\begin{equation}
 G^{(4)}(z_{i}, \, {\overline z}_{i}) \, \sim \, \frac{1}{
 \left| z_{12} \right|^{1/4} \, \left| z_{34} \right|^{1/4}}
 ~.
\label{eq:G4h}
\end{equation}
Comparing \calle{eq:G4g} and \calle{eq:G4h},
we conclude that $a=1/2$. Thus the final result is
$$
 G^{(4)}(z_{i}, \, {\overline z}_{i}) \,=\, \frac{1}{2} \,
 \left| \frac{1}{x \, z_{23} \, z_{41}} \right|^{1/4} \, 
 \left( \left| f_{1}(x) \right|^{2} \,+\, \left| f_{2}(x)
 \right|^{2} \right)~.
$$

\separator

\item
We can use the OPE in general to write
$$
 \phi_{n}(z, \overline{z}) \, \phi_{m}(0,0) \= \sum_{p} \,
 c_{nm}^{p} \, \chi_{nm}^{p}(z) \, \overline{\chi}_{nm}^{p}( \overline{z})
 \, z^{ \Delta_{p} - \Delta_{n} - \Delta_{m}} \, \overline{z}^{
 \overline{\Delta}_{p} - \overline{\Delta}_{n} - \overline{\Delta}_{m}}
 \, \phi_{p}(0,0)~,
$$
where
\beq
\label{eq:CMCC9}
 \chi_{nm}^{p}(z) \, \equiv \, \sum_{ \{ k \} } \, \beta_{nm}^{ p_{j} 
 \{ k \} } \, z^{\sum \, k_{i}} \, L_{-k_{1}} \, \ldots L_{-k_{n}}\,\ldots
\eeq
In the following,  we will drop the antiholomorphic part, and
concentrate on
\beq
\label{eq:CMCC6}
 \phi_{n}(z) \, \phi_{m}(0) \= \sum_{p} \,
 c_{nm}^{p} \, \chi_{nm}^{p}(z) 
 \, z^{ \Delta_{p} - \Delta_{n} - \Delta_{m}}
 \, \phi_{p}(0)~.
\eeq

Since
$$
 \phi_{n}(0) \, \ket{\emptyset} \= \ket{\Delta_{n}}~,
$$
applying \calle{eq:CMCC6} to the vacuum $\ket{\emptyset}$, we get
\beq
\label{eq:CMCC7}
 \phi_{n}(z)\ket{\Delta_{m}}\,=\, \sum_{p} \, c_{nm}^{p} \, \chi_{nm}^{p}(z)
 \, z^{\Delta_{p} - \Delta_{n} - \Delta_{m}} \, \ket{\Delta_{p}}
  ~.
\eeq
We define 
\beq
\label{eq:CMCC10}
 \ket{z,\,\Delta_{p}; \,nm} \, \equiv \, \chi_{nm}^{p}(z)\,\ket{\Delta_{p}}
 ~.
\eeq
Then for any $L_{s}$, $s>0$ we get
\beqn
 \lb L_{s}, \, \phi_{n}(z) \rb \, \ket{\Delta_{m}} \=
  L_s\phi_{n}(z) \ket{\Delta_{m}} 
   -\phi_n(z) L_s \ket{\Delta_{m}} 
 \=L_s\phi_n(z) \ket{\Delta_{m}} \nonumber \\
 \= \sum_{p} \,
 c_{nm}^{p} \, z^{ \Delta_p-\Delta_n-\Delta_m} \, L_s \,
 \ket{z, \, \Delta_p; \, nm}
 \nonumber \\
\Rightarrow \lb z^{s+1} \partial_{z} \,+\, (s+1) \, \Delta_n \, z^s \rb
 \, \phi_{n}(z) \, \ket{\Delta_{m}} 
  =  \sum_{p} \, c_{nm}^{p} \, z^{
 \Delta_{p} - \Delta_{n} - \Delta_{m}} \, L_{s} \,\ket{z, \, \Delta_{p};
 \, nm} 
  \nonumber \\
 \Rightarrow 
 \lb z^{s+1} \partial_{z} \,+\, (s+1) \, \Delta_{n} \, z^{s} \rb
 \, \sum_{p} \, c_{nm}^{p} \, z^{ \Delta_{p} - \Delta_{n} - \Delta_{m}}
 \, \ket{z, \, \Delta_{p}; \, nm} 
  \nonumber \\
 =  \sum_{p} \,
 c_{nm}^{p} \, z^{\Delta_{p}
 - \Delta_{n} - \Delta_{m}} \, L_{s} \, \ket{z, \, \Delta_{p}; \, nm}
 \nonumber \\
 \Rightarrow \lb z_{s} \, (\Delta_{p} \,-\, \Delta_{n} \,-\, \Delta_{m} \,+\,
 z^{s+1} \partial_{z} \,+\, (s+1) \, \Delta_{n} \, z^{s} \rb \,
 \ket{z, \, \Delta_{p}; \, nm}
 \nonumber \\
 = L_{s} \,\ket{z,\,\Delta_{p}; \, nm}~.~~~~~~~~
\label{eq:CMCC8}
\eeqn
These equations determine the state $\ket{z,\,\Delta_{p};\,nm}$. Actually,
the first two equations, with $s=1$ and $s=2$, are sufficient to determine
the state (this arises from the fact that the $L$'s
have to obey the Virasoro algebra). If we could determine the action of
$L_{s}$ on $\ket{z,\,\Delta_{p};\,nm}$ and solve 
\calle{eq:CMCC8}, then we would be able
to find the coefficients $ \beta_{nm}^{p; \, \{ k \} }$ using 
\calle{eq:CMCC9} and
\calle{eq:CMCC10}. Since we cannot proceed in this direct way,
we instead introduce the Taylor expansion
\beq
\label{eq:CMCC11}
 \ket{z,\,\Delta_{p};\,nm} \, \equiv \, \sum_{N=0}^{\infty} \, z^{N} \,
 \ket{N, \, \Delta_{p}; \, nm}~.
\eeq
We will show that the solution can now be
approximated up to any chosen order. From the definitions
\calle{eq:CMCC9}, \calle{eq:CMCC10}, and \calle{eq:CMCC11}, we see that
$$
 L_{0} \, \ket{N, \, \Delta_{p}; \, nm} \= (\Delta_{p} \,+\, N) \,
 \ket{N, \, \Delta_{p}; \, nm}~.
$$
Substituting \calle{eq:CMCC11} in \calle{eq:CMCC8}, we find
$$
 \lb N \,+\, (s+1) \Delta_{n} \,+\, \Delta_{p} \,-\, \Delta_{n} 
 \,-\, \Delta_{m} \rb \,\ket{N, \, \Delta_{p}; \, nm} \,=\, L_{s} \,
 \ket{N, \, \Delta_{p}; \, nm}~.
$$
Therefore
\bb
  L_{s} \, \ket{N, \, \Delta_{p}; \, nm} &=& 0~,~~~~~ N<s\\
 L_{s}\,\ket{N+s, \,\Delta_{p};\, nm} &=& 
 \lb N \,+\, s \Delta_{n} \,+\, \Delta_{p} \,-\, \Delta_{m} \rb
 \ket{N, \,\Delta_{p};\,nm} ~.
\ee
We have already mentioned that we need only the equations for $s=1$
and $s~=2$, which are now
\beqn
\label{eq:CMCC12}
 L_{1} \, \ket{ N+1} &=&
 (N \,+\, \Delta_{n} \,+\, \Delta_{p} \,-\, \Delta_{m}) \,\ket{N} ~,
             ~~~{\rm and}\\
\label{eq:CMCC13}
 L_{2} \, \ket{N+2}  &=&
 (N \,+\, 2\Delta_{n} \,+\, \Delta_{p} \,-\, \Delta_{m}) \,\ket{N} ~.
\eeqn
where we have simplified the notation for convenience.

Let us now compute
the state $\ket{z,\,\Delta_{p};\,nm}$ up to second order, 
i.e., determine the coefficients of the following expansion:
$$
 \ket{z,\,\Delta_{p}; \, nm} \= \lb 1 \,+\, z \, \beta_{nm}^{p;1} \,
 L_{-1} \, + \, z^{2} \,(\beta_{nm}^{p;1,1} \, L_{-1}^2
  + \beta_{nm}^{p;2} \, L_{-2}) \, + \, \ldots
 \rb \, \ket{ \Delta_{p}}~.
$$
This amounts to writing the states $\ket{N\,=\,0}$, $\ket{N\,=\,1}$,
and $\ket{N \,=\, 2}$ as linear combinations of states from the
Verma module of $\ket{\Delta_{p}}$.  One finds
\begin{eqnarray*}
  \ket{N=0} &=& \ket{\Delta_{p} }~, \\
  \ket{N=1} &=& \alpha L_{-1} \ket{\Delta_{p} }~,\\
  \ket{N=2} &=& (\gamma L_{-1}^{2} \,+\,
                         \delta L_{-2}) \ket{ \Delta_{p} }~,
\end{eqnarray*}
where
$$
 \alpha \, \equiv \, \beta_{nm}^{p;1} ~~,~~~ \gamma \, \equiv \,
 \beta_{nm}^{p;1,1} ~~,~~~{\rm and}~~~
 \delta \, \equiv \, \beta_{nm}^{p;2}
  ~.
$$
We present the calculation explicitly. Equation \calle{eq:CMCC12}
for $N=0$ implies
\begin{eqnarray*}
 ( \Delta_{n} \,+\, \Delta_{p} \,-\, \Delta_{m} ) \, \ket{\Delta_{p}} 
 &=& L_{1} \, \ket{N=1}\\
 &=& \alpha \, L_{1} \, L_{-1} \, \ket{ \Delta_{p} }\\
 &=& \alpha \, \lb L_{1}, \, L_{-1} \rb \, \ket{\Delta_{p} }\\
 &=& \alpha \, 2 L_{0} \, \ket{ \Delta_{p} }\\
 &=& 2\alpha\Delta_p \, \ket{ \Delta_{p} }~.
\end{eqnarray*}
Therefore,
$$
 \alpha \= \frac{ \Delta_{n} \,+\, \Delta_{p} \,-\, \Delta_{m} }{
 2 \Delta_{p}}~.
$$
From \calle{eq:CMCC13} for $N=0$, we  get
\begin{eqnarray*}
 (2 \Delta_{n} \,+\, \Delta_{p} \,-\, \Delta_{m} ) \, \ket{\Delta_{p}} 
 &=& L_{2} \, \ket{2}\\
 &=& L_{2} \, (\gamma \, L_{-1}^{2} \,+\, \delta \, L_{-2} ) \,
           \ket{ \Delta_{p} }\\
 &=& ( \gamma \, \lb L_{2}, \, L_{-1}^{2} \rb \,+\,
           \delta \, \lb L_{2}, \, L_{-2} \rb ) \, \ket{\Delta_{p}}\\
 &=& \lb \gamma \, 6 \Delta_{p} \,+\, \delta \, (4 \Delta_{p}
           \,+\, \frac{c}{2} ) \rb \, \ket{ \Delta_{p} }~,
\end{eqnarray*}
or, in other words,
\beq
\label{eq:CMCC14}
 6 \Delta_{p} \, \gamma \,+\, \lp 4 \Delta_{p} \,+\, \frac{c}{2}\rp \, \delta
 \= 2 \Delta_{n} \,+\, \Delta_{p} \,-\, \Delta_{m}
 ~.
\eeq
Again, from \calle{eq:CMCC12}, but now for $N=1$, we have
\beqn
 ( \Delta_{n} \,+\, \Delta_{p} \,-\, \Delta_{m} \,+\, 1) \, \ket{1}
 &=& L_{1} \, \ket{2}\nonumber \\
 \Rightarrow ( \Delta_{n} \,+\, \Delta_{p} \,-\, \Delta_{m} \,+\, 1)
  \, \alpha \,
 L_{-1} \, \ket{2} & \,=\, & (\gamma \, \lb L_{1}, \, L_{-1}^{2} \rb
 \,+\, \delta \, \lb L_{1}, \, L_{-2} \rb ) \, \ket{\Delta_{p}}
 \nonumber \\
 &=& L_1 (\gamma L_{-1}^{2} \,+\,
                         \delta L_{-2}) \ket{ \Delta_{p} }\nonumber \\
 &=& \lb \gamma \, 2(2 \Delta_{p} \,+\, 1) \,+\, 3 \delta )
 \rb \, L_{-1} \, \ket{\Delta_{p}}\nonumber \\
 \Rightarrow
 2 ( 2 \Delta_{p} \,+\, 1) \, \gamma \,+\, 3 \delta &=& 
 ( \Delta_{n} \,+\, \Delta_{p} \,-\, \Delta_{m} \,+\, 1) \, \alpha~.
\label{eq:CMCC15}
\eeqn
The system of \calle{eq:CMCC14} and \calle{eq:CMCC15} can be solved
explicitly for $\gamma$ and $\delta$.  This calculation gives
\begin{eqnarray*}
 \gamma &=& \frac{ (4\Delta_{p} + \frac{c}{2} ) \, 
                  ( \Delta_{n} + \Delta_{p} - \Delta_{m} + 1)
                  \, (\Delta_{n} + \Delta_{p} - \Delta_{m}) \,
                  -\, (6 \Delta_{n} + 3 \Delta_{p} - 3 \Delta_{m}
                  ) \, 2 \Delta_{p} }{
                  \lb 2 \Delta_{p} \, (8 \Delta_{p} - 5) \,+\,
                  c \, (1 + 2 \Delta_{p}) \rb \, 2\Delta_{p}}\\
 \delta &=& \frac{
                  (\Delta_{m} + \Delta_{n} - \Delta_{p}) \,+\,
                  2( \Delta_{n}  \Delta_{p} + \Delta_{m} 
                  \Delta_{p} + 3 \Delta_{n}  \Delta_{m} ) -
                  ( 3 \Delta_{n}^{2} + 3 \Delta_{m}^{2} -
                  \Delta_{p}^{2}) }{
                  2 \, \Delta_{p} \, ( 8  \Delta_{p} - 5) \,+\,
                  c \, (1 + 2  \Delta_{p}) }
\end{eqnarray*}

Now it is clear how we can obtain all the coefficients 
$\beta_{nm}^{p; \{ k \} }$
up to 
{\it n}-th order. All we have to do is to solve a linear algebraic
system of $P(n)$-th order. Then we obtain the initial coefficients
$c_{nm}^{p; \{ k \} \{ \overline{k} \} }$
by using equation \calle{eq:CMCC16}. This 
method is straightforward in principle, but discouraging
 in practice!

\separator

\item

In the amplitude expressed as
$$
   A \= \int\, dz_3 \,
        \average{c(z_1)c(z_2)c(z_4)V_1V_2V_3V_4} \,
    \= \int\, dz_3 \,
        \average{c(z_1)c(z_2)c(z_4)} \,
        \average{V_1V_2V_3V_4}~,
$$
one can calculate the correlation functions inside the integral easily
by using the bosonized version of $c(z)$ and the result of the problem
\ref{item:CMCC1}.  This enables us to re-write the
amplitude as
\bb
   A\= \int dz_3\, z_{12}z_{14}z_{24}
   \prod_{i<j} e^{k_i\cdot k_j\ln z_{ij}} 
   \= \int dz_3\, z_{12}z_{14}z_{24}
   \prod_{i<j} z_{ij}^{k_i\cdot k_j} ~ .
\ee
Since the result should not depend on the positions
at which we have placed the ghosts,
we can choose, without loss of generality, 
\bb
   z_1\rightarrow\infty~, ~~~ 
   z_2=1~, ~~~{\rm and}~~~
   z_4= 0~.
\ee 
Then
\bb
 P &\equiv&
 z_{12}^{k_1\cdot k_2+1}
 z_{13}^{k_1\cdot k_3}
 z_{14}^{k_1\cdot k_4+1}
 z_{23}^{k_2\cdot k_3}
 z_{24}^{k_2\cdot k_4+1}
 z_{34}^{k_3\cdot k_4} \\ 
 &{ {z_2=1,z_4=0} \atop =}& 
 (z_1-1)^{k_1\cdot k_2+1}
 z_{13}^{k_1\cdot k_3}
 z_1^{k_1\cdot k_4+1}
 (1-z_3)^{k_2\cdot k_3}
 1^{k_2\cdot k_4+1}
 z_3^{k_3\cdot k_4} ~.
\ee
Now momentum conservation gives
$$
  k_3\=-(k_1+k_2+k_4)~.
$$
Using this value of $k_3$ in the exponent of $z_{13}$ in $P$,
 we find
\bb
 P &=&
 (z_1-1)^{k_1\cdot k_2+1}
 z_{13}^{-k_1\cdot (k_1+k_2+k_4)}
 z_1^{k_1\cdot k_4+1}
 (1-z_3)^{k_2\cdot k_3}
 z_3^{k_3\cdot k_4} \\
 &=& {(z_1-1)^{k_1\cdot k_2+1}
 z_1^{k_1\cdot k_4+1}
 \over
 z_{13}^{k^2_1 + k_1\cdot k_2+k_1\cdot k_4} }
 \, (1-z_3)^{k_2\cdot k_3}
 z_3^{k_3\cdot k_4} \\
 &{z_1\rightarrow\infty \atop =}& 
 (1-z_3)^{k_2\cdot k_3} \, z_3^{k_3\cdot k_4} ~ ,
\ee
where we have used
the tachyon condition  $k_1^2=2$.
As a result, we end up with the {\bfseries Veneziano amplitude},
namely
\bb
   A\= \int_0^1\, dz_3\, 
   \lp 1-z_{3}\rp^{k_2\cdot k_3} z_3^{k_3\cdot k_4}
    \= B(k_2\cdot k_3+1,k_3\cdot k_4+1) ~ ,
\ee
where $B(x,y)$ stands for the well-known (beta)
 $B$-function\index{function!$B$-function}.

\end{enumerate}

\newchapter{OTHER MODELS IN  CFT}
\label{ch:OM}

\footnotesize
\noindent {\bfseries References}:
Orbifolds are reviewed in \cite{Ginz}, 
affine algebras in
\cite{Ginz,Fuchs2}. For information on the WZWN model,
see \cite{Fuchs2,Walton}. A review on the WZWN model from the
functional integral point of view is \cite{Gawe}.
\normalsize

\section{BRIEF SUMMARY}

\subsection{Orbifolds}

Let $M$ be a manifold and $\Gamma$ a discrete group acting on $M$,
i.e., a map $\sigma: \Gamma\times M\rightarrow M$ is given.  One
generally writes this as a group element $\gamma$ acting on
a point $x$ of the manifold. A point
$x\in M$ is a fixed point of this map provided 
that there is a $\gamma_0\in\Gamma$ such that
$\gamma_0\cdot x=x$ and $\gamma_0\ne {\rm id}$. 
The orbit of $x$ is  the set $\Gamma\cdot x$ generated by the action
of all the elements of $\Gamma$ on $x$. Then the quotient 
space $O=M/\Gamma$ is the set of all distinct orbits of $M$ under $\Gamma$.
Such a quotient space
is called an {\bfseries orbifold} (from {\bfseries orbi}t 
man{\bfseries ifold}). When the action $\sigma$ has no fixed points,
then $O$ is an ordinary manifold. When $\sigma$ has fixed points, then
$O$ is not exactly an ordinary manifold, because of the behavior at
the fixed points when the quotient space is constructed.

\begin{figure}[htb!]
\begin{center}
\psfrag{M}{$M$}
\psfrag{x}{$x$}
\psfrag{y}{$y$}
\psfrag{z}{$z$}
\psfrag{w}{$w$}
\psfrag{X}{$\lb x\rb$}
\psfrag{Y}{$\lb y\rb$}
\psfrag{Z}{$\lb z\rb$}
\psfrag{W}{$\lb w\rb$}
\psfrag{Gx}{$\Gamma\cdot x$}
\psfrag{Gy}{$\Gamma\cdot y$}
\psfrag{Gz}{$\Gamma\cdot z$}
\psfrag{Gw}{$\Gamma\cdot w$}
\psfrag{Mg}{$M\over \Gamma$}
\psfrag{Orb}{orbits}
\psfrag{After}{After}
\psfrag{identification}{identification}
\includegraphics[width=13cm]{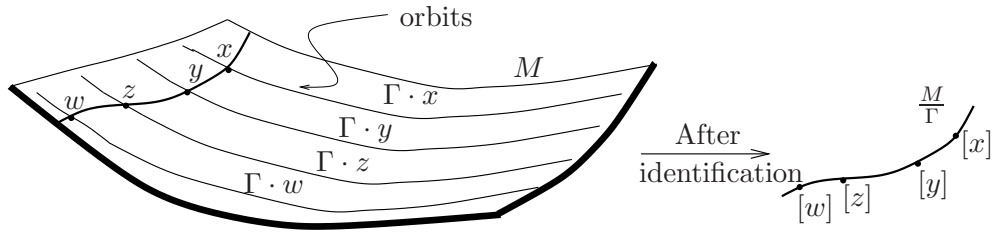}
\end{center}
\caption{The quotient space of a manifold $M$ by the group $\Gamma$.
         For ease of visualization, we have drawn continuous 
         (instead of discrete) orbits.
         }
\end{figure}

The concept of an orbifold thus generalizes that of a manifold.  This
is relevant for our purposes, as one can define CFTs on orbifolds.
In particular,  given a modular invariant conformal
field theory $\CT$
that admits a discrete symmetry group $\Gamma$, one can
construct an orbifold CFT (which is modular invariant)
by taking the original CFT $\CT$
and modding out by $\Gamma$. 

\begin{figure}[h!]
\begin{center}
\psfrag{S}{$S^1$}
\psfrag{SZ}{$S^1\over\BZ_2$}
\psfrag{f}{$\phi$}
\psfrag{-f}{$-\phi$}
\psfrag{Orbifolding}{orbifolding}
\psfrag{Fixed point}{Fixed point}
\includegraphics[height=5cm]{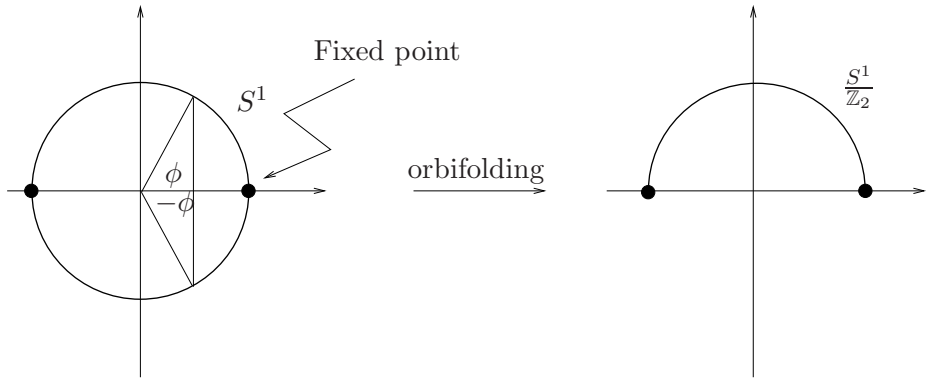}
\end{center}
\caption{The orbifold obtained from the circle by modding out
       by a $\BZ_2$ reflection group.
       The group $\BZ_2$ identifies points that are symmetric
       with respect to the horizontal axis, and therefore 
       the map has two fixed points.
       Compare this action with the action in $\BR P^1= S^1/\BZ_2\simeq S^1$,
       where $\BZ_2$ 
    identifies points symmetric with respect to origin of the coordinate system.
    In this case, the action on $S^1$ has no fixed points and the quotient space
       is a manifold, namely $S^1$.
         }
\end{figure}

One can adopt the notation
$$
   Z_{\CT}\=  \square
$$
to denote the partition function of $\CT$. Then the partition function of
$\CT/\Gamma$ is given by
$$
   Z_{\CT/\Gamma}\= {1\over|\Gamma|}\, 
    \sum_{ {\scriptscriptstyle g,h\in\Gamma}\atop
                   {\scriptscriptstyle gh=hg}  }
    ~~  
    \partition{g}{h} ~.
$$

\subsection{WZWN Model}

Given a group manifold G, we can define a 
corresponding $\sigma$-model, which is a theory with action
$$
   S_0 \= {k\over 16\pi}\, \iint\, d^2x\, {\rm tr}
   \lp \partial_a g(x) \partial^a g^{-1}(x) \rp~,
$$
where $g(x)$ is a field that takes its values in the group $G$.
This action is not conformally invariant. However, one can make
the theory conformally invariant by adding a {\bf Wess-Zumino term},
leading to the action 
\beq
  S \= S_0 + {k\over 24} \,
     \int\kern-8pt\int\kern-8pt\int\,d^3y \, f^{abc} \, {\rm tr}
    \lb  \lp  g(y) \partial_a g^{-1}(y) \rp
         \lp  g(y) \partial_b g^{-1}(y) \rp
         \lp  g(y) \partial_c g^{-1}(y) \rp \rb~,
\label{eq:WZWN}
\eeq
where $k$ is now an integer.  The Wess-Zumino term, which has been added
to the action, is topological. The integral is over a 3-dimensional space which
has the physical two-dimensional space as its boundary; the Wess-Zumino
term is the integral of a total derivative, and so ultimately only
depends on the physical two-dimensional space, not on the particular extension
into the three-dimensional space.  Single-valuedness of the action leads
to the requirement that $k$ be an integer.

The model \calle{eq:WZWN} has a local symmetry ${\rm G}\otimes\overline
{\rm G}$
which acts on the fields as\footnote{The groups G and $\overline{\rm G}$
are isomorphic; we use the overbar simply to indicate which side 
each group acts from.}
$$
   g(z,\overline z) \mapsto
   \Omega(z) g(z,\overline z) \overline\Omega(\overline z)~,
   ~~~~~\Omega\in {\rm G}, ~ \overline\Omega \in \overline{\rm G}~.
$$
This symmetry is generated by the currents
\bb
    J^a(z) &=& -{k\over 2}\, (\partial^a g)\, g^{-1}~~~{\rm and} \\
    \overline J^a(z) &=& -{k\over 2}\, g^{-1}\, (\partial^a g)~.
\ee
One has the associated OPEs
$$
   J^a(z)J^b(w)\= { k/2 \, \delta^{ab} \over (z-w)^2 }
                + { f^{ab}_{~~~c}\, J^c(w) \over z-w } 
                + {\rm reg} ~.
$$
The modes $J^a_n$ of $J^a(w)$,  
$$
     J^a(z) \= \sum_n \, {J^a_n \over z^{n+1} } ~,
$$ 
satisfy an affine $\hat{\mathfrak{g}}$ Kac-Moody algebra with level $k$:
$$
  \lb J^a_m, J^b_n \rb = i\, f^{ab}_{~~c}\, J^c_{m+n}
   + {k\over 2}\, m \, \delta^{a}\, \delta_{m+n,0}~.
$$  
The energy-momentum tensor is
$$
   T(z) = {1\over k+\tilde h_{\mathfrak{g}}} \,  
          \ddagger J_a(z) J_a(z) \ddagger~,
$$
where $\ddagger~\ddagger$ denotes normal ordering with respect to the
modes of $J^a(z)$, and $\tilde h_{\mathfrak{g}}$ is the dual Coxeter number
of $\hat{\mathfrak{g}}$.
Unitarity of the model
requires that $k$ be a positive integer (just as single-valuedness
of the action exponential does).
From the OPE of the energy-momentum tensor, we find that
$$
   c \= {k \, \dim\mathfrak{g} \over k + \tilde h_{\mathfrak{g}} }~.
$$ 
The heighest weight representations at level $k$ are labeled by a
vector $\bfp$ in the set
$$
   \{\bfp\in P ~|~ \bfp\cdot\bfom_i \ge 0~, ~\bfp\cdot\bfrh\le k~\}~,
$$
where $P$ is the weight lattice of $\hat{\mathfrak{g}}$, $\{\bfom_i\}$ the
corresponding fundamental
weights, and $\bfrh$ half the sum of the positive roots in 
$\hat{\mathfrak{g}}$.
The primary fields have dimensions
$$
    \Delta_{\bfp} \=
    {\bfp \cdot (\bfp+2\bfrh) \over k + \tilde h_{\mathfrak{g}} }~.
$$

\subsection{Parafermions}

The CFT UMM(3) is probably the simplest of the MMs;
it includes a
fermion field $\psi$ with weight $(1/2,0)$,
a spin variable $\sigma$, and a disorder
variable $\mu$. Together with the energy-momentum tensor
$T$, they build an algebra that enjoys a $\BZ_2$ symmetry.
This model is related to the Ising model of statistical physics.

One expects, correctly, that a similar construction is possible with other
cyclic groups. Indeed, an algebra that enjoys an analogous
$\BZ_3$ symmetry is constructed as follows. We start with two
fields $\psi$ and  $\psi^\dagger$ which both have weight $(2/3,0)$;
we call these {\bfseries parafermion}\index{Parafermion} fields. Although
such objects are non-local, together with the energy-momentum tensor,
they can be used to build a local
CFT algebra: 
\bb
   T(z)T(w) &=& {c/2\over (z-w)^4}+ {2T(w)\over (z-w)^2}
                + {\partial_w T(w)\over z-w} + \cdots ~,\\
   T(z)\psi(w) &=& {(2/3)\psi(w)\over (z-w)^2} + 
                   {\partial_w \psi(w)\over z-w} + \cdots ~,\\
   \psi(z)\psi(w) &=& {a\psi(w)\over (z-w)^{2/3}} 
                   + \cdots ~,\\
   \psi(z)\psi^\dagger(w) &=& {1\over (z-w)^{4/3}}\,
                  \lp 1+{8\over 9c}\, T(w) 
                   + \cdots\rp ~.
\ee
Associativity fixes the values of the constants in this
algebra to $c=4/5$ and $a=2/\sqrt{3}$.
Just as the $\BZ_2$ model is related to the Ising model,
this model is related to
the 3-state Potts model of statistical mechanics.

The above two cases are the first two models in the series
of parafermionic models. The series is parametrized by the
natural number $N\in\BN\smallsetminus\{0,1\}$. The $N$-th model
enjoys a $\BZ_N$ symmetry similar to the previous cases. 
In particular,  the $N$-th model contains $N-1$ parafermions
$\psi_1,\dots,\psi_{N-1}$  which are primary fields
with conformal weights
$$
 \Delta_k \= {k(N-k)\over N}~,~~~k=1,2,\dots,N-1~,
$$
which thus satisfy
$$
  T(z)\psi_k(w) \= { \Delta_k\psi_k(w)\over (z-w)^2} +
                   {\partial_w \psi_k(w)\over z-w} + \cdots ~.
$$
These fields also obey the conjugation relation
$$
   \psi_k^\dagger \= \psi_{N-k}~.
$$
The defining  OPEs of the algebra are:
\bb
   \psi_k(z)\psi_{k'}(w)&=& {c_{kk'}\over (z-w)^{2kk'/N}}\,
   \lb \psi_{k+k'}(w)+\CO(z-w)\rb~,~~~k+k'<N~, \\
   \psi_k(z)\psi^\dagger_{k'}(w)&=& {c_{k,N-k'}\over (z-w)^{2k(N-k')/N}}\,
   \lb \psi_{k-k'}(w)+\CO(z-w)\rb~,~~~k<k'~, \\ 
   \psi_k(z)\psi^\dagger_{k}(w)&=& {1\over (z-w)^{2k(N-k)/N}}\,
   \lb I+{2\Delta_k\over c}\, (z-w)^2 \, T(w)+\CO\lp(z-w)^3\rp\rb~.
\ee
Demanding associativity determines the central charge; its value is
$$
    c\= {2(N-1)\over N+2}~.
$$
The  model also contains a set of $N$ spin fields $\sigma$ with
weights 
$$
    \Delta(\sigma_k)\=  {l(N-l)\over 2N(N+2)}~,~~~~~l=0,1,\dots,N-1~.
$$
The theory includes $N$ different sectors 
differentiated according to the 
modes $A_n$ of $\psi_1$:
$$
  \psi_1\=\sum_n\,{A_n\over z^{n+{k(N-k)\over N}}}~.
$$

One of the most interesting results of the parafermionic models is
their relation to the SU$(2)_N$ WZWN model. One can show
that (as we will discuss in the problems)
\beq
\label{eq:OM20}
    \fbox{$\displaystyle
    \mbox{SU$(2)_N$ WZWN model}  
    ~~\simeq~~   
    \mbox{$\BZ_N$ parafermion model}
     \otimes
    \mbox{free boson}  
          $}
\eeq
The map between the two sides of this equivalence is given by:
\bb
     J_+(z) &=& \psi_1(z)\, : e^{i\sqrt{2\over N}\phi(z)}:~, \\
     J_-(z) &=& \psi^\dagger_1(z)\, : e^{-i\sqrt{{2\over N}}\phi(z)}:~, \\
     H(z) &=&  \sqrt{N}\, \partial \phi(z)~.
\ee
The primary fields of the WZWN model $G_l(z,\overline z)$ are constructed
as products of the spin fields $\sigma_l(z,\overline z)$ and the vertex 
operators of the free boson by
$$
  G_l(z,\overline z)\=\sigma_l(z,\overline z)\, 
  :e^{i{l\over2\sqrt{N}}\phi(z)}:~.
$$
Therefore any correlation function of WZWN fields will factorize 
into a correlation fuction of spin fields in the parafermion model
and a correlation function of vertex  operators for a free boson:
\bb
  \average{G_{l_1}(z_1,\overline z_1)
           G_{l_2}(z_2,\overline z_2)
           \dots
           G_{l_n}(z_n,\overline z_n)}\=
  \average{\sigma_{l_1}(z_1,\overline z_1)
           \sigma_{l_2}(z_2,\overline z_2)
           \dots
           \sigma_{l_n}(z_n,\overline z_n)} \\
   ~
  \times \average{V_{l_1}(z_1,\overline z_1)
           V_{l_2}(z_2,\overline z_2)
           \dots
           V_{l_n}(z_n,\overline z_n)}
   ~.
\ee
This identity can be used for the computation of all correlators
in  the parafermionic theory.

\newpage
\section{EXERCISES}

\begin{enumerate}

\item
We say that two elements $g,h\in\Gamma$ are conjugate, and write $g\sim h$,
if there is a $k\in\Gamma$ such that $g=khk^{-1}$. The set
$$
   \lb g\rb \eq \{h\in\Gamma ~|~ h\sim g\}
$$ 
is called the {\bfseries conjugacy class}\index{Class!Congugacy --}
of $g$. Let $N(g)$ be the set
$$
   N(g)\eq\{h\in\Gamma ~|~ hg=gh\}~.
$$
This set is called the 
{\bfseries stabilizer group}\index{Group!Stabilizer --}
or the
{\bfseries isotropy group}\index{Group!Isotropy --}
of $g$. 

Show that
$$
     |\Gamma|\=|N(g)| \, |\lb g\rb|~, ~~~~~\forall g~,
$$ 
where the notation $|S|$ stands for the number of elements in $S$.
Then show that the partition function of an orbifold can
be rewritten in the form:
$$
   Z_{\CT/\Gamma} \= \sum_{\lb h\rb\in\Gamma/\sim} \,
                      { 1\over |\lb h\rb| }\,
                     \sum_{g\in N(h)}\, \partition{g}{\lb h\rb}~.
$$

\item
The free boson and the SU(2) WZWN model at level 1 both have
$c=1$. In fact, these CFTs are equivalent, and we can express
the currents of the WZWN model in terms of the free boson field.

(a) Find the bosonized expressions for the currents
 $J^{\pm}(z)$ and  $H(z)$
such that their OPEs satisfy the  SU(2) Kac-Moody algebra at level 1.

(b) The level 1 SU(2) WZWN model has a single primary field in the spin 1/2
representation. Find the bosonized expression for this field.

\item
The SU(2) WZWN model at level 2 has $c=3/2$, as does the CFT
consisting of a free boson  $\phi$ plus a free fermion $\psi$.
As in the previous problem, one can show an equivalence between
these two CFTs.

(a) For suitable constants $a$ and $b$, one can represent
the Kac-Moody currents in the form
$$
 J^{\pm}(z)\=\psi\exp(\pm ia\phi)~~~{\rm and}
 ~~~~H(z)\=ib\,\partial_{z}\phi~.
$$
Find these constants.

(b) Find the representation of the primary fields (spin 1/2 and 1)
 of the WZWN theory 
 in terms of the free boson and fermion.

\item

Derive the following Ward identity (where $\Phi^j$ is a primary
field of spin $j$):
$$
 \lp\frac{\partial}{\partial z_{i}}\,-\,\frac{1}{k+2}\sum_{j\neq i}^{N}
 \frac{t_{i}^{a}t_{j}^{a}}{z_{i}-z_{j}}\rp\,
 \average{\Phi^{j_{1}}(z_{1})\ldots
 \Phi^{j_{N}}(z_{N})}\=0
$$
by using the null state
$$
 \chi\=\lp L_{-1}\,-\,\frac{1}{k+2}J_{-1}^{a}t^{a}\rp \Phi^{j}~.
$$

\item
Using only the parafermion algebra, calculate a recursion relation for the
correlation functions
$$
 \average{\psi_1(z_1)\dots\psi_1(z_n)
          \psi^\dagger(z'_1)\dots\psi^\dagger_1(z'_n)}~.
$$

\item
Using the map \calle{eq:OM20}, rederive the recursion relation
found in
the previous problem.

\item
Show that the $\BZ_k$ parafermion algebra is only consistent with the
conformal Ward identities if the central charge is given by $c=2(k-1)/(k+2)$.

\end{enumerate}

\newpage
\section{SOLUTIONS}

\begin{enumerate}

\item
(a) We can address the first part of the problem by
studying the general case of $\Gamma$
acting on the finite set $M$:
$$
  \Gamma\times M  \rightarrow M ~:~ (g,x)\mapsto g\cdot x~.
$$
Let $\Gamma \cdot x$ be the orbit of $x\in M$, and let $\Gamma_x$
be the 
isotropy group of $x$, i.e.,
$$
   \Gamma_x\=\la g\in\Gamma ~|~ g\cdot x=x \ra~.
$$
We will show that
\beq
\label{eq:OM31}
   |\Gamma| \= |\Gamma_x| \, |\Gamma \cdot x| ~.
\eeq

The group $\Gamma$ has $|\Gamma|$ elements. When $\Gamma$ acts on $x$,
each element takes $x$ to some point on its orbit.  The action of
$\Gamma$ on $x$ generates $|\Gamma|$ images of $x$. (Here, we are counting
degeneracies; for example, if a point on the orbit is produced by the
action of two distinct group elements, we count this as two images.)

We will compute this number again, but using another method, and then
compare the two results. 
Let $y$ be a point in the orbit $\Gamma \cdot x$. Then there exists an
element $\gamma\in\Gamma$ such that
$$
   \gamma\cdot x\= y~.
$$
Let $h\in\Gamma_x$, so that $h\cdot x=x$. Then
the group element
$$
    g\=\gamma h\in\Gamma~,
$$
has the property
$$
  g\cdot x\= (\gamma h)\cdot x\= \gamma (h\cdot x)\=\gamma\cdot x\= y~.
$$
Therefore for each point $y$ of the orbit $\Gamma\cdot x$, there are exactly
$|\Gamma_x|$ elements of $\Gamma$ that map $x$ to $y$.
Since the orbit has $|\Gamma\cdot x|$ \emph{distinct} points, 
the point $x$ has 
$|\Gamma_x| \, |\Gamma \cdot x|$ images under the action of the
group.
Equating our two results for this quantity gives us
the desired identity \calle{eq:OM31}.  In the case that $M$ is itself
the group $\Gamma$, this becomes the result we wished to derive.

(b) Rewriting the partition function in the form asked is straightforward
if we use the identity proved above:
\bb
 Z_{\CT/\Gamma} &=& 
   {1\over|\Gamma|}\,
    \sum_{{\scriptscriptstyle g,h\in\Gamma} \atop {\scriptscriptstyle gh=hg} }
    ~~
    \partition{g}{h}\\
   &=& \sum_{h\in\Gamma} ~
    {1\over |N(h)| \, |\lb h\rb| }\,
    \sum_{g\in N(h)}
    ~~ \partition{g}{h}\\
    &=& \sum_{h\in\Gamma/\sim} ~
    {1\over |N(h)| \, |\lb h\rb| }\,
    |N(h)| \sum_{g\in N(h)}
    ~~ \partition{g}{\lb h\rb}\\
    &=& \sum_{\lb h\rb\in\Gamma/\sim} \,
           { 1\over |\lb h\rb| }\,
       \sum_{g\in N(h)}\, \partition{g}{\lb h\rb}~.
\ee
To go from the second to the third equality, we noticed that $N(h)$ depends
only on the conjugacy class and not $h$ itself.

\separator

\item
(a) Using our general normalization
$$
 \average{ \phi(z) \, \phi(w)} \= - \frac{1}{2 g} \, \ln (z-w)~,
$$
we will prove that the fields
\beq
\label{eq:OM10}
 J^{\pm}(z) \, \equiv \, e^{\pm 2 i \sqrt{g} \phi(z)} ~~~{\rm and}~~~
 H(z) \, \equiv \, 2 i \, \sqrt{g} \, \partial_{z} \phi(z)
\eeq
satisfy the defining OPEs of the $\mathfrak{su}(2)$ Kac-Moody algebra
\beq
 J^{a}(z) \, J^{b}(w) \,=\, \frac{k \, q^{ab}}{(z \,-\, w)^{2}} \,
 + \, \frac{f^{ab}_c}{z \,-\, w} \, J^{c}(w) + \mbox{reg}
\label{eq:OM11}
\eeq
at level $k \,=\, 1$. We use the following conventions for the
structure constants:
\bb
 f^{0 \, +}_{+} &=& - \, f^{+ \,0}_{+} \= f^{- \, 0}_{-} \=
 - \, f^{0 \, -}_{-} \= 2~,\\
 f^{+ \, -}_{0} &=& - \, f^{- \, +}_{0} \= 1~.
\ee
The Killing form, consequently,  will be
$$
 \frac{1}{2} \, q^{0 \, 0} \= q^{+ \, -} \= q^{- \, +} 
 \= 1~.
$$
We can write the current-current OPEs \calle{eq:OM11}
explicitly for $k=1$:
\bb
 J^{+}(z) \, J^{+}(w) &=& {\rm reg} ~,\\
 J^{-}(z) \, J^{-}(w) &=& {\rm reg}~, \\
 H(z) \, H(w) &=& \frac{2}{(z-w)^{2}} + {\rm reg}~, \\
 J^{+}(z) \, J^{-}(w) &=& \frac{1}{(z-w)^{2}} +
    \frac{1}{z \,-\, w} \, H(w) + {\rm reg}~, \\
 H(z) \, J^{\pm}(w)  &=& \frac{\pm \, 2}{z-w} + {\rm reg}~.
\ee
We now  check that  the fields defined by \calle{eq:OM10} satisfy the previous
OPEs.
\bb
 J^{+}(z) \, J^{+}(w) & = &
 : e^{2i \sqrt{g} \phi(z)}: \, :e^{2i \sqrt{g} \phi(w)}:\\
 &=&
 (z-w)^{2} \, : e^{4i \sqrt{g} \phi(z)} : + \cdots
 \= {\rm reg}~, \\
 J^{-}(z) \, J^{-}(w) &=& 
 : e^{-2i \sqrt{g} \phi(z)} : \, : e^{-2i \sqrt{g} \phi(w)}:
 \\
 &=&
 (z-w)^{2} \, :e^{-4i \sqrt{g} \phi(w)}: + \cdots
 \= {\rm reg}~, \\
 H(z) \, H(w) & = & 2 i \, \sqrt{g} \, \partial_{z}
  :\phi(z): \,  2 \, i \, \sqrt{g} \, :\partial_{w} \phi(w):
 \\
 &=&
 - \, 4 g \, \partial_{z}  \partial_{w}  \phi(z) \phi(w) 
 + \mbox{reg}\\
 &=&
 - \, 4 g \, \lp - \, \frac{1}{2 g} \, \frac{1}{(z-w)^{2}}
 \rp  + \mbox{reg} \\
 &=&
 \frac{2}{(z-w)^2} + \mbox{reg}~, \\
 J^+(z) \, J^-(w) & = & : e^{ 2i \sqrt{g}
 \phi(z)}: \, : e^{-2i \sqrt{g} \phi(w)}:\\
 &=&
 \frac{1}{(z - w)^2} + \frac{2 \, i \, \sqrt{g} \,
 \partial_{w} \, \phi(w)}{z-w} + \mbox{reg}\\
 &=&
 \frac{1}{(z - w)^2} + \frac{1}{z-w} \, H(w)
 + {\rm reg}~,\\
 H(z) \, J^{\pm}(w) &=& 2 i \, \sqrt{g} \,
     \partial_{z} \phi(z) \, e^{\pm 2i \sqrt{g} \phi(z)}\\
 &=&
 2 i \, \sqrt{g} \, \frac{-i \, (\pm 2  \sqrt{g})}{
     2  g \, (z-w)} \, e^{\pm 2i \sqrt{g} \phi(w)}
     + \mbox{reg}\\
 &=&
 \frac{\pm 2 \, e^{\pm 2i \sqrt{g} \phi(z)}}{z - w} +
 \mbox{reg}\\
 &=&
 \frac{\pm 2}{z \,-\, w} \, J^{\pm}(w)  + \mbox{reg}
  ~.
\ee

(b) Let $ \Phi_{m}^{(1/2)}$, $m \,=\, \pm 1/2$, be the WZWN primary field.
Using the general formula
$$
 \Delta \left( \Phi^{(j)} \right) \,=\, \frac{j \, (j \,+\, 1)}{
 k \,+\, 2}~,
$$
we see that 
$$
 \Delta \lp \Phi^{(1/2)} \rp \= \frac{1}{4}~.
$$
To find a  bosonized expression for $\Phi_{m}^{(1/2)}$ we write
$$
 \Phi_{m}^{(1/2)}(z) \= e^{im \alpha \phi(z)}~.
$$
The constant $\alpha$ is chosen
so that
$$
  \Delta \lp \Phi^{(1/2)}_m \rp\=
 \frac{m^{2}  \alpha^{2}}{4 g} \= \frac{1}{4} ~~
 \Rightarrow ~~~ \alpha \= 2 \, \sqrt{g} ~.
$$
We will show now explicitly that
 the expression
$$
 \Phi_{m}^{(1/2)}(z) \= e^{2i \sqrt{g} m \phi(z)}
$$
is a representation of the algebra $\mathfrak{su}(2)$.
Recall that, in general, if $\phi_{(r)}$ transforms as
a representation $(r)$ of the algebra $\mathfrak{g}$, then
$$
 J^{(a)}(z) \, \phi_{(r)}(w) \, \sim \, \frac{t_{(r)}^{a}}{z-w}
 \, \phi_{(r)}(w)~,
$$
where the $t_{(r)}^{a}$ are representation matrices of the algebra
$\mathfrak{g}$.
In our case, it is easy to use the bosonic OPEs, as we did in part (a),
to find: 
\bb
 J^+(z) \, \lb \begin{array}{c} \Phi_{-1/2}^{(1/2)}\\ 
                                     \Phi_{+1/2}^{(1/2)}
                    \end{array} \rb
 &\sim& {1\over z-w}\, \lb \begin{array}{cc}0 & 1\\ 0 & 0\end{array} \rb\,
 \lb \begin{array}{c} \Phi_{-1/2}^{(1/2)}  \\ \Phi_{+1/2}^{(1/2)}
                    \end{array} \rb ~, \\
 J^-(z) \, \lb \begin{array}{c} \Phi_{-1/2}^{(1/2)}\\ \Phi_{+1/2}^{(1/2)}
                    \end{array} \rb
 &\sim&  {1\over z-w}\, \lb\begin{array}{cc} 0 & 0\\ 1 & 0\end{array} \rb
 \lb \begin{array}{c} \Phi_{-1/2}^{(1/2)}\\
                                     \Phi_{+1/2}^{(1/2)}
                    \end{array} \rb ~, \\
 H(z) \, \lb \begin{array}{c} \Phi_{-1/2}^{(1/2)}\\
                                     \Phi_{+1/2}^{(1/2)}
                   \end{array} \rb
 &\sim&  {1\over z-w}\,  \lb \begin{array}{cc} -1 & 0\\  0 & 1
                          \end{array} \rb
 \, \lb \begin{array}{c} \Phi_{-1/2}^{(1/2)}\\ \Phi_{+1/2}^{(1/2)}
                    \end{array} \rb~.
\ee

\separator

\item

(a) We will prove the map \calle{eq:OM20} for the general case.
We thus seek  a representation
of the Kac-Moody algebra at a generic  level of the form
\bb
 J^+(z) & = & a \, \psi \, e^{+i \alpha \phi(z)}~, \\
 J^-(z) & = & a \, \psi^{\dagger} \,
                 e^{-i \alpha \phi(z)}~, \\
 H(z) & = & i  b \, \partial_z \phi(z)~.
\ee
Without loss of generality, we choose
 $\psi_1$ as the representative of the
parafermion fields. Then
\bb
 J^+(z) & = & a \, \psi_1 \, e^{+i \alpha \phi(z)}~,\\
 J^-(z) & = & a \, \psi_1^{\dagger} \,
                 e^{-i \alpha \phi(z)}~.
\ee
The parameter $\alpha$ is fixed by dimensional
arguments:
$$
 1 \= \Delta (J^{\pm}) \= \Delta (\psi_{1}) +
 \Delta (e^{\pm i \alpha \phi(z)}) \= \lp 1 \,-\, \frac{1}{N}
 \rp \,+\, \frac{\alpha^2}{4g}~.
$$
Hence
$$
 \alpha \= 2 \, \sqrt{\frac{g}{N}}~.
$$
The constants $a$ and $b$ cannot be fixed by dimensional arguments;
instead, we use the requirement that $J^{\pm}(z)$ and $H(z)$ form a
representation of the Kac-Moody algebra.
Using Wick's theorem, we see that
\bb
   J^+(z) J^-(w) &=& a^2\, :\psi_1(z)e^{i\alpha\phi(z)}:
                        ~ :\psi_1(w)e^{-i\alpha\phi(w)}: \\
                 &=& a^2\, \lowerpairing{\psi_1(z)\psi}_1(w)
                  \, \lowerpairing{e^{i\alpha\phi(z)}e}^{-i\alpha\phi(w)}
                 +\cdots \\
                 &=& a^2\, (z-w)^{-2+2/N} \, (z-w)^{-2/N} + \cdots\\   
                 &=& {a^2\over (z-w)^2} + \cdots~.
\ee
Comparing this with the OPE for the Kac-Moody algebra at level $N$,
$$
  J^+(z) J^-(w)\= {N\over (z-w)^2} + \cdots~,
$$
we conclude that equivalence will require that $a=\sqrt{N}$. 
Similarly, the other OPEs require that
$b=2\sqrt{gN}$.
Therefore, we arrive at the map
\bb
 J^+(z) &=& \sqrt{N} \, \psi_{1} \, e^{2i \sqrt{\frac{g}{N}}
                \phi(z)}~,  \\
 J^-(z) &=& \sqrt{N} \, \psi_{1}^{\dagger} \, e^{-2i
                \sqrt{\frac{g}{N}} \phi(z)}~, \\
 H(z) &=& 2 i \, \sqrt{g N} \, \partial_{z} \phi(z)~,
\ee
as given in the introduction of the model.

(b) Now we confine ourselves to the case of level $N=2$. We have
two primary fields, $\Phi_{m}^{(1)}(z)$ and $\Phi_{m}^{(1/2)}(z)$,
to write in terms of a boson field $\phi(z)$, a fermion field $\psi(z)$,
and a spin field $\sigma(z)$. We notice that
the weights of the fields are rational numbers:
$$
 \Delta(\Phi^{(1)}) \= \frac{1}{2} ~~~{\rm and}~~~~ \Delta(\Phi^{(1/2)})
 \= \frac{3}{16}~.
$$
Since 
$$ 
   \Delta(\phi)\=0
$$
 and derivatives increase the
dimension by one, in the bosonized form of the $\Phi$'s, the boson
must enter in the expression in the form of a vertex
operator.
Thus we will have expressions of the
form $e^{i m \alpha \phi(z)}$, for suitable values
of $m$. Taking into account that
$$
 \Delta(\psi) \=\frac{1}{2} ~~~{\rm and}~~~ \Delta(\sigma) \,=\, \frac{1}{
 16} ~,
$$
we see that the only possibility is
\bb
 \Phi^{(1)}(z) & \equiv &
    \lb \begin{array}{c}
              \Phi_{-1}^{(1)}(z)\\
              \Phi_{0}^{(1)}(z)\\
              \Phi_{1}^{(1)}(z)\\ 
    \end{array} \rb
 \=
    \left[ \begin{array}{c}
              e^{-i \sqrt{2g} \phi(z)}\\
              \psi(z)\\
              e^{i \sqrt{2g} \phi(z)}\\
    \end{array} \rb ~, \\
 \Phi^{(1/2)}(z) & \equiv &
    \lb \begin{array}{c}
              \Phi_{-1/2}^{(1/2)}(z)\\
              \Phi_{1/2}^{(1/2)}(z)\\
    \end{array} \rb
 \=
    \lb \begin{array}{c}
              \sigma(z) \, e^{-i \sqrt{\frac{g}{2}} \phi(z)}\\
              \sigma(z) \, e^{i \sqrt{\frac{g}{2}} \phi(z)}\\
    \end{array} \rb ~.\\
\ee
The reader may be surprised that dimensional analysis is so
effective, so we think it useful to present the OPEs of the $\Phi$'s
with the generators of the algebra; notice that the matrices which
appear are indeed the 2-dimensional and 3-dimensional representations
of $\mathfrak{su}(2)$ with which we are familiar from quantum mechanics.
The OPEs are:
\bb
 J^+(z) \, \Phi^{(1)}(w) & \sim & {1\over z-w}\,
   \lb
         \begin{array}{ccc} 0 & 1 & 0\\
                            0 & 0 & 1\\
                            0 & 0 & 0
         \end{array}
         \rb      \, \Phi^{(1)}(w) ~, \\
 J^-(z) \, \Phi^{(1)}(w) & \sim & {1\over z-w}\,
   \lb
         \begin{array}{ccc} 0 & 0 & 0\\
                            1 & 0 & 0\\
                            0 & 1 & 0
         \end{array}
         \rb      \, \Phi^{(1)}(w)~, \\
 H(z) \, \Phi^{(1)}(w) & \sim & {1\over z-w}\,
  \lb
         \begin{array}{ccc} -1 & 0 & 0\\
                             0 & 0 & 0\\
                             0 & 0 & 1
         \end{array}
         \rb      \, \Phi^{(1)}(w)~,  \\
 J^+(z) \, \Phi^{(1/2)}(w) & \sim & {1\over z-w}\,
   \lb
         \begin{array}{cc} 0 & 1\\
                           0 & 0
         \end{array}
         \rb      \, \Phi^{(1/2)}(w)~, \\
 J^-(z) \, \Phi^{(1/2)}(w) & \sim & {1\over z-w}\,
   \lb
         \begin{array}{cc} 0 & 0\\
                           1 & 0
         \end{array}
         \rb     \, \Phi^{(1/2)}(w)~, \\
 H(z) \, \Phi^{(1/2)}(w) & \sim & {1\over z-w}\,
   \lb
         \begin{array}{cc} -1 & 0\\
                            0 & 1
         \end{array}
         \rb      \, \Phi^{(1/2)}(w)~.
\ee
We have given each of the above expressions up to a constant,
the determination of which is left as an  exercise to the reader.

\separator

\item

In the following, the notation $\Phi_{(r)}^{(j)}$ stands for the
primary field $\Phi^{(j)}(z)$ of spin $j$ in the representation
$(r)$. Since
$$
 \chi \= \lp L_{-1} \,-\, \frac{1}{k \,+\, 2} \, J_{-1}^{a}
 \, t^{a}_{(r_{i})} \rp \, \Phi_{(r_{i})}^{(j_{i})} (z_{i}) \, \equiv
 \, \CL_{-1} \, \Phi_{(r_{i})}^{(j_{i})} (z_{i})
$$
is a null state,
\begin{equation}
\label{eq:OM21}
 \lan \lp L_{-1} \,-\, \frac{1}{k \,+\, 2} \, J_{-1}^{a}
 \, t_{(r_i)}^a \rp \, \Phi_{(r_1)}^{(j_1)} (z_1)
 \, \ldots \, \Phi_{(r_i)}^{(j_i)} (z_i) \, \ldots \,
 \Phi_{(r_N)}^{(j_N)} (z_{N}) \ran \= 0~,
\end{equation}
where the operator $\CL_{-1}$ is acting on
$\Phi_{(r_{i})}^{(j_{i})} (z_{i})$. We have seen though that
\beqn
  \lan \Phi_{(r_{1})}^{(j_1)} (z_1) \, \ldots \,
 \CL_{-1} \, \Phi_{(r_i)}^{(j_i)} (z_i) \, \ldots \,
 \Phi_{(r_N)}^{(j_N)} (z_N) \ran
          & = &
 - \, \sum_{n \neq i}^{N} \, \frac{\partial}{\partial z_n} \,
 \lan \Phi_{(r_{1})}^{(j_{1})} (z_{1}) \, \ldots \, 
 \Phi_{(r_{N})}^{(j_{N})} (z_{N}) \ran
  \nonumber \\
          & = &
 \frac{\partial}{\partial z_i} \, \lan
 \Phi_{(r_{1})}^{(j_{1})} (z_{1}) \, \ldots \,
 \Phi_{(r_{N})}^{(j_{N})} (z_{N}) \ran
\label{eq:OM22}
\eeqn
On the other hand,
\beqn
 && \lan \Phi_{(r_{1})}^{(j_{1})} (z_{1}) \, \ldots \, J_{-1}^{a}
 \, t^{a}{(r_{i})} \, \Phi_{(r_{i})}^{(j_{i})} (z_{i}) \, \ldots \,
 \Phi_{(r_{N})}^{(j_{N})} (z_{N}) \ran
  \nonumber \\
  && \hspace{2cm}      \= 
 \lan \Phi_{(r_{1})}^{(j_{1})} (z_{1}) \, \ldots \,
 \ointleft \, \frac{dw}{2 \pi i} \, \frac{J^{a}(w)}{w \,-\, z_{i}} \,
 t_{(r_{i})}^{a} \, \Phi_{(r_{i})}^{(j_{i})} (z_{i}) \, \ldots \,
 \Phi_{(r_{N})}^{(j_{N})} (z_{N}) \ran
 \nonumber  \\
  && \hspace{2cm}      \=
 \ointleft \, \frac{dw}{2 \pi i} \, \frac{1}{w \,-\, z_{i}} \, \lan
 J^{a}(w) \, \Phi_{(r_{1})}^{(j_{1})} (z_{1}) \, \ldots \,
 \Phi_{(r_{N})}^{(j_{N})} (z_{N}) \ran \, t_{(r_{i})}^{a}
 \nonumber \\
  && \hspace{2cm}      \= 
 \ointleft \, \frac{dw}{w \,-\, z_{i}} \, \sum_{n\ne i} \,
 \frac{t_{(r_{n})}^{a}}{w \,-\, z_{n}} \, \lan
 \Phi_{(r_{1})}^{(j_{1})} (z_{1}) \, \ldots \,
 \Phi_{(r_{N})}^{(j_{N})} (z_{N}) \ran \, t_{(r_{i})}^{a}
 \nonumber \\
  && \hspace{2cm}      \= 
 \sum_{n \ne i} \, \frac{t_{(r_{n})}^{a} \, t_{(r_{i})}^{a}}{
 z_{i} \,-\, z_{n}} \, \lan \Phi_{(r_{1})}^{(j_{1})} (z_{1}) \,
 \ldots \, \Phi_{(r_{N})}^{(j_{N})} (z_{N}) \ran ~.
\label{eq:OM23}
\eeqn
Using \calle{eq:OM21}--\calle{eq:OM23}, we thus arrive at the equation
$$
 \lp \frac{\partial}{\partial z_{i}} \,-\, \frac{1}{k \,+\, 2} \,
 \sum_{n \neq i} \, \frac{t_{(r_{n})}^{a} \, t_{(r_{i})}^{a}}{
 z_i-z_n}\rp \, \lan
 \Phi_{(r_{1})}^{(j_{1})} (z_{1}) \, \ldots \,
 \Phi_{(r_{N})}^{(j_{N})} (z_{N}) \ran \= 0~.
$$

\separator

\item
We start by recalling the classic  Mittag-Leffler theorem.

\vspace{3mm}

{\bf Theorem} \lbrack Mittag-Leffler\rbrack\\
{\small
Let $f(z)$ be a complex function of a complex variable such that:

(i) $f(z)$ is analytic at $z=0$;

(ii) $f(z)$ has only simple poles $a_1$, $a_2$, $\ldots$
on the finite $z$-plane (without loss of generality, these are given
in order of increasing magnitude) with corresponding residues $b_1$, $b_2$, 
$\ldots$; 

(iii) there exists a sequence of positive  numbers $R_N$ such that 
      $R_N\rightarrow+\infty$ when $N\rightarrow+\infty$ and the circle
      $C_N\=\{ |z|=R_N\}$ does not pass through any pole of $f(z)$ for
      any values of $N$;

(iv) there exists a real number $M$ such that 
   $|f(z)|<M$ on the circles 
   $C_N$ for all $N$.

Then if $z$ is not a pole of $f(z)$:
\bb
 f(z) \= f(0) \,+\, \sum_k \, b_k \,
 \la \frac{1}{z \,-\, a_k} \,+\, \frac{1}{a_k} \ra~.
\ee
}%

\vspace{3mm}

\begin{figure}[htb!]
\begin{center}
\includegraphics[height=6cm]{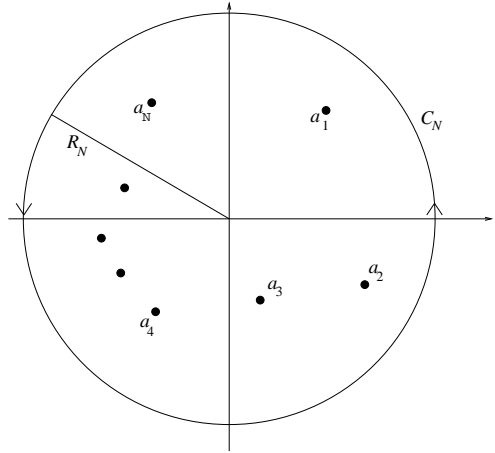}
\end{center}
\caption{The set of poles $a_1, a_2, \dots$ of a function $f(z)$ for which the
Mittag-Leffler theorem is applicable may be finite or countably infinite.
The circle $C_N$ does not pass through any pole.}
\end{figure}

The above formulation of the Mittag-Leffler theorem is simple, yet very
useful. There are several other equivalent formulations. 
Here is an example taken from an introductory modern text \cite{GreeneKrantz}
on complex variables:

\vspace{3mm}
{\bfseries Theorem} \lbrack Mittag-Leffler\rbrack\\
{\small                                       
Let $U\subseteq\BC$ be any open set. Let $a_1, a_2, \dots$ be a finite or
countably infinite set of distinct elements of $U$ with no accumulation point
in $U$. Suppose, for each $j$, that $U_j$ is a neighborhood of $a_j$ and 
$a_j\not\in U_k$ if $k\ne j$. Further assume, for each $j$, that $f_j$ is
a meromorphic function defined on $U_j$ with a pole at $a_j$ and no other
poles. Then there exists a meromorphic function $f$ on $U$ such that
$f-f_j$ is holomorphic on $U_j$ for every $j$ and which has no poles
other than those at the $a_j$.
}%

\vspace{3mm}
Now let us state and prove\footnote{We rush to make the folowing comment:
In textbooks of complex analysis, one never finds more than the
Mittag-Leffler theorem. This is because this theorem is all one needs.
If $f(z)$ has a double pole at $z=a$, then $(z-a)f(z)$ has a simple pole
at $z=a$, and one can continue from this point. However, we will make
use of the `generalized' Mittag-Leffler theorem since it appears, at
least superficially, simpler to apply in our case.} a
`generalized' Mittag-Leffler theorem.

\vspace{3mm}
{\bfseries Theorem}\\
{\small
Let $f(z)$ be a complex function of a complex variable such that: 

(i) $f(z)$ is analytic at $z=0$;

(ii) $f(z)$ has only double poles $a_1$, $a_2$, $\ldots$, on
the finite $z$-plane with corresponding residues $b_1$, $b_2$,
$\ldots$. Furthermore, let $c_1$, $c_2$, $\ldots$ be the
coefficients of the terms $1/(z-a_1)^2$, $1/
(z-a_2)^2$, $\ldots$ in the Laurent expansion of $f(z)$.
Without loss of generality we may assume that the poles are ordered
with increasing magnitude; 

(iii) there exists a sequence of positive  numbers $R_N$ such that
      $R_N\rightarrow+\infty$ when $N\rightarrow+\infty$ and the circle
      $C_N\=\{ |z|=R_N \}$ does not pass through any pole of $f(z)$ for
      any values of $N$;

(iv) there exists a real number $M$ such that
   $|f(z)|<M$ on the circles
   $C_N$ for all $N$.

Then if $z$ is not a pole of $f(z)$: 
\beq
\label{GMLT}
 f(z) \= f(0) \,+\, \sum_{k} \, b_{k} \, \lb \frac{1}{z \,-\,
 a_{k}} \,+\, \frac{1}{a_{k}} \rb \,+\, \sum_{k} \, c_{k} \,
 \lb \frac{1}{(z \,-\, a_{k})^{2}} \,-\, \frac{1}{a_{k}^{2}}
 \rb~.
\eeq
}%

\vspace{3mm}

\underbar{Proof}

We choose a point $J$ which is not a pole of $f(z)$. Then the function
$$
 F(z)  \eq \frac{f(z)}{z \,-\, J}
$$
has double poles at $a_{1}$, $a_{2}$, $\ldots$, and a simple pole at
$J$. By using the Residue Theorem, we can write
$$
 \ointleft_{C_N} \, \frac{dz}{2 \pi i} \, \frac{f(z)}{z \,-\, J} \,=\,
 \residue{z=J} \, F(z) \,+\, \sum_{k=1}^{N} \, \residue{z=a_{k}}
 \, F(z)~,
$$
where we have assumed that $C_N$ contains the poles $a_1$, $\ldots$,
$a_N$. Now,
\begin{eqnarray*}
 \residue{z=a_{k}} \, F(z) & = & \lim_{z \rightarrow a_{k}} \,
 \frac{d}{dz} \, (z \,-\, a_{k})^{2} \, F(z)\\ & = & \lim_{z 
 \rightarrow a_{k}} \, \left[ \frac{1}{z \,-\, J} \, \frac{d}{dz} \,
 (z \,-\, a_{k})^{2} \, f(z) \,-\, \frac{(z \,-\, a_{k})^{2}}{
 (z \,-\, J)^{2}} \, f(z) \right]\\
 & = & \frac{b_{k}}{a_{k} \,-\, J} \, - \, \frac{c_{k}}{(a_{k} \,-\,
 J)^2}~,
\end{eqnarray*}
and
$$ 
   \residue{z=J} \, F(z) \= f(J)~. 
$$
Consequently,
\beq
 \ointleft_{C_{N}} \, \frac{dz}{2 \pi i} \, \frac{f(z)}{z \,-\, J} \,=\,
 f(J) \,+\, \sum_{k=1}^{N} \, \lb  \frac{b_{k}}{a_{k} \,-\, J} \,-\,
 \frac{c_{k}}{(a_{k} \,-\, J)^{2}} \rb
  ~.
\label{eq:OM25}
\eeq
Since $f(z)$ is analytic at $z=0$, setting $J=0$ in the previous
relation gives
\beq
 \ointleft_{C_N} \, \frac{dz}{2 \pi i} \, \frac{f(z)}{z} \=
 f(0) \,+\, \sum_{k=1}^{N} \, \lb \frac{b_{k}}{a_{k}} \,-\,
 \frac{c_{k}}{a_{k}^{2}} \rb~.
\label{eq:OM26}
\eeq
By subtracting \calle{eq:OM26} from \calle{eq:OM25}, we get
\begin{eqnarray}
 J \, \ointleft_{C_N}\,\frac{dz}{2 \pi i} \, \frac{f(z)}{z \, (z \,-\, J)}
 \= f(J) \,-\, f(0) & + & \sum_{k=1}^{N} \, b_{k} \, \lb
 \frac{1}{a_{k} \,-\, J} \,-\, \frac{1}{a_{k}} \rb
 \nonumber \\
 & + & \sum_{k=1}^{N} \, c_{k} \, \lb \frac{1}{a_{k}^{2}}
 \,-\, \frac{1}{(a_{k} \,-\, J)^{2}} \rb~.
\label{eq:OM27}
\end{eqnarray}
Let us notice now that
$$
 \left| \, \ointleft_{C_N} \, \frac{dz}{2 \pi i} \, \frac{f(z)}{
 z \, (z \,-\, J)} \, \right| \, \leq \, \frac{M}{R_{N} \,
 (R_{N} \,-\, J)} \, \rightarrow \, 0 ~~,~~~~~ \mbox{as} ~~
 N \rightarrow +\infty~.
$$
This implies that, in the limit $N \rightarrow +\infty$, \calle{eq:OM27} gives
our advertised formula, namely
\bb
 f(J) \= f(0) \,+\, \sum_{k} \, b_{k} \, \lb \frac{1}{J \,-\,
 a_{k}} \,+\, \frac{1}{a_{k}} \rb \,+\, \sum_{k} \, c_{k} \,
 \lb \frac{1}{(J \,-\, a_{k})^{2}} \,-\, \frac{1}{a_{k}^{2}}
 \rb
 ~. \hspace{19mm}\blacksquare
\ee

We can derive a corollary of the above result for a function $f(z)$ 
that vanishes asymptotically.  Suppose
$$
   \lim_{|z|\rightarrow+\infty} f(z) = 0~.
$$
Then taking the limit $|z|\rightarrow+\infty$ of \calle{GMLT}, we find
$$
   0\= f(0) + \sum_k \, {b_k \over a_k}
            - \sum_k \, {c_k \over a_k^2}~.
$$
Using this result, \calle{GMLT} simplifies to
$$
   f(z) \=  \sum_k \, {b_k \over z-a_k}
         +  \sum_k \, {c_k \over (z-a_k)^2}
   ~.
$$

We will use the above result to derive a recursion relation for the
function
$$
   g(z_1)\eq
   \average{\psi_1(z_1)\dots\psi_1(z_n)
          \psi^\dagger(z'_1)\dots\psi^\dagger_1(z'_n)}~.
$$
(We are suppressing the dependence of $g$ on variables other than $z_1$
for simplicity, as we will be using $g$ to define another function that
does in fact depend only on $z_1$.)
Now $g(z_1)$, as defined, has branch cuts, and the Generalized Mittag-Leffler
Theorem we obtained
above is not directly applicable. However, we can define a new function
$f(z_1)$ which is a product of $g(z_1)$ and some factors that remove the branch 
cuts. Towards this end, we observe that the branch cuts in the variable
$z_1$ come from the OPEs:
\bb
   \psi_1(z)\psi_1(z_i) &=& {c_{11}\over (z_1-z_i)^{2/N} }\,
            \lb \psi_2(z_i)+\CO(z_1-z_i)\rb~~~{\rm and}\\
   \psi_1(z)\psi_1^\dagger(z'_i) &=& {1\over (z_1-z'_i)^{2-2/N} }\,
            \lb 1+\CO\lp (z_1-z_i)^2\rp \rb~.
\ee
Therefore, the function $f(z_1)$ should be defined as
$$
   f(z_1)\eq
          \lb \prod_{i=1}^n (z_1-z_i)^{2/N} \rb ~
          \lb \prod_{i=1}^n (z_1-z'_i)^{-2/N} \rb ~ g(z_1)~.
$$
The first factor removes any branch cuts or poles at the points
$z_1=z_i$. The second factor removes the branch cuts at the points
$z_1=z'_i$, but leaves double poles at these points.
Moreover,
$$
   \psi_1(z) ~~{\buildrel |z|\gg 1\over  \simeq }~~
   {1\over z^{2-2/N} } ~~
   {\buildrel {|z|\rightarrow+\infty} \over  \rightarrow}
   ~~ 0~.
$$
This guarantees that $f(z_1)\rightarrow0$ as $|z_1|\rightarrow+\infty$.
Therefore, $f(z_1)$ can be expressed in the form
\beq
\label{eq:OM28}
   f(z_1) \=  \sum_k \, {b_k \over z_1-z'_k}
         +  \sum_k \, {c_k \over (z_1-z'_k)^2}
   ~,
\eeq
where $b_k=b_k(z_2,\dots,z_n;z'_1,\dots,z'_n)$
and $c_k=c_k(z_2,\dots,z_n;z'_1,\dots,z'_n)$.
To finish the problem we must compute these functions.
In fact, it is easily seen that
\bb
  b_k &=&
       \lim_{z_1 \rightarrow z'_k} \, {d\over dz_1} \lb
       (z_1-z'_k)^2 \, f(z_1) \rb~,~~~{\rm and} \\
  c_k &=&
       \lim_{z_1 \rightarrow z'_k} \,  \lb
       (z_1-z'_k)^2 \, f(z_1)\rb ~.
\ee
 In the limit $z_1 \rightarrow z'_k$, the contraction
$\lowerpairing{\psi_1(z_1)\psi}_1(z'_k)$ will be dominant;
therefore, the second of the above equations gives:
\bb
 c_k &=&
       \lim_{z_{1} \rightarrow z_{k}'} 
       (z_1-z_{k}')^2 \, \lb \prod_{i=2}^{n} \,
       (z_1-z_i)^{2/N} \rb \, \lb \prod_{i=1}^{n}
       \, (z_1-z_{i}')^{-2/N} \rb \\
     &&
       \times\, \average{
       \lowerpairing{\psi_1(z_1)\psi_2(z_2)\ldots\psi_1^\dagger(z'_1)
       \dots\psi}_1^\dagger(z_{k}'\dots)
        	} \\
     &=&
       \lim_{z_{1} \rightarrow z_{k}'}
       (z_1-z_{k}')^2 \, \lb \prod_{i=2}^{n} \,
       (z_1-z_i)^{2/N} \rb \, \lb \prod_{i=1}^{n}
       \, (z_1-z_{i}')^{-2/N} \rb \, {(-1)^{2(n+k-2)/N}\over (z_1-z'_k)^{2-2/N}}
       \\
     &&
       \times\, \average{
       \hat\psi_1(z_1)\psi_2(z_2)\ldots\psi_1^\dagger(z'_1)
       \dots\hat\psi_1^\dagger(z_{k}'\dots)
                } \\
     &=&
       \lim_{z_{1} \rightarrow z_{k}'} (-1)^{2(n+k-2)/N}
       \lb \prod_{i=2}^{n} \,
       (z_1-z_i)^{2/N} \rb \, 
      \Big\lbrack \prod_{{\scriptscriptstyle i=1\atop i\ne k} }^n
       \, (z_1-z_{i}')^{-2/N} \Big\rbrack
       \\
     &&
       \times\, \average{
       \hat\psi_1(z_1)\psi_2(z_2)\ldots\psi_1^\dagger(z'_1)
       \dots\hat\psi_1^\dagger(z_{k}'\dots)
                } \\
     &=&
        (-1)^{2(n+k-2)/N}
       \lb \prod_{i=2}^{n} \,
       (z'_k-z_i)^{2/N} \rb \,
      \lb\prod_{{\scriptscriptstyle i=1\atop i\ne k}}^n
       \, (z'_k-z_{i}')^{-2/N} \rb
       \\
     &&
       \times\, \average{
       \hat\psi_1(z_1)\psi_2(z_2)\ldots\psi_1^\dagger(z'_1)
       \dots\hat\psi_1^\dagger(z_{k}'\dots)
                } \\
&=&
       \lb \prod_{i=2}^{n} 
       (z'_k-z_i)^{2/N} \rb \,
      \lb\prod_{i=1}^{k-1}
       \, (z'_i-z'_k)^{-2/N} \rb\,
      \lb\prod_{i=k+1}^n
       \, (z'_k-z_{i}')^{-2/N} \rb
       \\
     &&
       \times\, \average{
       \hat\psi_1(z_1)\psi_2(z_2)\ldots\psi_1^\dagger(z'_1)
       \dots\hat\psi_1^\dagger(z_{k}'\dots)
                } ~,
\ee
where the factors $(-1)^{2/N}$ came from the analytic continuation of  a
parafermion as one encircles another, i.e., from
$$
   \psi_1(\pi_C(x))  \psi_1(y) \= (-1)^{2/N}\,
   \psi_1(y)  \psi_1(x) ~,
$$
and where we have used the notation that a caret (`hat') above a 
quantity indicates that this quantity is to be omitted.
Analogous calculations determine also that
\bb
 b_k &=&
      {d\over dz'_k} \,
      \lb \prod_{i=2}^{n}
       (z'_k-z_i)^{2/N} \rb \,
      \lb\prod_{i=1}^{k-1}
       \, (z'_i-z'_k)^{-2/N} \rb\,
      \lb\prod_{i=k+1}^n
       \, (z'_k-z_{i}')^{-2/N} \rb
       \\
     &&
       \times \, \average{
       \hat\psi_1(z_1)\psi_2(z_2)\ldots\psi_1^\dagger(z'_1)
       \dots\hat\psi_1^\dagger(z_{k}'\dots)
                } \\ 
     &=&
       \frac{2}{N} \, \lb \sum_{i=2}^{n} \, \frac{1}{z_{k}' \,-\,
       z_i} \,-\, \sum_{i=1, i\neq k} \, \frac{1}{z_{k}' \,-\,
       z_i'} \rb \, 
       \lb \prod_{i=2}^{n} \, (z_{i} \,-\, z_{k}')^{2/N} \right] \,
       \lb \prod_{i=1}^{k-1} \, (z_{i}' \,-\, z_{k}')^{-2/N}
       \rb \\
       &   &
       \times \, \lb \prod_{i=k+1}^{n} \, (z_{k}' \,-\, z_{i}')^{-2/N} \rb
       \, \average{\hat{\psi}_{1}(z_{1}) \, \ldots \, \hat{\psi}_{1}^{
       \dagger}(z_k') }~.
\ee
Substituting these results for $b_k$ and $c_k$
in \calle{eq:OM28},  we find the recursion relation first written
by Zamolodchikov and Fateev:
\bb
 \average{ 
   \psi_1(z_{1}) \, \ldots \, \psi_1(z_n) \,
   \psi_1^\dagger(z_1') \, \ldots \, \psi_1^\dagger
   (z_{n}') }  
 \=
     \lb \prod_{i=2}^n \, (z_1 \,-\, z_{i})^{-2/N} \rb
     \, \lb \prod_{i=1}^{n} \, (z_{1} \,-\, z_{i}')^{2/N} \rb\\
     \times\,\sum_{k=1}^{n} \lb \frac{1}{(z_{1} \,-\, z_{k})^{2}}
     \,+\, \frac{2/N}{z_{1} \,-\, z_{k}'} \, \lp \sum_{l=2}^{n} \,
     \frac{1}{z_{k}' \,-\, z_{l}} \,-\, \sum_{l=1,l \neq k}^{n} \,
     \frac{1}{z_{k}' \,-\, z_{l}} \rp \rb\\
     \times\,\lb \prod_{i=2}^{n} \, (z_{i} \,-\, z_{k}')^{2/N} \rb \,
     \lb \prod_{i=1}^{k-1} \, (z_{i}' \,-\, z_{k}')^{-2/N} \rb \,
     \lb \prod_{i=k+1}^{n} \, (z_{k}' \,-\, z_{i}')^{-2/N} \rb\\
     \times\, \average{  
       \hat{\psi}(z_{1}) \, \ldots \, \psi_{1}(z_{n}) \, \psi_{1}^{
       \dagger}(z_{1}') \, \ldots \, \hat{\psi}_{1}^{\dagger}(z_{k}')
       \, \ldots \, \psi_{1}^{\dagger}(z_{n}') } ~.
\ee

\separator

\item

The mapping between the WZWN model and the parafermion model
lets us write the currents as
\bb
 J^{+}(z) & = & \sqrt{N} \, \psi_{1}(z) \, e^{2i \, \sqrt{
 \frac{g}{N}} \, \phi(z)} ~, \\
 J^{-}(z) & = & \sqrt{N} \, \psi_{1}^{\dagger}(z) \, e^{-2i \, \sqrt{
 \frac{g}{N}} \, \phi(z)}~,~~~{\rm and}\\
 H(z) & = & 2 i \, \sqrt{g N} \, \partial_{z} \, \phi(z)~.
\ee
Then the  correlation function
$$
 \average{\psi_1(z_1) \cdots \psi_1(z_n) \,
   \psi_1^\dagger(z_1') \cdots \psi_1^\dagger(z_2')} 
$$
can be computed from the identity
\bb
  \average{
  J^+(z_1) \cdots J^+(z_n)  J^-(z_1') \cdots J^-(z_n')} 
  &=& N^n \,
  \average{\psi_1(z_1) \cdots \psi_1(z_n) \,
   \psi_1^\dagger(z_1') \cdots \psi_1^\dagger(z_2')} ~ \\
  && \hspace{-3.1cm} \times\, 
  \average{ e^{2i \, \sqrt{\frac{g}{N}} \, \phi(z_1)}
            \dots e^{2i \, \sqrt{\frac{g}{N}} \, \phi(z_n)}
            e^{-2i \, \sqrt{\frac{g}{N}} \, \phi(z'_1)}
            \dots e^{-2i \, \sqrt{\frac{g}{N}} \, \phi(z'_n)} }
   \\
  &=& N^n \,
  \average{\psi_1(z_1) \cdots \psi_1(z_n) \,
    \psi_1^\dagger(z_1') \cdots \psi_1^\dagger(z_2')} ~ \\
  &&  \hspace{-3.1cm} \times\,  \lb \prod_{i<j} (z_i-z_j)^2\rb \,
     \lb \prod_{i<j} (z'_i-z'_j)^2\rb \,
     \lb \prod_{i,j} (z_i-z'_j)^{2/N}\rb ~.
\ee
The computation of the correlation function of the parafermions
 is thus reduced to
a computation of a correlation function of currents in the WZWN model.
For simplicity, we will concentrate on the 4-point correlation
function for the remainder of this solution; the computation can
be handled in the same way for higher point correlation functions. 
Therefore, let us consider the
4-point function
\bb
  \average{
  J^+(z_1)  J^+(z_2)  J^-(z_1')  J^-(z_2')}~. 
\ee
First, we recognize that
\bb
  \average{
 J^+(z_1) \, J^+(z_2) \, J^-(z_1') \, J^-(z_2')}
    & = & \average{
  \lowerpairing{J^+(z_1) J}^-(z_1')  J^+(z_2)  J^-(z_2')}
  \\
    &   & + \, \average{
 \lowerpairing{J^+(z_1) \, J}^-(z_2')  J^+(z_2)  J^-(z_1')}
  \\
    & = & \average{
  \lowerpairing{J^+(z_1 J}^-(z_1')  J^+(z_2)  J^-(z_2') }
 + \, (z_1' \, \leftrightarrow \, z_2')~.
\ee
One then computes
\bb
 \average{\lowerpairing{J^+(z_1)   J}^-(z_1') J^+(z_2)    J^-(z_2')}
 &=& \average{\lb \frac{N}{(z_{1} \,-\, z_{1}')^{2}} +
 \frac{H(z)}{z_{1} \,-\, z_{1}')} \rb \, J^{+}(z_{2}) \, 
 J^{-}(z_{2}')} \\
 && \hspace{-3cm} =
 \frac{N}{(z_1 - z_{1}')^{2}} \,
 \average{J^+(z_2) \, J^-(z_{2}')}
 +  \frac{1}{z_{1} - z_{1}'} \,
 \average{ H(z_{1}') \, J^+(z_{2})
 \, J^{-}(z_{2}')}\\
 & & \hspace{-3cm} =
 \frac{N}{(z_{1} - z_{1}')^{2}} \,
 \average{J^+(z_2) \, J^-(z_2')} \\
 && \hspace{-3cm} +  \frac{1}{z_1-z_1'}
 \, \lb \average{
 \lowerpairing{H(z_1') \, J}^+(z_2) \, J^-(z_2')} +
 \average{\lowerpairing{H(z_1') \, J}^-(z_2') \, J^+(z_2)} \rb \\
 && \hspace{-3cm} =
 \frac{N}{(z_1 \,-\, z_1')^2} \,
 \average{J^+(z_2) \, J^-(z_2')}
 \\
 &&  \hspace{-3cm}
 + \frac{1}{z_{1} \,-\, z_{1}'} \, \lb \frac{2}{z_{1}' \,-\, z_{2}}
 \,-\, \frac{2}{z_{1}' \,-\, z_{2}'} \rb \,
 \average{J^{+}(z_{2}) \, J^{-}(z_{2}')}\\
 & & \hspace{-3cm} =
 N \, \la \frac{1}{(z_{1} \,-\, z_{1}')^{2}} \,+\, \frac{2/N}{
 z_{1} \,-\, z_{1}'} \, \left[ \frac{1}{z_{1}' \,-\, z_{2}} \,-\,
 \frac{1}{z_{1}' \,-\, z_{2}')} \rb \ra \,
 \average{J^+(z_2) \, J^-(z_{2}')} ~.
\ee
Finally, we have
$$
 \average{J^+(z_2) \, J^-(z_2')} \= 
 (z_2 \,-\, z_2')^{-2/N} \, \average{\psi_1(z_2) \, 
 \psi_1^{\dagger}(z_2')} ~.
$$
Putting everything together, we obtain the relation
\beqn
 && \average{
 \psi_1(z_1) \, \psi_1(z_2) \, \psi_1^{\dagger}(z_1')
 \, \psi_{1}^{\dagger}(z_{2}')} \= \hspace{2cm}  \no 
 & & (z_1 \,-\, z_2)^{-2/N} \, (z_1 \,-\, z_1')^{2/N} \,
 (z_1 \,-\, z_2')^{2/N} 
 \, (z_2 \,-\, z_1')^{2/N} \, (z_{1}' \,-\, z_{2}')^{-2/N} \no
 &&~~~\times\,\lb\frac{1}{(z_{1} \,-\, z_{1}')^{2}} \,+\, \frac{2/N}{z_{1} \,-\,
 z_1'} \, \lp \frac{1}{z_{1}' \,-\, z_{2}} \,-\, \frac{1}{z_{1}'
 \,-\, z_{2}'} \rp \rb
 \, \average{\psi_1(z_2) \, \psi_{1}^{\dagger}(z_{2}')}
 \nonumber \\ & &
 + \, (z_2 \,-\, z_2')^{2/N} \, (z_{1}' \,-\, z_{2}')^{-2/N} \no
 &&~~~\times\, \lb \frac{1}{(z_1 \,-\, z_1')^2} \,+\, \frac{2/N}{
 z_1 \,-\, z_2'} \lp \frac{1}{z_2' \,-\, z_2} \,-\,
 \frac{1}{z_2' \,-\, z_1'} \rp \rb
 \, \average{\psi_{1}(z_{2}) \, \psi_1^{\dagger}(z_1') }~~~~
\label{eq:OM41}
\eeqn
which is exactly the relation we found in the previous problem.

\separator

\item
\def\expect#1{{ \langle\,{#1}\,\rangle }}
\def\bracket#1#2{{ \langle {#1} | {#2} \rangle }}
\def\p{{\psi}}
\def\o#1{{\lim_{{#1'}\to{#1}}}}
\def\c#1#2{{\p_{#1}(#2)\,\p_{#1}^{\dag}(#2')}}
\def\P#1{{\Psi_k^{(#1)}}}
\def\T#1{{\expect{T({#1}_1)T({#1}_2)}}}
\def\f#1#2{{{1 \over (#1-#1')^{2-2\D_{#2}}}}}
\def\cor{{{\expect{\p_1(z_1)\p_1(z_2)\p_1^{\dag}(z_1')\p_1^{\dag}(z_2')}}}}
\def\N{{2 \over N}}
\def\e{{\epsilon}}

The central charge may be found using the 2-point
 correlation function of the energy-momentum tensor, since
$$
 \expect{T(z)\,T(w)} \= {{c/2} \over {(z-w)^4}}~.
$$
Therefore,
let us first find an expression for the energy-momentum tensor
in terms of the parafermions.
To motivate the definition which is about to follow, let
us think initially about
 the well-understood $\BZ_2$
case. For this model, we have seen that\footnote{The normalization in this
 problem is that of Zamolodchikov and Fateev.}
\beq
   T(z)~=~-\,{1 \over 2}\,:\p(z)\,\partial\p(z):~.
\label{eq:4.1}
\eeq
Notice that \calle{eq:4.1} can be written as
\beqn
 T(z)&=&- {1\over2}\,\lb\,:\,\p (z)\p (z)\,:~+~:
 \,{1 \over {(z'-z)}}\,\p (z)~(z'-z)\partial\p (z)\,:\,\rb
 \nonumber \\
 &=&-\,{1\over2} ~\o{z} \,:\,{1 \over {(z'-z)}}\,\p (z)\,\lb\,\p (z)\+
  (z'-z)\partial\p (z)\+ \CO\lp{(z'-z)}^2\rp\,\rb\,:
 \nonumber \\  
 &=&{1\over2}~ \o{z} \,{1 \over (z-z')} \,:\,\p(z)\p(z')\,: ~. 
\label{eq:4.2}
\eeqn
Furthermore, since
$$
 \p^{\dag}\=\p~,
$$
equation \calle{eq:4.2}
can be written in the more convenient (for our purposes) form
\beq
  T(z)\=\o{z}\,:\,{1 \over z-z'}\,\psi(z)\psi^\dagger(z')\,:~. 
\label{eq:4.3}
\eeq
This equation is very suggestive: the energy-momentum tensor for the 
free fermion can be extracted from the OPE of the fermion with its
 hermitian conjugate. We shall  generalize equation \calle{eq:4.3}
 to the $\BZ_N$ case. For any parafermion $\psi_k$, we have the OPE
$$
 \c{k}{ z} \= {1 \over (z-z')^{2\Delta_k}}\,\lb\,1\+
  \alpha\,(z-z')\,\Psi_{k}^{(1)}\+\beta\,(z-z')^2\,
  \Psi_k^{(2)}\+\dots \,\rb~.
$$ 
Obviously, the conformal dimensions of the fields
$\Psi_k^{(1)},\Psi_k^{(2)},\dots$ are
\bb
  \Delta (\P{1})\=1~&,&~~~\overline{\Delta}(\P{1})\=0~,\\
             \Delta (\P{2})\=2~&,&~~~\overline{\Delta}(\P{2})\=0~,\\
             \dots~&,&~~~\dots~,
\ee
and therefore the corresponding spins are
$$
   s(\P{1})\=1~,~s(\P{2})\=2,~\dots~.
$$
A spin 1 field creates a U(1) symmetry (compare with QED), for which
$$
        \BZ_N \subset U(1)~.
$$
Since our parafermion theory should not have this larger
symmetry group, we impose the condition
$$
       \P{1}(z)~\equiv~0~
$$ 
on our algebra.
On the other hand, there is only one spin 2 field in the theory,
so we must identify this with the energy-momentum tensor, i.e.,
$$
   \P{2}(z)~\equiv~T(z)~.
$$
Using the method we have described in Exercise \ref{item:CMCC2}
of Chapter \ref{ch:CMCC}, we find
$$
 \beta\={2\Delta_k \over c}~.
$$

Thus, we define the energy-momentum tensor for the parafermion
theories as follows:
\beqn
  T(z)&=&{1 \over \beta}\, \o{z} \, {1 \over (z-z')^2}\,
         \lb\,-1\+(z-z')^{2\Delta_k}\,\c{k}{z}\rb\,
  \nonumber \\
      &=&{c \over 2\Delta_k}\,:\,\o{z} 
 \,{1 \over (z-z')^{2-2\Delta_k}}\,\c{k}{z}\,:~.
\label{eq:4.5}
\eeqn
Given this equation for the energy-momentum tensor, we have the OPE
\bb
    \T{z}&=&\o{z_1}\o{z_2}\,{c^2 \over 4\Delta_k^2}\,
{1 \over (z_1-z'_1)^2}
{1 \over (z_2-z'_2)^2}
     \\
 && \times\, \lb\,-1\+
(z_1-z'_1)^{2\Delta_k}\,
(z_2-z'_2)^{2\Delta_k}\,
\expect{\,\c{k}{z_1}\c{k}{z_2}\,}\,\rb~.
\ee
This result is true for any parafermion $\psi_k$. However,
since we have already studied the  correlation functions 
$$
  \expect{\p_1(z_1)\dots\p_1(z_n)\p_1^{\dag}(z_1')\dots\p_1^{\dag}(z_n')}~,
$$
 we are going to use the fields $\p_1$ and $\p_1^{\dag}$ to find $c$. 
In this case, then,
\bb
 \T{z}
&=&\o{z_1}\o{z_2}\,
{c^2 \over 4\Delta_1^2}
\, {1 \over (z_1-z'_2)^2}\,{1 \over (z_2-z'_2)^2} \\
 && \times\,\lb\,-1\+(z_1-z'_1)^{2\Delta_1}\,(z_2-z'_2)^{2\Delta_1}
 \expect{\,\c{1}{z_1}\c{1}{z_2}\,}\,\rb~\\
&=&\o{z_1}\o{z_2}\,{c^2 \over 4\Delta_1^2}\,
 {1 \over (z_1-z'_2)^2}\,{1 \over (z_2-z'_2)^2} \\
 && \times\, \lb\,-1\+(z_1-z'_1)^{2\Delta_1}\,(z_2-z'_2)^{2\Delta_1}\,(-1)^\N \, G\,\rb~,
\ee
where in the second equality we used the non-local behavior of parafermions
$$
  \psi_1(z_2)\psi_1^\dagger(z_1')\=
  (-1)^{2\Delta_1}\, \psi_1^\dagger(z_1') \psi_1(z_2)\=
  (-1)^{-2/N}\, \psi_1^\dagger(z_1') \psi_1(z_2)~,
$$
and then set
$$
  G\eq
 \average{\psi_{1}(z_{1}) \, \psi_{1}(z_{2}) \, \psi_{1}^{\dagger}(z_{1}')
 \, \psi_{1}^{\dagger}(z_{2}')}~.
$$
Inserting
$$ 
  \average{
 \psi_{1}(z) \, \psi_{1}^{\dagger}(z')} \,=\, {1\over (z-z')
 ^{2\Delta_1}} \= {1\over(z-z')^{2-2/N} }
$$
in equation \calle{eq:OM41}, we get
\bb
 G &=&
 (z_1-z_2)^{-2/N} \, (z_1-z'_1)^{2/N} \,
 (z_1-z'_2)^{2/N} \, (z_2-z'_1)^{2/N} \,
 (z'_1-z'_2)^{-2/N} \, (z_2-z'_2)^{2/N}\\
 &&
 ~\times\,\la \lb {1\over(z_1-z'_1)^2} +
 {2/N\over z_1-z'_1} \, \lp {1\over z'_1-z_2}
 -{1\over z'_1-z'_2} \rp \rb\,
 {1\over (z_2-z'_2)^2} \right.   \\
 && ~\left.
 \qquad + \, \lb {1\over (z_1-z'_2)^2} + {2/N\over z_1-z_2'} 
  \lp {1\over z'_2-z_2} -
 {1\over z'_2-z'_1} \rp \rb \,  {1\over (z_2-z'_2)^2}
  \ra  ~.
\ee
The calculations become more transparent if we use the variables
\bb
  \xi\eq{ (z'_1-z'_2)(z_1-z_2)\over (z_1-z'_2)(z_2-z'_1)}~,
  ~~~\epsilon\eq z_1-z'_1~,
  ~~~{\rm and}~~~\eta\eq z_2-z'_2~.
\ee
The prefactor in the  expression of $G$ is then just 
$ \xi^{-2/N} \, \epsilon^{2/N} \, \eta^{2/N}$.
The quantity in the curly bracket can be writen in the form
$$
 \la \, \ldots \, \ra \= \frac{1}{\epsilon^2}
 \, \frac{1}{\eta^2} \, \lp a \,+\, b\, \xi \,+\, d \, \xi^2
 \rp ~,
$$
with
$$
 a \= b \= \frac{2}{N} \, (N \,-\, 1) \= 2 \, \Delta_1~,
 ~~~{\rm and}~~~ d \= 1 ~.
$$
Then
$$
 G \=
 \epsilon^{-2 \Delta_1} \, \eta^{-2 \Delta_1} \, \xi^{-2/N} \,
 \lp a \,+\, b \, \xi \, + \, d \, \xi^2 \rp~,
$$
and
\beq 
   \T{z}\=\lim_{\e\to 0}\,\lim_{\eta \to 0}\,
{c^2 \over 4\Delta_1^2}\,
{1 \over\e^2}\, {1 \over \eta^2} \,
[\,-1\+(-1)^\N\,\xi^{-\N}\,(\,a\+ b\xi\+ d\xi^2\,)\,]~.
\label{eq:4.12}
\eeq
If the limit in equation \calle{eq:4.12}
 exists when $\e$ and $\eta$ approach zero 
independently, it must also exist when $\e=\eta$.
 In this case (see the appendix of this problem solution),
$$
   \xi\= - \lb \,
      1\+ {\e^2 \over y^2}
      \+ {\e^4 \over y^4}\+\CO(\e^6)\, \rb ~,
$$
where $y=z'_1-z'_2$.
Notice that because of the factor $1/\e^4$ which appears in equation
\calle{eq:4.12},
in the limit $\e \to 0$, terms of order $\CO(\e^k)$ for $k>4$ do not contribute.
Putting everything together, we find
\bb
   \T{z}&=&\lim_{\e\to 0}\,
          {c^2 \over 4\Delta_1^2}\,
          {1 \over\e^4} \,\Biggl\{ \,-1\+(-1)^\N\,
           (-1)^{-\N} \, 
      \lp \,
      1+ {\e^2 \over y^2}  + {\e^4 \over y^4}+\CO(\e^6)\, \rp ^{-\N}\, \\
     && \times\,\Biggl[                                    
    a- b\, \lp \, 1+{\e^2 \over y^2}+ {\e^4 \over y^4}+\CO(\e^6)\, \rp 
    + d\, \lp\, 1+ {\e^2 \over y^2}
      +{\e^4 \over y^4}+\CO(\e^6)\, \rp ^2\, \Biggr] \Biggr\} \\
    &=&\lim_{\e\to 0}\,
    {c^2 \over 4\Delta_1^2}\,
    {1 \over\e^4} \,\Biggl\{ \,-1\+
      \lp\,
      1\- {\N}\, {\e^2 \over y^2}
      {-\N {\e^4 \over y^4}+{{N+2} \over N^2}} {\e^4 \over y^4}\, \rp \\
    &&\times\,\Biggl[                                    
    a- b\,
      \lp\,
      1+ {\e^2 \over y^2}
      + {\e^4 \over y^4}\, \rp
    + d\,
      \lp\,
      1+ 2\,{\e^2 \over y^2}
      + 2\,{\e^4 \over y^4}+ {\e^4 \over y^4}\, \rp
     \, \Biggr] \+ \CO(\e^6)\, \Biggr\} \\
   &=&\lim_{\e\to 0}\,
   {c^2 \over 4\Delta_1^2}\,
   {1 \over\e^4} \,\Biggl\{ \,(-1+a-b+d) 
      + {\e^2 \over y^2} \lb \, -b+2d-\N\,(a-b+d) \, \rb \\
  &&+ {\e^4 \over y^4}\lb \,-b+3d-\N(-b+2d)+
  \lp {-1 \over N}+{2 \over N^2}\,(a-b+d)\, \rp
   \, \rb + \CO(\e^6)\, \Biggr\} \\
  &=&\lim_{\e\to 0}\,
      {c^2 \over 4\Delta_1^2}\,
     {1 \over\e^4} \, 
       \lb {\e^4 \over y^4}\,{(N^2+N-2) \over N^2}     
    ~+~ \CO(\e^6)\, \rb~, 
\ee
i.e.,
$$
 \T{z}\=
{c^2 \over 4\Delta_1^2}\,
        {(N^2+N-2) \over N^2}    
\,{1 \over y^4}~, 
$$
     which gives   
$$
    {c \over 2}\= {c^2 \over 4\Delta_1^2}\,{(N+2)(N-1) \over N^2}
    ~\Rightarrow~c\={2(N-1)\over N+2}~.
$$

\footnotesize
{\bf APPENDIX}

Considering the cross ratio $\xi$
 as a function of $z_{1}$ and $z_{2}$,
we write down the Taylor expansion of $\xi$ around the point
$z_1=z_1'$ and $z_2=z_2'$:
\begin{equation}
 \xi \, (z_{1}, \, z_{2}) \,=\, \sum_{n=0}^{+ \infty} \, \frac{1}{n!}
 \, \left( \epsilon \, \partial_{z_{1}} \,+\, \eta \, \partial_{
 z_{2}} \right)^{n} \, \xi (z'_1, \, z'_2) ~.
\label{eq:OM42}
\end{equation}
One can easily see that
\bb
 \partial_{z_{1}} \, \xi(z_{1}, \, z_{2}) & = &
 \frac{(z_{1}' \,-\, z_{2}') \, (z_{2} \,-\, z_{2}')}{
 (z_{2} \,-\, z_{1}') \, (z_{1} \,-\, z_{2}')^{2}}~~~{\rm and}\\
 \partial_{z_{2}} \, \xi(z_{1}, \, z_{2}) & = &
 \frac{(z_{1}' \,-\, z_{2}') \, (z_2' \,-\, z_1)}{
 (z_{1} \,-\, z_{2}') \, (z_{2} \,-\, z_{1}')}~,
\ee
and thus
$$
 \partial_{z_{1}} \, \xi(z'_1, \, z'_2) \= 
 \partial_{z_{2}} \, \xi(z'_1, \, z'_2) \= 0~.
$$
We also notice that
$$
 \xi (z_{1}', \, z_{2}') \= -1~.
$$
We can see further that
\bb
 && \partial_{z_{1}}^{k} \, \xi (z_{1}, \, z_{2}) \,=\,
 \frac{(z_{1}' \,-\, z_{2}') \, (z_{2} \,-\, z_{2}')}{
 (z_{2} \,-\, z_{1}') \, (z_{1} \,-\, z_{2}')^{k+1}} \,
 (-1)^{k+1} \, k!\\
 &\Rightarrow&\partial_{z_{2}} \,
 \partial_{z_{1}}^{k} \, \xi (z_{1}, \, z_{2}) \,=\,
 \frac{(z_{1}' \,-\, z_{2}') \, (z_{2}' \,-\, z_{1}')}{
 (z_{2} \,-\, z_{1}')^{2} \, (z_{1} \,-\, z_{2}')^{k+1}} \,
 (-1)^{k+1} \, k!\\
 &\Rightarrow& \partial_{z_{2}}^{m} \,
 \partial_{z_{1}}^{k} \, \xi (z_{1}, \, z_{2}) \,=\,
 \frac{-(z_{1}' \,-\, z_{2}')^{2}}{
 (z_{1} \,-\, z_{2}')^{k+1} \, (z_{2} \,-\, z_{1}')^{m+1}} \,
 (-1)^{k+1} \, k! (-1)^{m+1} \, m!\\
 &\Rightarrow& \partial_{z_{2}}^{m} \,
 \partial_{z_{1}}^{k} \, \xi (z'_1, \, z'_2) \,=\,
 \frac{k! \, m! \, (-1)^{k}}{(z_{1}' \,-\, z_{2}')^{k+m}}
 ~~,~~~ mk\neq0 ~.
\ee
For $ m k= 0$,  we have
\bb
 \partial_{z_{2}}^{m} \, \partial_{z_{1}}^{k} \,
 \xi (z_{1}', \, z_{2}') \= 0 ~, ~~~mk\=0~.
\ee
Thus, \calle{eq:OM42} is written
\begin{eqnarray*}
 \xi (z_{1}, \, z_{2}) & = & \sum_{n=0}^{+ \infty} \, \frac{1}{n!}
 \, \left( \epsilon \, \partial_{z_{1}} \,+\, \eta \, \partial_{z_{2}}
 \right)^{n} \, \xi (z_{1}', \, z_{2}')\\
 & = &
 -1 \, + \, \sum_{n=2}^{+ \infty} \,
 \left( \epsilon \, \partial_{z_{1}} \,+\, \eta \, \partial_{z_{2}}
 \right)^{n} \, \xi (z_{1}', \, z_{2}')\\
 & = &
 -1 \,+\, \sum_{n=2}^{+ \infty} \, \frac{1}{n!} \, \sum_{k=0}^{n}
 \, \pmatrix{n\cr k\cr} \, \epsilon^{k} \, \eta^{n-k} \, \partial_{z_{1}}^{k} \,
 \partial_{z_{2}}^{n-k} \, \xi(z_{1}', \, z_{2}')\\
 & = &
 -1 \,+\, \sum_{n=2}^{+ \infty} \, \frac{1}{n!} \, \sum_{k=1}^{n-1}
 \, \pmatrix{n\cr k\cr} \, \epsilon^{k} \, \eta^{n-k} \, \partial_{z_{1}}^{k} \,
 \partial_{z_{2}}^{n-k} \, \xi(z_{1}', \, z_{2}')\\
  & = &
 -1 \,+\, \sum_{n=2}^{+ \infty} \, \frac{1}{n!} \, \sum_{k=1}^{n-1}
 \, \frac{n!}{k! \, (n \,-\, k)!} \,  \epsilon^{k} \, \eta^{n-k} \,
 \frac{k! \, (n \,-\, k)! \, (-1)^{k}}{(z_{1}' \,-\, z_{2}')^{n}}\\
 & = &
 -1 \,+\, \sum_{n=2}^{+ \infty} \, \frac{1}{(z_{1}' \,-\, z_{2}')^{n}}
 \, \left[ \sum_{k=1}^{n-1} \, (-1)^{k} \, \epsilon^{k} \,
 \eta^{n-k} \rb \\
 &=&  -1 \, + \, \sum_{n=2}^{+ \infty} \, 
 \frac{1}{y^{n}} \, \left[ \sum_{k=1}^{n-1} \, (-1)^{k} \,
 \epsilon^{k} \, \eta^{n-k} \rb~.
\ee
If $\epsilon=\eta$, then
\begin{eqnarray*}
 \sum_{k=1}^{n-1} \, (-1)^{k} \, \epsilon^{k}\, \eta^{n-k} = 
 \epsilon^{n} \, \sum_{k=1}^{n-1} \, (-1)^{k}
  = 
 \cases{  - \, \epsilon^{n}~,& if $n$ is  even~, \cr
                         0~, & if $n$ is  odd~,  \cr}
\end{eqnarray*} 
and so 
$$
 \xi (z_{1}, z_{2}) \,=\, -1 \,-\, \sum_{k=1}^{+ \infty} \, 
 \left( \frac{\epsilon}{y} \right)^{2k} ~.
$$

\normalsize

\end{enumerate}

\newchapter{CONSTRUCTING NEW MODELS IN CFT}

\footnotesize
\noindent {\bfseries References}:
The Coulomb gas formulation (CGF) is reviewed in \cite{Dotsenko}.
For discussions of the coset construction,
see \cite{GodOli,Fuchs2}. These topics are
also treated in the 
CFT textbooks \cite{DiFMS,Ketov}.
\normalsize

\section{BRIEF SUMMARY}

\subsection{Coulomb Gas Construction or Formulation}

The Coulomb Gas Construction (CGC) is a standard method of
generating CFT minimal models which have particular symmetries,
starting from a set of bosonic fields. Sometimes the
models generated this way were previously known by other methods;
in this case perhaps it makes more sense to refer to
the CGF of these models, i.e., to view
this as a bosonized description of these models.
We will give an
overview of the method using the MMs of Chapter \ref{ch:GP}.

Consider a free boson in the presence of a background charge, normalized 
such that
$$
  T\=-{1\over2}\, (\partial\phi)^2+i\alpha_0\sqrt{2}\partial^2\phi~.
$$
Then the central charge is
$$
   c\=1-24 \, \alpha^2_0~,
$$
and the conformal weight of the vertex operator
$$
  V_\alpha(z)\= e^{i\sqrt{2}\alpha\phi(z)}
$$
is given by
$$
   \Delta_\alpha\=\alpha(\alpha-2\alpha_0)~.
$$
For the theory of
a boson with a background charge,  there is a
conserved U$(1)$ current
$$
   J(z) \= {1\over\sqrt{2}}\, i\partial\phi(z) ~.
$$
The corresponding charge $Q$ measures the `strength' of the vertex,
and is given by 
$$
     Q\=  {1\over\sqrt{2}}\, \int\,{dz\over 2\pi i}\, i\partial\phi~.
$$
The vertex \calle{vert1} has $\alpha$ units of  charge:
\bb
    \lb Q,V_\alpha\rb\=\alpha\,V_\alpha~.
\ee
Due to the background charge, a correlation function
$$
   \average{V_{\alpha_1}(z_1)V_{\alpha_2}(z_2)...V_{\alpha_n}(z_n)}
$$
vanishes by the $U(1)$ symmetry unless the net charge of the 
averaged operator  is $2\alpha_0$, i.e., unless
$$
  \sum_{k=1}^n\alpha_k\=2\alpha_0~.
$$ 
Correlation functions that do not satisfy this neutrality condition 
can be made non-vanishing 
using {\bfseries screening charges}\index{Charge!Screening --}.
A screening charge is an operator
$$
   S_\pm\=\int\, {dz\over 2\pi i}\, J_\pm(z)~,
$$ 
which has the effect of reducing the total charge
without affecting the conformal properties of the
function. 
The last condition requires
$$
  J_\pm(z)\= e^{i\sqrt{2}\alpha_\pm\phi(z)}~,~~~~~
  \alpha_\pm\= \alpha_0\pm\sqrt{\alpha_0^2+1}~.
$$
Now,  a correlation function that
does not satisfy the neutrality condition, such as
$$
   \average{ V_{2\alpha_0-\alpha}(z_1)
   V_{\alpha}(z_2)V_{\alpha}(z_3)V_{\alpha}(z_4)}~,
$$
may be transformed into one that does satisfy the neutrality condition
by inserting a sufficient number of
screening operators.  Suppose, therefore, 
that we insert some screening
operators, and construct the correlation function 
$$
   \average{ V_{2\alpha_0-\alpha}(z_1)
   V_{\alpha}(z_2)V_{\alpha}(z_3)V_{\alpha}(z_4)S_+^{n-1}S_-^{m-1}}
   ~.
$$
For a given $n$ and $m$, this will satisfy the neutrality
condition for certain values of the charge $\alpha$; 
one finds that $\alpha$ must take one of the discrete values
\beq
\label{amn}
       \alpha_{mn}\={1\over 2}\,\lb (1-m)\alpha_- + (1-n)\alpha_+\rb~,~~~
    n,m\in\mathbb{N}^*~.
\eeq
The corresponding primary fields are then
\beq
\label{fmn}
  V_{\alpha_{mn}}(z)\=e^{i\sqrt{2}\alpha_{mn}\phi(z)}\eq \phi_{mn}(z)~,
\eeq
with their conformal weights equal to
\beq
\label{hmn}
    \Delta_{mn}\={1\over 4}\,\lb (\alpha_-\, m+\alpha_+\,n)^2
    -(\alpha_++\alpha_-)^2\rb~.
\eeq
The above construction does not impose any constraint on the allowed values of
the central charge $c$, or equivalently, on the value of $\alpha_0$.
However, if we require that the number of primary fields be finite,
say $1\le n\le s-1$ and $1\le m\le r-1,$ where
$r,s\in \mathbb{N}\smallsetminus\{0,1\}$, one must impose
\beq
\label{a+a-}
       \alpha_+\=\sqrt{{r\over s}}~,~~~~~
       \alpha_-\=-\sqrt{{s\over r}}~.
\eeq
Obviously, $r$ and $s$ must be relatively 
prime, as we can always cancel their common
factors.
When the condition \calle{a+a-} is met,
we recover the MMs of CFT we described in the previous chapter.

\subsection{Coset Construction}

As we have seen in the WZWN model, an affine algebra
$\hat\mathfrak{g}$ generated by currents $J_a$
that satisfy OPEs
$$
   J^a(z)J^b(w)\= { k/2 \, \delta^{ab} \over (z-w)^2 }
                + { f^{ab}_{~~~c}\, J^c(w) \over z-w } 
                + {\rm reg} 
$$
gives rise to a Virasoro algebra with  energy-momentum tensor 
\beq
\label{eq:Suga}
   T(z) = {1\over k+\tilde h_{\mathfrak{g}}} \,  
          \ddagger J_a(z) J_a(z) \ddagger~,
\eeq
and central charge
$$
   c \= {k \, \dim\mathfrak{g} \over k + \tilde h_{\mathfrak{g}} }~.
$$ 
When $T$ is written as in equation \calle{eq:Suga}, we say that
it has been expressed in the  {\bf Sugawara construction}.

Given a subalgebra $\mathfrak{h}\subset\mathfrak{g}$,
one can construct a new conformal theory which corresponds to
the coset $\mathfrak{g}/\mathfrak{h}$ as follows.
Let $T_{\mathfrak{g}}$ and
$T_{\mathfrak{h}}$ be the energy-momentum tensors of
the CFTs from the the Sugawara construction
for $\mathfrak{g}$ and $\mathfrak{h}$, respectively. Then their difference
$$
   T_{\mathfrak{g}/\mathfrak{h}}\=
   T_{\mathfrak{g}}
   -T_{\mathfrak{h}}
$$
commutes with the $\mathfrak{h}$ current algebra and the
Sugawara construction on $\mathfrak{g}$, i.e.
\bb
  T_{\mathfrak{g}/\mathfrak{h}}(z) J_A(w) &=& \rm{reg}~,
  ~~~~~A=1,2,\dots,\dim\mathfrak{h}~,\\
  T_{\mathfrak{g}/\mathfrak{h}}(z)T_{\mathfrak{g}}(w) &=& \rm{reg}~.
\ee
The commutativity properties of 
the energy-momentum tensor $T_{\mathfrak{g}/\mathfrak{h}}$
imply that the Sugawara construction 
$T_{\mathfrak{g}}= T_{\mathfrak{g}/\mathfrak{h}}+T_{\mathfrak{h}}$
is a tensor product CFT, formed by tensoring the coset CFT with the
CFT on $\mathfrak{h}$. The conformal anomaly of the coset CFT
is
$$
 c_{\mathfrak{g}/\mathfrak{h}}\=
 c_{\mathfrak{g}}-c_{\mathfrak{h}}~.
$$

\newpage
\section{EXERCISES}

\begin{enumerate}

\item

Consider the bosonic theory  of
Exercise \ref{item:CNM1} of Chapter \ref{ch:GP}, with
$\beta=ie_0/4\pi$, defined on a
Riemann surface $M$ with no boundary and with genus $h$. 
Show that the correlation function
$$
 \average{\prod_i e^{i\alpha_i\Phi(\xi_i)}}
$$
can be non-vanishing only if 
$$
   \sum_i \alpha_i \= e_0\, (1-h)~.
$$

\item
Consider the free boson theory with background charge.  The energy
momentum tensor is
$$
 T(z)\,=\,-g\,\partial_{z}\phi\partial_{z}\phi+i
 e_{0}\partial_{z}^{2}\phi~.
$$
Expand the boson field in modes 
according to $i\partial_{z} \phi=\sum_{n}\phi_{n}\,z^{n-1}$. 

(a) Compute the commutation relations of the  generators
$L_{n}$ with the modes $\phi_{m}$.

(b) Find the $\phi_n^\dagger$.

(c) Define a $U(1)$ charge operator $Q$
such that 
$$
   \lb Q, V_\alpha(w) \rb \= \alpha\, V_\alpha(w)~.
$$
in terms of the modes.
Then show that for a generic operator $O$ with charge $q$,
$$
 \lb Q,O \rb \=q\, O~,
$$
and that the expectation value  $\average{O}$
can be non-zero only if $q=2e_0$.

\item
Construct the screening charges for the theory of
a free boson with a background charge.
Do this for generic normalization.

\item
\label{item:CNMX}
One way to construct irreducible representations of the Virasoro algebra
is as follows. A module $\lb \Phi_\Delta\rb$ is called degenerate if it contains
at least one state $\ket{\chi}$ with the properties
\bb
   L_n\ket{\chi} &=& 0~, ~~~~~n>0~, \\
   L_0\ket{\chi} &=& (\Delta+N)\ket{\chi}~,
\ee 
for some $N\in\BN^*$, which is called the 
{\bf degeneration level}\index{Level!Degeneration --}. Such states,
when found, must be set equal to zero 
(null~states\index{State!Null --}), as otherwise the representation would 
be reducible. This amounts to factorization of $\lb \Phi_\Delta\rb$
by $\lb \chi \rb$.

To construct null vectors for the module 
$\lb V_\alpha\rb=\lb V_{2\alpha_0-\alpha}\rb$,
we start with a highest weight state
$\ket{V_{2\alpha_0-\alpha-n\alpha_\pm}}$, and use screening operators to
create another highest weight state $\ket{\chi}$, which has charge
 $2\alpha_0-\alpha$
and weight $\Delta_\alpha+N$. Such a state,
being  in $\lb V_\alpha\rb$, must
must be  null, from which condition we
can derive the allowed spectrum of charges
and the degeneration level.

Use this method to  derive the spectrum of the highest weight operators
in the case of the CFT of a boson in the presence of a background charge.

\item
With the help of the screening charges, derive the fusion relations of the
MMs.

\item
Derive a contour integral representation for the correlation function
of Exercise \ref{ising1} from Chapter \ref{ch:CMCC} 
by using the CGF. 

\item
A {\bf character} for a Verma module built over the field $\Phi_\Delta$
with conformal weight $\Delta$ is defined by
$$
  \chi_\Delta(q) = q^{-c/24}\, {\rm tr}\lp q^{L_0} \rp~.
$$
Since for any state in the Verma module, the eigenvalue of $L_0$
has the form  $\Delta+N$, the character takes the form
$$
 \chi_\Delta(q) = q^{\Delta-c/24}\,\sum_{N=0}^{+\infty} \, p(N)\, q^N~~.
$$
The coefficient $p(N)$ in the expansion
counts the number of states at level $N$.

(a)  Show that if there are no null states, then
$$
 \chi_\Delta(q) \= {q^{\Delta-c/24}\over\prod\limits_{n=1}^{+\infty} (1-q^n)}~.
$$

(b) Show that if there is a null state at level $N$, then
$$
 \chi_\Delta(q) \= {q^{\Delta-c/24}\,(1-q^N)
 \over\prod\limits_{n=1}^{+\infty} (1-q^n)}~.
$$

(c) Explain how to generalize
this to cases with null states within null states.

\item
\label{item:CNM2}
For  the primary field $\Phi_{mn}$ of the minimal model MM$(r,s)$,
prove the {\bfseries Rocha-Caridi formula}
\beq
\label{eq:CNM21}
  \chi_{mn} \= {q^{-c/24}\over\prod\limits_{n=1}^{+\infty} (1-q^n)}\,
            \sum_{k=-\infty}^{+\infty}\, 
            \lp q^{a_{mn}(k)} - q^{b_{mn}(k)} \rp ~,
\eeq
where
\bb
     a_{mn}(k) &=& {(2rsk+sn-mr)^2 - (r-s)^2 \over 4rs}~,~~~{\rm and}\\
     b_{mn}(k) &=& {(2rsk+sn+mr)^2 - (r-s)^2 \over 4rs}~. 
\ee

\end{enumerate}
\newpage
\section{SOLUTIONS}

\begin{enumerate}

\item
Using the path integral representation, the correlation 
function given in the problem can be written as
\bb
   \average{\prod_i e^{i\alpha_i\Phi(\xi_i)}}
   &=&
   \Big\lmoustache d\mu\lb\Phi\rb \, 
   e^{-\alpha\iint d^2\xi\sqrt{g}g^{ab}\partial_a\Phi\partial_b\Phi}\,
   e^{-i{e_0\over2\pi}\iint d^2\xi\sqrt{g} R\Phi}\,
   \prod_i e^{i\alpha_i\Phi(\xi_i)}~.
\ee
This must be invariant under a shift in the bosonic field
$$
   \Phi ~\mapsto~ \Phi+ \delta~.
$$
However, introducing this transformation in the path integral, we find
\bb
  \average{\prod_i e^{i\alpha_i\Phi(\xi_i)}} ~\mapsto~
  \average{\prod_i e^{i\alpha_i\Phi(\xi_i)}}\,
  e^{i\lp\sum_i\alpha_i-{e_0\over2\pi}\iint d^2\xi\sqrt{g} R\rp\delta}
  ~.
\ee
Thus the correlation function is invariant only if
$$
   \sum_i\alpha_i-{e_0\over2\pi}\iint d^2\xi\sqrt{g} R\=0~.
$$
If this equation is not satisfied, then the correlation function 
vanishes.

On the other hand, the Gauss-Bonnet theorem states that
for a manifold without boundary
$$
   \iint d^2\xi\sqrt{g} R\= 4\pi\, (1-h)~.
$$
Thus, for the correlation function to be non-zero, it is
necessary to have
$$
   \sum_i\alpha_i\= 2e_0\, (1-h)~.
$$

\separator

\item
(a) We define the modes of $T(z)$ and $\phi(z)$ using the standard relations
\bb
 T(z) &\equiv& \sum_{n} \, L_{n} \, z^{-n-2}~,\\
 i \, \partial_{z} \phi (z) &\equiv& \sum_{n} \, \phi_{n} \,
 z^{-n-1}~.
\ee
From Cauchy's formula, we then have the explicit formul\ae:
\bb
 L_n &=& \ointleft \, \frac{dz}{2\pi  i} \, z^{n+1} \, T(z)~,~~~{\rm and}\\
 \phi_n &= & \ointleft \, \frac{dz}{2\pi i} \, z^{n} \, i \,
 \partial_z \phi(z)~.
\ee
Repeating the standard argument, we find that
\beq
 \lb L_n, \, L_m \rb \= \ointleft \, \frac{dw}{2\pi i}
 \, \ointleft \, \frac{dz}{2\pi  i} \, z^{n+1} \, w^m \, i \,
 \CR \left( \, T(z) \, \partial_{w} \phi(w) \rp~.
\label{eq:CNM8}
\eeq
In order to proceed, we need to find the OPE $\CR\lp T(z)\partial_{w}
\phi (w) \rp$. Notice that
$$
 T(z) \= T_{{\rm old}}(z) + i e_{0} \, \partial^{2} \phi(z)~.
$$
Then we find that
\bb
 \CR \lp \, T(z) \, \partial_w \phi(w) \rp &=& 
 \CR \lp\, T_{{\rm old}}(z) \, \partial_{w} \phi(w) \rp +
 i \, e_0 \, R \lp \partial_z^2 \phi(z) \, \partial_w
 \phi(w) \rp\\
 &=& \frac{\partial_w \phi(w)}{(z- w)^{2}} +
 \frac{\partial_w \, (\partial_w \phi(w))}{z- w} 
 -\frac{i e_0}{g} \, \frac{1}{(z - w)^3} +\mbox{reg}~.
\ee
Substituting in \calle{eq:CNM8},
\begin{equation}
 \lb L_n, \, \phi_m \rb \= -m \, \phi_{m+n} +
 \frac{e_0}{2g} \, n \, (n - 1) \, \delta_{n,-m}~.
\label{eq:CNM9}
\end{equation}
The above relation is true for any $n$ and $m$. 

(b) Let 
$$
   \phi(z) \= x_0 -i\phi_0\, \ln z+i\sum_{n\ne 0} {\phi_n\over n}\, z^{-n}~,
$$ 
for which the expansion of modes
$$
   i\partial\phi(z) \= \sum_n \phi_n\, z^{-n-1}
$$
follows.
As we have seen, for a field $\Phi(z)$ with weight $\Delta$, the adjoint
is the original field evaluated at the point $1/z$
and multiplied by $z^{-2\Delta}$ (see the solution of 
Exercise \ref{item:CMCC2} of Chapter \ref{ch:CMCC}): 
$$
   \Phi^\dagger(z) \= z^{-2\Delta} \, \Phi'\lp {1\over z}\rp ~.
$$
Therefore, for the bosonic field in the presence of background charge,
recalling \calle{eq:GP32}, we must have
\bb
  \phi^\dagger(z) &=& \phi\lp {1\over z}\rp +{ie_0\over g}\, \ln{1\over z}
                    +{ie_0\over 2g}\,\ln(-1)  \\
             &=& \lp x_0 +{ie_0\over 2g}\,\ln(-1)\rp 
                 +\lp -i\phi_0+{ie_0\over g}\rp \, \ln{1\over z}
                 +i\, \sum_{n\ne 0} {\phi_n\over n}\, z^n ~.
\ee
Then
\bb
   -i\partial\phi^\dagger(z) &=& \lp \phi_0-{e_0\over g}\rp \,{1\over z}
                 + \sum_{n\ne 0} \phi_{-n}\, z^{-n-1} ~.
\ee
From the last relation, we see that
\bb
    \phi^\dagger_0 &=& \phi_0-{e_0\over g}~, \qquad{\rm and}\\
    \phi_n^\dagger &=& \phi_{-n}~~,~~n\ne 0~.
\ee

(c) From the OPE
$$
   \CR\bigl(\partial\phi(z)e^{ia\phi(w)}\bigr)\=
   {-ia/2g\over z-w}\, e^{ia\phi(w)} + \mbox{reg}~,
$$
we see that the operator
$$
   \phi_0 \eq \ointleft {dz\over 2\pi i}\, i\partial\phi(z)~,
$$
satisfies the commutation relation
$$
  \lb \phi_0, V_a(w)\rb \=\ointleft {dz\over 2\pi i}\, 
  \CR\lp i\partial\phi(z)V_a(w)\rp  \= {a\over 2g}\, V_a(w)~.
$$
This implies that the charge operator we are looking for is
$$
   Q\eq 2g\, \phi_0~.
$$

As we have seen in Exercise \ref{item:GP4} of Chapter \ref{ch:GP},
$$
  L_0\= 2g\sum_{n=1}^{+\infty}\phi_{-n}\phi_n +g\,\phi_0^2-e_0\,\phi_0~.
$$
The vacuum state $\ket{\emptyset\,;\,e_0}$ must be annihilated by
this operator, so
\bb
  0 &=& L_0 \ket{\emptyset\, ;\, e_0} \= 
       (gx^2-e_0x)\, \ket{\emptyset\, ;\, e_0}~,
\ee
where we also used the fact that for $n>0$, $\phi_n\ket{\emptyset\, ;\, e_0}=0$,
and that $\phi_0\ket{\emptyset\, ;\, e_0}=x\ket{\emptyset\, ;\, e_0}$.
The last equation implies that $x=e_0/g$.

The commutation relation
$$
 \lb Q, \, O \rb \= q \, O
$$
now implies
\bb
  q \, \bra{\emptyset\, ;\, e_0} \, O \, \ket{\emptyset\, ;\, e_0} &=&
 2g \bra{\emptyset\, ;\, e_0} \, \phi_0 \, O \, \ket{\emptyset\, ;\, e_0} -
 2g \bra{\emptyset\, ;\, e_0} \, O \, \phi_0 \, \ket{\emptyset\, ;\, e_0} \\
 &=&  2g \bra{\emptyset\, ;\, e_0} \lp\phi_0^\dagger+{e_0\over g}\rp 
  \, O \, \ket{\emptyset\, ;\, e_0} -
 2g \bra{\emptyset\, ;\, e_0} \, O \, \phi_0 \, \ket{\emptyset\, ;\, e_0} \\
 &=& 2g \lp\frac{e_0}{g}+\frac{e_0}{g}\rp \,
 \bra{\emptyset\, ;\, e_0} \, O \, \ket{\emptyset\, ;\, e_0}
  - 2g \frac{e_0}{g} \,
 \bra{\emptyset\, ;\, e_0}  \, O \, \ket{\emptyset\, ;\, e_0}~,
\ee
or
\bb
 \lp q -2e_0 \rp \,
 \bra{\emptyset\, ;\, e_0} \, O \, \ket{\emptyset\, ;\, e_0} \= 0~.
\ee
From this, it follows immediately  that
$$
 q \neq  2e_0 ~\Rightarrow~
  \average{O}\= 0~,
$$
while $\average{O}$ can be non-zero if and only if $q=2e_0$.

\separator

\item
A screening charge
$$
   S\=\ointleft\, dz\, J(z)
$$
should be an operator that does not change the conformal properties
if inserted in the correlation functions.
This implies that
$$
     \Delta_J\=1~,
$$
since in this case $S$ commutes with the  Virasoro generators.
 To see this,
we first notice that for unit conformal weight 
\beq
\label{eq:CNM22}
  \CR \lp T(z)J(w)\rp \= \partial_w\lp {J(w)\over z-w}\rp +\mbox{reg} ~.
\eeq
Then 
\bb
   \lb L_n, J(z)\rb
   &=& \ointleft{dw\over 2\pi i} \, w^{n+1}\, \CR\lp T(w)J(z)\rp
   \=  \partial_z\lp T(z) \ointleft_z {dw\over 2\pi i}
       \, {w^{n+1}\, \over w-z}\rp \\
   &=& \partial_z \lp z^{n+1} J(z) \rp ~,~~~\forall n~.
\ee
For $n\ge-1$, the function 
$\partial_z(z^{n+1} J(z))=(n+1)z^nJ(z)+z^{n+1}\partial J(z)$
is analytic, and 
\beq
  \lb L_n, S\rb \= \ointleft{dz\over 2\pi i} \, \partial_z \lp z^{n+1} J(z) \rp
   \= 0~.
\label{eq:CNM51}
\eeq
This result says, in particular, that the screening operator $S$
commutes with the generators $L_{-1}$, $L_0$, and $L_1$ which 
generate the global
conformal group, as well as with all
the $L_k,~k>0,$ which annihilate the null states
(see the next problem). What do you think happens for $n< -1$? 

For a free boson, $J(z)$ may be of the form
$\partial\phi$ or $e^{i\alpha\phi}$. The first choice gives us the
U(1) charge
$$
   \ointleft\,{dz\over2\pi i}\, i\partial\phi \= \phi_0~.
$$
Therefore we consider the second 
possibility, $J(z)=e^{i\alpha\phi(z)}$.
Then we must have
$$
  1\= \Delta_J \= {\alpha(\alpha-2e_0)\over4g}
  \Rightarrow \alpha^2-2e_0\alpha-4g\=0~,
$$
the solution of which is
$$
   \alpha\eq e_\pm\= e_0\pm\sqrt{e_0^2+4g}~.
$$
Therefore, there are two screening charges:
$$
   S_\pm\=  \ointleft\,dz\,e^{i e_\pm\phi}~.
$$

\separator

\item

Given the vertex operator $V_{2\alpha_0-\alpha-n\alpha_+}(w)$, we can form
the operator
\beqn
   \chi^+  %
           = \ointleft_{C_1} {dz_1\over 2\pi i}
                \ointleft_{C_2} {dz_2\over 2\pi i}
                \cdots
                \ointleft_{C_n} {dz_n\over 2\pi i}
                 J_+(z_1)J_+(z_2)\cdots J_+(z_n)
                 V_{2\alpha_0-\alpha-n\alpha_+}(w)~,~~
\label{eq:CNM24}
\eeqn
carrying $2\alpha_0-\alpha$ units of charge.

Using equation \calle{eq:CNM51}, we can show that
the operator $\chi^+$ defined above is a  highest weight operator
\bb
   L_0 \chi^+ &=& (\Delta_\alpha+N)\, \chi^+~, \\
   L_k \chi^+ &=& 0~, ~~~~~k>0~,
\ee
where $\Delta_\alpha$ stands for the conformal weight of 
$V_\alpha$ and $N$ is a number that will be computed below.
To prove this, we examine the properties of the  integrand in equation
\calle{eq:CNM24}. Let us concentrate on the variable $z_1$ (similar
considerations are valid for any of the variables). Using
the identity
$$
   e^{i\sqrt{2}a\phi(z)} e^{ib\sqrt{2}\phi(w)}\=
   (z-w)^{2ab}\,
   :e^{ia\sqrt{2}\phi(z)} e^{ib\sqrt{2}\phi(w)}: ~,
$$
we see that the  factor
$$
   (z_1-z_2)^{2\alpha_+^2}\,
   (z_1-z_3)^{2\alpha_+^2}\,
   \dots (z_1-z_n)^{2\alpha_+^2}\,
    (z_1-w)^{2\alpha_+(2\alpha_0-\alpha-n\alpha_+)}
$$
will apppear. The branch cuts are avoided if
\beq
\label{eq:CNM31}
    2\alpha_+^2(n-1)+2\alpha_+(2\alpha_0-\alpha-n\alpha_+)\=-m-1~,
\eeq
where $m=1,2,\dots$. The value $m=0$ is excluded, as it gives
$V_{2\alpha_0-\alpha}$.
We thus have
\bb
  L_k\chi^+&=&
             \ointleft_{C_1} {dz_1\over 2\pi i}
              \ointleft_{C_2} {dz_2\over 2\pi i}
              \cdots
               \ointleft_{C_n} {dz_n\over 2\pi i}
                 L_k J_+(z_1)J_+(z_2)\cdots J_+(z_n)
                 V_{2\alpha_0-\alpha-n\alpha_+}(w) \\
             &=& \ointleft_{C_1} {dz_1\over 2\pi i}
              \ointleft_{C_2} {dz_2\over 2\pi i}
              \cdots
               \ointleft_{C_n} {dz_n\over 2\pi i}
                  J_+(z_1) L_k J_+(z_2)\cdots J_+(z_n)
                 V_{2\alpha_0-\alpha-n\alpha_+}(w) \\
            &=& \ointleft_{C_1} {dz_1\over 2\pi i}
              \ointleft_{C_2} {dz_2\over 2\pi i}
              \cdots
               \ointleft_{C_n} {dz_n\over 2\pi i}
                  J_+(z_1) L_k J_+(z_2)\cdots J_+(z_n)
                 V_{2\alpha_0-\alpha-n\alpha_+}(w) \\
             &=& \cdots\cdots \\
               &=& \ointleft_{C_1} {dz_1\over 2\pi i}
              \ointleft_{C_2} {dz_2\over 2\pi i}
              \cdots
               \ointleft_{C_n} {dz_n\over 2\pi i}
                  J_+(z_1)  J_+(z_2)\cdots J_+(z_n)
                 L_k V_{2\alpha_0-\alpha-n\alpha_+}(w) \\
             &=& \cases{ 0~, &if$~k>0~,$\cr
           \Delta_{2\alpha_0-\alpha-n\alpha_+}\chi^+~, &if$~k=0~,$\cr}
             ~.
\ee 
We have thus proved that $\chi^+$ is a highest weight operator and thus
a null operator in $\lb V_\alpha\rb$.

The spectrum of the theory is determined directly from what has been 
written above. In particular, 
equation \calle{eq:CNM31}, when we use the facts that
$$
  2\alpha_0\=\alpha_++\alpha_-~~~{\rm and}~~~
 \alpha_+\alpha_-\=-1~,
$$
 gives
$$
   \alpha\= {1-m\over 2}\, \alpha_-~, ~~~~~m=1,2,\dots~.
$$
We could repeat the same line of reasoning
for the case of $J_-$. Doing so leads to the result
$$
   \alpha\= {1-n\over 2}\, \alpha_+~,~~~~~n=1,2,\dots~.
$$
Combining the two results, we have thus found that the spectrum
of primary operators $V_\alpha$ is in one-to-one
correspondence with the charges
$$
    \alpha_{mn} \= {1-m\over 2}\, \alpha_- + {1-n\over 2}\, \alpha_+~,
    ~~~~~m,n=1,2,\dots~.
$$

To assist us later, we make some additional comments here. The charge
$2\alpha_0-\alpha_{mn}$ corresponds to $\alpha_{-m,-n}$, while
the charge $2\alpha_0-\alpha_{mn}-n\alpha_+$ corresponds
to $\alpha_{-m,n}$. The conformal weights have the properties
$\Delta_{mn}=\Delta_{-m,-n}$ and
$$
   \Delta_{-m,n}\=\Delta_{mn}+mn~.
$$
From this we conclude that the module $\lb V_{mn}\rb$
admits a null state at level $nm$.

Finally, notice that if $\alpha_+\=\sqrt{r/s}$, then
$$
   \Delta_{mn}\=\Delta_{r-m,s-n}~.
$$
This implies that the spectrum
in this case is restricted to a finite number of
primary fields with $m\le r-1$ and $n\le s-1$.
\separator

\item

To study the fusion rules
$$
   \lb \phi_{m_1,n_1}\rb \times
   \lb \phi_{m_2,n_2}\rb~~,
$$
we must study the 3-point correlation functions
$$
   \average{ \phi_{m_1,n_1} \phi_{m_2,n_2} \phi_{m_3,n_3} }~~,
$$
and find for what combination of the indices these
correlation functions do not vanish.
According to the CGF, this correlation function is computed by
\beq
\label{eq:CNM26}
 \average{V_{2\alpha_0-\alpha_1}V_{\alpha_2}V_{\alpha_3} S_+^{N_+} S_-^{N_-}}~,
\eeq
where $\alpha_i$ is the charge corresponding to the pair of integers
$m_i$ and $n_i$. Charge neutrality in the last equation requires
$$
  (2\alpha_0-\alpha_1)+\alpha_2+\alpha_3+N_+ \alpha_+ +N_- \alpha_-
  \= 2\alpha_0~,
$$
or in other words,
$$
   -\alpha_1+\alpha_2+\alpha_3+N_+ \alpha_+ +N_- \alpha_-\=0~.
$$
Substituting the charges with their equivalent expressions
from equation \calle{amn}, we find
$$
  \lp {1+m_1-m_2-m_3\over2}+N_+\rp\, \alpha_+ +
  \lp {1+n_1-n_2-n_3\over2}+N_-\rp\, \alpha_- \= 0~.
$$
Let us consider first the cancellation of the $\alpha_+$ charges.
Since $N_+\ge0$, the quantity $1+m_1-m_2-m_3$ must be an even
negative integer:
\beq
\label{eq:CNM25}
   1+m_1-m_2-m_3\=\mbox{even}~, ~~~~~1+m_1-m_2-m_3\le0~.
\eeq

Instead of \calle{eq:CNM26},
one could  have used either of the following two correlators:
\bb
 \average{V_{\alpha_1}V_{2\alpha_0-\alpha_2}V_{\alpha_3}S_+^{M_+}S_-^{M_-}}~,\\
 \average{V_{\alpha_1}V_{\alpha_2}V_{2\alpha_0-\alpha_3}S_+^{L_+}S_-^{L_-}}~.
\ee
In these cases, one finds analogous constraints, respectively:
\beqn
\label{eq:CNM27}
  1-m_1+m_2-m_3\=\mbox{even}~, ~~~~~1-m_1+m_2-m_3\le0~,~~~ {\rm or}\\
\label{eq:CNM28}
  1-m_1-m_2+m_3\=\mbox{even}~, ~~~~~1-m_1-m_2+m_3\le0~.
\eeqn
Adding the first relations of \calle{eq:CNM25}, \calle{eq:CNM27},
and \calle{eq:CNM28} gives:
$$
   m_1+m_2+m_3\= \mbox{odd}~.
$$
The second relation of \calle{eq:CNM28} implies that
$$
   m_3 \le  m_1+m_2 -1~,
$$
while the second relations of \calle{eq:CNM25} and \calle{eq:CNM27} 
imply that
$$
  \left.
  \begin{array}{c} m_3\ge m_2-m_1+1 \\ m_3\ge m_1-m_2 +1\end{array}
  \ra \Rightarrow  m_3\ge |m_1-m_2| +1 ~.
$$

Identical results are obtained for the second index using the
cancellation of the $\alpha_-$ charges.
We can thus write the fusion rules as follows:
$$
   \lb \phi_{m_1,n_1}\rb \times
   \lb \phi_{m_2,n_2}\rb 
   \= \sum^{m_1+m_2-1}_{m_3=|m_1-m_2| +1\atop  m_1+m_2+m_3=
   \mbox{\tiny odd}}
      ~~
    \sum^{n_1+n_2-1}_{n_3=|n_1-n_2| +1\atop  n_1+n_2+n_3= 
   \mbox{\tiny odd}}
   \lb \phi_{m_3,n_3}\rb  ~.
$$

For the MMs,
$$
  \Phi_{mn}\=\Phi_{r-m,s-n}~~~,
$$
and therefore
$$
   \lb \phi_{m_1,n_1}\rb \times \lb \phi_{m_2,n_2}\rb \=
   \lb \phi_{r-m_1,s-n_1}\rb \times \lb \phi_{r-m_2,s-n_2}\rb ~.
$$
Applying the formula we found above, we see
\bb
   \sum^{m_1+m_2-1}_{m_3=|m_1-m_2| +1\atop  m_1+m_2+m_3=
   \mbox{\tiny odd}}
      ~~
    \sum^{n_1+n_2-1}_{n_3=|n_1-n_2| +1\atop  n_1+n_2+n_3= 
   \mbox{\tiny odd}}
   \lb \phi_{m_3,n_3}\rb  \=
   \sum^{2r-m_1-m_2-1}_{m_3=|m_1-m_2| +1\atop  m_1+m_2+m_3=
   \mbox{\tiny odd}}
      ~~
    \sum^{2s-n_1-n_2-1}_{n_3=|n_1-n_2| +1\atop  n_1+n_2+n_3= 
   \mbox{\tiny odd}}
   \lb \phi_{m_3,n_3}\rb ~,
\ee
or
\bb
   \sum_{m_3>\min\{m_1+m_2-1,2r-m_1-m_2-1\}\atop  m_1+m_2+m_3=
   \mbox{\tiny odd}}
      ~~
    \sum_{n_3>\min\{n_1+n_2-1,2s-n_1-n_2-1\}\atop  n_1+n_2+n_3=
   \mbox{\tiny odd}}
   \lb \phi_{m_3,n_3}\rb \= 0~.
\ee
So, finally, the fusion rules for the MMs are
$$
   \lb \phi_{m_1,n_1}\rb \times
   \lb \phi_{m_2,n_2}\rb
   \=
 \sum^{\min\{m_1+m_2-1,2r-m_1-m_2-1\}}_{m_3=|m_1-m_2|+1\atop  m_1+m_2+m_3=
   \mbox{\tiny odd}}
      ~~
  \sum^{\min\{n_1+n_2-1,2s-n_1-n_2-1\}}_{n_3=|n_1-n_2|+1\atop  n_1+n_2+n_3=
   \mbox{\tiny odd}}
   \lb \phi_{m_3,n_3}\rb  ~.
$$

\separator

\item
For the Ising model $c=1/2$. This means that, using the CGF,
one has $1-24\alpha_0^2=1/2$, which sets the value of 
the background charge to be
$$ 
   \alpha_0 \= -{1\over 4\sqrt{3}}~.
$$
Then
$$
   \alpha_+\={\sqrt{3}\over2}~~~{\rm and}~~~
   \alpha_-\= -{2\over\sqrt{3}}~.
$$
The field $\sigma(z)$ has conformal dimension $1/16$, and corresponds to
$\phi_{12}$ with charge $\alpha_{12}=-\alpha_+/2=-\sqrt{3}/4$.

Using the prescription of the CGF, the correlation function
\bb
 G^{(4)}(z_i) \= \average{\sigma(z_1) \, \sigma(z_2) \,
 \sigma(z_3) \, \sigma(z_4)}
\ee
maps to the bosonic correlator
\bb
 G^{(4)}(z_{i}) &=& \average{\phi_{12}(z_{1}) \, \phi_{12}(z_{2}) \,
 \phi_{12}(z_{3}) \, \phi_{12}(z_{4})} \\
  &=& \ointleft_{C} \, dv \,
  \average{ 
    V_{\alpha_{12}}(z_{1}) \, V_{\alpha_{12}}(z_{2}) \,
    V_{\alpha_{12}}(z_{3}) \, V_{2\alpha_0-\alpha_{12}}(z_{4}) \,
    J_+(v)} ~,
\ee
which, upon using the identity
$$
 \lan \prod_{i=1}^{n} \, V_{\alpha_i}(z_i) \ran \=
 \prod_{i=1}^{n} \, \prod_{j=i+1}^{n} \, (z_i - z_j)^{2\alpha_i \alpha_j} ~,
$$
takes the form 
\bb
 G^{(4)}(z_i) &=& \phantom{\times} (z_1-z_2)^{2\alpha_{12}^2}
       \, (z_1 - z_3)^{2\alpha_{12}^2}
       \, (z_1 - z_4)^{2\alpha_{12}(2\alpha_0-\alpha_{12})}\\
                & &
       \times (z_2 - z_3)^{2\alpha_{12}^2}
       \, (z_2 - z_4)^{2\alpha_{12}(2\alpha_0-\alpha_{12})}
       \, (z_3 - z_4)^{2\alpha_{12}(2\alpha_0-\alpha_{12})}\\
                &&
       \times\ointleft_C dv \, (z_1-v)^{2\alpha_{12} \alpha_+ }
                \, (z_2-v)^{2\alpha_{12} \alpha_+ }
                \, (z_3-v)^{2\alpha_{12} \alpha_+ }
       \, (z_4-v)^{2\alpha_+(2\alpha_0-\alpha_{12} )}
       ~.
\ee
Recall from equation \calle{eq:G4b} that
the (holomorphic) correlation function can be
expressed as
$$
  G^{(4)} \= (xz_{23}z_{41})^{-1/8}\, f(x)
  ~\Rightarrow~
  f(x) \= (xz_{23}z_{41})^{1/8}\, G^{(4)}~.
$$ 
We can take advantage of conformal invariance to
fix the values  of the points $z_1$,
$z_2$, $z_3$, and $z_4$. We choose
\bb
 z_1 \=0~,~~~ z_2\=x~,~~~
 z_3 \=1~,~~~{\rm and}~~~ z_4 \rightarrow \infty~.
\ee
Then, using all of the above, we have
$$
  f(x) \=  x^{1/2} \, (1-x)^{1/2}\,
           \ointleft_C \, dv \, v^\alpha \, (x-v)^ \alpha  \, (1-v)^\alpha~,
$$
where
$$
  \alpha\= 2\alpha_{12} \alpha_+\=-3/4~.
$$
Now we have to choose the contour $C$ of integration; we have
two independent choices, which are clearly depicted in the
figure below:
\begin{center}
\psfrag{0}{$0$}
\psfrag{1}{$1$}
\psfrag{z}{$x$}
\psfrag{inf}{$\infty$}
\includegraphics[width=10cm]{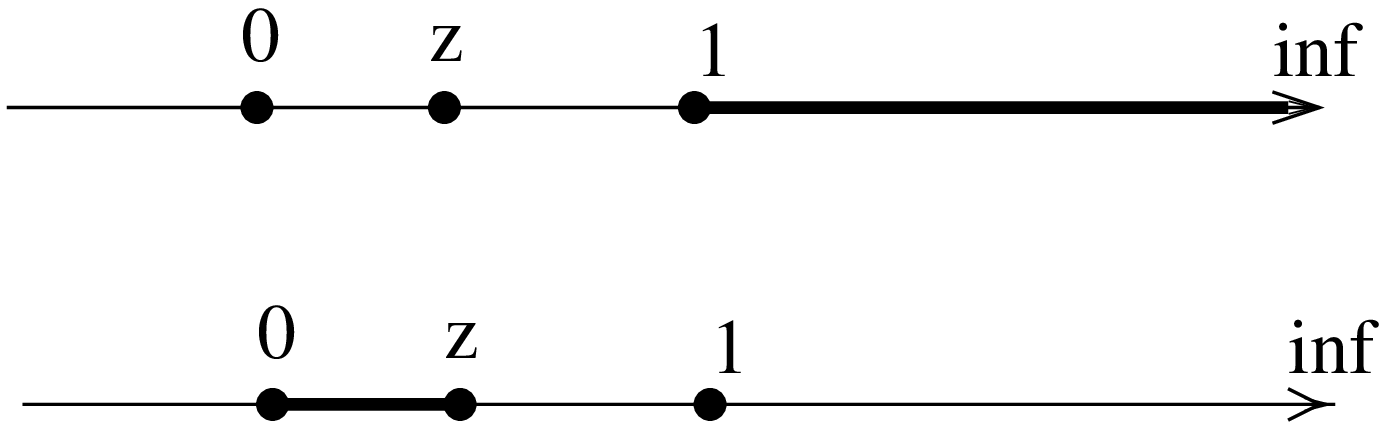}
\end{center}

Other choices are not independent. This means that we have two
independent solutions:
\beqn
   f_1(x) &=&  x^{1/2} \, (1-x)^{1/2}
            \, \int_1^\infty \, dv \,
    v^{\alpha} \, (x - v)^{\alpha} \, (1 - v)^\alpha~,~~~{\rm and}
\label{eq:CNM13} \\
   f_2(x) &=& x^{1/2} \, (1-x)^{1/2}
           \, \int_0^x \, dv \,
           v^\alpha \, (x - v)^{
           \alpha} \, (1 - v)^\alpha~.
\label{eq:CNM14} 
\eeqn
The function $f(x)$ will be a linear combination of $f_1(x)$ and 
$f_2(x)$.

Now recall that the hypergeometric function has the following
integral representation:
\begin{equation}
 F \lp a, \, b; \, c; \, x \rp \eq
 {\Gamma(c)\over\Gamma(b) \, \Gamma(c-b)} \, \int_0^1
 du \, u^{b-1} \, (1 \,-\, u)^{c-b-1} \, (1 \,-\, xu)^{-a}
  ~,
\label{eq:CNM15}
\end{equation}
for $|x|<1$, ${\rm Re} \, c \, > \, {\rm Re} \, b \, > \, 0$. For other
values, we can define $F$ by analytic continuation. 

By making the
substitutions
$v=1/u$ and $v=xu$
in the integrals of equations \calle{eq:CNM13}
and \calle{eq:CNM14}, respectively,
we find
\bb
    f_1(x) &=& x^{1/2}\,(1-x)^{1/2}\,\int_0^1 du \, u^{-3 \alpha -2} \, (1-xu)^
    \alpha \, (1 -u)^\alpha~,~~~{\rm and} \\
    f_2(x) &=& \phantom{x^{1/2}}\,(1-x)^{1/2}\,
    \int_0^1 du \, u^\alpha \, (1-xu)^\alpha \, (1 \,-\, u)^\alpha~.
\ee
Using the representation of the hypergeometic
function \calle{eq:CNM15}, we see
\bb
    f_1(x) &=& x^{1/2}\,(1-x)^{1/2}\,
    F \lp -\alpha, \, -3 \alpha \,-\,1; \, -2 \alpha; \, x \rp
   \= x^{1/2}\,(1-x)^{1/2}\, F\lp{3\over4},{5\over4};{3\over2};x\rp~,
   \\
    f_2(x) &=& \phantom{x^{1/2}}\,(1-x)^{1/2}\,
    F \lp -\alpha, \, \alpha+1; \, 2 \alpha+2; \, z \rp
   \= x^{1/2}\,(1-x)^{1/2}\, F\lp{3\over4},{1\over4};{1\over2};x\rp~,
\ee
There is an identity for the hypergeometric function that is useful
in the present context:
$$
  F\lp {1\over2}+a,a;2a;x\rp\= 2^{2a-1}\,(1-x)^{-1/2}\,
  \lp 1+\sqrt{1-x}\rp^{1-2a}~.
$$
Substituting the values $a=3/4$ and $\alpha=1/4$ in this identity gives,
respectively,
\bb
    F\lp{3\over4},{5\over4};{3\over2};x\rp &=&
   {1\over\sqrt{2}}\,(1-x)^{-1/2}\,{1\over\sqrt{1+\sqrt{1-x}}}~,~~~\mbox{and}\\
     F\lp{3\over4},{1\over4};{1\over2};x\rp  &=&
   \sqrt{2}\,(1-x)^{-1/2}\,\sqrt{1+\sqrt{1-x}}~.
\ee
Thus
\bb
   f_1(x) &=&  {1\over\sqrt{2}}\, {\sqrt{x}\over\sqrt{1+\sqrt{1-x}}}~,\\
   f_2(x) &=&  \sqrt{2}\, \sqrt{1+\sqrt{1-x}}~.
\ee
The function $f_1(x)$ can easily be rewritten in an equivalent form:
\bb
  f_1(x) &=&  {1\over\sqrt{2}}\, {\sqrt{x}\over\sqrt{1+\sqrt{1-x}}}
              {\sqrt{1-\sqrt{1-x}}\over\sqrt{1-\sqrt{1-x}}}
         \=   {1\over\sqrt{2}}\, {\sqrt{x}
               \sqrt{1-\sqrt{1-x}}\over \sqrt{1-(1-x)} } \\
         &=& {1\over\sqrt{2}}\,\sqrt{1-\sqrt{1-x}}~.
\ee
We have thus found  the same functions as in  Exercise \ref{ising1}
 of Chapter \ref{ch:CMCC}.

\separator

\item
(a) A state at level $N$ has the form
$$
   L_{-N}^{n_N}\dots L_{-2}^{n_2}L_{-1}^{n_1}\ket{\Delta}~,
$$
with the level $N$ given by
$$
   N\= N\, n_N  + (N-1)\,  n_{N-1} +\dots + 2\, n_2 + n_1 ~.
$$
Therefore, there are as  many states at level $N$ as
there are partitions $p(N)$
of $N$ into $k$ parts $(n_1,n_2,\dots,n_k)$ with $k=1,2,\dots,N$
according to
$$
   p(N) \= \sum_{n_1,n_2,\dots,n_N\atop n_1+2n_2+\dots+Nn_N=N} 1~.
$$
Consequently,
\bb
   \sum_{N=0}^{+\infty} p(N) q^N 
   &=& \sum_{N=0}^{+\infty} 
       \sum_{n_1,n_2,\dots,n_N\atop n_1+2n_2+\dots+Nn_N=N}
        q^{n_1+2n_2+\dots+Nn_N} \\
  &=&  \sum_{n_1,n_2,\dots, n_N, \dots }
        q^{n_1+2n_2+\dots+N n_N + \dots} \\
  &=&  \sum_{n_1=0}^{+\infty} q^{n_1} 
       \sum_{n_2=0}^{+\infty} q^{2n_2} 
        \cdots 
       \sum_{n_N=0}^{+\infty} q^{Nn_N} 
        \cdots \\ 
  &=&  {1\over 1-q} \, {1\over 1-q^2} \, {1\over 1-q^3}\cdots \\
  &=&  {1\over \prod_{k=1}^{+\infty} (1-q^k)}~. 
\ee
Hence, without null states,
$$
   \chi_\Delta(q)\= {q^{\Delta-c/24}\over \prod_{k=1}^{+\infty} (1-q^k)}~.
$$

(b) We saw above that, without a null state, one has
$$
   \chi_{\mbox{\tiny naive}}(q)\= 
  {q^{\Delta-c/24}\over \prod_{k=1}^{+\infty} (1-q^k)}~.
$$
However, once there are null states, this expression overcounts the
states. To compensate, we must subtract the null states .

If there is a null state $\ket{\Delta+N}$ at level $N$, then  all
states of the form 
$$
   L_{-k}^{n_k}\dots L_{-2}^{n_2}L_{-1}^{n_1}\ket{\Delta+N}
$$
are also null. They form a Verma module over  $\ket{\Delta+N}$.
Therefore the corresponding character is given by
$$
   \chi_{\mbox{\tiny null}}(q)\=
   {q^{\Delta+N-c/24}\over \prod_{k=1}^{+\infty} (1-q^k)}~.
$$

The correct counting of states is therefore given by
$$
 \chi_\Delta(q)\=\chi_{\mbox{\tiny naive}}-\chi_{\mbox{\tiny null}}(q)
  \= {q^{\Delta-c/24} \over \prod_{k=1}^{+\infty} (1-q^k)}
  (1-q^N)~. 
$$

(c) Referring to part (b), if the Verma module over the null
state $\ket{\Delta+N}$ itself has a null state at level $M$,
then we must first subtract the null states from this module
to find the correct number of states that we should subtract from
$\chi_{\mbox{\tiny naive}}$. This gives
$$
   \chi_{\mbox{\tiny naive}}-\lb \chi_{\mbox{\tiny null}} - 
   \chi_{\mbox{\tiny null of null}} \rb ~,
$$
or 
$$
  {q^{\Delta-c/24} \over \prod_{k=1}^{+\infty} (1-q^k)}
 \, (1-q^N+q^{N+M})~.
$$
It now becomes obvious how to extend this in cases
where the null states  within the null states continue to a deep
level, or even {\it ad infinitum}:
$$
   \chi_{\mbox{\tiny naive}}
   -\la \chi_{\mbox{\tiny null}} 
   -\lb \chi_{\mbox{\tiny null of null}} 
   -\Big( \chi_{\mbox{\tiny null of null of null }}
   -(~~\cdots~~)
     \Big) \rb \ra ~,
$$
Also, it is straightforward to generalize these results
to cases in which the Verma modules have more than one null state
at each subtraction step, and which can then overlap (see  problem
\ref{item:CNM2}). 

\separator

\item
Before we proceed to the solution, we introduce some notation
to simplify our formul\ae.  Let
\bb
    a(k) &\equiv& \Delta_{m,n+2sk}\=\Delta_{m-2rk,n}~~,~~~{\rm and}\\
    b(k) &\equiv& \Delta_{-m,n+2sk}\=\Delta_{-m-2rk,n}~,
\ee
where we have also used the properties
$$
  \Delta_{mn}\=
  \Delta_{r-m,s-n}\=
  \Delta_{-m,-n}\=
  \Delta_{r+m,s+n}~.
$$
We also let
$$
  \lb a(k) \rb \= \lb V_{m,n+2sk}\rb~~~{\rm and}~~~
  \lb b(k) \rb \= \lb V_{-m,n+2sk}\rb~.
$$

To solve this problem, we directly apply the results from the previous
problems. In Exercise \ref{item:CNMX}, we saw that the  module
$\lb V_{mn}\rb=\lb a(0)\rb$ of the MMs admits a 
null state  with weight $\Delta_{-m,n}=b(0)$.
Because of the symmetry $\Delta_{mn}=\Delta_{r-m,s-n}$, the same module
also admits a second null state with weight 
$\Delta_{-(r-m),s-n}=\Delta_{r-m,-s+n}=\Delta_{-m,-2s-n}=b(-1)$.
Therefore, to compute the number of states for $\lb a(0)\rb$
we must subtract the states of $\lb b(0)\rb$ and 
$\lb b(-1)\rb$:
$$
  \lb a(0)\rb - \lb b(0)\rb -\lb b(-1)\rb~.
$$
However, this is not the end, since, by the same argument, the modules 
$ \lb b(0)\rb$ and $\lb b(-1)\rb$ have null states themselves. Continuing
this way, we discover an infinite tower of null states embedded within
null states. Using simple arguments as above, we find the following embeddings:
\begin{itemize}
\item The module $\lb a(k)\rb, ~k>0$ has null states with weights
      $b(k)$ and $b(-k-1)$.
\item The module $\lb a(-k)\rb,~k>0$ has null states with weights
      $b(k)$ and $b(-k-1)$.
\item The module $\lb b(k)\rb,~k>0$ has null states with weights
      $a(k+1)$ and $a(-k-1)$.
\item The module $\lb b(-k)\rb,~k>0$ has null states with weights
      $a(k)$ and $a(-k)$.
\end{itemize}
The following diagram makes it easy to appreciate the series of embeddings:
\begin{center}
\psfrag{a(0)}{$a(0)$}
\psfrag{b(0)}{$b(0)$}
\psfrag{a(1)}{$a(1)$}
\psfrag{b(1)}{$b(1)$}
\psfrag{a(-1)}{$a(-1)$}
\psfrag{b(-1)}{$b(-1)$}
\psfrag{b(-2)}{$b(-2)$}
\psfrag{dots}{\dots}
\includegraphics[height=2.5cm]{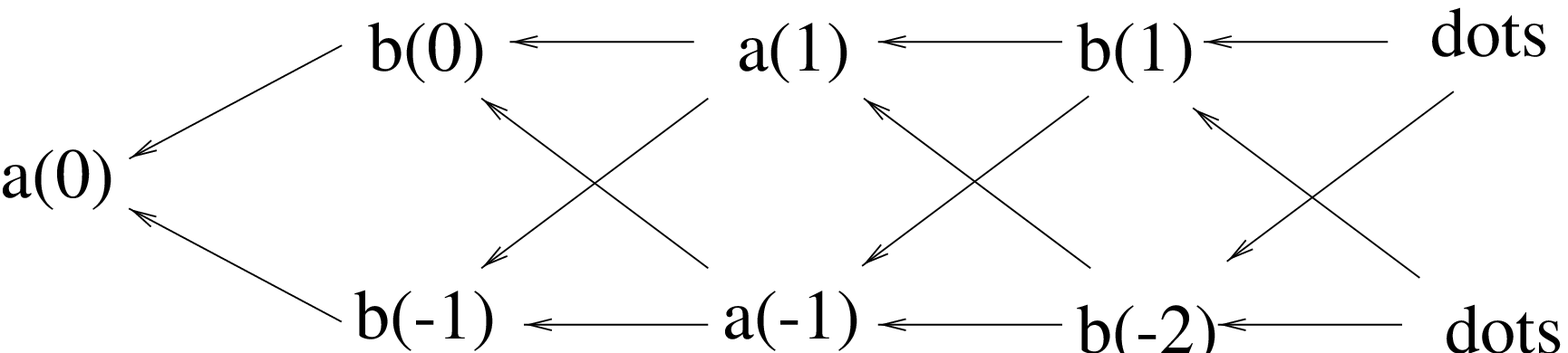}
\end{center}

Since there is overlapping among the modules of the  null states, 
we must be careful to subtract each null
state only once. Notice, too, that
the various modules are entering in the series with alternating
signs:
\begin{center}
\psfrag{a(0)}{$a(0)$}
\psfrag{b(0)}{$b(0)$}
\psfrag{a(1)}{$a(1)$}
\psfrag{b(1)}{$b(1)$}
\psfrag{a(-1)}{$a(-1)$}
\psfrag{b(-1)}{$b(-1)$}
\psfrag{b(-2)}{$b(-2)$}
\psfrag{dots}{\dots}
\psfrag{+}{+}
\psfrag{-}{$-$}
\includegraphics[height=4.6cm]{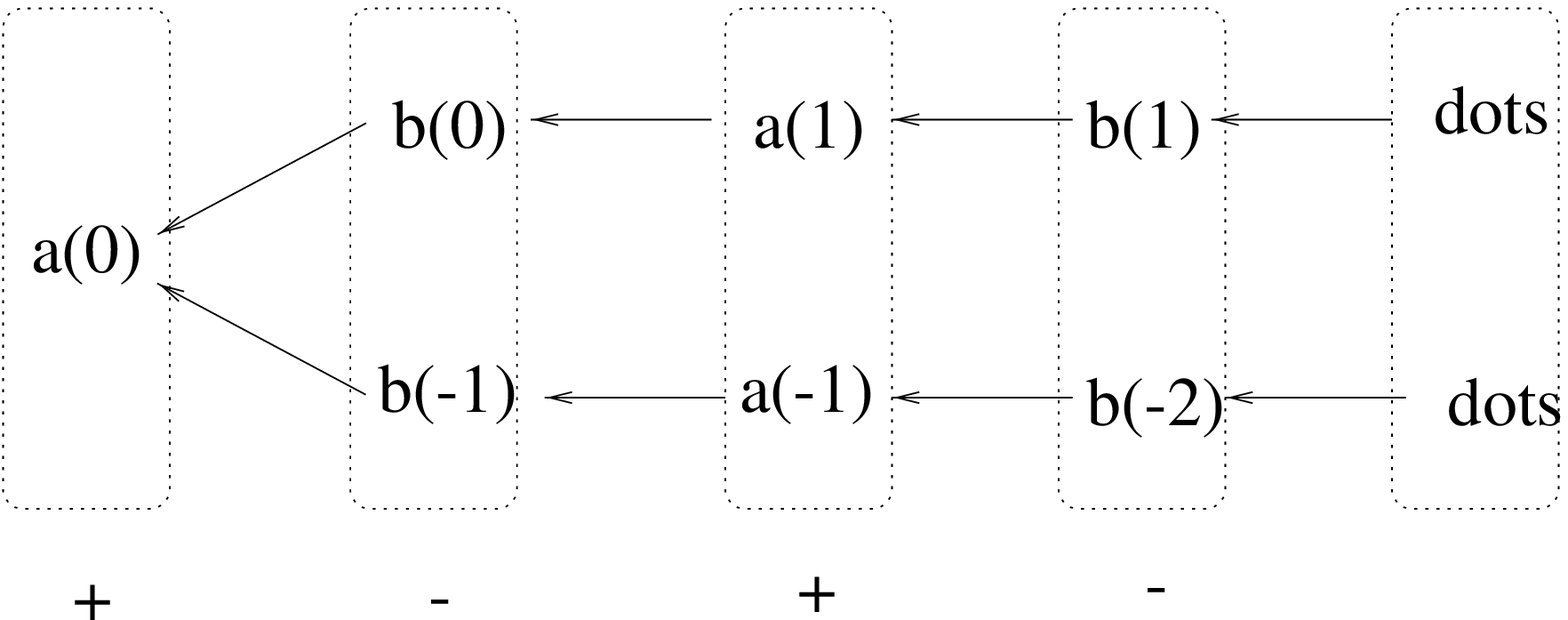}
\end{center}
Therefore,
$$
  \chi_{mn}(q)\= {q^{-c/24}\over\prod_{n=1}^{+\infty}(1-q^n)}\,
   \sum_{k\in\BZ} \lp q^{a(k)}-q^{b(k)} \rp~.
$$
The product $\prod_{n=1}^{+\infty}(1-q^n)$ appears often when
one considers characters and related objects.  It is standard
and convenient to define 
the {\bfseries Detekind $\eta$-function}\index{Function!Detekind --}
$$
   \eta(q)\eq q^{1/24}\, \prod_{n=1}^{+\infty} (1-q^n)~,
$$
which we will encounter again often.

\end{enumerate}

\newchapter{MODULAR INVARIANCE}
\label{ch:MI}

\footnotesize
\noindent {\bfseries References}:
The mathematical literature on the topics covered in this chapter is vast.
However, most of the main tools and techniques can also be found
in physics-oriented papers and books,
such as \cite{Belokolos,DiFMS,Ginz,Ketov}. A popular reference
on Riemann surfaces is \cite{Farkas}, and one
on $\theta$-functions is \cite{Mumford}.
A recent short introduction to the theory
of functions on compact surfaces is
\cite{Korotkin}. More results on algebraic geometry may be found in the classic
book \cite{Griffiths}.
\normalsize

\section{BRIEF SUMMARY}

\subsection{Riemann $\vartheta$-function}

On the sphere (or, equivalently, on the
compactified plane), the monomials
$z-z_i$ can be viewed as the fundamental building blocks out
of which functions are built.\footnote{See, for example,
the Mittag-Leffler theorem in Chapter \ref{ch:OM}.} 
The 
analogous building blocks for functions
defined on higher genus Riemann surfaces are the so-called 
$\vartheta$-functions discussed in this chapter. 

The {\bf Riemann} $\vartheta$-{\bf function} 
is defined by the expression
\bb
  \vartheta(u_1,u_2,\dots,u_g | \tau_{ij} ) \eq
  \sum_{m_1=-\infty}^{+\infty}\,
  \sum_{m_2=-\infty}^{+\infty}\,
  \dots
  \sum_{m_g=-\infty}^{+\infty}\,
  e^{ 2\pi i \sum\limits_{j=1}^g u_jm_j}\,
  e^{ \pi i \sum\limits_{j,k=1}^g m_i \tau_{ij} m_j}~.
\ee
The convergence of the Riemann $\vartheta$-function is guaranteed if
$\mbox{Im}\tau_{ij}>0$.
Usually the following compact notation is used:
\beq
\label{modular3}
  \vartheta(\bfu | \bfta) \=
  \sum_{\bfm\in\BZ^g}\, e^{ 2\pi i \bfu \bfm}\,
                e^{ \pi i \bfm^T \bfta \bfm}~,
\eeq
where $\bfu$ and $\bfm$ 
are column matrices, and $\bfta$ is a $g\times g$
symmetric  matrix:
\bb
  \bfu\eq\lb u_i\rb~,~~~~~
  \bfm\eq\lb m_i\rb~,~~~~~
  \bfta\eq\lb \tau_{ij}\rb~.
\ee
When there is simply a dot product such as $u_im_i$, to avoid
clutter we will write this as $\bfu\bfm$ rather than the
more proper $\bfu^T\bfm$.
When the $\vartheta$-function is related to a Riemann surface, 
the parameter $g$ is the genus of the surface and $\bfta$ is
the period matrix.

One can also introduce the Riemann 
{\bf $\vartheta$-function with characteristics}:
\beqn
\label{eq:MI1}
  \vartheta\lb\matrix{\bfa\cr\bfb\cr}\rb(\bfu | \bfta) &\equiv&
  \sum_{\bfm\in\BZ^g}\, e^{ 2\pi i(\bfm+\bfa)(\bfu+\bfb)}\,
                e^{ \pi i (\bfm+\bfa)^T \bfta (\bfm+\bfa)}
   \\
\label{eq:MI2}
  &=& e^{ 2\pi i\bfa(\bfu+\bfb)}\,
  e^{ \pi i \bfa^T \bfta \bfa}\,
  \vartheta(\bfu+\bfta\bfa+\bfb  | \bfta) ~,
\eeqn
where $\bfa, \bfb\in\BR^g$.

For $g=1$, we introduce the simplified  notation
$$
  q\equiv e^{2i\pi\tau}~,~~~~~
  z\equiv e^{2i\pi u}~.
$$
In addition, in mathematical physics,
the following notation is typically used:
\bb
  \vartheta_1(u | \tau) &\eq&
  \sum_{m=-\infty}^{+\infty}\, (-1)^m\, z^{m+1/2}\, q^{(m+1/2)^2/2}
  \= -i\vartheta\lb\matrix{1/2\cr 1/2\cr}\rb (u|\tau)~;\\
  \vartheta_2(u | \tau) &\eq&
  \sum_{m=-\infty}^{+\infty}\,  z^{m+1/2}\, q^{(m+1/2)^2/2}
  \= \vartheta\lb\matrix{1/2\cr 0\cr}\rb (u|\tau)~;\\
  \vartheta_3(u | \tau) &\eq&
  \sum_{m=-\infty}^{+\infty}\,  z^m\, q^{m^2/2}
  \= \vartheta\lb\matrix{0\cr 0\cr}\rb (u|\tau)~;~~~{\rm and} \\
  \vartheta_4(u | \tau) &\eq&
  \sum_{m=-\infty}^{+\infty}\, (-1)^m\, z^m\, q^{m^2/2}
  \= \vartheta\lb\matrix{0\cr 1/2\cr}\rb (u|\tau)~.
\ee
The above four functions are
called the {\bf Jacobi} $\vartheta_i$-{\bf functions}
($i=1,2,3,4$).

In $\BC^g$, we construct the lattice $L_{\bfta}$:
$$
    L_{\bfta}\=\BZ^g+\bfta\, \BZ^g \= 
     \{ \bfr\in\BC^g ~|~ \exists\bfl,\bfn\in\BZ^g,~\bfr=\bfl+\bfta\bfn\}~.
$$ 
The $\vartheta$-function is completely characterized by its behavior
under shifts in the above lattice:
\beqn
\label{modular1}
  \vartheta\lb\matrix{\bfa\cr\bfb\cr}\rb(\bfu+\bfta\bfn+\bfl | \bfta) &=&
                e^{ -\pi i \bfn^T_k \bfta \bfn}\,
                e^{ - 2\pi i \bfn(\bfu+\bfb) }\,
                e^{ 2\pi i \bfa \bfl }\,
  \vartheta\lb\matrix{\bfa\cr\bfb\cr}\rb(\bfu | \bfta) 
\eeqn
In other words, the $\vartheta$-function is a section of a (holomorphic line)
bundle on the complex torus
$$
    J(\Sigma)={\BC^g\over L_{\bfta}}~.
$$
$J(\Sigma)$ is called the {\bf Jacobian}  of the Riemann surface $\Sigma$.

One sees in addition that
\beq
\label{modular2}
  \vartheta\lb\matrix{\bfa+\bfn\cr\bfb+\bfl\cr}\rb(\bfu | \bfta) =
                e^{ 2\pi i \bfa \bfl }\,
  \vartheta\lb\matrix{\bfa\cr\bfb\cr}\rb(\bfu | \bfta) ~.
\eeq

\subsection{The Modular Group}

For a surface $\Sigma$, let ${\rm Diff}^+(\Sigma)$ be the group of all 
orientation preserving diffeomorphisms of $\Sigma$,
and let ${\rm Diff}^+_0(\Sigma)$ be the (normal) subgroup of all 
such diffeomorphisms connected to the identity. The quotient group
$$
   \CM(\Sigma) \= { {\rm Diff}^+(\Sigma) \over {\rm Diff}^+_0(\Sigma) }
$$
is called the {\bf modular group} (or {\bf mapping class group}).
A non-trivial class in $\CM(\Sigma)$ (i.e., element of $\CM(\Sigma)$)
is called a
{\bf modular transformation}.
For any such class, we can always choose a representative given by
a {\bf Dehn twist} $D_\gamma$ defined as follows. Let $\gamma$ be a loop
on $\Sigma$. We now cut the surface $\Sigma$ along the loop
and twist one of the edges thus produced by $2\pi$ while
leaving the other unaltered. Finally, we glue back the two
edges. A complete set of generators is produced when we consider
all Dehn twists along curves which wind around a single handle
or at most two handles.

A matrix representation $M(D_\gamma)$ of the Dehn twists
$D_\gamma$ can be found
by considering their action on the homology basis.
The intersection matrix is invariant under diffeomorphisms,
and so the action of $\CM(\Sigma)$ on $H_1(\Sigma,\BZ)$ 
must also preserve the intersection matrix, and therefore
$M(D_\gamma)\in {\rm Sp}(2g,\BZ)$. In fact, it is known that 
the matrices $M(D_\gamma)$ generate all of
${\rm Sp}(2g,\BZ)$.

A Dehn twist about a  homologically trivial cycle does not
affect the homology class of any curve\footnote{Be careful!
This twist is still non-trivial.}; it is thus represented by 
a unit matrix. All such twists generate a subgroup
$\CT(\Sigma)$ of $\CM(\Sigma)$ known as the
{\bf Torelli group}. One can show that
$$
   {\rm Sp}(2g,\BZ) \= {\CM(\Sigma)\over\CT(\Sigma)}~.
$$

\begin{figure}[htb]
\begin{center}
\psfrag{O1}{$\omega_1$}
\psfrag{O2}{$\omega_2$}
\includegraphics[height=4cm]{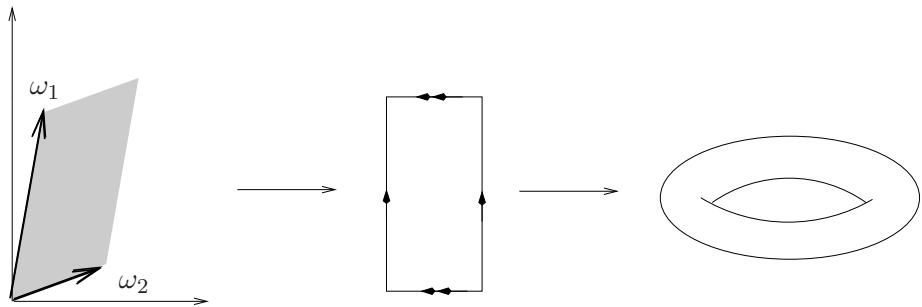}
\end{center}
\caption{The torus is the quotient space of the plane divided by a
         2-dimensional lattice.}
\end{figure}

\subsection{Partition Functions and Modular Invariance}

Mapping the plane into a cylinder, the dilation operators on
the plane become the translation operators on the cylinder:
$$
  L_{-1}^{\rm cyl} \= {2\pi i\over\omega_1}\,
  \lp L_0^{\rm pl}-{c\over 24}\rp ~~~{\rm and}~~~
  \overline L_{-1}^{\rm cyl} \= - {2\pi i\over\overline\omega_1}\,
  \lp \overline L_0^{\rm pl}-{c\over 24}\rp~.
$$
The constant $-c/24$ arises as a Casimir effect due to the periodic boundary
conditions. Assuming periodic boundary conditions on the bases of the
cylinder, we create a toroidal geometry, with cycles $\omega_1$ and
$\omega_2$. The partition function is thus given by
\bb
   Z &=& {\rm tr} \lp e^{\omega_2 L_{-1}^{\rm cyl} +
               \overline\omega_2 \overline L_{-1}^{\rm cyl} }\rp \\
     &=& 
 e^{-2\pi i{c\over 24}
    \lp{\omega_2\over\omega_1}+{\overline\omega_2\over\overline\omega_1}\rp}
 \, {\rm tr}
  \lp e^{2\pi i {\omega_2\over\omega_1} L_0^{\rm pl} } 
 e^{2\pi i {\overline\omega_2\over\overline\omega_1}\overline L_0^{\rm pl}}
  \rp~. 
\ee
Defining
$$
  \tau\eq {\omega_2\over\omega_1}~~~{\rm and}~~~
  q\eq e^{2\pi i \tau}~,
$$
and taking into account the decomposition of the Hilbert space into
holomorphic and antiholomorphic pieces, we can write the
partition function as
$$
   Z \= \sum_{\Delta,\overline\Delta} \, \CN_{\Delta\overline\Delta}
        \, \chi_\Delta(\tau)\,
           \overline\chi_{\overline\Delta}(\tau)~,
$$
where $\chi_\Delta$ is the character for the Verma module built over
the field with dimension $\Delta$.
The partition function contains the whole operator content of a model.

Mathematically speaking, the torus is the quotient of the complex plane
divided by the lattice $\Lambda$ generated by the vectors
$\omega_1$ and $\omega_2$. Physical intuition
suggests that the partition function may depend on the geometry
of the torus parametrized by $\tau$, but it should be invariant
under different choices of the lattice basis $\omega_1,\omega_2$.
In other words, the transformations
$$
   \lb\matrix{\omega'_1\cr\omega'_2\cr}\rb \=
   \lb\matrix{a & b\cr c & d \cr}\rb \,
   \lb\matrix{\omega_1\cr\omega_2\cr}\rb ~,
   ~~~~a,b,c,d\in\BZ~,~~~ad-bc=1~
$$
should not affect the partition function.  These
transformations constitute the group SL$(2,\BZ)$, and,
since they do not change $\tau$, we call this the
modular group of the torus.
The parameter $\tau$ is called the 
{\bfseries modular parameter} (or simply the {\bfseries modulus})
of the torus.

SL$(2,\BZ)$ can be generated by the following two transformations
of $\tau$:
\bb
   T: \tau &\mapsto& \tau+1 \Rightarrow 
      (\omega_1,\omega_2)\mapsto (\omega_1,\omega_1+\omega_2)~,\\
   S: \tau &\mapsto& -{1\over\tau}\Rightarrow 
      (\omega_1,\omega_2)\mapsto(\omega_2,-\omega_1)~.
\ee
The requirement of modular invariance of the partition
function reduces, therefore, to the conditions
$$
   Z(\tau+1)\=Z(-1/\tau)\=Z(\tau)~.
$$

Let $\chi(\tau)$ be a vector that contains the characters
for all the representations present in a particular model.
The partition function then takes the form
$$
   Z(\tau) \=\chi^\dagger \CN \chi~.
$$
Different models at fixed $c$ corresponds to different matrices $\CN$.
Modular invariance implies
$$
   \CN\=T \CN T^\dagger~~~{\rm and}~~~ \CN\=S \CN S^\dagger~.
$$
The first equation simply requires $\Delta-\overline\Delta\in\BZ$ 
--- only integer spins are allowed. The second equation is a 
Diophantine equation that strongly restricts the allowed matrices $\CN$.

There is a remarkable formula found by E. Verlinde and proved by
Moore and Seiberg. The matrices $\bfN_i=\lb N^k_{ij}\rb$ that determine
the fusion rules
$$
  \lb\phi_i\rb \,\times\, \lb \phi_j\rb\=N^k_{ij}\, \lb\phi_k\rb~,
$$
can be written in terms of the modular $S$-matrix as follows:
$$
 N^k_{ij}\=\sum_n \, {S^n_j \, S^n_i \, S^k_n\over S^k_0}~.
$$
This equation is known as the {\bfseries Verlinde formula}\index{formula!Verlide --}.
A corollary of this formula is that the modular $S$-matrix diagonalizes the
matrices $\lb \bfN_i\rb$:
$$
      \bfN_i \= S\, \bfD_i \, S^\dagger~,
$$
where
$$
    \bfD_i \= \mbox{diag}\lb {S_i^n\over S_0^n} \rb~.
$$

\newpage
\section{EXERCISES}

\begin{enumerate}

\item
 (a) From the definition \calle{eq:MI1}, prove equation \calle{eq:MI2}
for the Riemann $\vartheta$-function with characteristics. 

(b) Now use equation \calle{eq:MI2} to prove properties \calle{modular1} 
    and \calle{modular2}. 

\item
(a) For a function $g(x_1,x_2,...,x_g)$ let
\beqn
 F_1 &\equiv& \sum_{m_1=-\infty}^{+\infty}
 \sum_{m_2=-\infty}^{+\infty}
 \dots
 \sum_{m_g=-\infty}^{+\infty}
 g(m_1,m_2,...,m_g)~,~~~{\rm and}
 \nonumber \\ 
 F_2 &\equiv& \sum_{n_1=-\infty}^{+\infty}
 \sum_{n_2=-\infty}^{+\infty}
 \dots
 \sum_{n_g=-\infty}^{+\infty} 
  \int\limits_{-\infty}^{+\infty}\,dy_1
 \int\limits_{-\infty}^{+\infty}\,dy_2
 \dots
 \int\limits_{-\infty}^{+\infty}\,dy_g
  \, g(y_1,y_2,...,y_g)\, e^{-2\pi i \sum_{j=1}^g n_j y_j}~.
  \nonumber
\eeqn

Show that
$$
   F_1 \= F_2 ~.
$$
This formula is known as the {\bf Poisson resummation formula}. 

(b) Use the Poisson resummation formula to prove that 
the Riemann function \calle{modular3} can also be written in the form:
$$
  \vartheta(\bfu | \bfta) = {i^{g/2}\over \sqrt{|\det\bfta|}}\,
                e^{ -\pi i \bfu^T\bfta^{-1}\bfu }\,
                \sum_{\bfn}\,
                e^{ 2\pi i \bfn^T\bfta^{-1} \bfu }\,
                e^{ -\pi i \bfn^T \bfta^{-1} \bfn}~.
$$

\item
Prove that Jacobi's $\vartheta_3$-function has only one zero,
which is located at ${1\over 2}\tau+{1\over 2}$.

\item
In this problem, you are to prove {\bf Jacobi's triple identity} 
\beq
 \prod_{n=1}^{+\infty}\, (1-q^n)\,
                         (1+w\,q^{n-1/2})\,
                         (1+w^{-1}\,q^{n-1/2})
  \= \sum_{n=-\infty}^{+\infty}\, w^n\, q^{n^2/2}~,
\label{Jacobitriple}
\eeq
for $|q|<1$ and $w\not=0$.
You should do this two different ways:

(a) Prove the result using physical arguments \cite{Ginz}, starting by 
    writing down  the grand canonical partition
    function for a free system of fermion and antifermion oscillators.

(b) Prove the results purely mathematically \cite{And86}.  Define
\beqn
      P(q,w)&\equiv& \prod_{n=1}^{+\infty}\,
                   (1+w\,q^{n-1/2})\,
                   (1+w^{-1}\,q^{n-1/2})~~~{\rm and}
    \nonumber \\
      Q(q,w)&\equiv& {1\over\prod\limits_{n=1}^{+\infty}(1-q^n)}\,
                \sum_{n=-\infty}^{+\infty}\, w^n\, q^{n^2/2}~,
    \nonumber
\eeqn
and then prove that both satisfy the functional relation
$$
     X(q,w) = q^{1/2}\, w \, X(q, qw)~.
$$
Use this to establish \calle{Jacobitriple}.

\item
Derive {\bfseries Euler's pentagonal number theorem}
\bb
   \prod_{k=1}^{+\infty}\,
                   (1-q^k) \= \sum_{k=-\infty}^{+\infty}\,
                   (-1)^k\,q^{{k(3k-1)\over 2}}~,
\ee
using the following two methods:

(a) Use Jacobi's triple identity.

(b) Use the Rocha-Caridi formula for the minimal model
   with $p=2$ and $q=3$.

\item
\label{item:MI6}

(a) Find an infinite product expression for each of the Jacobi 
$\vartheta_i$-functions.

(b) Let 
$$
   \theta_i(\tau)\eq\vartheta_i(0|\tau)~.
$$
Show that
$$
   \eta^3(\tau) \= {1\over 2}\, \theta_2(\tau)\theta_3(\tau)\theta_4(\tau)~.
$$

\item
Write down the matrix representation of the Dehn twists for 
a genus two surface.

\item
Compute the partition function for a free boson. Then verify that
it is modular invariant.

\item
Compute the partition function for a compactified free boson, i.e.,
a free boson restricted to move on a circle. Then verify that
this partition function is modular invariant.

\item
\label{item:MI1}
(a) Find the characters $\chi_0$, $\chi_{1/2}$, and
    $\chi_{1/16}$ of the free fermion theory by direct enumeration
    of the states.

(b)  Determine the
     modular transformation matrix $\bfS$ acting on the
     character basis of part (a). 

(c) Show that $\bfS$ diagonalizes the fusion rules of the theory.

\item
Generalizing part (b) of Exercise \ref{item:MI1}, show that under the
$T$ and $S$ transformations,  the character vector of the MMs transforms
as
$$
  T\chi(\tau) \= \chi(\tau+1)~~~{\rm and}~~~
  S\chi(\tau) \= \chi(-1/\tau)~,
$$
and derive the exact form of the matrices $T$ and $S$.

\item
For a theory with dyons $(q,g)$ (i.e., a particle with electric
charge $q$ and magnetic charge $g$), the 
{\bfseries Dirac-Schwinger-Zwanziger} (DSZ) {\bfseries condition} says that
the charges of any two dyons satisfy the quantization equation
\beq
\label{DSZquant}
   q_1 g_2 - q_2 g_1 \=  2\pi n_{12}~,~~~~~n_{12}\in\BZ~.
\eeq

(a) Show that the general solution of this equation takes the
    form
\bb
     q &=& q_0\, \lp n + {\theta\over2\pi}\, m\rp~,\\
     g &=& g_0\, m~,
\ee
where $n,m\in\BZ$; $q_0$ and $g_0$ are constants; and $\theta$ is a
parameter with a value $0\le\theta< 2\pi$.

(b) Show that the parameter $\theta$ is related to CP violation.
    In particular, show that for $\theta=0$ and $\theta=\pi$, the theory
    is CP invariant, while for all other values of $\theta$ between
    $0$ and $2\pi$,
    the theory violates CP invariance.

(c) Rewrite the solution of part (a) in the form
$$
   q+ig \= q_0\, (n+m\, \tau) ~.
$$
 Explain how this form shows that the solutions to the DSZ condition
 \calle{DSZquant}
 satisfy an
 SL$(2,\BZ)$ invariance. Discuss the transformation properties of $q+ig$ 
   under this group.
   
\end{enumerate}

\newpage
\section{SOLUTIONS}

\begin{enumerate}

\item

(a) From the definition \calle{eq:MI1}, we have:
\bb
   \vartheta\lb\matrix{\bfa\cr\bfb\cr}\rb(\bfu | \bfta) &\equiv&
  \sum_{\bfm}\, e^{ 2\pi i(\bfm+\bfa)(\bfu+\bfb)}\,
                e^{ \pi i (\bfm+\bfa)^T \bfta (\bfm+\bfa)} \\
   &=& 
   \sum_{\bfm}\,
                e^{ 2\pi i\bfm^T(\bfu+\bfb)}\,
                e^{ 2\pi i\bfa(\bfu+\bfb)}\,
                e^{ \pi i \bfm^T \bfta \bfm}\, 
                e^{ \pi i \bfm^T \bfta \bfa} \,
                e^{ \pi i \bfa^T \bfta \bfm}\,
                e^{ \pi i \bfa^T \bfta \bfa} ~.
\ee  
Since the matrix $\bfta$ is symmetric, one has
\bb
     e^{ \pi i \bfm^T \bfta \bfa} \,
                e^{ \pi i \bfa^T \bfta \bfm}\=
     e^{2 \pi i \bfm^T \bfta \bfa}~.
\ee
Hence,
\bb
   \vartheta\lb\matrix{\bfa\cr\bfb\cr}\rb(\bfu | \bfta) &=&
    \sum_{\bfm}\,
                e^{ 2\pi i\bfm^T(\bfu+\bfb)}\,
                e^{ 2\pi i\bfa(\bfu+\bfb)}\,
                e^{ \pi i \bfm^T \bfta \bfm}\,
                e^{2 \pi i \bfm^T \bfta \bfa}\,
                e^{ \pi i \bfa^T \bfta \bfa} \\
    &=&         e^{ 2\pi i\bfa(\bfu+\bfb)}\, e^{ \pi i \bfa^T \bfta \bfa}\,
                \sum_{\bfm}\,
                e^{ 2\pi i\bfm^T(\bfu+ \bfta \bfa+\bfb)}\,
                 e^{ \pi i \bfm^T \bfta \bfm}~,
\ee
which gives the formula sought:
\bb
   \vartheta\lb\matrix{\bfa\cr\bfb\cr}\rb(\bfu | \bfta) &=&
   e^{ 2\pi i\bfa(\bfu+\bfb)}\, e^{ \pi i \bfa^T \bfta \bfa}\,
    \vartheta(\bfu+\bfta\bfa+\bfb | \bfta)~.
\ee

(b) Before we proceed to prove equations \calle{modular1}
and \calle{modular2}, we first obtain, as an intermediate step,
an expression for 
$\vartheta(\bfu+\bfta\bfn+\bfl | \bfta)$
in terms of
$\vartheta(\bfu | \bfta)$. 
One observes that
\bb
  \vartheta(\bfu+\bfta\bfn+\bfl | \bfta)
  &=& \sum_{\bfm\in\BZ^g}\, e^{ 2\pi i (\bfu+\bfta\bfn+\bfl) \bfm}\,
       e^{ \pi i \bfm^T \bfta \bfm} \\
  &=&  \sum_{\bfm\in\BZ^g}\,
      e^{ 2\pi i \bfu \bfm}\,
      e^{ 2\pi i (\bfta\bfn)\bfm}\,
      e^{ \pi i \bfm^T \bfta \bfm} ~,
\ee
since the dot 
product $\bfl\bfm$ is an integer, i.e., $e^{ 2\pi i\bfl\bfm}=1$.
Now we make the change of variables $\bfm=\bfk-\bfn$, which yields
\bb
    \vartheta(\bfu+\bfta\bfn+\bfl | \bfta) 
  &=& \sum_{\bfk\in\BZ^g}\,
      e^{ 2\pi i \bfu(\bfk-\bfn)}\,
      e^{ 2\pi i (\bfta\bfn)(\bfk-\bfn)}\,
      e^{ \pi i (\bfk-\bfn)^T\bfta(\bfk-\bfn)}\\ 
   &=& 
      e^{ -2\pi i \bfu\bfn}\,
      e^{ -2\pi i (\bfta\bfn)\bfn}\,
      e^{ \pi i \bfn^T\bfta\bfn}\,
      \sum_{\bfk\in\BZ^g}\,
      e^{ 2\pi i \bfu\bfk}\,
      e^{ \pi i \bfk^T\bfta\bfk}\,
      e^{ 2\pi i (\bfta\bfn)\bfk}\,
      e^{ -\pi i \bfn^T\bfta\bfk}\,
      e^{ -\pi i \bfk^T\bfta\bfn} \\
   &=&
      e^{ -2\pi i \bfu\bfn}\,
   e^{ -\pi i (\bfta\bfn)\bfn}
      \sum_{\bfk\in\BZ^g}\,
      e^{ 2\pi i \bfu\bfk}\,
      e^{ \pi i \bfk^T\bfta\bfk}~.
\ee
Thus,
$$
  \vartheta(\bfu+\bfta\bfn+\bfl | \bfta)\=
   e^{ -2\pi i \bfu\bfn}\,
   e^{ -\pi i (\bfta\bfn)\bfn}
   \,\vartheta(\bfu | \bfta)~.
$$

Now we are ready to prove formul\ae \calle{modular1}
and \calle{modular2}. To do this, we will make use of
equation \calle{eq:MI2} (which we proved in part (a)
of this problem solution)
and the equation we just derived. In particular,
\bb
   \vartheta\lb\matrix{\bfa\cr\bfb\cr}\rb(\bfu+\bfta\bfn+\bfl | \bfta) &=&
   e^{ 2\pi i \bfa (\bfu+\bfta\bfn+\bfl+\bfb)}\,
   e^{ \pi i \bfta^T\bfta\bfa}
   \vartheta(\bfu+\bfta\bfn+\bfl+\bfta\bfa+\bfb | \bfta) \\
   &=&
   e^{2\pi i \bfa (\bfu+\bfta\bfn+\bfl+\bfb)}\,
   e^{ \pi i \bfta^T\bfta\bfa}
   \vartheta(\bfu+\bfta\bfal+\bfb|\bfta) \,
   e^{ -2\pi i (\bfu+\bfta\bfb)\bfn}\,
   e^{ -\pi i (\bfta\bfn)\bfn} \\
   &=& 
   e^{ -\pi i (\bfta\bfn)\bfn} \,
   e^{ -2\pi i (\bfu+\bfb)\bfn}\,
   e^{ 2\pi i \bfa\bfl}\,
   e^{ 2\pi i \bfa(\bfu+\bfb)}\,
   e^{ \pi i \bfa^T\bfta\bfa}\,
   \vartheta(\bfu+\bfta\bfa+\bfb|\bfta) \\
   &=&
   e^{ -\pi i (\bfta\bfn)\bfn} \,
   e^{ -2\pi i (\bfu+\bfb)\bfn}\,
   e^{ 2\pi i \bfa\bfl}\,
    \vartheta\lb\matrix{\bfa\cr\bfb\cr}\rb(\bfu | \bfta)~.
\ee
Additionally, 
\bb
  \vartheta\lb\matrix{\bfa+\bfn\cr\bfb+\bfl\cr}\rb(\bfu | \bfta) &=&
  e^{ 2\pi i(\bfa+\bfn)(\bfu+\bfb+\bfl)}\, 
  e^{ \pi i (\bfa+\bfn)^T \bfta (\bfa+\bfn)}\,
    \vartheta(\bfu+\bfta\bfa+\bfta\bfn+\bfb+\bfl | \bfta)\\
  &=&
  e^{ 2\pi i(\bfa+\bfn)(\bfu+\bfb+\bfl)}\,
  e^{ \pi i (\bfa+\bfn)^T \bfta (\bfa+\bfn)}\,
   \vartheta(\bfu+\bfta\bfa+\bfb | \bfta)\,
  e^{ -2\pi i(\bfu+\bfta\bfa+\bfb)\bfn}\,
  e^{ -\pi i (\bfta\bfn)\bfn}~.
\ee
Since $e^{2\pi i\bfn\bfl}=1$ and some
of the exponentials cancel each other,
we obtain, in the end,
\bb
  \vartheta\lb\matrix{\bfa+\bfn\cr\bfb+\bfl\cr}\rb(\bfu | \bfta) &=&
    e^{ 2\pi i\bfa\bfl}\,
    e^{ 2\pi i\bfa(\bfu+\bfb)}\,
    e^{ \pi i\bfa^T\bfta\bfa}\,
   \vartheta(\bfu+\bfta\bfa+\bfb | \bfta) \\
   &=& e^{ 2\pi i\bfa\bfl}\,
   \vartheta\lb\matrix{\bfa\cr\bfb\cr}\rb(\bfu | \bfta)~.
\ee

\separator

\item
(a) We define the function
\beq
\label{eq:MI3}
  f(\bfx) \eq \sum_{\bfm\in\BZ^g} g(\bfx+\bfm)~.
\eeq
Clearly, the function $f$ is periodic, and so
it can be expanded in a Fourier
series.  We thus write
\beq
\label{eq:fFourier}
  f(\bfx) \= \sum_{\bfn} a_{\bfn}\, e^{2\pi i\bfb\cdot\bfx}~,
\eeq
with 
\bb
   a_{\bfn} &=& \int_0^1\, f(\bfy)\, e^{-2\pi i\bfn\cdot\bfy}\, d\bfy \\
   &=& \sum_{\bfm} \int_0^1\, g(\bfy+\bfm)\, e^{-2\pi i\bfn\cdot\bfy}\, d\bfy \\
   &=& \sum_{\bfm} \int_{\bfm}^{\bfm+{\bf1}}\, g(\bfw)\, 
       e^{-2\pi i\bfn\cdot(\bfw-\bfm)}\, d\bfw ~,
\ee
where in the second equality we 
have used the definition of $f$, and in the
third equality we have made the change of variables $\bfw=\bfy+\bfm$.
In addition, {\bf 1} stands for a vector with 
all its components equal to 1.
To proceed further, we notice that 
$$
   e^{-2\pi i\bfn\cdot\bfm} \= 1~.
$$
Then
\bb
  a_{\bfn} &=& \sum_{\bfm} \int_{\bfm}^{\bfm+{\bf1}}\, g(\bfw)\, 
       e^{-2\pi i\bfn\cdot\bfw}\, d\bfw \\
   &=& \sum_{\bfm} \int_{-\infty}^{+\infty}\, g(\bfw)\, 
       e^{-2\pi i\bfn\cdot\bfw}\, d\bfw ~.
\ee
Inserting this result into the expression \calle{eq:fFourier},
we find
\bb
  f(\bfx) 
  \=   \sum_{\bfm} e^{-2\pi i\bfn\cdot\bfx}
       \int_{-\infty}^{+\infty}\, g(\bfw)\,
       e^{-2\pi i\bfn\cdot\bfw}\, d\bfw ~,
\ee
or, in other words,
\bb
   \sum_{\bfm} g(\bfx+\bfm)
  \=   \sum_{\bfm} e^{-2\pi i\bfn\cdot\bfx}
       \int_{-\infty}^{+\infty}\, g(\bfw)\,
       e^{-2\pi i\bfn\cdot\bfw}\, d\bfw ~.
\ee
For the special case $\bfx=0$, we find the Poisson resummation formula
\beq
\label{eq:MI4}
  \sum_{\bfn} g(\bfm)
  \=   \sum_{\bfm} 
       \int_{-\infty}^{+\infty}\, g(\bfw)\,
       e^{-2\pi i\bfn\cdot\bfw}\, d\bfw ~.
\eeq

(b) 
Comparing the definition \calle{modular3} of 
the $\vartheta$-function with that of the
auxiliary function \calle{eq:MI3} we used in part (a), we see that
$$
  g(\bfu)\= e^{2\pi i\bfu\cdot\bfm}
            \, e^{\pi i\bfm^T\bfta\bfm}~.
$$
Inserting this in the identity \calle{eq:MI4}, we can rewrite
the $\vartheta$-function as
\bb
  \vartheta(\bfu|\bfta) &=& \sum_{\bfn}\int\,d\vec y\,
   g(\vec y)\, e^{-2\pi i\vec n\cdot\vec y} \\
  &=& \sum_{\bfn}\int\,d\vec y\,
   e^{2\pi i\vec y\cdot\vec n} \, e^{\pi i\vec n\bfta\vec n}
   \, e^{-2\pi i\vec n\cdot\vec y} \\
  &=& \sum_{\bfn}\int\,d\vec y\,
   e^{\pi i\vec n\bfta\vec n}
   e^{-2\pi i(\vec u-\vec n)\cdot\vec y} \\
  &=& \sum_{\bfn}\int\,d\vec y\,
   e^{-\pi \vec N\bfT\vec N}
   e^{2\pi (\vec U-\vec N)\cdot\vec y} ~,
\ee
where we have set
$$
  \vec u\eq i\vec U~, ~~~
  \bfn\eq i\bfN~, ~~~{\rm and}~~~
  \bfT\eq i\bfta~.
$$
 The integral that has appeared is Gaussian, and so
 is easily evaluated according to the standard result
$$
  \int\,d\vec y\,e^{-\vec y \bfA \vec y}\, e^{\bfB\vec y}\=
  {\pi^{g/2}\over\sqrt{\det\bfA}}\, e^{{1\over4}\bfB^T\bfA^{-1}\bfB}~.
$$
Therefore
\bb
  \vartheta(\bfu|\bfta) &=& \sum_{\bfn}
   \, {\pi^{g/2}\over\sqrt{\det(\pi\bfT)}}  \,
   e^{-{1\over4}4\pi^2(\vec U-\vec N){\bfT^{-1}\over\pi}(\vec U-\vec N)}
   \\
  &=&  \sum_{\bfn}
   \, {1\over(-i)^{g/2}\sqrt{\det\bfta}}  \,
   e^{-i\pi(\vec u-\vec n)\bfta^{-1}(\vec u-\vec n)}
   \\
  &=&
    {i^{g/2}\over\sqrt{\det\bfta}}  \,
    \sum_{\bfn}
   e^{-i\pi\vec u\bfta^{-1}\vec u}
   e^{+i\pi(\vec u\bfta^{-1}\vec n+\vec n\bfta^{-1}\vec u)}
   e^{-i\pi\vec n\bfta^{-1}\vec n}
   \\
  &=&
    {i^{g/2}\over\sqrt{\det\bfta}}  \,
   e^{-i\pi\vec u\bfta^{-1}\vec u}  \,
    \sum_{\bfn}
   e^{2i\pi\vec n\bfta^{-1}\vec u}
   e^{-i\pi\vec n\bfta^{-1}\vec n}~,
\ee
which completes the proof.

\separator

\item

From the general properties of the $\vartheta$-function,
 we can see that
\bb
   \vartheta_3(u+a\tau+b|\tau)&=&
   e^{-\pi i a^2\tau}\, e^{-2\pi i a u}\, \vartheta_3(u|\tau)~,~~~{\rm and}\\
   \vartheta_3(-u|\tau)&=&\vartheta_3(u|\tau)~,
\ee
where $a,b\in\BZ$.
We can differentiate the first of these equations, which produces
the identity
\bb
   \vartheta'_3(u+a\tau+b|\tau)&=&
   e^{-\pi i a^2\tau}\, e^{-2\pi i a u}\,
   \lb \vartheta'_3(u|\tau)-2\pi i a\, \vartheta_3(u|\tau)\rb~.
\ee

\noindent\begin{minipage}{6.5cm}
Using the above equations, we find that
\bb
   {\vartheta'_3(u+1)\over\vartheta_3(u+1|\tau)} &=&
   {\vartheta'_3(u)\over\vartheta_3(u|\tau)} ~,~~~{\rm and} \\
    {\vartheta'_3(u+\tau)\over\vartheta_3(u+\tau|\tau)} &=&
    {\vartheta'_3(u)\over\vartheta_3(u|\tau)}-2\pi i~.
\ee
\end{minipage}
\begin{minipage}{8cm}
\psfrag{x}{$\displaystyle x$}
\psfrag{y}{$y$}
\psfrag{t}{$\tau$}
\psfrag{1}{$1$}
          \begin{flushright}
          \includegraphics[width=6cm]{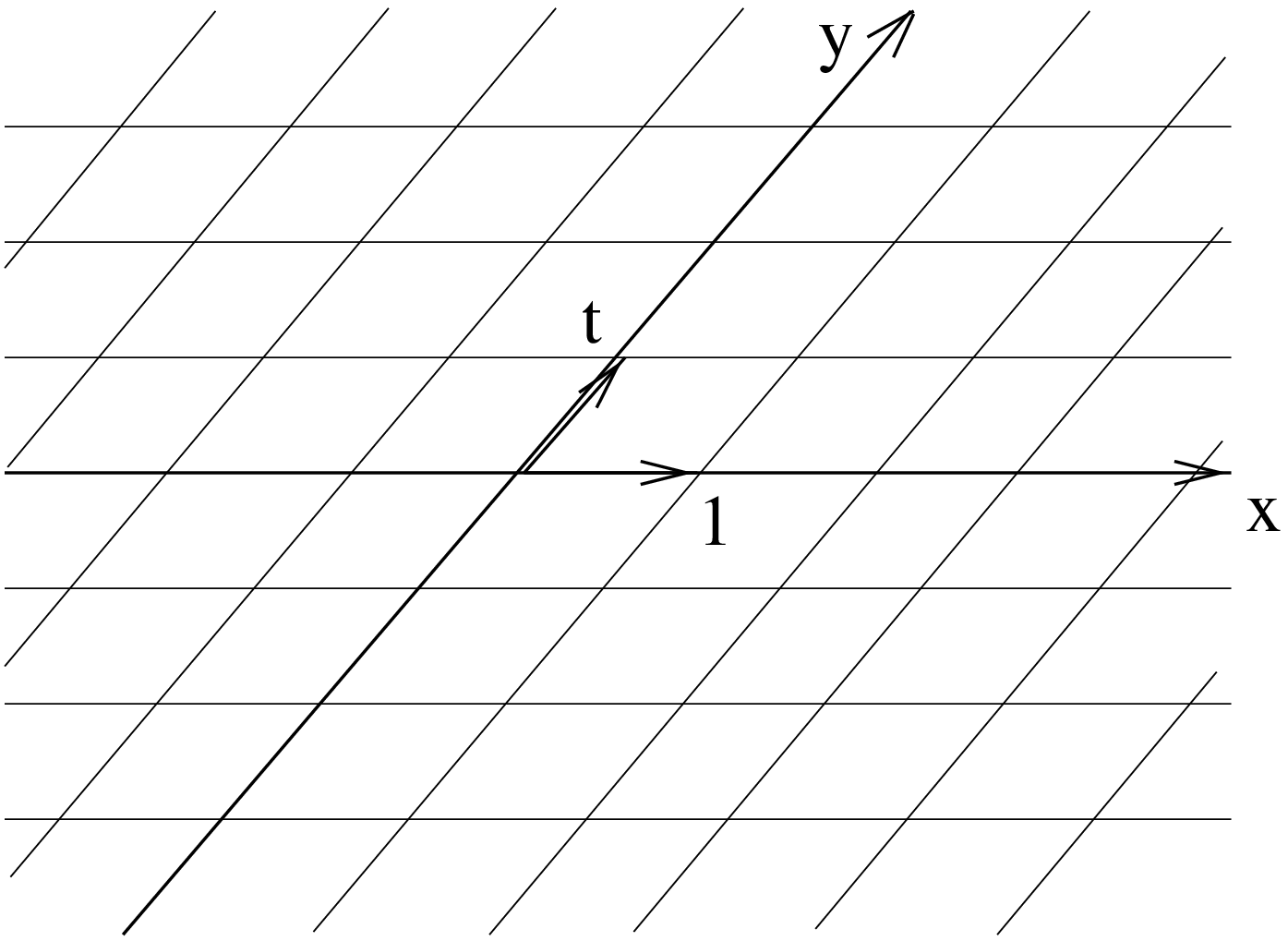} 
          \end{flushright}
\end{minipage}

\vspace{-5mm}
\noindent
Now from complex analysis, we know that
\bb
   N_{roots}-N_{poles} &=&
   {1\over2\pi i}\,\ointleft_{unit~cell}
   {\vartheta'_3(u)\over\vartheta_3(u)} \, du \\
   &=&{1\over2\pi i}\,\int_0^1
   {\vartheta'_3(x)\over\vartheta_3(x)} \, dx
   +{1\over2\pi i}\,\int_0^\tau
   {\vartheta'_3(1+iy)\over\vartheta_3(1+iy)}\, idy \\
   && +{1\over2\pi i}\,\int_1^0
   {\vartheta'_3(x+\tau)\over\vartheta_3(x+\tau)}\, dx
   +{1\over2\pi i}\,\int_\tau^0
   {\vartheta'_3(iy)\over\vartheta_3(iy)} \, idy \\
   &=&{1\over2\pi i}\,\int_0^1
   {\vartheta'_3(x)\over\vartheta_3(x)} \, dx
   +{1\over2\pi i}\,\int_0^\tau
   {\vartheta'_3(iy)\over\vartheta_3(iy)}\,  idy \\
   && -{1\over2\pi i}\,\int_0^1
   \lp {\vartheta'_3(x)\over\vartheta_3(x)}-2\pi i\rp\,  dx
   -{1\over2\pi i}\,\int_0^\tau
   {\vartheta'_3(iy)\over\vartheta_3(iy)} \, idy \\
   &=& 1 ~.
\ee
In the same way, we can show that
\bb
   N_{poles} \= {1\over2\pi i} \,\ointleft_{unit~cell}
   \vartheta_3(u) du \= 0~.
\ee
Therefore, $\vartheta_3$ has no poles and  only one root in the unit cell.

From the properties of $\vartheta_3$, we see that if $u_0$ is a
root, so are $u_0+a\tau+b$ and $-u_0$. This means that the set of
all roots of
$\vartheta_3$ form a periodic set with respect to the lattice
$\BZ+\tau\BZ$ and  is also symmetric around 0. So symmetry
dictates that the single  root of $\vartheta_3$  must be located at
one of the
points $0$, $1/2$, $\tau/2$, or $(\tau+1)/2$.
This is the point ${\tau+1\over2}$ as can be
checked\footnote{One can also verify explicitly that the other values,
$0$, $1/2$, and $\tau/2$, render the $\vartheta_3$-function
non-vanishing.
In particular, from its definition, it is straightforward to see that
$$
  \vartheta_3(0|\tau)=\sum_{m=-\infty}^{+\infty} \, q^{m^2/2} \ne 0~.
$$
Then from the properties
\bb
   \vartheta_3(u+1/2|\tau) &=& \vartheta_4(u|\tau)~,\\
   \vartheta_3(u+\tau/2|\tau) &=& e^{-i\pi u}\, q^{-1/8}\,
                                   \vartheta_2(u|\tau)~,
\ee
for $u=0$ we find
\bb
 \vartheta_3(1/2|\tau) &=& \vartheta_4(0|\tau)~,\\
 \vartheta_3(\tau/2|\tau) &=&  q^{-1/8}\,\vartheta_2(0|\tau)~.
\ee
Using the definition of  $\vartheta_2$-function
$$
  \vartheta_3(\tau/2|\tau) =  q^{-1/8}\,
  \sum_{m=-\infty}^{+\infty} \, q^{(m+1/2)^2/2} \ne 0~.
$$
Also, by consulting the results of problem \ref{item:MI6} of this
chapter,
one can immediately see that $\vartheta_4(0|\tau)$ is non-zero,
and thus $\vartheta_3(1/2|\tau)$ is   non-zero.
}%
    explicitly from the definition of $\vartheta_3$:
\bb
  \vartheta_3\lp u+{1\over2}\tau+{1\over2}\rp &=& q^{-1/8}\,
  \sum_{m=-\infty}^{+\infty} (-)^m z^m q^{(m+1/2)^2/2}
\ee
which, for $u=0$, gives
\bb
  \vartheta_3\lp {1\over2}\tau+{1\over2}\rp &=& q^{-1/8}\,
  \sum_{m=-\infty}^{+\infty}  (-1)^m q^{(m+1/2)^2/2}\\
  &=& q^{-1/8}\,\sum_{m=-\infty}^{-1} (-1)^m q^{(m+1/2)^2/2}
      +q^{-1/8}\,\sum_{m=0}^{+\infty}  (-1)^m q^{(m+1/2)^2/2}\\
  &=& q^{-1/8}\,\sum_{n=1}^{n=+\infty} (-1)^n q^{(n-1/2)^2/2}
      +q^{-1/8}\,\sum_{m=0}^{+\infty}  (-1)^m q^{(m+1/2)^2/2}\\
  &{n=l+1\atop=}& -q^{-1/8}\,\sum_{l=0}^{l=+\infty} (-1)^l
q^{(l+1/2)^2/2}
      +q^{-1/8}\,\sum_{m=0}^{+\infty}  (-1)^m q^{(m+1/2)^2/2}\\
  &=& 0~.
\ee

\separator

\item
(a) The grand canonical partition function of the system is
$$
     \Xi = \sum_{\rm configurations}\, e^{-\beta E +\mu\beta N}~,
$$
where $N$ is the total fermion number $N=N_f-N_{\bar f}$, which
takes values
in $\BZ$. If $n_k$ and $\bar n_k$ are the occupation numbers for the
$k$-th fermion and antifermion modes, respectively, then
$$
  N_f\= \sum_k\, n_k~,~~~{\rm and}~~~
  N_{\bar f}\= \sum_k\, \bar n_k~.
$$
The Hamiltonian of the system is
$$
     H= E_0 \, \sum_k\,\lp k-{1\over 2}\rp\lp n_k+\bar n_k\rp ~.
$$
We will
use the symbols $w$ and $q$ for the fugacity and `fundamental'
 Boltzman weight, respectively, i.e.,
$$
    w\eq e^{\beta\mu} ~~~{\rm and}~~~ q\eq e^{-\beta E_0}~.
$$

We now calculate the grand canonical partition function in two distinct ways.
First, we write the grand canoncial partition function 
as a simple infinite product as follows:
\beqn
  \Xi &=& \sum_{\rm configurations}\, 
  q^{\sum\limits_k\,\lp k-{1\over 2}\rp\lp n_k+\bar n_k\rp}
  w^{\sum\limits_k\,\lp n_k+\bar n_k\rp}\nonumber\\
   &=& \sum_{\rm configurations}\,
  q^{\sum\limits_k\, (k-1/2)\, n_k }\,
  w^{\sum\limits_k\, n_k}~
  q^{\sum\limits_k\, (k-1/2)\, \bar n_k }\,
  w^{\sum\limits_k\, \bar n_k}\nonumber\\
  &=& \sum_{ \{n_k\},\{\bar n_k\} }\,
  \prod_k\lp q^{k-1/2} w\rp^{n_k}~
  \prod_k\lp q^{k-1/2} w^{-1}\rp^{\bar n_k}\nonumber\\
  &=& \prod_{k=1}^{+\infty} 
   \sum_{n_k=0}^1 \lp q^{k-1/2} w\rp^{n_k}~
  \sum_{\bar n_k=0}^1\lp q^{k-1/2} w^{-1}\rp^{\bar n_k}\nonumber\\
  &=& \prod_{k=1}^{+\infty} 
  \lp 1+ q^{k-1/2} w\rp  ~
  \lp 1+ q^{k-1/2} w^{-1}\rp ~.
\label{eq:Xi1}
\eeqn

The second way we calculate the grand
canoncial partition function begins by expressing $\Xi$
as an infinite sum involving canonical partition functions.
We see that
\beqn
   \Xi &=& \sum_{\rm configurations}\, e^{-\beta E} w^N \nonumber \\
       &=& \sum_{N=-\infty}^{+\infty} w^N \, 
   {{\displaystyle\sum\nolimits^\prime}\atop\mbox{\footnotesize configurations}}
           \, e^{-\beta E} \nonumber\\
       &\equiv& \sum_{N=-\infty}^{+\infty} w^N \, Z_N(q)~, 
\label{eq:Xi2}
\eeqn
where
$Z_N(q)$ is the partition function of the system for a fixed
fermion number $N$.
At fixed $N$,  the energy $E$ of the system is a sum of the energy 
$E_N$ for
the first $N$ levels filled plus the energy $\Delta E$ for zero
fermion number excitations. The energy $E_n$ is given by
$$
  E_N \= E_0\,  \sum_{k=1}^N \, (k-1/2) = {E_0 \, N^2\over 2}~.
$$
The zero fermion number excitations have equal numbers of fermions and
anti-fermions, i.e., $n_k=\bar n_k$. In terms of $n_k$, then, we have
$$
   \Delta E \= E_0\, \sum_{k=1}^m\, k\, n_k ~,
$$
up to some level $m$. However, there can be excitations for all possible values
of $m$, thus implying that
\bb
  Z_N(q) &=&
 \sum_{\rm excitations}   \, e^{-\beta E_N} \, e^{-\beta\Delta E}\\
  &=& q^{N^2\over 2}\, \sum_{m=0}^{+\infty}\, q^{\sum\limits_{k=1}^m\, k\, n_k} \\
  &=& q^{N^2\over 2}\, \prod_{n=1}^{+\infty} 
  \lp \sum_{k_n=0}^{+\infty} \, q^{n\, k_n} \rp \\
  &=& q^{N^2\over 2}\, \prod_{n=1}^{+\infty} {1\over 1-q^n}~.
\ee
Substituting this in \calle{eq:Xi2}, we get
$$
       \Xi\= \prod_{n=1}^{+\infty} {1\over 1-q^n}~
     \sum_{N=-\infty}^{+\infty} w^N \,
       q^{N^2\over 2}~.
$$
Combining this result with \calle{eq:Xi1} proves Jacobi's triple identity.

(b) For $P(q,w)$ we have
\bb
  P(q,qw) &=& \prod_{n=1}^{+\infty}\, (1+w q^{n+1-1/2}) (1+w^{-1} q^{n-1-1/2}) 
     \\
          &=& \prod_{n=1}^{+\infty}\, (1+w q^{n+1-1/2}) 
              \prod_{n=1}^{+\infty}\, (1+w^{-1} q^{n-1-1/2}) ~.
\ee
In the first product, we make the change of variables $k=n+1$,
while
in the second product, we make the change of variables $l=n-1$.
This yields
\bb
  P(q,qw)  
          &=& \prod_{k=2}^{+\infty}\, (1+w q^{k-1/2}) 
              \prod_{l=0}^{+\infty}\, (1+w^{-1} q^{l-1/2}) \\
          &=& {1+w^{-1}q^{-1/2}\over 1+wq^{1/2}}\,
              \prod_{k=1}^{+\infty}\, (1+w q^{k-1/2}) 
              \prod_{l=1}^{+\infty}\, (1+w^{-1} q^{l-1/2}) \\
          &=& {1+w^{-1}q^{-1/2}\over 1+wq^{1/2}}\, P(q,w)
          = q^{-1/2}w^{-1}\, P(q,w)~.
\ee
  
Similarly, for $Q(q,w)$, we find
\bb
  Q(q,qw) &=& {1\over\prod_{n=1}^{+\infty}(1-q^n)} \,
              \sum_{l=-\infty}^{+\infty}\, q^{l^2/2}\, q^l \, w^l \\
          &=& {1\over\prod_{n=1}^{+\infty}(1-q^n)} \,
               q^{-1/2}\, \sum_{l=-\infty}^{+\infty}\, q^{(l+1)^2/2}\,  w^l ~.
\ee
Making the change of variables $k=l+1$, we finally arrive at the
desired result, namely
\bb
  Q(q,qw) 
          &=& q^{-1/2}\, {1\over\prod_{n=1}^{+\infty}(1-q^n)} \,
               \sum_{k=-\infty}^{+\infty}\, q^{k^2/2}\,  w^{k-1}\\
          &=& q^{-1/2}w^{-1}\,Q(q,w)~.
\ee
This last result shows that
$$
    P(q,w)  = {\rm const}\, Q(q,w)  ~.
$$
To complete the proof, we must calculate this constant. This is simplified
by comparing the coefficients of  $w^0$.  

To this end, we first make a small digression on partitions.
Any partition of $n$ into parts $n=p_1+p_2+\dots+p_n$,
$0\le p_1\le p_2\le \dots \le p_n\le n$ can be rearranged in a
matrix 
\beq
\label{p-graph}
   \lp\matrix{ a_1 & a_2 & \dots & a_r \cr
               b_1 & b_2 & \dots & br  \cr}\rp~,
\eeq
called the {\bf Frobenius symbol}, such that
$$
    0 \le a_1 < a_2 < \dots a_r ~, ~~~
    0 \le b_1 < b_2 < \dots b_r ~, ~~~
    n = r + \sum_{i=1}^r a_i +\sum_{i=1}^r b_r~. 
$$
To understand how this map can be created, let us examine a specific example.
Consider the following partition of the number  20:
$$
   20 = 5 + 5 + 3 + 3 + 2 + 1 + 1 ~.
$$
This can be represented graphically:
$$
  \matrix{ \bullet & \bullet & \bullet & \bullet & \bullet \cr
           \bullet & \bullet & \bullet & \bullet & \bullet \cr
           \bullet & \bullet & \bullet &         &         \cr
           \bullet & \bullet & \bullet &         &         \cr
           \bullet &         &         &         &         \cr
           \bullet &         &         &         &         \cr
           \bullet &         &         &         &         \cr}
$$
Notice that the diagonal has $r=3$ dots. We erase them. Then the rows
to the right of the diagonal give $a_1$, $a_2$, and $a_3$,
and the columns to the left
of the diagonal give $b_1$, $b_2$, and $b_3$:
$$
  \matrix{   & * & * & * & * & & \leftarrow & a_1\cr
           \star &   & * & * & * & & \leftarrow & a_2\cr
           \star & \star &   &   &   & & \leftarrow & a_3\cr
           \star & \star & \star &   &   & \cr
           \star & \star &   &   &   & \cr
           \star &   &   &   &   & \cr
           \star &   &   &   &   & \cr
             &   &   &   &   & \cr
           \uparrow  & \uparrow  & \uparrow   &   &   & \cr
           b_1 & b_2  & b_3  &   &   & \cr}
$$
That is, we have rewritten the partition as
$$
     20 = 3 + ( 4 + 3 + 0) + (6 + 3 + 1) ~,
$$
giving us the Frobenius symbol
$$
  \lp\matrix{  4  & 3  & 0 \cr
               6  & 3  & 1\cr}\rp~.
$$
For each partition of $n$, there is exactly one Frobenius symbol.
Turning this around, for each $n$, we can construct $p(n)$ 
Frobenius symbols, where  as usual
$$
  \sum_{n=0}^{+\infty} \, p(n)\, q^n = \prod_{n=1}^{+\infty}\, 
  {1\over 1-q^n}~.
$$

Now, returning to calculating the coefficient of the constant term
in $P(q,w)$, we observe that we get  a contribution with 
weight 1 every time we have a product of $r$ terms $wq^{n-1/2}$ with
$r$ terms $w^{-1}q^{m-1/2}$:
$$
     q^{-r+n_1+n_2+\dots+n_r+m_1+m_2+\dots+m_r}~.
$$
This term can be mapped to the Frobenius symbol
\beq
   \lp\matrix{ n_1-1 & n_2-1 & \dots & n_r-1 \cr
               m_1-1 & m_2-1 & \dots & m_r-1  \cr}\rp~,
\eeq
since the entries satisfy all the required properties.
Therefore, the sum of contributions is exactly the number of
Frobenius symbols. This determines that const=1, so that
$P(q,\omega)=Q(q,\omega)$, and the
proof is complete.

\separator

\item

(a) In Jacobi's triple identity, we use $w=-p^{-1/2}$ and $q=p^3$
to obtain
$$
   \prod_{n=1}^{+\infty}\, (1-p^{3n}) \, (1-p^{3n-2}) \, (1-p^{3n-1})
   \= \sum_{n=-\infty}^{+\infty}\, (-1)^n\, p^{-n/2} \, p^{3n^2/2}~.
$$
Since for every $k\in\BZ$, there exists $n\in\BZ$ such that
$$
   k=3n~,~~~
   k=3n-1~,~~~\mbox{or}~~~
   k=3n-2~,
$$
when $n$ runs over all $\BZ$, $k$ also runs over all $\BZ$. Therefore,
the product in the previous equation is exactly
$$
   \prod_{k=1}^{+\infty} \, (1-q^n)~.
$$
Putting everything together, we arrive at Euler's pentagonal
number theorem.

(b) For $p=2$ and $q=3$, the central charge is $c=0$, and the model
has only the identity operator. Therefore, $\chi_{1,1}=1$. 
From the Rocha-Caridi formula
$$
   \chi_{1,1} = {1\over\prod\limits_{n=1}^{+\infty} (1-q^n)}\,
                 \sum_{k=-\infty}^{+\infty}\,
                \lp q^{6k^2+k} - q^{6k^2+5k+1} \rp~.
$$
From this we conclude that
\bb
   \prod_{n=1}^{+\infty} \, (1-q^n) &=&
                 \sum_{k=-\infty}^{+\infty}\,
                \lp q^{6k^2+k} - q^{6k^2+5k+1} \rp \\
    &=&
                 \sum_{k=-\infty}^{+\infty}\,
                \lb (-1)^{2k}\, q^{3 (2k)^2+(2k)\over 2} 
                   + (-1)^{2k-1}\, q^{3 (2k-1)^2+(2k-1)\over 2} \rb\\
    &=&        \sum_{l=-\infty}^{+\infty}\,
                 (-1)^l\, q^{3 l^2+l\over 2}~,
\ee
where to go from the first equality to the second, we just rewrite the
various quantities in an equivalent form, and to go from the second 
equality to the third, we just notice that the two terms are the
even and odd partial sums of one sum.
Setting $l=-n$ in the last equation, we find again the advertised relation.

\separator

\item

(a) In Jacobi's triple identity, we set $w=z$ or $w=-z$. 
In these two cases, we find, respectively,
\bb
   \vartheta_3(u|\tau) &=& \prod_{n=1}^{+\infty}
        (1-q^n)(1+zq^{n-1/2})(1+z^{-1}q^{n-1/2})~,~~~{\rm and} \\
   \vartheta_4(u|\tau) &=& \prod_{n=1}^{+\infty}
        (1-q^n)(1-zq^{n-1/2})(1-z^{-1}q^{n-1/2})~.
\ee
For $\vartheta_2$, we set $w=zq^{1/2}$ in Jacobi's triple identity,
which gives 
\bb
   \sum_{n=-\infty}^{+\infty} z^n q^{n^2+n\over 2} &=& \prod_{n=1}^{+\infty}
        (1-q^n)(1+zq^n)(1+z^{-1}q^{n-1})~.
\ee
The l.h.s. can be rearranged to give
\bb
   \sum_{n=-\infty}^{+\infty} z^n q^{n^2+n\over 2} &=& 
   \sum_{n=-\infty}^{+\infty} z^n q^{ {(n+1/2)^2\over 2}} \, q^{-1/8}\\
  &=& z^{-1/2}q^{-1/8}\sum_{n=-\infty}^{+\infty} z^{n+1/2} q^{(n+1/2)^2\over 2}
    \\
     &=& z^{-1/2}q^{-1/8}\, \vartheta_2(u|\tau)~,
\ee
and therefore
\bb
   \vartheta_2(u|\tau)\= z^{1/2}q^{1/8}\,
    \prod_{n=1}^{+\infty}(1-q^n)(1+zq^n)(1+z^{-1}q^{n-1})~.
\ee
In the same way, setting $w=-zq^{1/2}$ in Jacobi's trible identity,
we find that
\bb
   \vartheta_1(u|\tau)\= z^{1/2}q^{1/8}\,
    \prod_{n=1}^{+\infty}(1-q^n)(1-zq^n)(1-z^{-1}q^{n-1})~.
\ee

(b) From the results of part (a), we have 
\beqn
   \theta_2(\tau) &=& q^{1/8}\, \prod_{n=1}^{+\infty}(1-q^n)(1+q^n)(1+q^{n-1})
                      \nonumber \\
                  &=&  2q^{1/8}\, \prod_{n=1}^{+\infty}(1-q^n)(1+q^n)^2~;
                       \label{eq:MI16} \\
   \theta_3(\tau) &=& \prod_{n=1}^{+\infty}(1-q^n)(1+q^{n-1/2})(1+q^{n-1/2})
                      \nonumber \\
                  &=& \prod_{n=1}^{+\infty} (1-q^n)(1+q^{n-1/2})^2~;
             ~~~ {\rm and}
                       \label{eq:MI17} \\
   \theta_4(\tau) &=& \prod_{n=1}^{+\infty}(1-q^n)(1-q^{n-1/2})(1-q^{n-1/2})
                      \nonumber \\
                  &=& \prod_{n=1}^{+\infty}  (1-q^n)(1-q^{n-1/2})^2 ~.
                       \label{eq:MI18} 
\eeqn
Multiplying these results together, we find
\bb
   \theta_2(\tau) \theta_3(\tau) \theta_4(\tau)  &=& 2q^{1/8}\,
   \lb \prod_{n=1}^{+\infty}(1-q^n)\rb^3 \, 
   \lb \prod_{n=1}^{+\infty}(1+q^n)(1+q^{n-1/2})(1-q^{n-1/2})\rb^2 \\
   &=& 2\, \eta^3(\tau)\, f^2(q)~,
\ee
where
\beqn
  f(q) &=& \prod_{n=1}^{+\infty}(1+q^n)(1+q^{n-1/2})(1-q^{n-1/2})\nonumber \\
       &=& \prod_{n=1}^{+\infty}(1+q^n)(1-q^{2n-1})~.
\label{eq:MI11}
\eeqn
To complete the proof, we must show that $f(q)=1$. Let us break the 
infinite product of terms $1+q^n$ into terms with odd and even powers.
Then,
\bb
  f(q) &=& \prod_{n=1}^{+\infty}(1+q^{2n})(1+q^{2n-1})(1-q^{2n-1})\\ 
       &=& \prod_{n=1}^{+\infty}(1+q^{2n})(1-q^{2(2n-1)}) \\
       &{\calle{eq:MI11}\atop=}& f(q^2)~.
\ee
From this identity we can compute the derivatives of $f(q)$ at $q=0$.
These are
\bb
   f'(q)    &=& 2q f'(q^2)~,\\
   f''(q)   &=& 2 f'(q^2) + 4q^2 f''(q^2)~,\\
   f'''(q)   &=& 4q f''(q^2) + 8q f''(q^2) + 8q^3 f'''(q^2)~,\\
  \dots\dots
\ee
from which we see that $0=f'(0)=f''(0)=\dots$. Then from the Taylor expansion 
$$
  f(q)\=f(0)+ f'(0) \, q + {f''(0)\over 2!}\, q^2 + \cdots ~,
$$
we find $f(q)=f(0)=1$, which completes our proof.

\separator

\item
For the canonical homology basis $\{a_1,a_2,b_1,b_2\}$,
we consider the Dehn twists along the cycles $a_1$, $a_2$, 
$b_1$, $b_2$, and $a_1^{-1}a_2$. 

\begin{figure}[htb]
\begin{center}
\includegraphics[height=3cm]{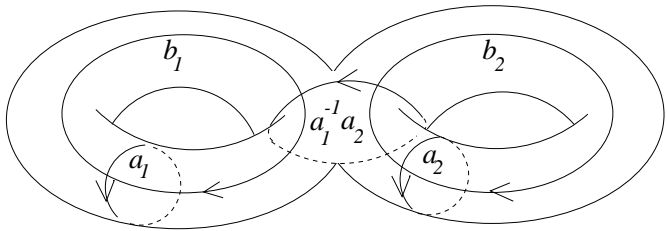}
\end{center}
\caption{The cycles for a genus 2 Riemann surface.}
\label{fig:cyclesT2}
\end{figure}

After twisting along the cycle $a_1$ the new cycles will
look as in figure \ref{fig:Dehna1}.

\begin{figure}[htb]
\begin{center}
\includegraphics[height=3cm]{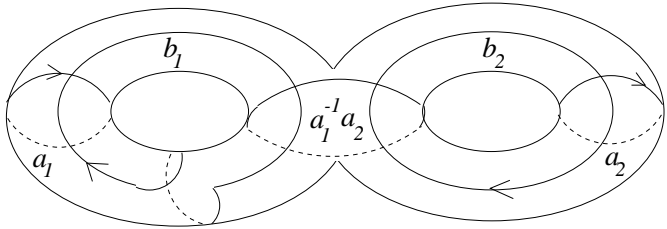}
\end{center}
\caption{A Dehn twist around the $a_1$ cycle.}
\label{fig:Dehna1}
\end{figure}

From this we conclude that
$$
  M(D_{a_1}) \=
  \lb\matrix{  1 & 0 & 0 & 0 \cr
               0 & 1 & 0 & 0 \cr
               1 & 0 & 1 & 0 \cr
               0 & 0 & 0 & 1 \cr }\rb ~.
$$
Similarly, we can find the following matrices:
\bb
  M(D_{b_1}) &=&
  \lb\matrix{  1 & 0 & 1 & 0 \cr
               0 & 1 & 0 & 0 \cr
               0 & 0 & 1 & 0 \cr
               0 & 0 & 0 & 1 \cr }\rb ~, ~~~~~
  M(D_{b_2}) \=
  \lb\matrix{  1 & 0 & 0 & 0 \cr
               0 & 1 & 0 & 1 \cr
               0 & 0 & 1 & 0 \cr
               0 & 0 & 0 & 1 \cr }\rb ~, \\
  M(D_{a_2}) &=&
  \lb\matrix{  1 & 0 & 0 & 0 \cr
               0 & 1 & 0 & 0 \cr
               0 & 0 & 1 & 0 \cr
               0 & 1 & 0 & 1 \cr }\rb ~, ~~~~~
  M(D_{a_1^{-1}a_2}) \=
  \lb\matrix{  1 & 0 & 0 & 0 \cr
               0 & 1 & 0 & 1 \cr
               -1 &1 & 1 & 0 \cr
               1 & -1 & 0 & 1 \cr }\rb ~.
\ee

\separator

\item

The trace 
$$
   Z\= q^{-c/24}\, \overline q^{-c/24}\,
       \mbox{tr}\lp q^{L_0}\overline q^{\overline L_0}\rp
$$
is a product of independent contributions from different oscillators
$\alpha_{-n}$ and momentum $p$. The contribution from the states
$$
    \la \ket{k,n}\equiv\alpha_{-n}^k\ket{\emptyset}~, ~k=0,1,2,\dots \ra~,
$$
which have 
$$ 
   L_0\ket{k,n}\= kn \, \ket{k,n}~,
$$
is
$$
  \sum_{k=0}^{+\infty} q^{kn}\= 1+q^n+q^{2n}+\dots \= {1\over 1-q^n}~.
$$
 The contribution from the states
$$
    \la \ket{p},~ p\in\BR \ra~,
$$
which have 
$$
   L_0\ket{p}\= {p^2\over 2}\, \ket{p}~,
$$
is
$$
  \int_{-\infty}^{+\infty} dp\, q^{p^2/2}\,\overline q^{p^2/2}
  \=  \int_{-\infty}^{+\infty} dp\, e^{-2\pi\mbox{Im}\tau\, p^2}\=
  \sqrt{1\over 2\mbox{Im}\tau}~.
$$
Therefore, putting all the factors together, the partition function
for the  free boson is
\bb
  Z\={1\over\sqrt{2\mbox{Im}\tau}}\, {1\over\eta(q)\eta(\overline q)}~,
\ee
where $\eta(q)$
is the Dedekind $\eta$-function.
Its modular properties (see the appendix) are
\bb
  \eta(\tau+1) &=& e^{i\pi/12}\, \eta(\tau)~,\\
  \eta(-1/\tau) &=& \sqrt{-i\tau} \, \eta(\tau)~.
\ee
Therefore, under the $T$-transformation,
\bb  
  Z \mapsto Z' &=& {1\over\sqrt{2\mbox{Im}(\tau+1)}}\,
               {1\over\eta(\tau+1)\eta(\overline \tau+1)} \\
            &=& {1\over\sqrt{2\mbox{Im}\tau}}\,
               {1\over  e^{i\pi/12}\eta(\tau)\,
                         e^{-i\pi/12}\eta(\overline \tau)} \\
            &=& {1\over\sqrt{2\mbox{Im}\tau}}\,
                {1\over\eta(\tau)\eta(\overline \tau)} \= Z~.
\ee
Likewise, under the $S$-transformation,
\bb  
  Z \mapsto Z' &=& {1\over\sqrt{2\mbox{Im}{-1\over\tau}}}\,
               {1\over\eta(-1/\tau)\eta(-1/\overline \tau)} \\
            &=& {1\over\sqrt{2\mbox{Im}{-1\over\tau}}}\,
               {1\over\eta(-1/\tau)\eta(-1/\overline \tau)} \\
            &=& {1\over\sqrt{\mbox{Im}\tau\over \tau\overline \tau}}\,
               {1\over \sqrt{-i\tau}\eta(\tau)\,
               \sqrt{i\overline\tau}\eta(\overline \tau)}\= Z~.
\ee

\footnotesize
{\bf APPENDIX}

From the definition of the Dedekind $\eta$-function
$$
   \eta(q)\eq q^{1/24}\, \prod_{n=1}^{+\infty} (1-q^n)~,
$$
we see that under a $T$-transformation, 
$$
   \eta(q)\mapsto q^{1/24} e^{i\pi/12}\, 
          \prod_{n=1}^{+\infty} (1- e^{2\pi i n}q^n) \=
         q^{1/24} e^{i\pi/12}\, \prod_{n=1}^{+\infty} (1-q^n)
         \= e^{i\pi/12}\,\eta(q)~.
$$

To determine how $\eta$ behaves under an $S$-transformation, we use the
identity
$$
    \eta^3(\tau)\={1\over2}\,\theta_2(\tau)\theta_3(\tau)\theta_4(\tau)~.
$$
From the properties of the $\vartheta$ function, it is easy to show that
\beqn
   \theta_2(-1/\tau) &=& \sqrt{-i\tau}\, \theta_4(\tau)~,
   \label{eq:MI12} \\
   \theta_3(-1/\tau) &=& \sqrt{-i\tau}\, \theta_3(\tau)~,~~~{\rm and}
   \label{eq:MI13} \\
   \theta_4(-1/\tau) &=& \sqrt{-i\tau}\, \theta_2(\tau)~.
   \label{eq:MI14} 
\eeqn
Thus we see that
\beq
   \eta(-1/\tau)\=  \sqrt{-i\tau}\,\eta(\tau)~.
   \label{eq:MI15} 
\eeq

\normalsize

\separator

\item

The compactification of the boson on a circle of radius $R$ has two 
consequences. First, the momenta $p$ are quantized, restricted to
the values
$$
   p\={m\over R}~, ~~~m\in\BZ~.
$$
Second, there are new winding states, because of the boundary conditions
$$
   \Phi(\sigma+2\pi) ~\sim~ \Phi(\sigma)+2\pi R\, w~, ~~~w\in\BZ~.
$$
The boson decomposes into two chiral components:
\bb
  \phi(z) &=& x-i p \, \ln z +i\sum_{n\ne 0} {\alpha_n\over n}\, z^{-n}~~~
   {\rm and}\\
  \overline\phi(\overline z) &=& \overline x-i \overline p\, \ln\overline z 
      +i\sum_{n\ne 0}{\overline\alpha_n\over n}\,\overline z^{-n}~.
\ee
The eigenvalues of $p$ and $\overline p$ are, respectively,
$$
  p\={m\over R}+{w R\over 2}~~~{\rm and}~~~
  \overline p\={m\over R}-{w R\over 2}~.
$$
We see that the contribution of the bosonic modes
in the present case is identical to the contribution of these
modes in the previous problem. However, the integral over
the momentum states becomes a discrete sum:
$$
  \int_{-\infty}^{+\infty} dp\, q^{p^2/2}\,\overline q^{p^2/2}
  \longrightarrow
  \sum_{m,w}  q^{\lp{m\over R}+{w R\over 2}\rp^2/2}
           \, \overline q^{\lp{m\over R}-{w R\over 2}\rp^2/2}~.
$$
Therefore,
$$
  Z\= \sqrt{2\,\mbox{Im}\tau} Z_{\mbox{\footnotesize boson}}\,
      \sum_{m,w}  q^{\lp{m\over R}+{w R\over 2}\rp^2/2}
           \, \overline q^{\lp{m\over R}-{w R\over 2}\rp^2/2}~,
$$
where $Z_{\mbox{\footnotesize boson}}$ stands for the partition function of the
uncompactified boson.

Under the $T$-transformation
\bb
  Z \mapsto Z' &=& \sqrt{2\,\mbox{Im}\tau} Z_{\mbox{\footnotesize boson}}\,
      \sum_{m,w}  q^{\lp{m\over R}+{w R\over 2}\rp^2/2}
           \, \overline q^{\lp{m\over R}-{w R\over 2}\rp^2/2}
                  e^{-i\pi\lp{m\over R}+{w R\over 2}\rp/2}\, 
                  e^{i\pi\lp{m\over R}-{w R\over 2}\rp/2} \\
               &=& \sqrt{2\,\mbox{Im}\tau} Z_{\mbox{\footnotesize boson}}\,
      \sum_{m,w}  q^{\lp{m\over R}+{w R\over 2}\rp^2/2}
           \, \overline q^{\lp{m\over R}-{w R\over 2}\rp^2/2}
               \,   e^{-2i\pi mw}\\
               &=& \sqrt{2\,\mbox{Im}\tau} Z_{\mbox{\footnotesize boson}}\,
      \sum_{m,w}  q^{\lp{m\over R}+{w R\over 2}\rp^2/2}
           \, \overline q^{\lp{m\over R}-{w R\over 2}\rp^2/2} \= Z~.
\ee
               
To study the transformation of $Z$ under the $S$-transformation, we 
first rewrite $Z$ in an equivalent form using the Poisson resummation
formula.  We see that we can write $Z$ as
$$
  Z\= \sqrt{2\,\mbox{Im}\tau} Z_{\mbox{\footnotesize boson}}\,
      \sum_{m,w}  f(m,w)
   \= \sqrt{2\,\mbox{Im}\tau} Z_{\mbox{\footnotesize boson}}\,
      \sum_{\vec x} \tilde f(\vec x)~,
$$
where
\bb
     f(m,w) &\equiv& q^{\lp{m\over R}+{w R\over 2}\rp/2} \\ 
            &=& \exp\lb-i\pi\tau\lp{m^2\over R^2}+{w^2 R^2\over4}+wm\rp
            +i\pi\overline\tau\lp{m^2\over R^2}+{w^2 R^2\over4}-wm\rp\rb\\
            &=& \exp\lb-i\pi \lp 
                (\tau-\overline\tau){m^2\over R^2}
                +(\tau-\overline\tau){w^2 R^2\over4}
                +(\tau+\overline\tau)wm\rp \rb \\
            &=& \exp\lb-i\pi \lp 
                2i\mbox{Im}\tau\,{m^2\over R^2}
                +i\mbox{Im}\tau\,{w^2 R^2\over2}
                +2\mbox{Re}\tau\,wm\rp \rb \\
            &=& e^{-i\pi \vec m^T A(\tau) \vec m}~,         
\ee
and
\bb
   \tilde f(\vec x) &=& \int d\vec x e^{-2\pi i \vec x\cdot \vec m}\,
                         f(\vec m)
     \= \int d\vec x e^{-2\pi i \vec x\cdot \vec m}\, 
           e^{-\pi\vec m^T A\vec m} \\
     &=& {\pi\over\sqrt{\det(\pi A)}}\, 
        e^{-{1\over4}(2\pi i\vec x)^T (\pi A)^{-1} (2\pi i\vec x)}\\
     &=& {1\over\tau\overline\tau}\, e^{\pi \vec x^T A^{-1}(\tau)\vec x}~,
\ee
where, for convenience, we have defined
\bb
                \vec m\=\lb\matrix{m\cr w\cr}\rb~~~{\rm and}~~~
           A(\tau)\=\lb\matrix{
                            {2\mbox{Im}\tau\over R^2}& -i\mbox{Re}\tau\cr
                            -i\mbox{Re}\tau & {R^2\mbox{Im}\tau\over2}\cr
                              }\rb~,
\ee
and
\bb
       A^{-1}(\tau)\=\lb\matrix{ {R^2\mbox{Im}\tau\over2}& i\mbox{Re}\tau\cr
                                i\mbox{Re}\tau & {2\mbox{Im}\tau\over R^2}\cr}
                    \rb~.
\ee                    
Therefore, the partition function is
\bb
  Z \= {\sqrt{2\,\mbox{Im}\tau}\over\sqrt{\tau\overline\tau}}\,
       Z_{\mbox{\footnotesize boson}}(\tau)\,
      \sum_{\vec x}  e^{\pi \vec x^T A^{-1}(\tau)\vec x}~.
\ee
Under an $S$-transformation, we now see that
\bb
   Z\mapsto Z' &=& \sqrt{2\,\mbox{Im}{-1\over\tau}} \,\sqrt{\tau\overline\tau}\,
       Z_{\mbox{\footnotesize boson}}(\tau)\,
      \sum_{\vec x}  e^{\pi \vec x^T A^{-1}(-1/\tau)\vec x}\\
    &=& {\sqrt{2\,\mbox{Im}\tau}\over\sqrt{\tau\overline\tau}} \,
        \sqrt{\tau\overline\tau}\,
       Z_{\mbox{\footnotesize boson}}(\tau)\,
      \sum_{\vec x}  e^{\pi \vec x^T P A^{-1}(\tau)P\vec x}\\
   &=& \sqrt{2\,\mbox{Im}\tau} \,
       Z_{\mbox{\footnotesize boson}}(\tau)\,
      \sum_{\vec x}  e^{\pi \vec x^T P A^{-1}(\tau)P\vec x}~,
\ee
where we have taken advantage of the fact that
\bb
    A^{-1}(-1/\tau) &=&
   \lb\matrix{ {R^2\mbox{Im}\tau\over2}& 
               -i\mbox{Re}\tau\cr
               -i\mbox{Re}\tau &
                {2\mbox{Im}\tau\over  R^2}\cr}\rb \\
      &=& P A^{-1}(\tau)P~, ~~~~~
     P\=\lb\matrix{0&1\cr1&0\cr}\rb~.
\ee
If we now make the change of variables $\vec n\=P\vec x$ (which amounts
simply
to a permutation of the components of $\vec x$), then at last we
see that
$$
  Z\mapsto Z'\= \sqrt{2\,\mbox{Im}\tau} \,
       Z_{\mbox{\footnotesize boson}}(\tau)\,
      \sum_{\vec n}  e^{\pi \vec n^T  A^{-1}(\tau)\vec n}\= Z~.
$$

\separator

\item
\def\zero{\chi_0(\tau)}
\def\one{\chi_{1 \over 2}(\tau)}
\def\two{\chi_{1 \over 16}(\tau)}
\def\tthree{\sqrt{\theta_3(\tau) \over \eta(\tau)}}
\def\ttwo{\sqrt{\theta_2(\tau) \over \eta(\tau)}}
\def\tfour{\sqrt{\theta_4(\tau) \over \eta(\tau)}}
\def\sq{{{(\tau)~\to~\sqrt{-i\tau}\,}}}
\def\h{{1 \over 2}}
\def\r{{1 \over \sqrt{2}}}

(a) The map 
$$
   w ~\mapsto~ z\=e^w
$$
maps the $w$-cylinder to the $z$-plane. The fermion on the cylinder is
$$
   \psi_{\mbox{cyl}}\= z^{1/2}\, \psi(z)\= \sum_n\, e^{-nw}\, \psi_n~.
$$
After going around the cycle of the cylinder, $w\mapsto w+2\pi i$, the fermion
may return to its initial value or take an opposite value. Thus $n$ can take
values either in $\BZ$ (the P sector, with periodic boundary
conditions) or  in $\BZ+1/2$ (the A sector, with antiperiodic boundary
conditions).
The $L_0$ operator is given by 
$$
  L_0 \= \sum_n n\psi_{-n}\psi_n + c_0~,
$$
where 
$c_0=1/24$ in the P sector and $c_0=-1/48$ in the A sector.
(For the derivation of these
values, see Exercise \ref{item:FSFT1} of Chapter \ref{ch:FSFT}.)

Starting with the vacuum state $\ket{\emptyset}$, one can build a 
tower of states for the free fermion using the operators $\psi_{-n},~n>0$:
$$
  \psi_{-n_k}\dots\psi_{-n_1}\ket{\emptyset}~.
$$
The calculation of $\mbox{tr}q^{L_0}$ is elementary. First we notice that
\bb
 \mbox{tr}q^{L_0}  &=& q^{c_0}\,\mbox{tr}\lp q^{\sum_{n>0} n\psi_{-n}\psi_n}\rp
  \\
                    &=& q^{c_0}\,\mbox{tr}\lp\bigotimes_{n>0}
                        q^{n\psi_{-n}\psi_n}\rp ~.
\ee
Now we notice that the subspace generated by the the $n$-mode has a
very simple basis, given by
\bb
  \la \ket{\emptyset},~\psi_{-n}\ket{\emptyset}\ra~.
\ee
As a result, the trace we compute above becomes
\bb
   \mbox{tr}q^{L_0}  &=& q^{c_0}\, \mbox{tr} \lp \bigotimes_{n>0}
                         \lb\matrix{1&0\cr 0& q^n\cr}\rb  \rp \\
                     &=& q^{c_0}\, \prod_{n>0}\mbox{tr}
                         \lb\matrix{ 1&0\cr 0& q^n\cr}\rb \\
                     &=& q^{c_0}\, \prod_{n>0} (1+q^n) ~.
\ee

There is a related trace that it is useful to
compute. The operator $(-1)^F$
measures the number of fermions mod 2 in a state (it takes the value $-1$
for a fermionic state, and $+1$ for a bosonic state). Repeating the previous 
calculation,
we find that
\bb
    \mbox{tr}\lp(-1)^Fq^{L_0}\rp  &=& q^{c_0}\, \prod_{n>0}\mbox{tr}
                         \lb\matrix{ 1&0\cr 0& -q^n\cr}\rb \\
                     &=& q^{c_0}\, \prod_{n>0} (1-q^n) ~.
\ee

The traces we have found do not give immediately the characters of 
the primary 
fields. Let us study the A sector first. Table \ref{table:MI1} gives
some of the states of this sector. We see that the representation is
actually the sum $\lb 0\rb\oplus\lb 1/2\rb$. The module $\lb 0\rb$
contains states with an even number of fermions, while the module
$\lb 1/2\rb$ contains states with an odd number of fermions. In other words,
the two modules can be separated by using the projection operators
$$
   P_\pm \= {1\pm (-1)^F\over   2}~.
$$
\begin{table}[h]
\begin{center}
\begin{tabular}{|c|c|}\hline\hline
 $L_0$ &  state \\ \hline\hline
0 & $\ket{\emptyset}$ \\ \hline
1/2 & $\psi_{-1/2}\ket{\emptyset}$ \\ \hline
3/2 & $\psi_{-3/2}\ket{\emptyset}$ \\ \hline
2 & $\psi_{-3/2}\psi_{-1/2}\ket{\emptyset}$ \\ \hline
5/2 & $\psi_{-5/2}\ket{\emptyset}$ \\ \hline
3 & $\psi_{-5/2}\psi_{-1/2}\ket{\emptyset}$ \\ \hline
7/2 & $\psi_{-7/2}\ket{\emptyset}$ \\ \hline
\end{tabular}
\caption{The first states of the A sector of the free fermion.}
\label{table:MI1}
\end{center}
\end{table}

Therefore  
\bb
  \chi_0 &=& \mbox{tr}_{\mbox{A}}\lp P_+ q^{L_0}\rp\\
         &=& {1\over2}\,\mbox{tr}_{\mbox{A}} q^{L_0} +
           {1\over2}\, \mbox{tr}_{\mbox{A}}\lp(-1)^Fq^{L_0}\rp \\
         &=& {1\over2}\,  q^{-1/48}\lp \prod_{n=1}^{+\infty}
             (1+q^{n-1/2})+\prod_{n=1}^{+\infty} (1-q^{n-1/2})\rp\\ 
         &=& {1\over2}\, \lb \sqrt{\theta_3(\tau)\over\eta(\tau)}
             +\sqrt{\theta_4(\tau)\over\eta(\tau)}\rb~,
\ee
and
\bb
  \chi_{1/2} &=& \mbox{tr}_{\mbox{A}}\lp P_- q^{L_0}\rp \\
         &=& {1\over2}\,\mbox{tr}_{\mbox{A}} q^{L_0} -
           {1\over2}\, \mbox{tr}_{\mbox{A}}\lp(-1)^Fq^{L_0}\rp \\
         &=&  q^{-1/48}\lp \prod_{n=1}^{+\infty}
        (1+q^{n-1/2})-\prod_{n=1}^{+\infty} (1-q^{n-1/2})\rp\\ 
         &=&  {1\over2}\, \lb \sqrt{\theta_3(\tau)\over\eta(\tau)}
             -\sqrt{\theta_4(\tau)\over\eta(\tau)}\rb~.
\ee

In the P sector, the anticommutation relations
\bb
   \la \psi_0,\psi_n\ra &=& 0~,~~~n\ne0~, \\
   \la \psi_0,\psi_0\ra &=& 1~,~~~{\rm and}~~~
   \la \psi_0, (-1)^F \ra \= 0~
\ee
require two degenerate ground states $\ket{1/16;\pm}$. The modules
built over them have, obviously, equal numbers of states, and therefore we can
conclude immediately that
\bb
  \chi_{1/16} &=& \mbox{tr}_{\mbox{P}}q^{L_0}\\
         &=&  q^{1/24} \prod_{n=1}^{+\infty} (1-q^n)\\
         &=& {1\over\sqrt{2}}\, \sqrt{\theta_2(\tau)\over\eta(\tau)}~.
\ee

(b) By definition, when one performs the modular transformation
$$ 
        S \, : \, \tau~ \mapsto ~\tau' \=-~{1 \over \tau}~,
$$ 
the characters transform according to a matrix $\bfS$ as
$$ 
     \chi_i(\tau)~\mapsto~\chi_i(\tau') \=\bfS_{ij} \, \chi_j(\tau)~.
$$

Using the modular properties of the $\vartheta$-function
and $\eta$-function, given 
in equations \calle{eq:MI12}-\calle{eq:MI15}, we
find
\bb
 \zero  &\mapsto& \chi_0(\tau') \=  \\
          &=& {1 \over 2} \lp \, \tthree ~+~ \ttwo \, \rp  \\
          &=& {1 \over 2} \, \lb \, \ttwo  ~+~{1 \over 2}
              \lp \tthree~+~ \tfour \, \rp  ~+~ {1 \over 2} \,
              \lp \, \tthree ~-~ \tfour \, \rp   \rb  \\
          &=& {1 \over 2} \lb \,  \sqrt{2} \, \two ~+~ \zero~+~ \one \, 
              \rb  ~,  
\ee
i.e.,
\beq  
      \zero~\mapsto~
   {1 \over 2} \, \zero ~+~ {1 \over 2} \, \one ~+~ 
          {1 \over \sqrt{2}} \, \two ~.
\label{eq:3.1}
\eeq
In the same way, we find    
\beqn
    \one &\mapsto& 
   {1 \over 2} \, \zero~+~{1 \over 2} \, \one ~-~{1 \over   
    \sqrt{2}}\, \two  ~,
\label{eq:3.2} \\
    \two &\mapsto& 
  {1 \over \sqrt{2}}\,\zero~-~{1 \over \sqrt{2}}\, \one ~.     
\label{eq:3.3}
\eeqn
From equations \calle{eq:3.1}, \calle{eq:3.2}, and \calle{eq:3.3}, we see that
\beq
    \bfS \=\pmatrix
            { {1\over2} & {1\over2} & \r \cr
              {1\over2} & {1\over2} & -\r \cr
              \r & -\r & 0  \cr }~~. 
\label{eq:3.4}
\eeq
Notice that
$$ 
    \bfS^{-1} \= \bfS^\dagger \= \bfS ~.
$$
  Should we have expected this result?

(c) The fusion rules are written in general as
$$
  [\,\phi_i \,]~[\, \phi_j \,]~=~N_{ij}^k~[\,\phi_k\,]~,
$$
and the matrices to be diagonalized are
$$
   \bfN_i~\equiv~[\,N_{ij}^k\,]~.
$$
In the case of Ising model, the conformal families are
$$
   [\,\phi_i\,]~:~[\,1\,]~,~[\,\epsilon\,]~,~~{\rm and}~~[\,\sigma\,]~,
$$
and the non-trivial fusion rules are
\bb
    \lb\,\epsilon\,\rb~\lb\,\epsilon \,\rb &=& \lb\,1\,\rb~, \\
    \lb\,\epsilon\,\rb~\lb\, \sigma\,\rb &=& \lb\,\sigma\,\rb~,~~~{\rm and} \\
    \lb\,\sigma\,\rb~\lb\,\sigma\,\rb &=& \lb\,1\rb~+~\lb\,\epsilon\,\rb~. 
\ee
Hence, the corresponding $\bfN_i$ matrices are 
$$
    \bfN_1~=~\pmatrix{1~0~0\cr 0~1~0\cr 0~0~1\cr} ~,~~~  
             \bfN_\epsilon~=~\pmatrix{0~1~0\cr 1~0~0\cr 0~0~1\cr} ~,~~~{\rm and}~~~
             \bfN_\sigma~=~\pmatrix{0~0~1\cr 0~0~1\cr 1~1~0\cr}~. 
$$ 
It is a trivial matter now to show that
\bb
   \bfS\,\bfN_1 \,\bfS &=& {\rm diag}(\,1,1,1\,)~,\\
   \bfS\,\bfN_\epsilon\,\bfS&=&{\rm diag}(\,1,1,-1\,)~,~~{\rm and}\\
  \bfS\,\bfN_\sigma\,\bfS&=& {\rm diag}
   (\, \sqrt{2},-\sqrt{2},0 \,)~.
\ee

\separator

\item
Under the $T$ transformation, $T: \tau\mapsto \tau+1$, the
powers of the parameter $q$ transform as $q^a\mapsto q^a\, e^{2\pi ia}$,
and any character
$$
   \chi_\Delta\= q^{\Delta-c/24}\, 
   \sum_{N=0}^{+\infty}\, p(N)\, q^N
$$
transforms as
\bb
   \chi_\Delta(q) \mapsto\chi_\Delta(q') &=&
   q^{\Delta-c/24}\, e^{2\pi i (\Delta-c/24)}\,
    \sum_{N=0}^{+\infty}\, p(N)\, q^N
   \, e^{2\pi i N} \\
   &=&
   q^{\Delta-c/24}\, e^{2\pi i (\Delta-c/24)}\,
   \sum_{N=0}^{+\infty}\, p(N)\, q^N
    \\
    &=&e^{2\pi i (\Delta-c/24)}\, \chi_\Delta(q)~.
\ee
This is a general result. In particular, for the minimal 
model MM$(r,s)$, this becomes
$$
   \chi_\Delta(q) ~ {\buildrel T \over \mapsto} ~ 
    e^{-2\pi i \lp{1\over24}-{(mr-ns)^2\over4rs}\rp}\, \chi_\Delta(q)~.
$$

To find the transformation of the characters under the 
$S$ transformation, we write the characters in a more convenient form.
Let 
$$
  n_\pm\eq sn\pm mr~~~{\rm and}~~~
  N\eq 2rs~.
$$
Then
\bb
    a(k) &=& {(N k+n_-)^2-(r-s)^2\over 2N}~,~~~{\rm and}\\
    b(k) &=& {(N k+n_+)^2-(r-s)^2\over 2N}~.
\ee
Thus
\bb
    \chi(q)\= {1\over \eta(q)}\,
              \sum_{k=-\infty}^{+\infty} \lp
              q^ {(N k+n_-)^2\over 2N}-
              q^ {(N k+n_+)^2\over 2N} \rp~,
\ee
where, as usual,
$$
   \eta(q)\= q^{-1/24}\, \prod_{n=1}^{+\infty} (1-q^n)~.
$$

Define the function
$$
   \Theta_{\lambda,N}(\tau)\eq \sum_{k=-\infty}^{+\infty} 
              e^{i\pi \tau {(N k+\lambda)^2\over N}}~.
$$
With this definition, we can write
$$
    \chi_{m,n}(q)\= {1\over \eta(q)}\, \lb \Theta_{n_-,N}
    - \Theta_{n_+,N} \rb~.
$$

It is now straightforward to work out
the behavior of $\Theta_{\lambda,n}(\tau)$ under
the $S$ transformation:
\bb
   \Theta_{\lambda,N}(\tau) \mapsto \Theta_{\lambda,N}(-{1\over\tau})
    &=&  \sum_{k=-\infty}^{+\infty} 
              e^{-{\pi i\over\tau}{(N k+\lambda)^2\over N}}
    \\
     & {\calle{eq:MI4} \atop =} &  
     \sum_{k=-\infty}^{+\infty} 
     \int_{-\infty}^{+\infty}
     dy\, 
      e^{-{\pi i\over\tau}{(N y+\lambda)^2\over N}}
      e^{-2\pi i k y}
    \\
    &{ x=N y+\lambda\atop =} &
     {1\over N}\,
   \sum_{k=-\infty}^{+\infty}
      e^{2\pi i k\lambda\over N}
    \int_{-\infty}^{+\infty}
     dx\,
      e^{-{\pi i\over\tau}\,{x^2\over N}}
      e^{-{2\pi i kx\over N}}
    \\
    &=& {1\over N}\,  \sum_{k=-\infty}^{+\infty} e^{2\pi i k\lambda\over N}
         \sqrt{\tau\over iN}\,  e^{\pi i \tau k^2\over N}~.
\ee
To continue further, we observe that any integer $k$ can be written in the form
$k=mN+\lambda'$, where $\lambda'\in\{0,\dots,N-1\}$. Then the previous expression
can be written as
\beqn
  \Theta_{\lambda,N}(\tau) \mapsto \Theta_{\lambda,N}(-{1\over\tau})
    &=&
    {1\over N}\,  \sum_{m=-\infty}^{+\infty} 
    \,  \sum_{\lambda'=0}^{N} 
         e^{2\pi i {(mN+\lambda')\lambda\over N}}
         \sqrt{N\tau\over i}\,  e^{\pi i \tau {(mN+\lambda')^2\over N}}
    \nonumber \\
      &=&
    \sqrt{-i\tau\over N}\,\sum_{\lambda'=0}^{N-1}
    e^{2\pi i {\lambda'\lambda\over N}}
    \, \sum_{m=-\infty}^{+\infty}e^{\pi i \tau {(mN+\lambda')^2\over N}}
    \nonumber \\
     &=&
    \sqrt{-i\tau\over N}\,\sum_{\lambda'=0}^{N-1}
    e^{2\pi i {\lambda'\lambda\over N}}\,
    \Theta_{\lambda',N}(\tau)~.
\label{eq:MI5}
\eeqn
Notice also that the function $\Theta$ defined above has the 
properties
\beq
  \Theta_{-\lambda,N}\=
  \Theta_{\lambda,N}\=
  \Theta_{\lambda+N,N}~.
\label{eq:MI6}
\eeq

Using the property \calle{eq:MI5}, we can write the transformation
of the character under the $S$ transformation in the form
\bb
  \chi_{m,n}(q) \mapsto \chi_{m,n}(q')
  &=& {1\over\sqrt{N}}\,{1\over\eta(q)}\,
          \sum_{\lambda'=0}^{+\infty}\, 
    \sum_{\lambda'=0}^{N-1} \lp
    e^{2\pi i {\lambda' n_-\over N}} -e^{2\pi i {\lambda' n_+\over N}}\rp\,
    \Theta_{\lambda',N}(\tau)~.
\ee
Now, $\lambda$ runs over $0,\dots, 2rs-1$, or, equivalently,
over $-rs,\dots,rs$. Using the property \calle{eq:MI6}, we can thus
write 
\small
\bb
  \chi_{m,n}(q) \mapsto \chi_{m,n}(q')
  &=& {1\over\sqrt{N}}\,{1\over\eta(q)}\,
    \sum_{\lambda'=1}^{rs} \lp
    e^{2\pi i {\lambda' n_-\over N}} -e^{2\pi i {\lambda' n_+\over N}}
    +e^{-2\pi i {\lambda' n_-\over N}} -e^{-2\pi i {\lambda' n_+\over N}}\rp\,
    \Theta_{\lambda',N}(\tau) \\
  &=&{2\over\sqrt{N}}\,{1\over\eta(q)}\,
    \sum_{\lambda'=1}^{rs} \lb
    \cos\lp 2\pi  {\lambda' n_-\over N}\rp
    -\cos\lp 2\pi {\lambda' n_+\over N}\rp\rb  \,
    \Theta_{\lambda',N}(\tau) \\
  &=& {4\over\sqrt{N}}\,{1\over\eta(q)}\,
    \sum_{\lambda'=1}^{rs} \lb
    \sin\lp \pi  {\lambda' (n_++n_-)\over N}\rp
    \sin\lp \pi  {\lambda' (n_+-n_-)\over N}\rp\rb  \,
    \Theta_{\lambda',N}(\tau) \\
  &=&{4\over\sqrt{N}}\,{1\over\eta(q)}\,
    \sum_{\lambda'=1}^{rs} \lb
    \sin\lp \pi  {\lambda' n\over r}\rp
    \sin\lp \pi  {\lambda' m\over s}\rp\rb  \,
    \Theta_{\lambda',N}(\tau) ~,
\ee \normalsize
where we have used the identity
$$
   \cos A -\cos B\= 2\sin{A+B\over2}\,\sin{A-B\over2}~,
$$
as well as the relationships
$$
  {n_+ + n_-\over N}\={n\over r}~~~{\rm and}~~~
  {n_+ - n_-\over N}\={m\over s}~.
$$
Finally, we rewrite the sum over $\lambda'$ as a sum over
$m'$ and $n'$. To do this, we recognize that $\lambda'$ must 
take the values
$n'_\pm=sn'\pm rm'$. Then
\small
\bb
   \chi_{m,n}(q')
  &=&{4\over\sqrt{N}}\,{1\over\eta(q)}\,
    \sum_{n'_-} \lb
    \sin\lp \pi  {n_-' n\over r}\rp
    \sin\lp \pi  {n_-' m\over s}\rp\rb  \,
    \Theta_{n_-',N}(\tau) 
  \\
  &+& {4\over\sqrt{N}}\,{1\over\eta(q)}\,
    \sum_{n'_+} \lb
    \sin\lp \pi  {n_+' n\over r}\rp
    \sin\lp \pi  {n_+' m\over s}\rp\rb  \,
    \Theta_{n_+',N}(\tau)
  \\
  &=&{4\over\sqrt{N}}\,{1\over\eta(q)}\,
    \sum_{n'_-} 
    \sin\lp \pi{nn's\over r}-\pi m'n\rp
    \sin\lp \pi n'm+ \pi{mm'r\over s}\rp  \,
    \Theta_{n_-',N}(\tau)
  \\
  &+& {4\over\sqrt{N}}\,{1\over\eta(q)}\,
    \sum_{n'_+} 
    \sin\lp \pi{nn's\over r}+\pi m'n\rp
    \sin\lp \pi n'm- \pi{mm'r\over s}\rp  \,
    \Theta_{n_+',N}(\tau)
  \\
  &=&{4\over\sqrt{N}}\,{1\over\eta(q)}\,
    \sum_{n'_-} (-)^{m'n}\,
    \sin\lp \pi{nn's\over r}\rp \,
    (-)^{n'm}\, \sin\lp \pi{mm'r\over s}\rp  \,
    \Theta_{n_-',N}(\tau)
  \\
  &+& {4\over\sqrt{N}}\,{1\over\eta(q)}\,
    \sum_{n'_+} (-)\, (-)^{m'n}\,
    \sin\lp \pi{nn's\over r}\rp \, (-)^{n'm}\,
    \sin\lp \pi{mm'r\over s}\rp  \,
    \Theta_{n_+',N}(\tau)
  \\
   &=& {4\over\sqrt{N}}\,{1\over\eta(q)}\,
    \sum_{n',m'} (-)^{m'n+n'm}\,
    \sin\lp \pi{nn's\over r}\rp \, \sin\lp \pi{mm'r\over s}\rp\,
    \lp  \Theta_{n_-',N}(\tau)-\Theta_{n_+',N}(\tau)\rp 
  \\
    &=& {2\sqrt{2}\over\sqrt{rs}}\,
    \sum_{n',m'} (-)^{m'n+n'm}\,
    \sin\lp \pi{nn's\over r}\rp \, \sin\lp \pi{mm'r\over s}\rp\,
    \chi_{m',n'}(q)~.
\ee \normalsize

\separator

\item
(a) We can assume without loss of generality that
there are states with
electric charge but no magnetic charge.  (We can always choose
a basis in which this is true.)  Let $q_0>0$ 
be the smallest such positive electric charge. There is guaranteed
to be such a charge if there is magnetic charge, due to the 
DSZ quantization condition \cite{Dirac,Schw,Zwan}. 

The DSZ condition for the electric charge $q_0$ (with no
magnetic charge) and a dyon
$(q_i,g_i)$ implies that 
$$
   q_0 g_i \= 2\pi\, n_i \Rightarrow  g_i \= {2\pi\over q_0}\, n_i~,
$$
for some integer $n_i$. Therefore the smallest
positive magnetic charge
that can exist in this theory is
$$
    g_0 \= {2\pi\over q_0}\, m_0~,
$$
where $m_0=\min\{|n_j|\}_j$ and  depends on the details
of the theory. 

Next, the DSZ condition for two dyons
$(q_1,g_0)$ and $(q_2,g_0)$ implies
$$
   q_1-q_2\= {n\over m_0} \, q_0~,
$$
for some integer $n$. The integer $n$ must be a multiple
of $m_0$, or else it would be possible to construct a 
state with electric charge between $0$ and $q_0$.
This analysis leads to the conclusion, then, 
that the \textit{difference} of the electric charges is
quantized, although the actual values of the electric
charges may be arbitrary. We thus parametrize the electric charge as
$$
   q\=q_0 n + \Theta~.
$$
It proves convenient to write
$\Theta\=q_0\, {\theta\over 2\pi}$,
i.e.,
$$
    q\=q_0\, \lp n + {\theta\over 2\pi}\rp~.
$$
Since $\theta\mapsto\theta+2\pi$ is equivalent to $n\mapsto n+1$,
we can take $\theta$ to be an angular variable with values
in $\lb0,2\pi\rp$. 

We can repeat the previous argument for states of magnetic charge $m g_0$.
Then the result is
$$
    q\=q_0\, \lp n + m\, {\theta\over 2\pi}\rp~.
$$
This is usually referred to as the 
{\bfseries Witten effect}\index{Effect!Witten --}
\cite{Wittena,Wittenb}.

(b) A CP transformation on the electric and magentic charges
has the effect
$$
   (q,g)\mapsto (-q,g)~.
$$
Therefore, a CP invariant theory must have both states
 $(q,g)$ and $(-q,g)$ in its spectrum. 
We then apply the DSZ condition for $(q,g_0)$ and $(-q,g_0)$:
$$
    2qg_0\=2\pi n \Rightarrow 2q_0 \= q_0 n
    \Rightarrow q_0\in {1\over2}\BZ~.
$$
Comparing this with the general solution of the DSZ condition,
we see that CP invariant states require $\theta=0$ or
$\theta=\pi$.
Any other value of $\theta$ signals CP violation.

(c) The general solution 
\bb
    q &=& q_0 \, \lp n +q_0\, {\theta\over 2\pi} m\rp~,\\
    g &=& g_0\, m~,
\ee
of the DSZ condition can also be written in the form
\bb
    q+ig\= q_0\,\lp n+\theta\, \tau\rp~,
\ee
upon defining
$$
   \tau\eq {\theta\over 2\pi}+i\, {2\pi  m_0\over q^2_0}~.
$$
Physical states $\ket{n,m}$
with electric and magnetic charges $(q,g)$ are located
on a discrete 2-dimensional lattice with periods $q_0$ and
$q_0\tau$. We can change the basis of this lattice
with an SL$(2,\BZ)$ transformation
$$
  \tau\mapsto{a\tau+b\over c\tau+d}~~,~~~~\qquad a,b,c,d\in\BZ,~~ad-bc=1~,
$$
without changing the physics.  Under such a transformation,
there is a corresponding transformation of the electric
and magnetic charges, given by
$$
  \lb\matrix{n\cr m\cr}\rb \mapsto
  \lb\matrix{a & b \cr c & d\cr}\rb~
  \lb\matrix{n\cr m\cr}\rb ~.
$$
The effect of the transformation on the charges
can be studied by considering the effect
of the two generating transformations $T$ and $S$.
The effect of $T: \tau\mapsto\tau+1$ is trivial,
producing
$$
  \theta\mapsto \theta+2\pi\Rightarrow
  \ket{n,m}\mapsto\ket{n-m,m}~,
$$
The effect of $S:\tau\mapsto-1/\tau$ is less trivial --- it interchanges
the magnetic and electric roles!
For example, if $\theta=0$, then 
$ \ket{n,m}\mapsto\ket{m,-n}$.  This duality between electric
and magnetic charges is the simplest example of a phenomenon
that has become central to our understanding of string theory
and related fields 
\cite{AGH,Bilal2,Dabholkar,DHoker,TASI96,GodOli2,GHL,ZHDS,%
IntrSei,Itoyama1,Itoyama2,Ketov2,Klemm,Marshakov,DSzeta}.

\end{enumerate}

\newpage
\def\hodd{BIBLIOGRAPHY}
\def\heven{BIBLIOGRAPHY}
\footnotesize

\clearpage{\pagestyle{empty}\cleardoublepage}

\end{document}